\documentclass[]{dukedissertation2025}


\usepackage{placeins}
\usepackage[T1]{fontenc}
\usepackage{titlesec}
\usepackage{amsmath, amssymb, amsfonts, amsthm}
\usepackage{graphicx}
\usepackage[title,titletoc]{appendix}
\usepackage{apptools}
\usepackage{color}
\usepackage{multirow}
\usepackage{setspace}

\usepackage{newpxtext}
\usepackage{MnSymbol}
\usepackage{bm}
\usepackage{enumitem}
\usepackage{xfrac}
\usepackage{mathtools}
\usepackage{caption}
\usepackage{subcaption}
\usepackage{tabularx}
\usepackage{booktabs}
\usepackage{siunitx}
\usepackage{pdflscape}
\usepackage{listings}
\usepackage{xcolor}
\usepackage{pgfplots}
\usepackage{makecell}
\usepackage[super]{nth}
\usepackage{xr}
\usepackage{tabularx}
\usepackage{colortbl}
\usepackage{needspace}
\usepackage{microtype}
\usepackage{tikz}
\usetikzlibrary{matrix,arrows.meta,calc,positioning,backgrounds,shapes.geometric}

\makeatletter
\AtBeginDocument{%
  \let\originalBibitemShut\BibitemShut
  \def\BibitemShut#1{\unskip\originalBibitemShut{#1}}%
}
\makeatother

\setlist[itemize]{noitemsep,nolistsep}
\setlist[enumerate]{noitemsep,nolistsep}
\graphicspath{{./figures/}{./Chapter1}} 
\author{Andrew Gordeev}
\title{Constraining Light and Strange Flavor Equilibration in Relativistic Heavy-Ion Collisions}
\supervisor{Steffen A. Bass}
\department{Physics} 

\date{2026} 

\copyrighttext{All rights reserved except the rights granted by the \href{http://creativecommons.org/licenses/by-nc/3.0/us/}{Creative Commons Attribution-Noncommercial Licence}}

\defensedate{July 15, 2026}
\member{Berndt Mueller}
\member{Jian-Guo Liu}
\member{Anselm Vossen}
\member{Ashutosh V. Kotwal}

\usepackage[numbers,sort&compress]{natbib}

\definecolor{linkred}{RGB}{160,0,0}

\usepackage[
    pdfusetitle,
    plainpages=false,
    letterpaper,
    bookmarks,
    bookmarksnumbered,
    colorlinks,
    linkcolor=linkred,
    citecolor=linkred,
    urlcolor=linkred,
    linktoc=all
]{hyperref}
\usepackage[capitalise]{cleveref}
\creflabelformat{equation}{#2#1#3}
\makeatletter

\newcommand\getfontsizeofheading{The current font size is: \fontname\font \f@size pt}
\newcommand\getfontsizeforfontsizeclass[1]{{\string #1 is printed as #1  \fontname\font \f@size pt\par}}
\newcommand\printfontsizeforfontsizeclass[1]{{#1 \string #1 is printed in  \fontname\font \f@size pt\par}}
\makeatother

\makeatletter
\xpatchcmd{\NR@chapter}
  {\refstepcounter{chapter}}
  {\vspace*{-80pt}%
   \refstepcounter{chapter}%
   \vspace*{80pt}}
  {}
  {\PackageWarning{dissertation}{Could not patch chapter hyperlink anchor}}
\makeatother

\makeatletter
\xpatchcmd{\@schapter}
  {\Hy@raisedlink{%
     \hyper@anchorstart{\@currentHref}\hyper@anchorend
   }}
  {\vspace*{-80pt}%
   \Hy@raisedlink{%
     \hyper@anchorstart{\@currentHref}\hyper@anchorend
   }%
   \vspace*{80pt}}
  {}
  {\PackageWarning{dissertation}{Could not patch starred chapter hyperlink anchor}}
\makeatother



\makeatletter
\xpatchcmd{\tableofcontents}
  {\chapter*{\contentsname}}
  {\chapter*{\contentsname}%
   \Hy@writebookmark{}{\contentsname}{\@currentHref}{0}{toc}}
  {}
  {\PackageWarning{dissertation}{Could not patch Contents bookmark}}
\makeatother

\usepackage{fancyhdr}

\fancypagestyle{dissertation}{
    \fancyhf{}

\fancyhead[L]{%
  \rmfamily\scshape\nouppercase{\rightmark}}

    \fancyhead[R]{%
        \small\rmfamily\thepage%
    }

    \fancyfoot{}

    \renewcommand{\headrulewidth}{0.3pt}

    \renewcommand{\headrule}{%
  \hbox to\headwidth{%
    \color{linkred}\leaders\hrule height \headrulewidth\hfill
  }%
}
}

\fancypagestyle{plain}{
    \fancyhf{}
    \fancyhead[R]{\normalsize\rmfamily\thepage}
    \renewcommand{\headrulewidth}{0pt}
    
}

\renewcommand{\chaptermark}[1]{%
  \markboth{\thechapter\quad #1}{\thechapter\quad #1}%
}

\makeatletter

\renewcommand{\fnum@figure}{%
  \figurename\nobreakspace\thefigure%
}

\renewcommand{\fnum@table}{%
  \tablename\nobreakspace\thetable%
}

\long\def\@makecaption#1#2{%
  \vskip\abovecaptionskip
  \normalbaselines
  {\color{linkred} #1.} #2\par
  \vskip\belowcaptionskip
}

\makeatother

\newcommand{\chapterlabel}{Chapter}
\titleformat{\chapter}[display]
  {\normalfont\rmfamily}
  {\color{linkred}\LARGE\scshape \chapterlabel\ \thechapter}
  {0.5em}
  {\fontsize{23pt}{27pt}\selectfont}
  [\vspace{0.5em}{\color{linkred}\titlerule[0.6pt]}]

\titlespacing*{\chapter}
  {0pt}
  {-80pt}
  {4.5em}

\titleformat{\section}
  {\normalfont\rmfamily\Large\bfseries\color{linkred}}
  {\thesection}{0.65em}{}

\titleformat{\subsection}
  {\normalfont\rmfamily\large\bfseries\color{linkred}}
  {\thesubsection}{0.65em}{}

\titleformat{\subsubsection}
  {\normalfont\rmfamily\normalsize\bfseries\color{linkred}}
  {}{0pt}{}
  
\titlespacing*{\section}
  {0pt}
  {34pt plus 5pt minus 4pt}
  {12pt plus 2pt minus 1pt}

\titlespacing*{\subsection}
  {0pt}
  {24pt plus 4pt minus 3pt}
  {8pt plus 2pt minus 1pt}

\titlespacing*{\subsubsection}
  {0pt}
  {17pt plus 3pt minus 2pt}
  {6pt plus 1pt minus 1pt}

\newcolumntype{L}[1]{>{\raggedright\let\newline\\\arraybackslash\hspace{0pt}}b{#1}}
\newcolumntype{C}[1]{>{\centering\let\newline\\\arraybackslash\hspace{0pt}}b{#1}}
\newcolumntype{R}[1]{>{\raggedleft\let\newline\\\arraybackslash\hspace{0pt}}b{#1}}

\usepackage{subfiles}

\makeatletter

\newcommand{\arxivtitlepage}{%
  \begin{titlepage}
    \thispagestyle{empty}
    \normalbaselines

    \begin{center}

      \vspace*{0.16\textheight}

      {\fontsize{20pt}{30pt}\selectfont
       \rmfamily
       \@title\par}

      \vspace{1.5em}

      {\color{linkred}\rule{\textwidth}{1.0pt}\par}

      \vspace{10em}

      {\Large\rmfamily \@author\par}

      \vfill

      {\normalsize\rmfamily PhD Dissertation\par}

      \vspace{0.7em}

      {\normalsize\rmfamily
       Advisor: Steffen A. Bass\par}

      \vspace{0.7em}

      {\normalsize\rmfamily
       Department of Physics\par
       Duke University\par}

      \vspace{1.2em}

      {\normalsize\rmfamily 2026\par}

      \vspace*{0.06\textheight}

    \end{center}
  \end{titlepage}
}

\makeatother

\begin{document}

\hypersetup{pageanchor=false}
\arxivtitlepage

\pagenumbering{roman}
\hypersetup{pageanchor=true}
\pagestyle{dissertation}

\abstract

Relativistic heavy-ion collisions provide the only known means of creating and studying the quark-gluon plasma, a hot and dense state of QCD matter in which quarks and gluons are no longer confined within hadrons. Hydrodynamic modeling has established that this medium behaves as a strongly coupled fluid of very low viscosity, but it typically assumes that the quark and gluon composition reaches chemical equilibrium by the onset of the hydrodynamic phase. The validity of this assumption remains uncertain: gluon-dominated initial states suggest that quark production may continue well into the hydrodynamic stage, with potentially different equilibration rates for light and strange quarks.

This dissertation develops a dynamical framework for light and strange quark chemical equilibration. Incomplete equilibration of each flavor is described through time-dependent quark fugacities that modify both the equation of state and the particlization of the medium, so that its evolving flavor composition shapes both the medium's evolution and its final observables. Implemented within a multistage framework of fluctuating initial conditions, viscous relativistic hydrodynamics, particlization, and hadronic transport, the model is used to study the effects of chemical equilibration on the hydrodynamic evolution and on hadronic and electromagnetic observables.

To confront the model with experiment, this framework is embedded within a Bayesian analysis. Gaussian process emulators trained on the model calculations are used to infer the initial light and strange quark fugacities, their equilibration timescales, and selected transport coefficients from Au+Au collision data at RHIC. The strange sector is found to begin substantially undersaturated, with an initial fugacity well below its equilibrium value and more suppressed than that of the light quarks; the data further favor a slower approach to equilibrium for strangeness, though the equilibration timescales themselves remain weakly constrained. Taken together, these results favor a quark-gluon plasma whose flavor composition is still approaching chemical equilibrium during the hydrodynamic phase rather than fixed at its outset, and lay the groundwork for future analyses across collision systems of differing size and energy.

\tableofcontents

\listoftables	

\listoffigures	



%
%
%

\chapter{Introduction to QCD matter}
\label{chapter1}

The nature of matter has always been one of the most fundamental questions at the heart of physics. Although humanity's understanding of it has transformed dramatically over history, the Standard Model stands today as one of the greatest triumphs of modern physics: a robustly tested theory of the known elementary particles and their interactions, albeit with the glaring omission of gravity. Most of modern nuclear and particle physics rests on this foundation. 

Quantum chromodynamics (QCD), the theory of the strong interaction, stands out within the Standard Model as the explanation of what binds quarks into the protons and neutrons of atomic nuclei. The vast majority of ordinary matter's mass comes from QCD dynamics; it is, in a very real sense, the theory that describes most of us. Yet there is still much about it that is not well understood. Heat ordinary matter sufficiently, and the familiar protons and neutrons disintegrate into an exotic fluid: the quark-gluon plasma (QGP). 

This fluid has extraordinary properties. It is an incredibly strongly coupled medium composed entirely of quarks and gluons, the fundamental constituents of protons, neutrons, and many other hadrons, within which they are ordinarily confined and never seen in isolation. Furthermore, it has a remarkably low specific viscosity, because of which it is often said to flow as the ``most perfect'' fluid in the universe. And that title means less yet more at the same time if we rewind to the early universe, just a microsecond or so after the Big Bang, when the QGP was effectively the only fluid --- all matter, everywhere, was hot and dense enough to be in this state until the universe expanded and cooled.

Today, the QGP is not known to exist anywhere in nature, with the closest analog being related forms of quark matter which may occur inside neutron stars. If the Standard Model can be claimed as a great theoretical triumph, though, then the experiments validating it are just as meritorious. This has largely come about through generations of scattering and collider experiments, in which particles and nuclei are collided in different configurations to test their properties. Among the most important modern facilities for this purpose are the Large Hadron Collider (LHC) near Geneva and the Relativistic Heavy-Ion Collider (RHIC) on Long Island, which have been the leading experimental programs for creating and studying the QGP in the laboratory.

The experimental production of the QGP is frustratingly fleeting, however. With each collision, a small volume of QGP --- on the order of $10^{-14}$ m in length, comparable to the size of the atomic nuclei themselves --- is produced and decays on a timescale of $\sim\!10^{-23}$ seconds. This is far too brief to allow for direct observation; all we can actually see is the final shower of particles formed directly or indirectly from its decay, once they reach the detectors. Anything we want to know about the medium must then be inferred from these final particles. Worse still, it is not a simple process that neatly transitions from the QGP to observables. A single collision involves a rapid succession of phases that each encompass different physics: the initial impact between nuclei, the flow of the QGP, its reformation into particles, and their later interactions before detection must all be understood in tandem to meaningfully deduce what happened at earlier times.

Put simply, the study of the QGP is an inverse problem. If we could observe every stage of a collision in detail, then theoretical models could be compared directly to the intermediate evolution of the system, making it far easier to identify where our understanding succeeds or fails. An inverse problem is more insidious, because it precludes such direct comparisons. To work backward from the data toward the processes that produced them, we cannot simply run the collision in reverse. Many effects can overlap, compensate for one another, or imitate one another in the final observables, so the unknown properties of the system must be disentangled through systematic model-to-data comparison.

For such problems, statistical inference is not just a convenient tool, but what makes them quantitatively tractable in the first place. It allows for systematic model-to-data comparison in a landscape with too much theoretical freedom for reliable hand-tuning, while also providing uncertainty estimates for the inferred model parameters. It is from such inference --- Bayesian analyses in particular --- that some of the most quantitative constraints on key QGP properties, including its transport coefficients, have been obtained to date.

The central conceit of this work is to extend the current heavy-ion collision modeling framework to describe chemical equilibration within the QGP. Quarks of different flavors, or species, are produced and redistributed throughout the stages of a heavy-ion collision, but there remains substantial ambiguity surrounding their early-time chemical equilibration. The goal here is therefore to develop a model that allows these equilibration dynamics to be probed more directly, and to implement it within a Bayesian inference framework that can constrain the relative rates of light and strange quark equilibration from experimental data.

The outline of this dissertation is as follows. For the rest of chapter~\ref{chapter1}, I provide a broad overview of our theoretical and experimental understanding of the matter produced in heavy-ion collisions. In chapter~\ref{chapter2}, I detail the multistage model framework used to computationally simulate heavy-ion collisions. In chapter~\ref{chapter3}, I introduce this work's treatment of chemical equilibration and discuss its impact on the hydrodynamic evolution and experimental observables. In chapter~\ref{chapter4}, I provide an overview of the Bayesian inference methodology used for model-to-data comparison. Finally, in chapter~\ref{chapter5}, I present the resulting Bayesian constraints on QGP chemical equilibration and other medium properties.

\section{The strong interaction}
\label{section11}

Quantum chromodynamics is the gauge theory of the strong interaction, invariant under local SU($3$) gauge transformations whose associated charge is called color charge. Its matter fields are the quarks, which come in six flavors: up, down, strange, charm, bottom, and top. Each quark also carries one of three color charges, conventionally denoted red, green, or blue. The interactions between them are mediated by eight gauge bosons, the gluons. The quark masses span an enormous range, from a few MeV for the up and down quarks to over $170$ GeV for the top, as summarized in Table~\ref{tab:quark_properties}. Because of this hierarchy, at the temperatures reached in a heavy-ion collision, only the three lightest flavors are thermally produced in appreciable numbers. The far heavier charm, bottom, and top quarks cannot be created thermally to any significant degree, although charm and bottom quarks are produced in initial hard scatterings and can therefore serve as hard probes of the medium. The relative production of light and strange quarks, in particular, is central to this work and is taken up in section~\ref{section16}. Throughout this work, I will use “light quarks” to refer to up and down quarks, while treating strange quarks separately because their larger mass makes their chemical equilibration dynamics distinct.

\begin{table}[!t]
  \centering
  \caption[The six quark flavors and some of their basic properties]{The six quark flavors and some of their basic properties. Charges are in units of the elementary charge $e$. For the light quarks, the quoted masses are current-quark masses in the $\overline{\text{MS}}$ scheme at a renormalization scale of $2$~GeV. The charm and bottom masses are $\overline{\text{MS}}$ masses quoted at their own mass scales, while the top mass is extracted from event kinematics. Values are taken from the Particle Data Group~\cite{ParticleDataGroup:2024cfk}.}
  \label{tab:quark_properties}

  \renewcommand{\arraystretch}{1.12}
  \setlength{\tabcolsep}{12pt}

  \begin{tabular}{lccc}
    \toprule
    \textbf{Flavor}
    & \textbf{Generation}
    & \textbf{Charge $[e]$}
    & \textbf{Mass} \\
    \midrule

    up      & I   & $+2/3$ & $2.16 \pm 0.07$ MeV \\
    down    & I   & $-1/3$ & $4.70 \pm 0.07$ MeV \\
    strange & II  & $-1/3$ & $93.5 \pm 0.8$ MeV \\

    \addlinespace[4pt]

    charm   & II  & $+2/3$ & $1.273 \pm 0.005$ GeV \\
    bottom  & III & $-1/3$ & $4.183 \pm 0.007$ GeV \\
    top     & III & $+2/3$ & $172.57 \pm 0.29$ GeV \\

    \bottomrule
  \end{tabular}
  \vspace{16pt}
\end{table}

The dynamics of the theory are encoded in the QCD Lagrangian density,

\begin{equation}
  \mathcal{L}_{\text{QCD}} =
  \sum_f \bar{\psi}_f \left(i\gamma^\mu D_\mu - m_f\right)\psi_f
  - \frac{1}{4}G^a_{\mu\nu}G^{a\,\mu\nu},
  \label{eq:qcd_lagrangian}
\end{equation}
where $\psi_f$ is the quark field of flavor $f$, understood as a color triplet with color indices suppressed. The first term describes the propagation of the quarks and their coupling to the gluon field, both contained within the covariant derivative $D_\mu = \partial_\mu - i g A^a_\mu t^a$, where $A^a_\mu$ are the gluon fields, $t^a$ are the generators of SU(3) in the fundamental representation, and $g$ is the strong coupling. The quark masses of Table~\ref{tab:quark_properties} enter through the flavor-dependent mass terms $m_f$. The final term is built from the gluon field-strength tensor,
\nopagebreak[4]
\begin{equation}
  G^a_{\mu\nu} = \partial_\mu A^a_\nu - \partial_\nu A^a_\mu + g f^{abc} A^b_\mu A^c_\nu,
  \label{eq:gluon_field_strength}
\end{equation}
and governs the propagation and self-interaction of the gluons, with $f^{abc}$ the structure constants of SU($3$).

What most starkly distinguishes QCD from Abelian gauge theories such as quantum electrodynamics (QED), the quantum field theory of electromagnetism, is the self-interaction of the gluons. The photon is electrically neutral and does not couple to itself, so electromagnetic field lines spread freely through space. Gluons, by contrast, carry color charge, and the non-Abelian terms in the field-strength tensor lead to their mutual coupling through three- and four-gluon vertices. This single feature, a direct consequence of the non-Abelian structure of the gauge group, is central to the two defining properties of QCD: asymptotic freedom and confinement.

\subsection{Asymptotic freedom}
\label{section111}

The strength of the strong interaction is not fixed, but depends on the momentum scale $Q$ at which it is probed. This running of the coupling is set by the QCD $\beta$-function, whose leading term is negative, meaning that the effective coupling grows weaker as $Q$ increases and shorter distances are resolved. At one-loop order it takes the form
\begin{equation}
  \alpha_s(Q^2) = \frac{4\pi}{\beta_0\,\ln\!\left(Q^2/\Lambda_{\text{QCD}}^2\right)},
  \qquad \beta_0 = 11 - \frac{2}{3}n_f,
  \label{eq:running_coupling}
\end{equation}
where $\alpha_s = g^2/4\pi$ and $n_f$ is the number of active quark flavors. The two contributions to $\beta_0$ have opposite physical origins. The term proportional to $n_f$ comes from quark loops, which screen the color charge much as electron loops screen electric charge in QED. The constant term has no counterpart in QED; it comes from gluon loops and antiscreens the charge instead. For the physical number of active flavors, and indeed for any integer $n_f \le 16$, this gluonic antiscreening wins, and the coupling weakens at high energy.

\begin{figure}[!htbp]
  \vspace{12pt}
  \centering
  \includegraphics[width=\textwidth]{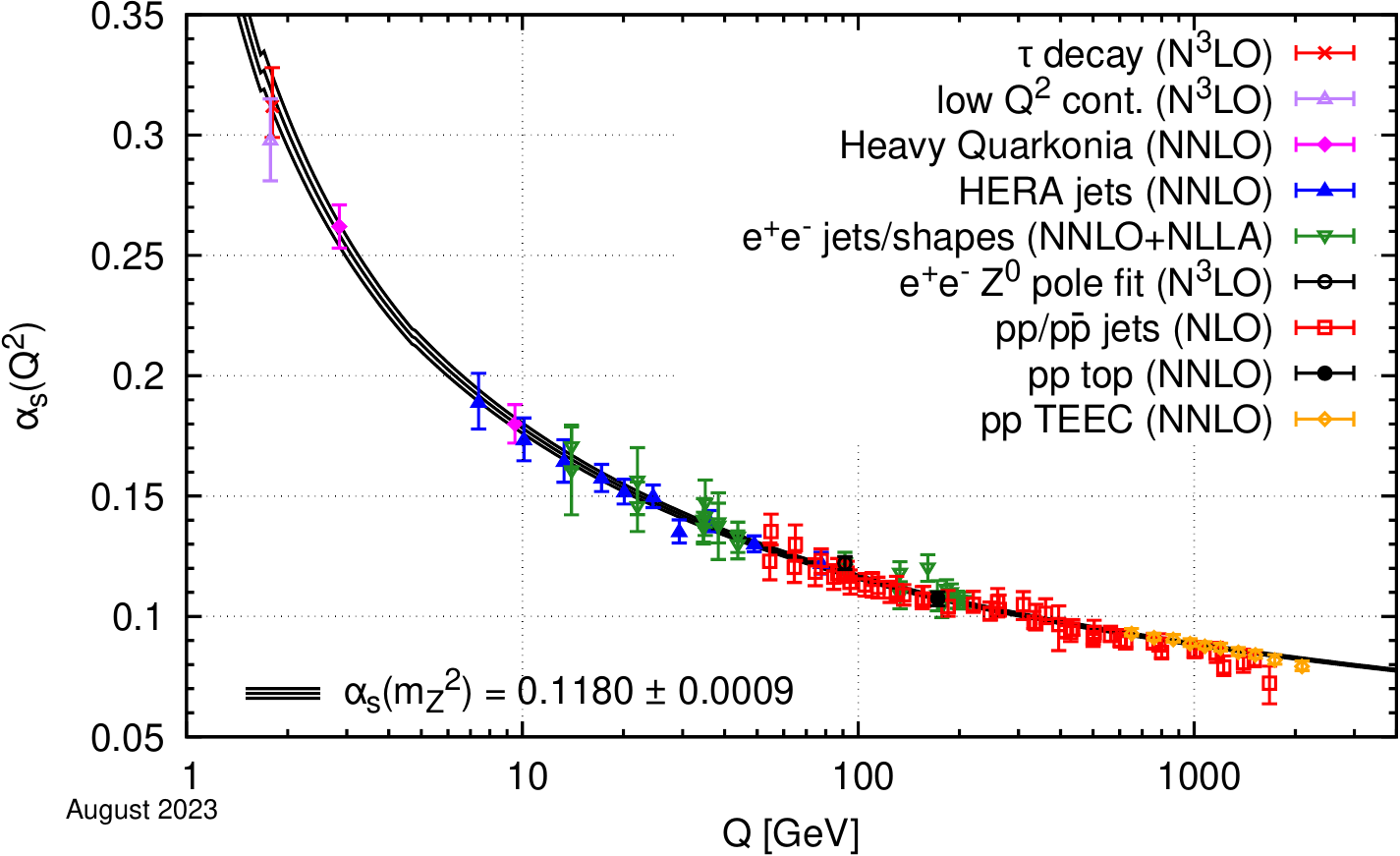}
  \caption[Strong coupling $\alpha_s$ as a function of the momentum-transfer scale $Q$]{The measured strong coupling $\alpha_s$ as a function of the momentum-transfer scale $Q$, illustrating its decrease toward higher energies. Reproduced from Ref.~\cite{ParticleDataGroup:2024cfk}.}
  \label{fig:alpha_s_running}
\end{figure}

This behavior is known as asymptotic freedom~\cite{Gross:1973id,Politzer:1973fx}. Experimental evidence for it was first established in deep inelastic scattering experiments, in which the quarks inside a proton were found to behave as quasi-free, pointlike constituents when probed at high momentum. At higher energies, quarks and gluons interact more weakly and behave increasingly as free particles. At sufficiently high energy scales, the coupling becomes small enough to allow for a controlled expansion in $\alpha_s$, making perturbation theory a reliable tool. The scale $\Lambda_{\text{QCD}} \sim 200$~MeV sets the characteristic nonperturbative scale of QCD; as $Q$ approaches this scale, the coupling becomes large and the perturbative description breaks down. This measured scale dependence of $\alpha_s$ is shown in Fig.~\ref{fig:alpha_s_running}.

\Needspace{4\baselineskip}
\subsection{Confinement}
\label{section112}

At low momenta, the running coupling becomes large enough that perturbation theory fails altogether. This nonperturbative regime is associated with confinement: quarks and gluons are never observed in isolation, but only within color-neutral hadrons. This is precisely why the colored constituents of matter are never seen directly under ordinary conditions; pulling a quark away from its partners only feeds energy into the color field until it becomes energetically favorable to create a new quark-antiquark pair, yielding more hadrons rather than a liberated color charge.

What becomes of confinement under extreme conditions? If hadronic matter is heated or compressed sufficiently, its hadrons begin to overlap and their color charges screen one another, until quarks and gluons are no longer bound within individual hadrons but propagate throughout the medium. This deconfined state of matter is the quark-gluon plasma.

One should note, however, that the matter created in heavy-ion collisions is not a weakly coupled gas of quarks and gluons. Only at asymptotically high temperatures, far beyond those reached in any experiment to date, does the coupling become weak enough for such a picture to hold. Near the transition temperature, where the plasma is actually produced and studied, the coupling remains large and the medium is strongly coupled, hence its remarkable fluid dynamical properties. Characterizing this strongly coupled matter is the subject of the latter half of this chapter.

\section{Relativistic heavy-ion collision experiments}
\label{section12}

As established in the previous section, the quark-gluon plasma is the deconfined phase of QCD matter, reached only when strongly interacting matter is heated or compressed past the deconfinement transition. Recreating it in a laboratory setting requires depositing an enormous energy density over a region large enough to form an appreciable volume of plasma. This is one of the central purposes of ultrarelativistic heavy-ion collisions, in which beams of heavy nuclei are accelerated to nearly the speed of light and brought into collision. Heavy nuclei such as gold and lead have been primarily used because their large size increases the number of nucleons that participate in the collision, producing a larger and longer-lived droplet of deconfined matter than lighter projectiles would. 

The two foremost facilities for this purpose, introduced at the start of this chapter, are the LHC at the European Organization for Nuclear Research (CERN) and RHIC at Brookhaven National Laboratory. Over a program spanning 2000 to early 2026, RHIC collided gold nuclei (Au+Au) at center-of-mass energies up to $\sqrt{s_{\text{NN}}} = 200$~GeV, together with proton-proton and smaller asymmetric systems used as references. Its principal detectors included STAR and PHENIX, with sPHENIX later succeeding PHENIX for the final stage of the RHIC program. The LHC heavy-ion program began with Pb+Pb collisions in 2010, and as of this writing, has just finished its third run, with planned operations for many more years to come following a several-year shutdown. It reaches substantially higher energies, colliding lead nuclei (Pb+Pb) at $\sqrt{s_{\text{NN}}} = 2.76$, $5.02$, and $5.36$~TeV to produce a hotter and longer-lived plasma, and studies heavy-ion physics primarily through the ALICE, ATLAS, and CMS experiments. Beyond these, the smaller BRAHMS and PHOBOS experiments at RHIC and the earlier fixed-target heavy-ion program at the CERN SPS played important roles in establishing the field.

A single collision is an inherently stochastic event, and no individual collision can be observed in enough detail to reveal the properties of the medium directly. Instead, experiments record enormous numbers of collisions and aggregate them into event-averaged distributions and correlations, which form the statistical objects against which theoretical models are compared. This statistical character is what makes a quantitative characterization of the QGP possible in the first place. The early goal of establishing whether a deconfined state of matter can be created in the laboratory was largely met over the first decade of RHIC operation, culminating in the now-famous conclusion that the medium behaves as a nearly perfect fluid~\cite{BRAHMS:2004adc,PHOBOS:2004zne,STAR:2005gfr,PHENIX:2004vcz}. Attention has since turned to characterizing it quantitatively, constraining its transport coefficients and equation of state through systematic comparison of models to data, as well as to mapping the QCD phase diagram in search of a critical point. These aims overlap with a broad range of related questions, including whether QGP-like behavior extends to the smallest collision systems, such as high-multiplicity proton-proton and proton-nucleus collisions; how energetic partons lose energy while traversing the medium; and how heavy quarks are transported through and hadronize from the medium. The present work falls under the quantitative characterization of the medium's bulk properties, with a particular focus on chemical equilibration, and does not pursue these other directions.

Two things specify a collision system at the most basic level: its energy and the species of the colliding nuclei. The collision energy is quoted as the center-of-mass energy per nucleon-nucleon pair, $\sqrt{s_{\text{NN}}}$, defined per pair so that systems of different mass number can be compared on a common footing. For a symmetric collider with beam energy $E$ per nucleon in the high-energy limit, $\sqrt{s_{\text{NN}}} \approx 2E$, so that RHIC's $100$~GeV per nucleon beams yield $\sqrt{s_{\text{NN}}} = 200$~GeV. Varying the energy changes the initial energy density and temperature of the medium, while varying the colliding nuclei changes the volume and lifetime of the matter created. The quark-gluon plasma is by now firmly established in large collision systems, such as Au+Au at RHIC and Pb+Pb at the LHC; more recently, light-ion collisions at the LHC in 2025 have provided strong evidence for QGP-like final-state effects in systems as light as oxygen-oxygen~\cite{CMS:2025bta, ALICE:2026zck}.

Even for a specified energy and collision system, individual collisions vary wildly. A large part of this is geometry. The degree of overlap between the two nuclei is characterized by the impact parameter, the transverse distance between their centers: a head-on collision with near-zero impact parameter is termed central, while a glancing one with large impact parameter is peripheral, as illustrated in Fig.~\ref{fig:collision_geometry} for three representative centralities. The impact parameter is not directly measurable, so collisions are instead classified by their centrality, typically defined operationally from the final-state multiplicity or deposited detector energy: a higher activity corresponds, on average, to a more central collision with greater nuclear overlap. Centrality is conventionally expressed as a percentile of the total inelastic cross section, with the $0$--$5$\% class denoting the most central collisions and larger percentiles increasingly peripheral ones.

\usetikzlibrary{arrows.meta}
\definecolor{partc}{HTML}{E8A04B}  
\definecolor{spec}{HTML}{CBD2D8}   
\newcommand{\collgeom}[2]{
  \fill[spec] (-#1,0) circle (0.9);
  \fill[spec] ( #1,0) circle (0.9);
  \begin{scope}
    \clip (-#1,0) circle (0.9);
    \fill[partc] (#1,0) circle (0.9);
  \end{scope}
  \draw[thick] (-#1,0) circle (0.9);
  \draw[thick] ( #1,0) circle (0.9);
  \fill (-#1,0) circle (1.2pt);
  \fill ( #1,0) circle (1.2pt);
  \draw[{Latex[length=1.6mm]}-{Latex[length=1.6mm]}, thin] (-#1,0) -- (#1,0);
  \node at (0.16,0.28) {$b$};
  \node[below] at (0,-1.35) {#2};
}

\begin{figure}[!htbp]
  \centering
  \begin{tikzpicture}
    \begin{scope}[xshift=0cm]   \collgeom{0.2}{central}      \end{scope}
    \begin{scope}[xshift=4.5cm] \collgeom{0.55}{semi-central} \end{scope}
    \begin{scope}[xshift=9cm]   \collgeom{0.8}{peripheral}   \end{scope}
    \begin{scope}[xshift=4.5cm]
      \draw[thin] (0.5,1.4) -- (0.12,0.5);
      \node[anchor=south] at (0.85,1.48) {participants};
      \draw[thin] (-1.15,1.4) -- (-0.82,0.42);
      \node[anchor=south] at (-1.42,1.48) {spectators};
    \end{scope}
  \end{tikzpicture}
  \caption[Schematic transverse view of the collision geometry for three
    centralities]{Schematic transverse view of the collision geometry for three
    centralities. As the impact parameter $b$ grows from central to peripheral
    collisions, the overlap region of participating nucleons (orange) shrinks and
    becomes increasingly anisotropic, while the non-overlapping spectator nucleons
    continue essentially undeflected. This almond-shaped spatial anisotropy of
    non-central collisions is what drives the anisotropic flow discussed in
    section~\ref{section132}.}
  \label{fig:collision_geometry}
\end{figure}
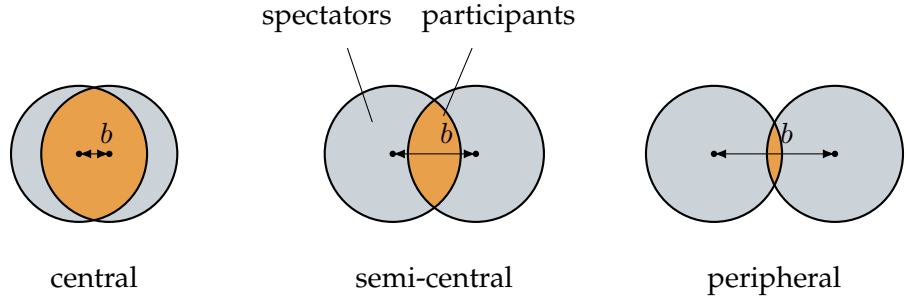

Because the beam direction is singled out by the collision geometry, particles are described in terms of their motion along and transverse to the beam axis. The longitudinal motion is naturally described by the rapidity,
\begin{equation}
  y = \frac{1}{2}\ln\!\left(\frac{E + p_z}{E - p_z}\right),
  \label{eq:rapidity}
\end{equation}
where $E$ and $p_z$ are the particle's energy and longitudinal momentum. Rapidity is the natural variable for this purpose because it is additive under Lorentz boosts along the beam axis: such a boost shifts every rapidity by a common constant, leaving rapidity differences invariant. Since rapidity requires knowledge of a particle's mass or identity, which is not always available, one often uses instead the pseudorapidity,
\begin{equation}
  \eta = -\ln\tan\!\left(\frac{\theta}{2}\right),
  \label{eq:pseudorapidity}
\end{equation}
which depends only on the polar emission angle $\theta$ and coincides with the rapidity in the ultrarelativistic limit, where the mass is negligible relative to the momentum. The region near zero rapidity, termed midrapidity, corresponds to particles emitted roughly transverse to the beams; in the center-of-mass frame of a symmetric collision this is where particle production is largest and where many measurements are concentrated. The transverse motion, in turn, is quantified by the transverse momentum $p_T$, the magnitude of the momentum projected into the plane perpendicular to the beam. Unlike the longitudinal momentum, $p_T$ is itself invariant under boosts along that axis, which together with the rapidity makes $(y, p_T)$ the standard pair for describing a particle's kinematics.

Beyond this kinematic role, the transverse plane carries the geometric structure of the collision. The initial nuclear overlap region is in general anisotropic, and these anisotropies --- together with their event-by-event fluctuations --- are imprinted on the momenta of the produced particles, driving the collective flow observables discussed in the next section.

\section{Experimental observables}
\label{section13}

As emphasized at the start of this chapter, the medium produced in a heavy-ion collision cannot be observed directly, and its properties must instead be inferred from the particles that reach the detectors. These particles are characterized in a few complementary ways: the abundances of the various species produced, the distributions of their momenta, and the correlations among their emission angles. A smaller population of rare, high-momentum particles and jets, known collectively as hard probes, provides a complementary view of the medium, as does the electromagnetic radiation emitted over its lifetime. Each class of observable described below may be measured differentially in centrality, transverse momentum, and rapidity or pseudorapidity, using the kinematic and geometric classifications described in the previous section.

\subsection{Hadron yields and spectra}
\label{section131}

The overwhelming majority of particles emerging from a collision are soft hadrons, produced in a clear hierarchy of abundances. Pions are the most numerous, followed by kaons, then protons, and then successively rarer species, such as multi-strange baryons. The most basic characterization of these particles is how many of them there are: their integrated yields, or multiplicities. For unidentified charged particles, this is commonly reported as the charged particle pseudorapidity density $dN_{\text{ch}}/d\eta$, a basic measure of the size of the produced final state. They can further be distinguished by their transverse momentum spectra, the distributions of a given species in $p_T$, often reported as $dN/dp_T$ or as invariant yields. The shape of the spectrum at low momentum carries information about both the collective expansion of the medium and the thermal motion of the particles near kinetic freeze-out, when they cease to interact.

\begin{figure}[!htbp]
  \centering
  \includegraphics[width=\textwidth]{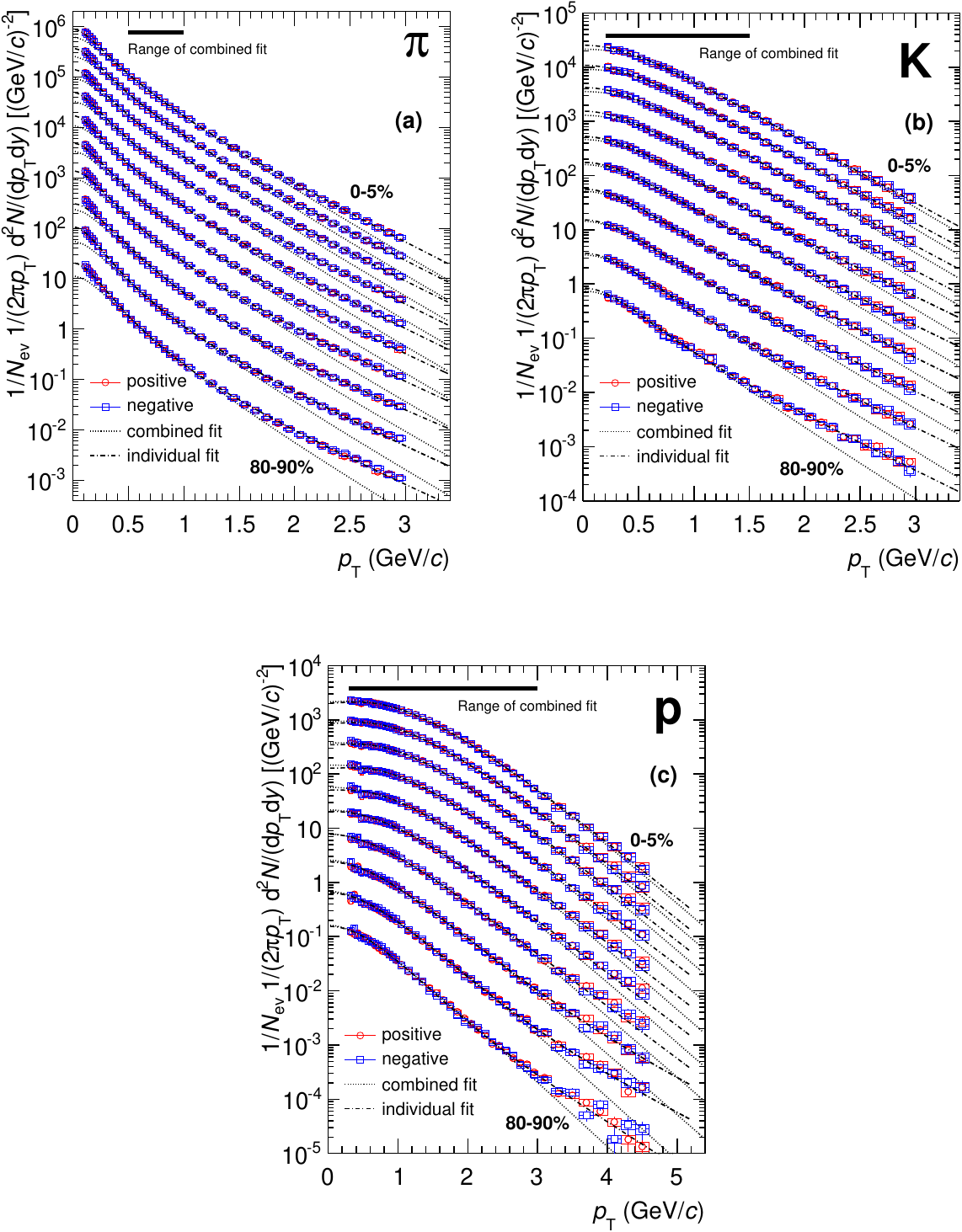}
  \caption[Identified particle transverse momentum spectra for pions, kaons, and protons at representative centralities in Pb+Pb collisions at $\sqrt{s_{\text{NN}}} = 2.76$~TeV]{Identified particle transverse momentum spectra for pions, kaons, and protons at representative centralities in Pb+Pb collisions at $\sqrt{s_{\text{NN}}} = 2.76$~TeV. Reproduced from Ref.~\cite{ALICE:2013mez}.}
  \label{fig:identified_spectra}
\end{figure}

At low and intermediate $p_T$, the spectra reflect a collective outward expansion of the medium known as radial flow. As the medium expands, its collective transverse velocity shifts the momenta of emitted particles to higher values. Because the momentum associated with a common flow velocity depends on a particle's mass, heavier species are pushed toward higher $p_T$, producing a characteristic mass ordering of the spectra, illustrated in Fig.~\ref{fig:identified_spectra}. This is an isotropic effect, shared equally by particles emitted in all azimuthal directions, and should be distinguished from the anisotropic flow discussed in the next subsection, which concerns the variation of the yield with azimuthal angle.

A convenient summary of a spectrum is its mean transverse momentum $\langle p_T \rangle$, which captures in a single number the characteristic hardness of the spectrum. It is governed by the same combination of radial flow and thermal motion, and is therefore sensitive to the expansion dynamics of the medium. In particular, larger bulk viscosity tends to suppress radial expansion and lower $\langle p_T \rangle$, a connection discussed further in section~\ref{section14}.

Beyond the spectra of individual species, their relative abundances are themselves informative. Particle ratios such as $K/\pi$ and $p/\pi$ are measured as functions of centrality and collision energy, and constrain the chemistry of the system at chemical freeze-out, when inelastic reactions cease to change the particle abundances. Of particular relevance to this work is the production of strange hadrons, which is enhanced in nucleus-nucleus collisions relative to a scaled proton-proton baseline. This strangeness enhancement was among the earliest proposed signatures of the quark-gluon plasma, and it is taken up in detail in section~\ref{section16}.
\Needspace{4\baselineskip}
\subsection{Anisotropic flow and collective behavior}
\label{section132}

The particles emerging from a collision are not distributed uniformly in azimuthal angle. This azimuthal anisotropy is conventionally quantified by expanding the angular distribution as a Fourier series,
\begin{equation}
  \frac{dN}{d\phi} =
  \frac{N}{2\pi}\left(1 + 2\sum_{n=1}^{\infty} v_n \cos\!\left[n(\phi - \Psi_n)\right]\right),
  \label{eq:flow_fourier}
\end{equation}
where $\phi$ is the azimuthal angle of an emitted particle, the coefficients $v_n$ are known as the flow harmonics, and the angles $\Psi_n$ give the corresponding event-plane orientations. The flow harmonics are among the most important observables in heavy-ion physics, as they provide some of the clearest evidence for the collective, fluid-like behavior of the medium~\cite{Heinz:2013th}.

The origin of this anisotropy lies in the collision geometry introduced in the previous section. The region of overlap between the two nuclei is generally spatially anisotropic, and the pressure gradients within the expanding medium are correspondingly uneven, being steepest along the short axis of the overlap region. The collective expansion therefore develops more rapidly in some directions than in others, converting the initial spatial anisotropy into an anisotropy of the final particle momenta. This conversion is efficient only when the medium behaves as a low viscosity fluid, which is why anisotropic flow is regarded as central evidence for the near perfect fluid picture and for the success of a hydrodynamic description.

\begin{figure}[!htbp]
  \centering
  \includegraphics[width=0.60\textwidth]{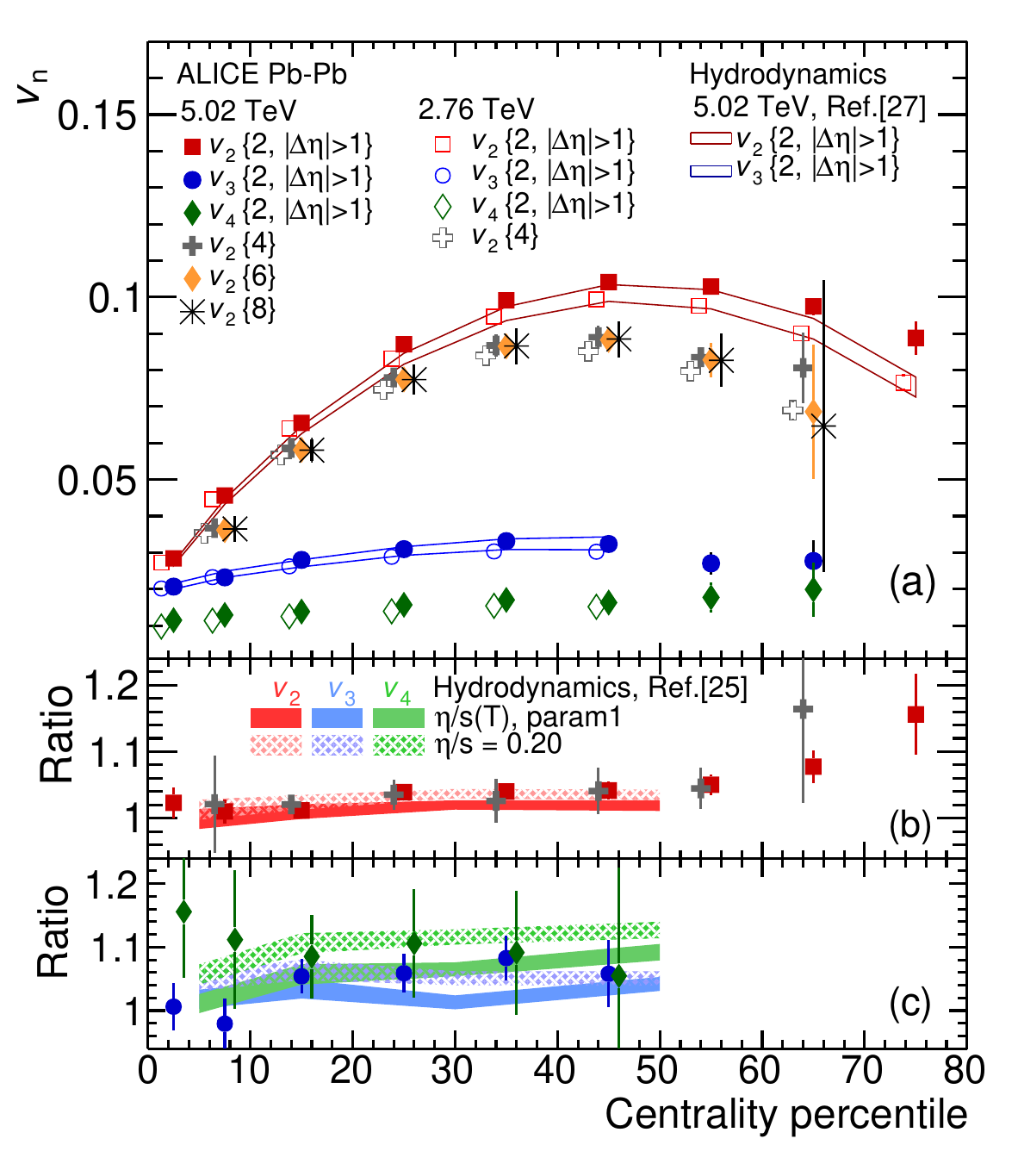}
  \caption[The flow harmonics $v_2$, $v_3,$ and $v_4$ for Pb+Pb at $2.76$ and $5.02$ TeV as a function of centrality]{The flow harmonics $v_2$, $v_3,$ and $v_4$ for Pb+Pb at $2.76$ and $5.02$ TeV as a function of centrality, from ALICE. Reproduced from Refs.~\cite{ALICE:2011ab,ALICE:2016ccg}. Also plotted are comparisons to hydrodynamic models from annotated references in the figure, corresponding to ~\cite{Niemi:2015voa,Noronha-Hostler:2015uye}.}
  \label{fig:vn_centrality}
\end{figure}

The individual harmonics carry distinct physical meaning. The second harmonic $v_2$, known as elliptic flow, is typically the largest in non-central collisions, where the average overlap region takes on a characteristic almond-like shape. The third harmonic $v_3$, or triangular flow, arises primarily from event-by-event fluctuations in the positions of the nucleons rather than from the average collision geometry. Higher harmonics receive contributions both from the corresponding higher-order initial anisotropies and from nonlinear coupling among lower-order flow harmonics. Common to all of these is a sensitivity to the shear viscosity of the medium: larger shear viscosity damps the conversion of spatial into momentum anisotropy and thereby suppresses the $v_n$, making anisotropic flow one of the primary experimental constraints on this quantity. The centrality dependence of $v_2$ and $v_3$ is shown in Fig.~\ref{fig:vn_centrality}.

At intermediate transverse momentum, the elliptic flow of different hadron species approximately follows constituent quark number scaling, an observation historically interpreted as evidence that collective flow develops at the partonic level~\cite{STAR:2003wqp}. In practice, because the orientation of the overlap region is not known experimentally on an event-by-event basis, the flow harmonics are not measured against a fixed plane but are instead extracted from azimuthal correlations among the produced particles. The flow harmonics build up over the course of the medium's hydrodynamic evolution, described in section~\ref{section23}.
\enlargethispage{2\baselineskip}
\Needspace{4\baselineskip}
\subsection{Hard and electromagnetic probes}
\label{section133}

A further class of observables comes not from the bulk of low-momentum hadrons but from rare, high-momentum particles produced in the earliest moments of a collision. Energetic partons and heavy quarks are created in early hard scatterings and must traverse the medium before fragmenting or hadronizing into observable particles. In doing so, they lose energy and momentum to the medium, a phenomenon known as jet quenching.

The resulting suppression is commonly quantified by the nuclear modification factor $R_{AA}$, the yield in nucleus-nucleus collisions divided by the proton-proton yield scaled by the expected number of binary nucleon-nucleon collisions. A value below unity at high transverse momentum indicates that energetic partons have lost energy in the medium, and is taken as evidence of its density and opacity~\cite{PHENIX:2001hpc}.

A related signature is the modification of quarkonium production, involving bound states of a heavy quark and its antiquark such as the $J/\psi$ and the $\Upsilon$. These states are expected to dissociate in a deconfined medium, with more loosely bound states melting more readily, although at high collision energies regeneration can occur as heavy quarks recombine. Such quarkonium modification has long been regarded as a signature of deconfinement~\cite{Matsui:1986dk}.

Distinct from these hard probes, electromagnetic radiation provides a penetrating view of the medium's interior throughout its evolution. Photons and dileptons (correlated lepton-antilepton pairs) are emitted at every stage of a collision, from initial hard scatterings through the thermal radiation of the quark-gluon plasma and the hadronic phase. Because they couple to the medium only electromagnetically, they escape essentially without final-state interactions. Their spectra therefore integrate over the entire spacetime evolution, with the thermal component carrying information about the temperature and chemical composition of the medium during its earliest, hottest stages~\cite{PHENIX:2008uif,ALICE:2015xmh}, although this signal must first be separated from a large background of photons and lepton pairs produced in hadron decays. This sensitivity to the medium's evolving composition is taken up in chapter~\ref{chapter3}.
\Needspace{4\baselineskip}
\section{Bulk properties of QCD matter}
\label{section14}

The observables described in the previous section are experimental handles on the collective properties of the medium itself. These bulk properties describe the thermodynamic and transport behavior of the soft, collective sector, rather than the properties of individual final-state hadrons directly. The most important of them for the present work are the phase structure of QCD matter, its equation of state, and the transport coefficients that govern its dissipative evolution.

\subsection{The QCD phase diagram}
\label{section141}

The phases of strongly interacting matter are conventionally organized in a phase diagram spanned by the temperature $T$ and the baryon chemical potential $\mu_B$, the latter controlling the net baryon density, or equivalently the excess of quarks over antiquarks. At low temperature and density lies ordinary confined matter, in which quarks and gluons are bound into hadrons, while at high temperature lies the deconfined quark-gluon plasma introduced in section~\ref{section11}.

At vanishing baryon chemical potential and physical quark masses, the transition between these regimes is known from lattice QCD to be a smooth crossover rather than a sharp phase transition~\cite{Aoki:2006we}. There is consequently no unique transition temperature, but rather a pseudocritical temperature, commonly quoted around $T_c \approx 155$--$158$~MeV~\cite{Borsanyi:2010bp,HotQCD:2018pds}, marking the approximate center of the crossover region. At larger baryon chemical potential, many theoretical scenarios predict a first-order transition line terminating at a critical point~\cite{Bzdak:2019pkr}, although neither the line nor the point has yet been established from first principles or experimentally. Lattice QCD, which provides the firmest theoretical handle at $\mu_B \approx 0$, cannot be applied straightforwardly at large real $\mu_B$ because of the numerical sign problem, leaving much of the high density region of the phase diagram uncertain.

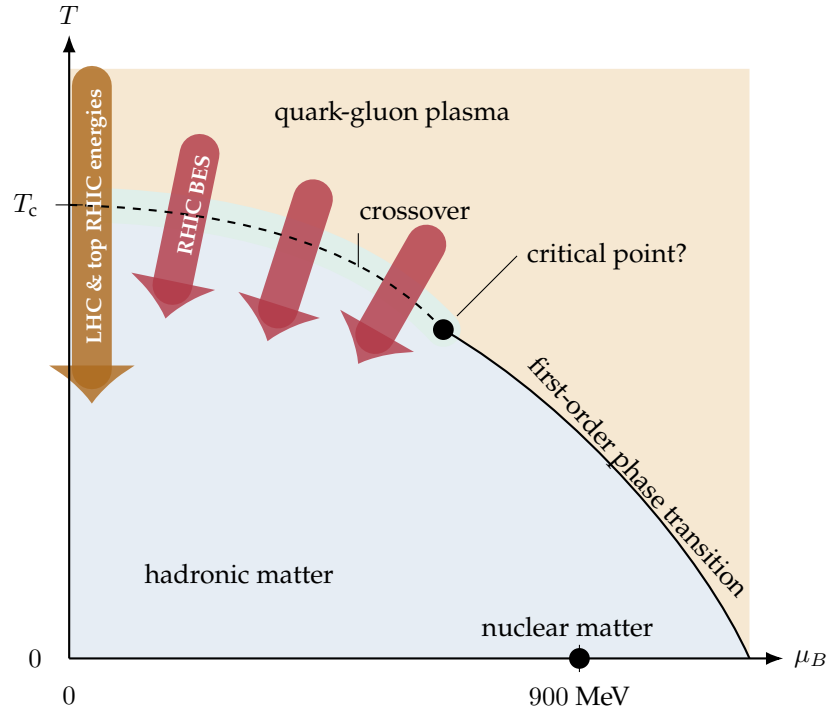
\begin{figure}[!t]
  \centering
    \definecolor{bandHRG}{HTML}{E6EDF4}  
    \definecolor{bandMID}{HTML}{E0EFEA}  
    \definecolor{bandQGP}{HTML}{F6E8D2}  
    \definecolor{arrHi}{HTML}{AE6B1F}    
    \definecolor{arrBES}{HTML}{B23A48}   
    \begin{tikzpicture}[>=Latex, scale=1.5, font=\small,
        traj/.style={-{Latex[length=5mm,width=12mm]}, line width=15pt, line cap=round, opacity=0.82}]
    
      \fill[bandQGP]
        (0,4.0) -- (0,5.2) -- (6.0,5.2) -- (6.0,0)
        .. controls (5.6,0.9) and (4.5,2.2) .. (3.3,2.9)
        .. controls (2.7,3.55) and (1.7,3.95) .. (0,4.0) -- cycle;
      \fill[bandHRG]
        (0,0) -- (0,4.0)
        .. controls (1.7,3.95) and (2.7,3.55) .. (3.3,2.9)
        .. controls (4.5,2.2) and (5.6,0.9) .. (6.0,0) -- (0,0) -- cycle;
      \draw[bandMID, line width=13pt, line cap=round]
        (0.2,4.0) .. controls (1.7,3.95) and (2.7,3.55) .. (3.3,2.9);
    
      \draw[traj, arrHi]  (0.2,5.05) -- (0.2,2.25);
      \draw[traj, arrBES] (1.15,4.45) -- (0.85,3.0);
      \draw[traj, arrBES] (2.15,4.05) -- (1.75,2.8);
      \draw[traj, arrBES] (3.15,3.65) -- (2.55,2.6);
    
      \draw[dashed, thick] (0,4.0) .. controls (1.7,3.95) and (2.7,3.55) .. (3.3,2.9);
      \draw[thick]         (3.3,2.9) .. controls (4.5,2.2) and (5.6,0.9) .. (6.0,0);
    
      \fill (3.3,2.9) circle (2.6pt);
      \draw[thin] (3.42,3.02) -- (3.95,3.55);
      \node[anchor=west] at (3.95,3.55) {critical point?};
    
      \fill (4.5,0) circle (2.6pt);
      \node[anchor=south west] at (3.55,0.1) {nuclear matter};
    
      \draw[->, thick] (0,0) -- (6.3,0) node[right] {$\mu_B$};
      \draw[->, thick] (0,0) -- (0,5.5) node[above] {$T$};
    
      \draw (-0.12,4.0) -- (0.12,4.0);   \node[anchor=east] at (-0.18,4.0) {$T_\text{c}$};
      \draw (4.5,-0.12) -- (4.5,0.12);   \node[anchor=north] at (4.5,-0.16) {$900$ MeV};
      \node[anchor=east]  at (-0.15,0) {$0$};
      \node[anchor=north] at (0,-0.16) {$0$};
    
      \node at (2.85,4.8) {quark-gluon plasma};
      \node at (1.5,0.75) {hadronic matter};
    
      \node[rotate=0, anchor=south] at (3.05,3.78) {crossover};
      \node[rotate=-48, anchor=south] at (4.85,1.45) {first-order phase transition};
      \draw[thin] (2.55,3.75) -- (2.55, 3.45);
      \node[rotate=90, anchor=center, white, font=\scriptsize\bfseries] at (0.24,3.9) {LHC \& top RHIC energies};
      \node[rotate=79, anchor=center, white, font=\scriptsize\bfseries] at (1.1, 4.0) {RHIC BES};
    \end{tikzpicture}
  \caption[Schematic phase diagram of QCD matter in the plane of temperature $T$ and baryon chemical potential $\mu_B$]{Schematic phase diagram of QCD matter in the plane of temperature $T$ and baryon chemical potential $\mu_B$, showing the hadronic and quark-gluon plasma regions, the crossover at low $\mu_B$, and the conjectured first-order transition line and critical point at larger $\mu_B$. The broad arrows indicate the schematic trajectories of high-energy collisions at the LHC and top RHIC energies and of the RHIC Beam Energy Scan (probing larger $\mu_B$).}
  \label{fig:phase_diagram}
\end{figure}

It is important to keep in mind that a heavy-ion collision does not sit at a single point in this diagram. The medium is a finite, rapidly evolving system, and as it expands and cools it traces out a dynamical trajectory across the phase diagram. The matter produced near midrapidity at top RHIC energy and at the LHC is created at high temperature and small baryon chemical potential, placing it close to the $\mu_B = 0$ crossover region where lattice QCD constraints are strongest. These features are summarized schematically in Fig.~\ref{fig:phase_diagram}.

\Needspace{5\baselineskip}
While the $(T,\mu_B)$ plane is the conventional representation of the QCD phase diagram, it is only a projection of a larger thermodynamic parameter space. In equilibrium, additional chemical potentials may be introduced for other conserved charges, such as the electric-charge and strangeness chemical potentials $\mu_Q$ and $\mu_S$, and heavy-ion applications often impose constraints such as strangeness neutrality and a fixed charge-to-baryon ratio. The two-dimensional phase diagram should therefore be understood as a useful projection, not as the complete thermodynamic description of QCD matter.

These conserved charge chemical potentials should also be distinguished from the non-equilibrium flavor fugacities considered later in this work. The latter do not represent conserved charges, but instead parameterize departures of the light and strange quark abundances from their chemical equilibrium values. This distinction is central to the extension of the equation of state developed in chapter~\ref{chapter3}. Since even the conventional $(T,\mu_B)$ plane remains incompletely understood, this broader parameter space leaves an expansive landscape for further theoretical and experimental exploration.

\subsection{The equation of state}
\label{section142}

Closely related to the phase structure is the equation of state, the relation among the equilibrium thermodynamic quantities of the medium: its pressure, energy density, and entropy density, expressed as functions of temperature and chemical potentials. The equation of state encodes how the medium responds to changes in temperature and density, and through doing so, governs how the internal pressure drives the collective expansion seen in the experimental observables.

The qualitative behavior of the equation of state follows the change in the underlying degrees of freedom across the crossover. At low temperature the relevant degrees of freedom are the massive hadrons, while at high temperature they are the far more numerous quarks and gluons. As a result, the energy and entropy densities rise steeply through the transition region as these additional degrees of freedom are liberated. The equation of state also softens in this region, meaning that the pressure responds less stiffly to changes in energy density than it does in either the hadronic or the high-temperature limit. This softening is significant because pressure gradients accelerate the fluid, so a softer equation of state leads to a more gradual collective expansion.

At vanishing chemical potential, the equation of state is known from first-principles lattice QCD calculations at high temperature and matches smoothly onto a hadron resonance gas description at low temperature. The detailed construction of this equilibrium equation of state is presented in section~\ref{subsection233}. A central aim of this work is to extend this equilibrium description to systems with chemically non-equilibrium flavor composition, in which the light and strange quark abundances depart from their equilibrium values. As noted above, this is implemented through flavor fugacities rather than through chemical potentials associated with exactly conserved charges. This flavor-dependent equation of state is developed in chapter~\ref{chapter3}.

\subsection{Transport coefficients}
\label{section143}

Beyond its equilibrium thermodynamics, the medium is also characterized by transport coefficients, which describe how it responds dissipatively to gradients in its flow. The two most important for the bulk evolution are the shear and bulk viscosities. These are commonly expressed in terms of the ratios $\eta/s$ and $\zeta/s$, where $\eta$ and $\zeta$ are the shear and bulk viscosities and $s$ is the entropy density. In natural units, dividing by the entropy density renders these ratios dimensionless, allowing them to be compared across different fluids and thermodynamic conditions. The ratio $\eta/s$ is known as the specific shear viscosity and $\zeta/s$ as the specific bulk viscosity.

Bulk viscosity describes the resistance of the medium to uniform expansion or compression. In an expanding fluid such as the QGP, bulk viscosity gives rise to a bulk pressure that acts as a correction to the equilibrium pressure and tends to suppress the radial expansion. A larger bulk viscosity therefore tends to lower the mean transverse momentum of the final-state hadrons, providing one of its main experimental signatures.

Shear viscosity describes the resistance of the medium to shearing, when adjacent layers of fluid flow past one another at different velocities. Its primary experimental signature is in anisotropic flow: larger shear viscosity damps the conversion of initial spatial anisotropy into final-state momentum anisotropy, thereby suppressing the flow harmonics $v_n$. The specific shear viscosity of the QGP is remarkably small, among the lowest of any known fluid, and it is this property that underlies the description of the medium as a nearly perfect fluid.

Both $\eta/s$ and $\zeta/s$ are in general functions of temperature and chemical potentials. The temperature dependence of $\zeta/s$ is not precisely known, but it is generally expected to peak near the crossover region and to fall off in either direction. The chemical potential dependence is more poorly constrained still, and is discussed further in chapter~\ref{chapter3}. Other transport coefficients, such as charge and baryon diffusion coefficients, become important at finite baryon density, but the present work focuses on the shear and bulk viscosities that control the dissipative evolution of the soft medium. The dissipative hydrodynamic framework in which these viscosities enter, together with the parameterized forms used to model their temperature dependence, is developed in section~\ref{section23}.

Taken together, the equation of state and the transport coefficients constitute the principal material inputs to a relativistic viscous hydrodynamic description of the medium's bulk evolution. Such a description is applicable because the matter produced in a collision appears to reach a hydrodynamic regime very rapidly and thereafter behaves collectively as a fluid. The dynamics of this approach to fluid behavior, including the processes of thermalization and hydrodynamization, are the subject of section~\ref{section15}, while the hydrodynamic framework used to model the subsequent evolution is developed in chapter~\ref{chapter2}.

\section{Kinetic equilibration of QCD matter}
\label{section15}

In the previous section, the equation of state and transport coefficients were presented as the material inputs to a viscous hydrodynamic description of the medium. Such a description is traditionally motivated by proximity to local thermodynamic equilibrium, but modern heavy-ion theory has shown that hydrodynamics can become applicable before full equilibration, provided the nonhydrodynamic degrees of freedom have decayed sufficiently. How the highly excited matter produced at the moment of impact reaches this fluid regime, and how quickly, is the subject of the present section.

We should first consider how a collision evolves in space and time, as sketched in Fig.~\ref{fig:spacetime}. The two nuclei, Lorentz-contracted into thin sheets, pass through one another and leave behind a highly excited state of matter that is initially far from equilibrium. Over a very short time this matter relaxes toward a hydrodynamic regime, forming a deconfined QCD medium that expands and cools as a relativistic fluid. As the temperature falls through the transition region, the matter hadronizes, its quarks and gluons binding into hadrons. These hadrons continue to scatter and decay for a time, until the system becomes dilute enough that the remaining particles cease to interact and stream freely outward to the detectors. The entire evolution, from impact to the final decoupling of the hadrons, spans a time of order $10$~fm/$c$ in large collision systems.

\definecolor{cPre}{HTML}{D4D6D3}  
\definecolor{cQGP}{HTML}{EFCE8E}  
\definecolor{cGas}{HTML}{BCD0E8}  
\definecolor{cNuc}{HTML}{4A5A6A}  
\def\stTmax{4}
\newcommand{\stageregion}[3]{(0,0) -- (-\stTmax,\stTmax) -- (-#2,\stTmax)
  -- plot[domain=-#2:#2,samples=#3] (\x,{sqrt(\x*\x+#1*#1)})
  -- (#2,\stTmax) -- (\stTmax,\stTmax) -- cycle}
  
\enlargethispage{2\baselineskip}
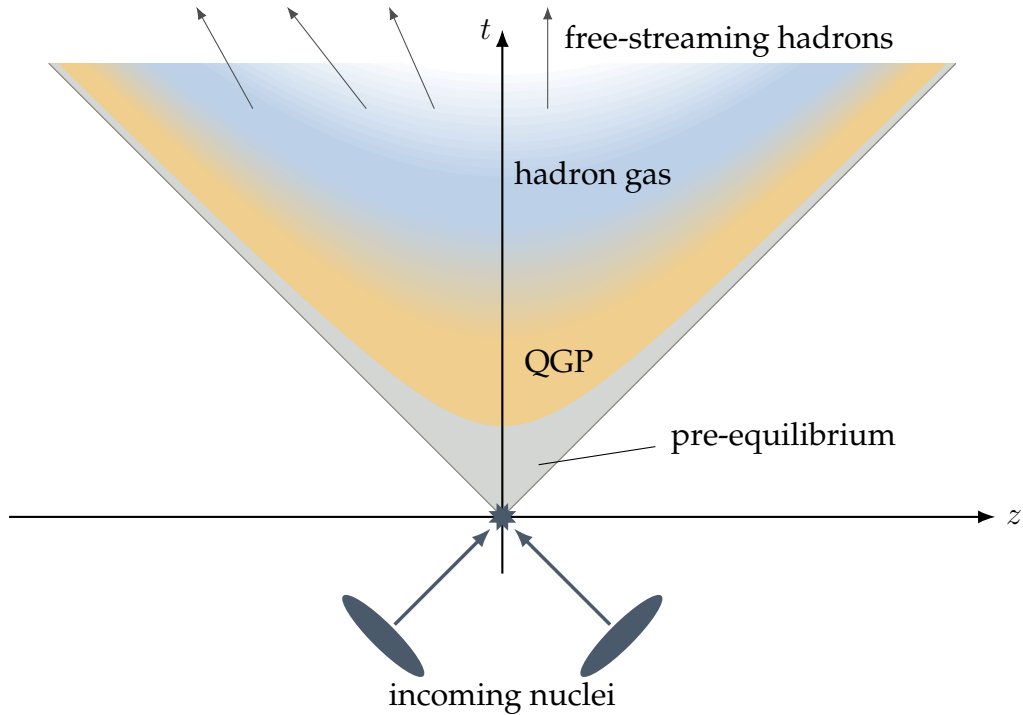
\begin{figure}[!htbp]
  \vspace{12pt}
  \centering
  \begin{tikzpicture}[>=Latex, scale=1.5, font=\large]
    \fill[white] (0,0) -- (-\stTmax,\stTmax) -- (\stTmax,\stTmax) -- cycle;
    \foreach \tau in {3.9,3.825,...,0.9} {
      \pgfmathsetmacro\z{sqrt(\stTmax*\stTmax-\tau*\tau)}
      \pgfmathsetmacro\bmix{max(0,min(100,100*(2.6-\tau)))}
      \pgfmathsetmacro\wmix{max(0,min(100,100*(4.0-\tau)))}
      \fill[cQGP!\bmix!cGas!\wmix!white] \stageregion{\tau}{\z}{48};
    }
    \fill[cPre] \stageregion{0.8}{3.919}{100};
    \draw[gray, thin] (0,0) -- (-\stTmax,\stTmax);
    \draw[gray, thin] (0,0) -- (\stTmax,\stTmax);
    \begin{scope}[shift={(-1.05,-1.05)}, rotate=45]
      \fill[cNuc] (0,0) ellipse [x radius=0.11, y radius=0.5];
    \end{scope}
    \begin{scope}[shift={(1.05,-1.05)}, rotate=135]
      \fill[cNuc] (0,0) ellipse [x radius=0.11, y radius=0.5];
    \end{scope}
    \draw[-{Latex[length=2.6mm]}, line width=1.3pt, cNuc] (-0.95,-0.95) -- (-0.1,-0.1);
    \draw[-{Latex[length=2.6mm]}, line width=1.3pt, cNuc] ( 0.95,-0.95) -- ( 0.1,-0.1);
    \node[font=\large] at (0,-1.6) {incoming nuclei};
    \draw[->, thick] (0,-0.5) -- (0,4.3) node[left] {$t$};
    \draw[->, thick] (-4.35,0) -- (4.35,0) node[right] {$z$};
    \node[star, star points=8, star point ratio=0.55, fill=cNuc, inner sep=4.0pt] at (0,0) {};
    \foreach \sx/\sy/\ex/\ey in {0.4/3.6/0.4/4.5, -0.6/3.6/-1.0/4.5,
        -2.2/3.6/-2.7/4.5, -1.2/3.6/-1.9/4.5}
      \draw[->, gray!60!black] (\sx,\sy) -- (\ex,\ey);
    \node[font=\large] at (2.0,4.2) {free-streaming hadrons};
    \node[align=center, fill=white, fill opacity=0.0, text opacity=1,
          inner sep=2.5pt, rounded corners=1.5pt] at (0.5,1.35) {QGP};
    \node[fill=white, fill opacity=0.0, text opacity=1,
          inner sep=2.5pt, rounded corners=1.5pt] at (0.8,3.0) {hadron gas};
    \draw[thin] (1.35,0.62) -- (0.32,0.46);
    \node[anchor=west] at (1.4,0.7) {pre-equilibrium};
  \end{tikzpicture}
  \caption[Schematic spacetime diagram of a heavy-ion collision in the plane
    of time $t$ and the longitudinal beam coordinate $z$] {Schematic spacetime diagram of a heavy-ion collision in the plane of time $t$ and the longitudinal beam coordinate $z$. The two Lorentz-contracted nuclei approach along the light cone and collide at the origin. The subsequent stages of pre-equilibrium, the hydrodynamic quark-gluon plasma, a hadron resonance gas, and free-streaming after freeze-out are separated approximately by curves of constant proper time $\tau = \sqrt{t^2 - z^2}$.}
  \label{fig:spacetime}
\end{figure}

It is often useful to describe this evolution using coordinates adapted to the longitudinal expansion. Instead of the laboratory time $t$ and beam coordinate $z$, one introduces the proper time
\begin{equation}
    \tau = \sqrt{t^2 - z^2}
\end{equation}
and the spacetime rapidity
\begin{equation}
    \eta_s = \frac{1}{2}\ln\left(\frac{t+z}{t-z}\right).
    \label{eq:spacetime_rapidity}
\end{equation}

These variables are the spacetime counterparts of the momentum-space rapidity introduced in section~\ref{section12}. Near midrapidity at top RHIC and LHC energies, the system is often approximated as boost invariant, meaning that its local properties depend primarily on $\tau$ and the transverse coordinates rather than on $\eta_s$.

The remainder of this section concerns the earliest part of this evolution, during which the system passes from its initial far-from-equilibrium state to a regime in which hydrodynamics becomes applicable. This is the problem of kinetic equilibration: the relaxation of the local momentum distribution, or equivalently the stress tensor, toward local equilibrium. It is distinct from the chemical equilibration of quark abundances, treated in the following section.

Kinetic equilibration in this sense encompasses several distinct milestones: the decay of the most rapidly varying nonhydrodynamic modes, the onset of a hydrodynamic description, the reduction of pressure anisotropy, and finally the approach to a locally thermal momentum distribution. These need not occur at the same time, and it is this broader process, rather than any single one of them, that the present section addresses. Of these, hydrodynamization carries particular phenomenological weight: it marks the switching point at which the hydrodynamic stage takes over, and so enters the multistage model directly as the time from which the medium is evolved hydrodynamically. Its relation to full thermalization --- in particular, that the former does not require the latter --- is taken up in the following subsection.
\Needspace{4\baselineskip}
\subsection{Thermalization and hydrodynamization}
\label{section151}

It has long been recognized that hydrodynamics describes heavy-ion data remarkably well when the fluid evolution is initialized very early, within roughly $1$~fm/$c$ of the initial impact~\cite{Busza:2018rrf}. This is at first sight surprising, especially from the perspective of weakly coupled quasiparticle scattering, where full thermal equilibration might be expected to take longer. The resolution lies partly in a distinction between two notions that are easily conflated.

Thermalization refers to the approach to local thermal equilibrium, in which the local momentum distribution becomes isotropic and approaches its equilibrium form, up to collective flow and the local values of the conserved charge densities. Hydrodynamization, by contrast, refers to the earlier point at which a hydrodynamic description becomes accurate for the subsequent evolution. The modern understanding is that these are not the same: a hydrodynamic description can become valid well before the system has fully thermalized, while it is still substantially anisotropic in momentum space~\cite{Busza:2018rrf}.

This anisotropy arises from the rapid longitudinal expansion of the medium along the beam direction, which initially stretches the system far faster than it expands transversely, driving the pressure along the beam axis well below the transverse pressure. A convenient measure of the resulting anisotropy is the ratio of the longitudinal pressure $P_L$ to the transverse pressure $P_T$. In local equilibrium, and in the absence of viscous corrections, the momentum distribution is isotropic and the two pressures are equal, so that $P_L/P_T = 1$, whereas at early times the longitudinal expansion can drive $P_L/P_T \ll 1$. Hydrodynamics becomes applicable while this ratio is still well below unity, with full isotropization $P_L/P_T \to 1$ corresponding to thermalization proper. Viscous hydrodynamics can describe this anisotropic early stage precisely because the difference between the longitudinal and transverse pressures is the kind of departure from equilibrium encoded in its dissipative corrections.

One way to understand the early applicability of hydrodynamics is through the existence of hydrodynamic attractors. In a variety of expanding systems, far-from-equilibrium solutions of the underlying evolution equations are found to converge toward a common trajectory, the attractor, before the system reaches equilibrium~\cite{Heller:2015dha}. Once on this attractor, the evolution becomes largely insensitive to the details of the initial state, which helps explain why a hydrodynamic description with relatively few parameters can be predictive even when applied at very early times. Detailed kinetic theory calculations find that this rapid hydrodynamization occurs on a timescale of about $1$~fm/$c$~\cite{El:2007vg, Kurkela:2015qoa, Keegan:2016cpi, Kurkela:2018vqr}, a conclusion supported by real-time classical-statistical lattice simulations~\cite{Epelbaum:2013ekf, Berges:2013fga, Schenke:2015aqa}. This is consistent with the strongly interacting character of the medium established earlier, for which a fast onset of collective behavior is expected.

\subsection{Pre-equilibrium dynamics}
\label{section152}

To understand the starting point of this evolution, one must consider the state of the matter immediately after the two nuclei pass through one another. At the high energies of interest, this initial state is dominated not by quarks but by gluons, which are produced in great abundance. The phenomenological success of initial-state models built on this picture supports the idea of a gluon-saturated initial state~\cite{Schenke:2012wb, Eskola:1999fc}, in which gluon production saturates as gluon splitting and recombination reach a dynamic balance~\cite{McLerran:1993ni, Gribov:1983ivg, Gelis:2010nm}. In the color-glass condensate picture, the gluon occupation numbers are high enough that the earliest configuration is better described as a strong classical color field than as a collection of individual particles. The initial state is therefore gluon-dominated and correspondingly quark-poor.

In weak-coupling descriptions, the subsequent evolution can be understood in stages that reflect the falling gluon density and occupation number. While the occupation numbers remain large, the dynamics are approximately those of classical color fields. As the system expands and dilutes, the occupation numbers fall and an effective kinetic theory, formulated in terms of quasiparticle scatterings, becomes the appropriate description. In the weakly coupled picture, gluon thermalization is expected to proceed rapidly through the bottom-up scenario~\cite{Baier:2000sb}. The same early estimates suggest that the quark and antiquark abundances equilibrate more slowly, owing to the smaller cross sections for producing quark pairs~\cite{Shuryak:1992wc, Biro:1993qt}, a difference in timescales that is central to the chemical equilibration discussed in the next section.

Underlying this evolution is a competition between two effects. Interactions among the constituents drive the system toward equilibrium, while the rapid longitudinal expansion continually dilutes it and acts to sustain the momentum-space anisotropy. For hydrodynamics to become applicable as early as it does, the equilibrating interactions must win this competition quickly. These pre-equilibrium descriptions carry the system up to the point at which it can be treated by a hydrodynamic description.

Once the medium has hydrodynamized, its evolution is described by the viscous fluid characterized in section~\ref{section14}. The residual departures from local equilibrium are what the shear and bulk viscosities quantify, and in a second-order hydrodynamic description the associated dissipative stresses relax toward their equilibrium values over finite timescales. In this sense the viscosities re-enter the dynamical description in the role of near-equilibrium relaxation, with the corresponding relaxation equations developed in section~\ref{section23}.

The relaxation of the momentum distribution is, however, only one aspect of the medium approaching equilibrium. The relative abundances of the different quark flavors equilibrate through their own, distinct processes, and as noted above this chemical equilibration is expected to lag hydrodynamization. The medium may therefore evolve hydrodynamically while its flavor content is still far from chemical equilibrium, with potentially significant consequences for its properties and evolution. Whether, and how completely, chemical equilibrium is reached is considerably less certain than the corresponding picture for momentum equilibration, and is the subject of the next section.

\section{Chemical equilibration of QCD matter}
\label{section16}

Approaching equilibrium involves more than the relaxation of the momentum distribution treated in the previous section. While kinetic equilibration concerns the shape of the momentum distribution, chemical equilibration concerns the flavor content of the medium: the abundances of the various quark species relative to their equilibrium values. The two are distinct, and a medium whose momenta have equilibrated need not have reached its equilibrium composition. 

The natural language for describing chemical composition is that of chemical potentials and fugacities. For a chemical potential $\mu_i$, the corresponding fugacity is
\begin{equation}
    \gamma_i = e^{\mu_i/T}.
    \label{eq:fugacity_basic}
\end{equation}

In equilibrium thermodynamics, such chemical potentials are usually associated with conserved charges such as baryon number, electric charge, and strangeness. The flavor fugacities considered later in chapter~\ref{chapter3} are different in interpretation: light and strange quark pair numbers are not conserved, so their effective fugacities parameterize departures from chemical equilibrium rather than exact conservation laws. Chemical equilibration is then the process by which these non-equilibrium fugacities relax toward their equilibrium values, $\gamma_i = 1$, or equivalently toward vanishing effective chemical potentials.

The initial state provides a natural reason to expect the medium to begin far from chemical equilibrium. As discussed in the previous section, saturation-based descriptions motivate an early-time state dominated by gluonic degrees of freedom and correspondingly undersaturated in quarks, with the gluon density characterized by the saturation scale $Q_s$. The quark abundances, and the strange quark abundance in particular, may therefore begin below their equilibrium values and must be built up dynamically as the collision proceeds. The degree to which they do so before hadronization and chemical freeze-out is the question this section addresses.

\subsection{Quark production and equilibration rates}
\label{section161}

The various quark flavors need not equilibrate at the same rate. One important scale is set by the quark masses listed in Table~\ref{tab:quark_properties}: the light up and down quarks are nearly massless on QCD scales and are pair-produced copiously, whereas the heavier strange quark is produced less readily and can equilibrate more slowly. Charm and heavier quarks are far too massive to be produced thermally in appreciable numbers in the bulk medium, and therefore play a negligible role in its chemical equilibration.

Strangeness occupies a special position in this hierarchy. The colliding nuclei contain no valence strange quarks, so the observed bulk strangeness must be generated dynamically during the collision. In a deconfined medium, the leading production channels are gluon fusion, $gg \to s\bar{s}$, and light quark-antiquark annihilation, $q\bar{q}\to s\bar{s}$. Because strangeness is the lightest of the flavors that must be generated dynamically and is slower to equilibrate, the strange sector serves as the natural probe of incomplete chemical equilibration.

Estimates of the relevant timescales have a long history. Early perturbative calculations, noted in the previous section, already found that quarks and antiquarks equilibrate more slowly than gluons, owing to the smaller cross sections for producing quark pairs~\cite{Shuryak:1992wc, Biro:1993qt}. Subsequent kinetic theory and two-stage equilibration analyses have placed the chemical equilibration of quarks and antiquarks on a timescale of several fm/$c$, in some estimates exceeding $3$~fm/$c$, later than hydrodynamization but still preceding full thermalization of the medium~\cite{Xu:2004mz, Kurkela:2018oqw, Kurkela:2018xxd}. Although the precise timescale remains uncertain, these studies establish a consistent picture in which the chemical equilibration of the quark flavors can lag the formation of the fluid, so that the medium evolves through much of its lifetime with a flavor composition still approaching equilibrium. Whether full chemical equilibrium, and strangeness equilibration in particular, is reached by the end of the collision depends on the size of the system and the collision energy.

\subsection{Strangeness enhancement and chemical freeze-out}
\label{section162}

The special role of strangeness has a direct experimental counterpart. In a deconfined medium, strangeness can be produced efficiently through partonic reactions such as gluon fusion, whereas in a hadron gas the corresponding strangeness-producing reactions typically have higher thresholds and proceed more slowly. This can be understood intuitively in terms of the relevant mass scales: producing strange quark-antiquark pairs in the high-temperature QGP depends on the strange quark mass of $\sim\!100$ MeV, while in the lower temperature hadron gas strangeness must instead be carried by strange hadrons — the lightest being kaons, at masses near $500$ MeV — which strangeness conservation requires to be produced in pairs. Strange hadron yields are therefore enhanced in nucleus-nucleus collisions relative to scaled proton-proton baselines, an effect that grows with the strangeness content of the hadron, so that the multi-strange $\Xi$ and $\Omega$ baryons are especially enhanced. This strangeness enhancement was among the earliest proposed signatures of quark-gluon plasma formation~\cite{Rafelski:1982pu}, and one of the key indicators of its experimental discovery.

\begin{figure}[!htbp]
  \centering
  \includegraphics[width=0.7\textwidth]{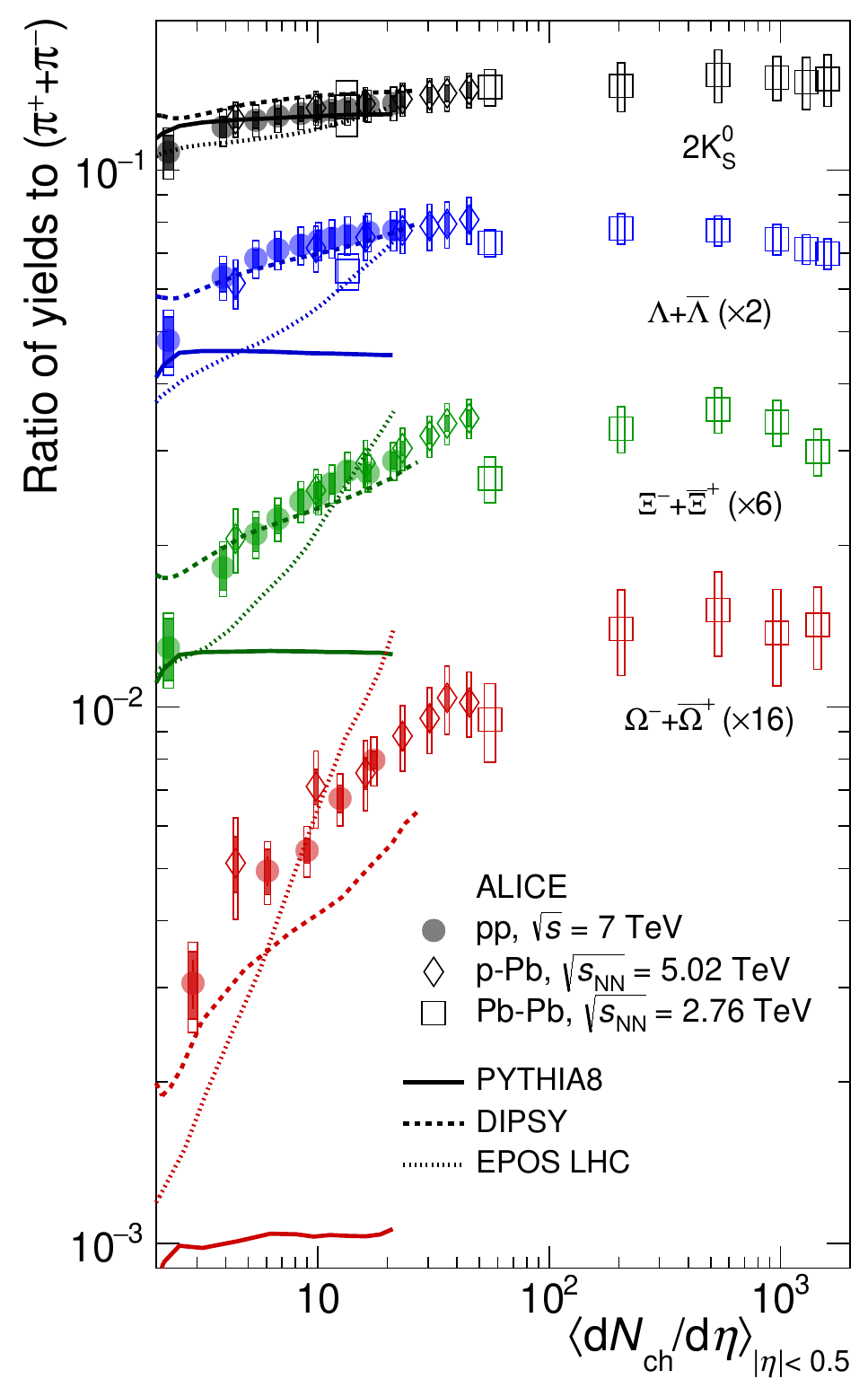}
  \caption[Yield ratios of strange hadrons to pions as a function of
    the average charged particle multiplicity $\langle dN_{\text{ch}}/d\eta \rangle$]{Yield ratios of strange hadrons to pions as a function of the average charged particle multiplicity $\langle dN_{\text{ch}}/d\eta \rangle$, rising smoothly from pp through p+Pb to Pb+Pb collisions at the LHC. Reproduced from Ref.~\cite{ALICE:2016fzo}.}
  \label{fig:strangeness_enhancement}
\end{figure}

The contemporary picture is more nuanced than a simple contrast between large and small systems. Strangeness enhancement is now seen to rise smoothly with the produced particle multiplicity across collision systems of very different size, with the suppression of strangeness in the smallest systems understood in part as a consequence of conservation laws enforced over a small volume~\cite{ALICE:2016fzo}. This smooth rise of strange hadron production with multiplicity, spanning collision systems from pp to Pb+Pb, is shown in Fig.~\ref{fig:strangeness_enhancement}.

The abundances of the produced hadrons are described with notable success by statistical hadronization models~\cite{Andronic:2017pug}. In this picture the integrated yields are described by only a few parameters, principally a temperature and chemical potentials, together with the volume of the system. The temperature is interpreted as the chemical freeze-out temperature, the stage at which inelastic, abundance-changing reactions cease and the relative yields of the various species become fixed. Chemical freeze-out is distinct from, and generally precedes, the later kinetic freeze-out, at which the elastic scatterings also cease and the momentum spectra stop evolving. A striking feature of these fits is that the extracted chemical freeze-out temperature, around $155$~MeV at top RHIC and LHC energies, closely coincides with the pseudocritical temperature of the QCD transition discussed in section~\ref{section141}, suggesting that the hadron abundances are established close to the transition itself as the medium hadronizes.

\FloatBarrier

Within the statistical model, incomplete strangeness equilibration is accommodated through a strangeness saturation or phase-space occupancy factor, $\gamma_s$. This is the strange quark fugacity introduced above, here in its established statistical-model role: it multiplies the equilibrium abundance of a hadron according to its total strange quark content, with multi-strange hadrons receiving higher powers of $\gamma_s$. Some formulations also introduce an analogous light quark factor, often denoted $\gamma_q$. I write this light quark factor as $\gamma_l$ here to emphasize its connection to the notation used later in this work. Schematically, for a hadron species $i$,
\begin{equation}
    N_i \propto
    \gamma_l^{n_l^i+n_{\bar l}^i}
    \gamma_s^{n_s^i+n_{\bar s}^i}
    N_i^{\text{eq}}(T,\mu_B,\mu_Q,\mu_S),
\end{equation}
where $n_l^i+n_{\bar l}^i$ counts the total number of light up and down quarks and antiquarks in hadron species $i$, while $n_s^i+n_{\bar s}^i$ counts the total number of strange and antistrange quarks. These saturation factors should not be confused with chemical potentials for conserved charges: $\gamma_s$ weights the total number of strange and antistrange quarks, not the net strangeness. A value of $\gamma_s = 1$ corresponds to full strangeness equilibration, while $\gamma_s < 1$ indicates an undersaturated strange sector. Fitted values below unity, found particularly in smaller systems and at lower collision energies, provide phenomenological evidence that strangeness need not reach full chemical equilibrium~\cite{Becattini:2005xt}. These saturation factors are static quantities, extracted at freeze-out, and condense the entire prior equilibration history into a single number for each flavor.

\subsection{Dynamical chemical equilibration}

The saturation factors capture the end result of chemical equilibration, but reveal little about its dynamics. How the flavor abundances evolve during the collision, and how that evolution feeds back on the medium itself, is far less well constrained. This feedback is the essential point: a chemically undersaturated medium has a different equation of state from one in full chemical equilibrium, and therefore expands and cools differently. The chemical composition is not merely a property to be read off at freeze-out, but one that can shape the entire evolution and, through it, the final observables.

A number of studies have begun to treat this evolution explicitly, modeling the hydrodynamic medium in partial chemical equilibrium with a time-dependent equation of state that interpolates from the gluon-dominated initial state toward full equilibrium~\cite{Vovchenko:2015yia, Vovchenko:2016ijt}. These works found the chemical composition of the early medium to leave an imprint on electromagnetic observables, namely photon and dilepton production, indicating that such probes are sensitive to the early stages of the collision. To date, however, this approach has not been carried over into a precision Bayesian study of hadronic observables, which are produced and measured in far greater abundance at RHIC and the LHC and may carry their own signatures of the medium's composition.

It is this gap that the present work addresses. Building on the partial chemical equilibrium picture, this work develops a framework in which the light and strange quark abundances are allowed to evolve dynamically out of chemical equilibrium over the course of the hydrodynamic evolution. The flavor fugacities introduced in this work generalize this idea from a static freeze-out saturation factor to a dynamical property of the evolving QGP. Rather than using $\gamma_s$ only to describe the final hadron abundances, the model allows the light and strange quark occupancies to evolve during the hydrodynamic stage, modifying both the equation of state and the eventual hadron yields. I then embed this framework within a Bayesian analysis that constrains the rates of light and strange quark equilibration from hadronic data. The model itself is developed in chapter~\ref{chapter3}, the Bayesian methodology is introduced in chapter~\ref{chapter4}, and the resulting constraints are presented in chapter~\ref{chapter5}.

\chapter{Modeling relativistic heavy-ion collisions}
\label{chapter2}

Relativistic heavy-ion collisions evolve through multiple distinct physical regimes in a way that defies any single, physically justifiable modeling framework. Instead, the standard practice when studying the bulk dynamics is to use multistage models that typically evolve through the following physical timeline: 

\begin{enumerate}[
    label={\arabic*.},
    leftmargin=2.2em,
    labelsep=0.6em,
    itemsep=0.8em,
    topsep=0.8em,
    parsep=0pt
]

    \item \textbf{Initial conditions:}
    The medium forms due to the constituents of the two nuclei interacting as the nuclei interpenetrate at the beginning of the collision.

    \item \textbf{Pre-equilibrium:}
    The early medium evolves in a highly non-equilibrium state. Multiple interactions drive the system toward equilibrium and the formation of a quark-gluon plasma.

    \item \textbf{Quark-gluon plasma:}
    The medium is well described hydrodynamically in the QGP phase as it expands and cools.

    \item \textbf{Hadronization:}
    As the medium cools, the deconfined partons convert into hadrons.

    \item \textbf{Hadronic transport:}
    The hadrons continue to interact through scattering and decays until freeze-out as they propagate out from the collision region.

    \item \textbf{Detection:}
    The hadrons reach the various experimental detectors, at which point we obtain data that can be analyzed to compute observables.

\end{enumerate}

\begin{figure}[!t]
  \centering
  \resizebox{\textwidth}{!}{%
  \begin{tikzpicture}[
      stage/.style={
          rectangle, rounded corners=2pt, draw=black!65, line width=0.7pt,
          text width=22mm, minimum height=17mm, align=center,
          inner sep=3pt, font=\small, fill=#1
      },
      ghost/.style={
          rectangle, rounded corners=2pt, draw=black, dashed, line width=0.7pt,
          text width=22mm, minimum height=10mm, align=center,
          inner sep=3pt, font=\scriptsize, text=black, fill=none
      },
      flow/.style={-{Stealth[length=2.8mm, width=2.2mm]}, line width=0.8pt, draw=black},
      hand/.style={font=\scriptsize, text=black, align=center, inner sep=2pt}
  ]
      \node[stage=orange!16]                          (trento) {\textbf{Initial condition}\\[2pt]{\scriptsize T\textsubscript{R}ENTo}};
      \node[stage=orange!14, right=22mm of trento]   (music)  {\textbf{Viscous hydro}\\[2pt]{\scriptsize MUSIC}};
      \node[stage=orange!12, right=18mm of music]    (is3d)   {\textbf{Particlization}\\[2pt]{\scriptsize iS3D}};
      \node[stage=orange!10, right=18mm of is3d]   (smash)  {\textbf{Hadronic transport}\\[2pt]{\scriptsize SMASH}};
      \node[stage=black!8, right=18mm of smash]    (obs)    {\textbf{Observables}\\[2pt]{\scriptsize $\langle N\rangle$,\, $\langle p_T\rangle$,\, $v_n$}};

      \draw[flow] (trento) -- node[hand, above]{$\varepsilon(x,y)$ at $\tau_0$} (music);

      \coordinate (mid1) at ($(trento.east)!0.5!(music.west)$);
      \node[ghost, below=9mm of mid1] (preq) {Pre-equilibrium};
      \draw[dotted, black, line width=0.6pt] (preq.north) -- (mid1);

      \draw[flow] (music)  -- node[hand, above]{hypersurface\\at $T_\text{part}$} (is3d);
      \draw[flow] (is3d)   -- node[hand, above]{sampled\\hadrons} (smash);
      \draw[flow] (smash)  -- node[hand, above]{final\\particles} (obs);

      \draw[-{Stealth[length=2.6mm]}, draw=black, line width=0.6pt]
          ([yshift=-15mm]trento.south west) -- ([yshift=-15mm]obs.south east)
          node[midway, below=1pt, font=\scriptsize, text=black]{Proper time $\tau$ };
  \end{tikzpicture}%
  }
  \caption[The multistage model pipeline]{The multistage model pipeline used in this work.}
  \label{fig:pipeline}
\end{figure}
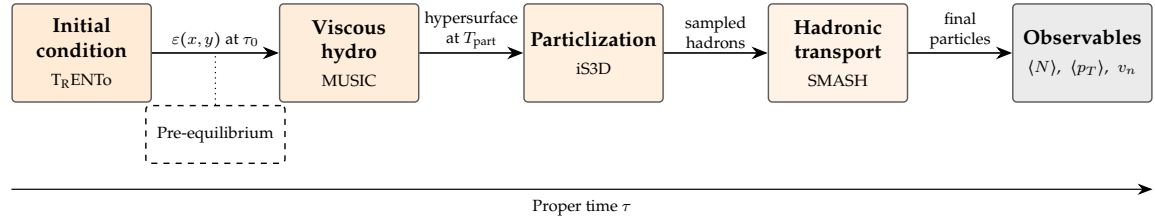

Multistage models vary greatly in both dynamics and form, and many do not explicitly treat each of the above steps. For example, the model pipeline presented in this work will, by design, largely sidestep the pre-equilibrium phase. This is shown schematically in Fig. \ref{fig:pipeline}. However, all of these stages must in some way be accounted for in order to construct a reasonable framework for model-to-data comparisons. The only direct experimental knowledge is at two points: the species and energies of the colliding nuclei, and the final ensemble of detected hadrons. Any model which cannot coherently link the two is necessarily incomplete.

The remainder of this chapter will loosely follow the above timeline, discussing some of the major models used for each step, as well as how the final-stage outputs are processed to allow for coherent comparison to experimental data. In some cases, these models provide only an effective description of the corresponding physical process rather than a direct microscopic treatment. Note that this discussion is restricted to the bulk dynamics of the medium, or soft-sector observables, and therefore does not provide a complete picture of relativistic heavy-ion collisions. Conspicuously omitted from this discussion are jets, heavy quarks, and other high-momentum particles that thermalize less readily with the medium but provide a rich category of observables in their own right. These typically necessitate a separate, parallel chain of models that couples to the medium at certain points but evolves the hard particles semi-independently.

\section{Initial conditions}
\label{section21}

The initial conditions are, in some sense, the most difficult part of a heavy-ion collision to model physically, as they are simultaneously the most extreme state reached and the farthest from the final hadrons that we can observe. One might naively expect that with QCD and knowledge of the structure and properties of the incoming nuclei, it should be possible to directly calculate the properties of the initial medium. While this is true to a very limited extent --- the previous chapter discussed, for example, the phenomenon of gluon saturation, which is accessible to weak-coupling methods --- the interactions during and just after the collision are highly nonperturbative, making direct calculations mostly intractable. This necessitates effective models for the initial state.

The primary quantity that needs modeling is the initial energy or entropy density profile, which is largely defined by the initial geometry. Recall from section \ref{section12} that this strongly depends on the nuclear geometry, the centrality of the collision, and, for non-spherical nuclei, the relative orientation of the nuclei, which altogether define the shape of the medium. It would not be sufficient to simply define the initial state as some ellipsoid, however, as matter is deposited around discrete nucleons rather than uniformly. Any robust initial condition model must thus account for the collision geometry, event-by-event fluctuations in the positions of nucleons, and nuclear structure --- in some cases, even sub-nucleonic. 

In this section, I will briefly review three classes of models: Glauber models, gluon saturation-based models, and T\textsubscript{R}ENTo, the last of which is used throughout the rest of this work.

\subsection{Glauber models}

The original Glauber model is perhaps the simplest and most intuitive picture of initial energy deposition that still finds widespread use in heavy-ion physics~\cite{Miller:2007ri}, primarily in the form of the Monte Carlo Glauber (MC-Glauber) model. It is motivated primarily by geometry: the two incoming nuclei are treated as collections of nucleons moving along straight-line trajectories, and the amount of matter produced in the collision is assumed to depend on the geometric overlap of the nuclei. 

In the optical Glauber model, each nucleus is described by a smooth density distribution $\rho_A(\mathbf{x})$, from which one defines the corresponding thickness function
\begin{align}
    T_{A,B}(\mathbf{x}_\perp) = \int dz\, \rho_{A,B}(\mathbf{x}_\perp,z).
\end{align}

The overlap of the two thickness functions then determines the transverse geometry of the collision at a given impact parameter. From this overlap, one can estimate quantities such as the density of participant nucleons and the density of binary nucleon-nucleon collisions.

In this analytic form, event-by-event fluctuations in the positions of individual nucleons are washed out. Realistic simulations seek to account for such fluctuations, necessitating a somewhat different approach. The MC-Glauber model instead places the two nuclei at a given impact parameter $b$, samples nucleon positions within each, and determines which pairs collide according to a probability that depends on their relative transverse distance and the inelastic nucleon-nucleon cross section $\sigma_{NN}^\text{inel}$. The resulting participant nucleons are then deposited according to Eq. \ref{eq:nucleon_sum} to construct the full profile.

The radial density of a spherical heavy nucleus is typically described by a Woods-Saxon distribution:
\begin{align}
    \rho(r) = \frac{\rho_0}{1 + \exp\left(\frac{r-R}{a}\right)}
\end{align}
where $\rho_0$ is the maximum density, $R$ is the nuclear radius, and $a$ is the surface thickness. This has the form of a sigmoidal function, where $R$ defines the distance at which the density drops off (i.e., the edge of the nucleus) and $a$ governs the smoothness of that drop-off (i.e., how sharply defined the edge of the nucleus is). 

In practice, $R$ and $a$ are determined by nuclear structure measurements for all the species of interest. Nucleon positions can then simply be sampled from the radial probability $P(r) \propto r^2 \rho(r)$, with angular orientations sampled isotropically. Neglecting nucleon substructure, one can write the density of each individual nucleon as a simple Gaussian
\begin{align}
    \rho_n(\mathbf{x}) = \frac{1}{(2\pi w^2)^{3/2}} \exp\left(-\frac{|\mathbf{x}|^2}{2w^2}\right)
    \label{eq:nucleon_gaussian}
\end{align}
for some nucleon width $w$. To match realistic multiplicity distributions, it is a common practice to multiply each nucleon density by a randomly sampled weight factor, such as from a gamma distribution with unit mean:
\begin{align}
    \gamma_i \sim f(\gamma) = \frac{k^k}{\Gamma(k)} \gamma^{k-1} e^{-k\gamma}.
\end{align}

The participant density of each nucleus in the center-of-mass frame then takes the form
\begin{align}
    \rho^{\text{part}}_{A,B}(\mathbf{x}) =
    \sum_{i=1}^{N_{\text{part}}^{A,B}}
    \gamma_i^{A,B}\,
    \rho_n\!\left(
        \mathbf{x}-\mathbf{x}_i^{A,B}
        \pm \frac{\mathbf{b}}{2}
    \right),
    \label{eq:nucleon_sum}
\end{align}
where $\gamma_i$ is the random weight, $\mathbf{x}_i$ is the position of the $i$th participant nucleon relative to its parent nucleon's center, and $\mathbf{b}$ is the impact parameter vector separating the centroids of the two nuclei; the $+$ ($-$) sign applies to nucleus $A$ ($B$), whose centers sit at $\mp \mathbf{b}/2$, and the full initial profile is the sum of the two nuclei's contributions.

\subsection{Gluon saturation-based models}

Recall from section \ref{section16} that the initial condition is believed to be gluon-saturated. Effective theories based on this principle describe the early regime of the medium as a color-glass condensate (CGC)~\cite{Gelis:2010nm} dominated by low-momentum gluons. There have been a number of successful models within this framework, most notably IP-Glasma~\cite{Schenke:2012wb}.

In the CGC picture, partons are organized by their Bjorken-$x$, the fraction of the parent nucleus's longitudinal momentum that they carry. The large-$x$ degrees of freedom in the incoming nuclei act as effective color sources for the softer, small-$x$ gluon fields that dominate the early dynamics. Because the gluon occupation numbers are large in the saturated regime, these fields can be treated approximately as classical color fields~\cite{McLerran:1993ni, McLerran:1993ka}. Immediately after the collision, the system is then described by a highly non-equilibrium state of longitudinal color fields known as the ``Glasma''.

In the IP-Glasma model, the impact-parameter-dependent saturation model (IP-Sat)~\cite{Kowalski:2003hm} is used to determine the fluctuating saturation scale and color charge structure of the incoming nuclei, after which the early-time Glasma fields are evolved using classical Yang-Mills dynamics.  When coupled to viscous hydrodynamics, IP-Glasma initial conditions have been shown to reproduce the measured anisotropic flow harmonics, simultaneously describing $v_2$ and $v_3$ in a way that earlier saturation-based initial conditions could not~\cite{Gale:2012rq}.

There have been a number of other saturation-based models. One notable example is the EKRT model~\cite{Eskola:1999fc, Niemi:2015qia}, which combines perturbative QCD minijet production with a saturation conjecture. This framework deposits energy entirely through minijets produced above a transverse momentum scale, fixed locally by the saturation conjecture; below this scale, nonlinear parton interactions are assumed to shut off further production. Although it is not a Glasma model in the same sense as IP-Glasma, it is similarly motivated by the expectation that perturbative gluon dynamics largely govern the early production of matter in heavy-ion collisions.

These models are appealing because of both their ability to reproduce experimental observables and their theoretical connections to perturbative QCD and gluon saturation physics. However, neither model --- nor any other in this category to date --- has been able to perfectly describe all observables simultaneously. Furthermore, both IP-Glasma and EKRT are computationally expensive and relatively inflexible for large-scale parameter studies. It is for these reasons that many multistage models prefer more flexible, parametric approaches. One of the most widely used examples of this is T\textsubscript{R}ENTo.

\subsection{\texorpdfstring{T\textsubscript{R}ENTo}{TRENTo}}

The initial condition model used in the rest of this work is the Reduced Thickness Event-by-event Nuclear Topology model, much more commonly known as T\textsubscript{R}ENTo~\cite{Moreland:2014oya}. This is a parametric model that does not assume any particular physical mechanism for entropy production in the initial state, and is designed to reproduce a broad range of features of the energy deposition while remaining flexible for data-driven calibration. For any two colliding nuclei, the steps of the model are as follows:

\begin{enumerate}[
    label=\arabic*.,
    leftmargin=2.2em,
    labelsep=0.6em,
    itemsep=0.8em,
    topsep=0.8em,
    parsep=0pt
]

    \item \textbf{Sample nucleon positions.}
    Within each nucleus, sample nucleon positions according to a Woods-Saxon distribution, or an alternative distribution for small or deformed nuclei, subject to a constraint that there is a minimum distance between any two nucleons $|\mathbf{x}_i - \mathbf{x}_j| > d_\text{min}$.

    \item \textbf{Determine inelastic collisions.}
    Determine which nucleons collide inelastically according to their positions and the inelastic nucleon-nucleon cross section $\sigma_{NN}^{\text{inel}}$.

    \item \textbf{Construct the participant densities.}
    For each nucleus, as with the Glauber model, assume each nucleon has a Gaussian density of the form in Eq.~\ref{eq:nucleon_gaussian} and determine the summed density of the participants using Eq.~\ref{eq:nucleon_sum}, where the weights $\gamma_i$ are taken from a gamma distribution with unit mean and a variance of $1/k$.

    \item \textbf{Construct the participant thickness functions.}
    For each nucleus, project onto the transverse plane by integrating over the beam axis:
    \begin{align}
        \tilde{T}_{A,B}(\mathbf{x}_\perp)
        = \int dz \rho^{\text{part}}_{A,B}(\mathbf{x}_\perp,z).
    \end{align}

    \item \textbf{Construct the reduced thickness.}
    Combine the participant thicknesses for the two nuclei as the reduced thickness, given by the generalized mean
    \begin{align}
        T_R(p; \tilde{T}_A, \tilde{T}_B)
        = \left(
            \frac{\tilde{T}_A^p + \tilde{T}_B^p}{2}
          \right)^{1/p},
    \end{align}
    where $p$ is a free parameter whose different values mimic different physical mechanisms for combining the contributions of the two nuclei. Up to some normalization constant, this reduced thickness is then presumed to be equivalent to either the energy or entropy density of the system.

\end{enumerate}

Note that in the infinite limits, $T_R = \max(\tilde{T}_A, \tilde{T}_B)$ as $p \to \infty$ and $T_R = \min(\tilde{T}_A, \tilde{T}_B)$ as $p \to -\infty$. Intermediate values of $p$ interpolate between familiar functional forms such as the arithmetic mean when $p = 1$, the geometric mean when $p = 0$, and the harmonic mean when $p = -1$.

One useful way to characterize the initial-state geometry is through its azimuthal anisotropy, quantified by the eccentricity harmonics $\varepsilon_n$. These measure the spatial deformation of the deposited energy density and are of central importance because, during the hydrodynamic evolution, pressure gradients convert this spatial anisotropy into the momentum-space anisotropy of the final-state particles --- the anisotropic flow harmonics $v_n$. The $n$th-order eccentricity is defined as
\begin{align}
    \varepsilon_n \, e^{i n \Phi_n} = -\frac{\int d^2x_\perp \, r^n \, e^{i n \phi} \, \varepsilon(\mathbf{x}_\perp)}{\int d^2x_\perp \, r^n \, \varepsilon(\mathbf{x}_\perp)},
    \label{eq:eccentricity}
\end{align}
where $\varepsilon(\mathbf{x}_\perp)$ is the transverse energy density and $r$ and $\phi$ are polar coordinates measured from the energy density centroid. The magnitude $\varepsilon_n$ sets the strength of the $n$th-order spatial anisotropy, while the phase $\Phi_n$ defines the corresponding participant-plane angle. The lowest harmonics dominate: the ellipticity $\varepsilon_2$ reflects the almond shape of the nuclear overlap region in non-central collisions, while the triangularity $\varepsilon_3$ arises almost entirely from event-by-event fluctuations in the nucleon positions.

\begin{figure}[!b]
  \vspace{12pt}
  \centering
  \includegraphics[width=\textwidth]{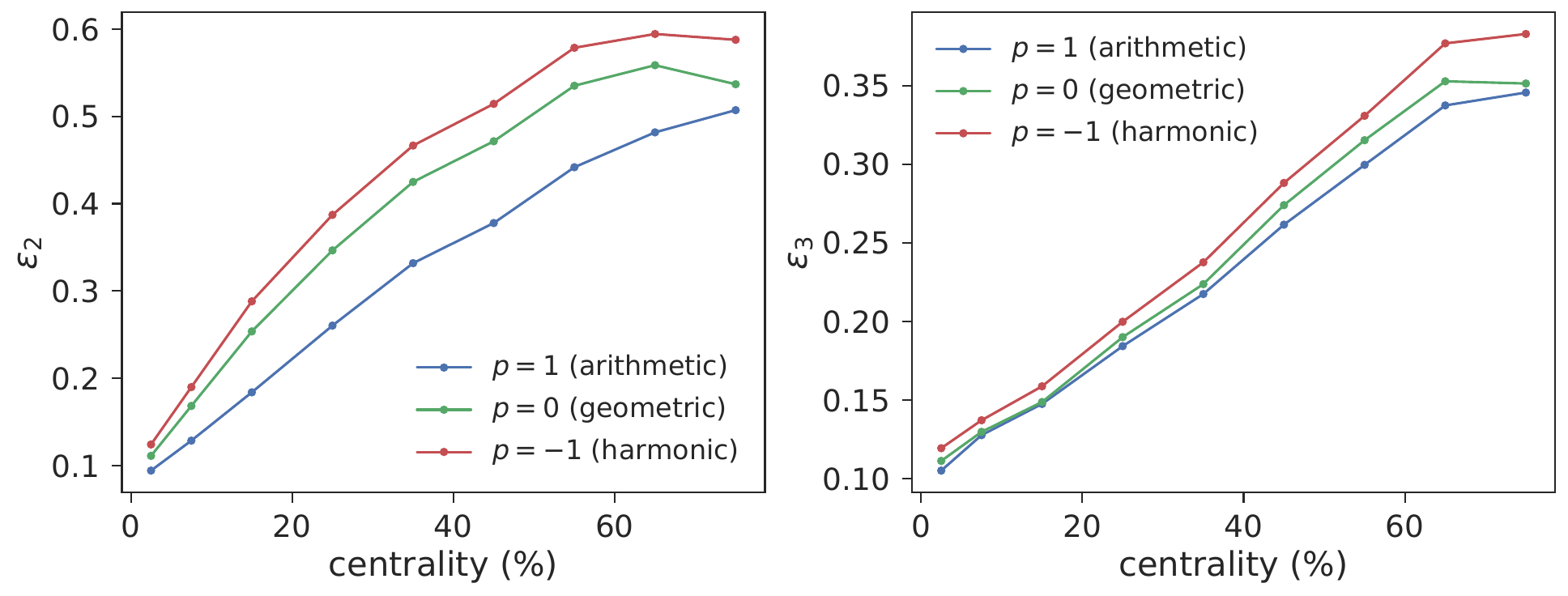}
  \caption[Initial-state eccentricities $\varepsilon_2$ and $\varepsilon_3$ versus centrality for three values of $p$]{Initial-state eccentricity harmonics $\varepsilon_2$ (left) and $\varepsilon_3$ (right) versus centrality for Au+Au collisions at $\sqrt{s_{NN}} = 200$~GeV, computed from T\textsubscript{R}ENTo events at three values of the reduced-thickness parameter $p$. Centrality is estimated from the impact parameter.}
  \label{fig:trento_eccentricity}
\end{figure}

The choice of $p$ carries a direct geometric consequence. Because the geometric and harmonic limits ($p \leq 0$) suppress regions where only one nucleus contributes, decreasing $p$ concentrates the deposited matter within the overlap region and enhances the eccentricity of the initial state. This is shown in Fig.~\ref{fig:trento_eccentricity}: the elliptic eccentricity $\varepsilon_2$ increases systematically as $p$ decreases, while the triangularity $\varepsilon_3$, driven primarily by event-by-event nucleon fluctuations, remains largely insensitive to $p$.

This scheme has several free parameters: the minimum nucleon-nucleon distance $d_\text{min}$, the nucleon width $w$, the gamma distribution shape parameter $k$, the reduced-thickness parameter $p$, and a final normalization constant. This intentionally flexible parameterization allows T\textsubscript{R}ENTo to mimic the entropy-deposition scaling of many other initial condition models, including the aforementioned MC-Glauber and EKRT models --- and, notably, IP-Glasma, whose scaling the geometric-mean limit ($p \approx 0$) was shown to reproduce~\cite{Moreland:2014oya}. This correspondence holds at the level of the average deposition scaling, however, not the full event-by-event geometry. T\textsubscript{R}ENTo cannot reproduce the detailed fluctuation structure of a model like IP-Glasma, whose sub-nucleonic color-charge fluctuations produce far spikier profiles than the smooth Gaussian nucleon ansatz.

\begin{figure}[!b]
  \vspace{12pt}
  \centering
  \includegraphics[width=\textwidth]{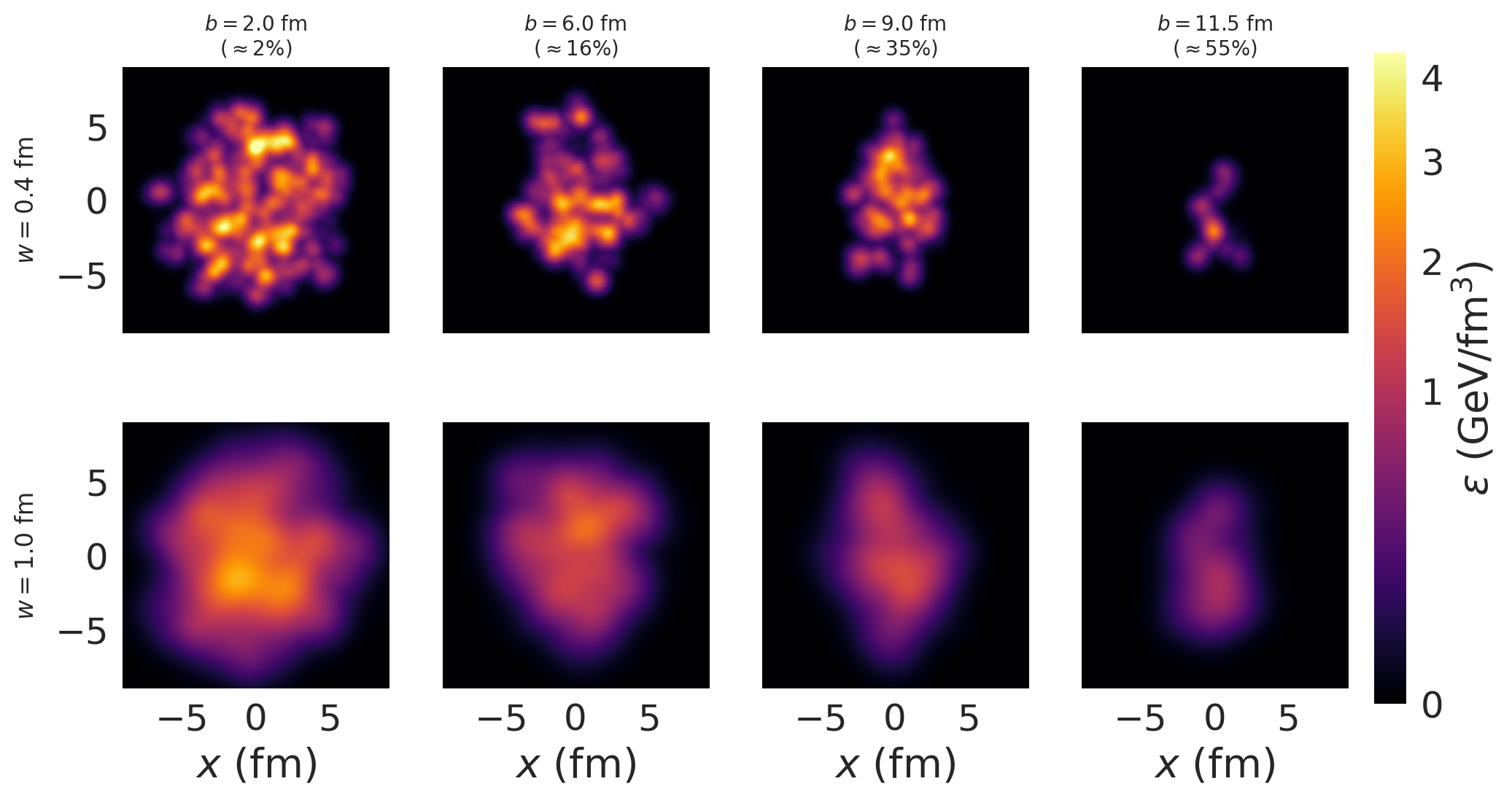}
  \caption[T\textsubscript{R}ENTo initial energy density profiles for two nucleon widths across a range of impact parameters]{T\textsubscript{R}ENTo initial energy density profiles for Au+Au collisions at $\sqrt{s_{NN}} = 200$~GeV, shown for two nucleon widths $w$ (rows) across a sequence of impact parameters (columns). Centralities are estimated from the impact parameter.}
  \label{fig:trento_width}
\end{figure}

The nucleon width $w$ sets the granularity of the deposited profile, as illustrated in Fig.~\ref{fig:trento_width}. A small width preserves the spiky structure of individual nucleon hotspots, revealing nucleon-scale event-by-event fluctuations, whereas a larger width smooths these features into a more continuous distribution. The same panels show the centrality dependence of the initial geometry: with increasing impact parameter, the overlap region shrinks and becomes progressively more elliptical.

\enlargethispage{\baselineskip}
The usefulness of T\textsubscript{R}ENTo lies in the fact that this simple generalized-mean ansatz can interpolate between several familiar initial condition scalings while remaining computationally cheap. This makes it particularly well suited for modern Bayesian calibration studies, where one needs to evaluate the full multistage model over many points in parameter space. For these reasons, T\textsubscript{R}ENTo will be the initial condition model used in the remainder of this work.

\section{Pre-equilibrium dynamics}
\label{section22}

As discussed in section \ref{section15}, the matter immediately after the collision is generally too far from local equilibrium for hydrodynamics to apply at arbitrarily early times. For this reason, a pre-equilibrium stage is often introduced as a bridge between the initial deposition of energy and the onset of hydrodynamics. Physically, this stage stands in for the evolution by which the far-from-equilibrium initial state hydrodynamizes, a process whose microscopic dynamics in QCD remain incompletely understood. In a multistage model, the practical role of this stage is to propagate the early system to a switching time $\tau_0$ and provide hydrodynamic initial conditions in the form of the local stress-energy tensor. During this phase, initial collective flow can develop, and the spatial profile and dissipative components of the stress-energy tensor can be reshaped before being fed into hydrodynamics.

In the simplest treatments, the pre-equilibrium period is neglected or absorbed into an effective hydrodynamic starting time. In this case, the model does not claim that no dynamics occur before $\tau_0$; rather, those dynamics are not treated as a distinct stage of the simulation. Their effects are instead folded into the chosen initial condition, the overall normalization, the initialization of flow and viscous stresses, and the choice of hydrodynamic starting time. This approach is less physically explicit, but it has the practical advantage of reducing the number of separate model components and free parameters.

One common approach is free streaming~\cite{Broniowski:2008qk}, in which the initial condition is allowed to evolve as a non-interacting ensemble of partons for some tunable free-streaming time $\tau_{\text{fs}}$. In some sense, this is counterintuitive: by the onset of hydrodynamics, the medium behaves as a strongly coupled fluid. But during free streaming, the opposite assumption is made: that all interactions vanish, only to be abruptly switched on at $\tau_{\text{fs}}$. Nevertheless, free streaming remains useful as a simple effective model of the pre-equilibrium stage, since it generates early flow and broadens the initial energy density profile without introducing the full complexity of a realistic microscopic description.

Kinetic theory provides a more realistic approach to modeling the system's approach to hydrodynamic behavior, as discussed in section \ref{section15}. Rather than assuming that the medium free streams without interactions, kinetic theory evolves the phase-space distributions of the underlying particles while explicitly accounting for collisions among them~\cite{Arnold:2002zm}. In this way, it can describe the gradual emergence of collective behavior and provide a more physically motivated bridge between the initial condition and the onset of hydrodynamics. Frameworks such as KoMPoST implement this idea efficiently by propagating the system using linear response around far-from-equilibrium backgrounds~\cite{Kurkela:2018wud, Kurkela:2018vqr}.

The primary outputs of the pre-equilibrium stage are the hydrodynamic fields at the switching time: the energy density, flow velocity, and dissipative stresses. These quantities are also affected by the initial condition itself, which means that the pre-equilibrium dynamics are, to some extent, degenerate with choices made in the initial state. For example, different assumptions about the duration or dynamics of the pre-equilibrium stage can change the amount of radial flow, the smoothing of the initial profile, and the eccentricities that later drive anisotropic flow. These effects can sometimes be compensated for by retuning parameters such as the initial normalization, nucleon width, or hydrodynamic starting time.

In this work, there is no distinct pre-equilibrium stage: hydrodynamics is initialized directly from a T\textsubscript{R}ENTo initial condition at $\tau_0 = 0.6$ fm/$c$. The effects of chemical pre-equilibrium will instead be incorporated into the initial conditions for hydrodynamics, through the framework detailed in chapter \ref{chapter3}.
\Needspace{4\baselineskip}
\section{Viscous relativistic hydrodynamics}
\label{section23}

Once the medium is sufficiently close to local thermodynamic equilibrium, its subsequent evolution can be described as that of a relativistic fluid. This is the primary means by which the evolution of the quark-gluon plasma is modeled directly, with the applicability of hydrodynamics itself having long been taken as a key QGP signature~\cite{Heinz:2013th}.

\subsection{Ideal hydrodynamics}

Relativistic hydrodynamics is fundamentally built on the conservation of energy and momentum together with a separation of scales that allows the microscopic dynamics to be smoothed out and replaced by continuous fields. In particular, hydrodynamics tends to apply when the mean free path of the interacting particles is much smaller than the characteristic length scale of the system --- in this case, the mean free paths of partons and the size of the QGP fireball, respectively. This is heuristically represented by the condition that the Knudsen number \(\text{Kn}\), given by the ratio of the mean free path $\lambda$ and length scale $L$, is much less than one:
\begin{align}
    \text{Kn} = \frac{\lambda}{L} \ll 1.
\end{align}

The primary conserved quantity in relativistic hydrodynamics is the stress-energy tensor, $T^{\mu\nu}$, whose conservation is expressed by
\begin{align}
    \partial_\mu T^{\mu\nu} = 0.
    \label{eq:dmuTmunu}
\end{align}

In heavy-ion collisions, $T^{\mu\nu}$ is typically decomposed into an ideal part plus dissipative corrections associated with shear and bulk viscosities. In the absence of any viscosities, we are in the ideal limit, where
\begin{align}
    T^{\mu\nu}_{\text{ideal}} = \varepsilon u^\mu u^\nu - P (g^{\mu\nu} - u^\mu u^\nu),
\end{align}
for the energy density $\varepsilon$, flow velocity $u^\mu$, pressure $P$, and metric tensor $g^{\mu\nu}$. Throughout this work, I will use the mostly-minus convention $g^{\mu\nu} = \mathrm{diag}(1,-1,-1,-1)$. Note that the pressure has, to this point, not been mentioned as a hydrodynamic field. This is because it is not an independent quantity; the pressure is given as a function of the energy density as an equation of state, which is an input to hydrodynamics. This will be explained in section~\ref{subsection233}, but for now it suffices to say that, at vanishing chemical potential, we have some function $P(\varepsilon)$. The latter term in $T^{\mu\nu}_{\text{ideal}}$ is often written in terms of a projection operator onto the space orthogonal to the flow velocity $\Delta^{\mu \nu} = g^{\mu \nu} - u^\mu u^\nu$, yielding
\begin{align}
    T^{\mu\nu}_{\text{ideal}} = \varepsilon u^\mu u^\nu - P \Delta^{\mu \nu}.
\end{align}

Each component of the stress-energy tensor can be understood as the flux of the four-momentum component $p^\mu$ across a surface of constant coordinate $x^\nu$. This by itself may seem quite abstract, so let us first consider $T^{\mu\nu}_{\text{ideal}}$ in the local rest frame, when $u^\mu = (1,0,0,0)$. In this case, we have
\begin{align}
    T^{\mu\nu}_{\text{ideal,LRF}} =  
    \begin{pmatrix}
        \varepsilon & 0 & 0 & 0 \\
        0 & P & 0 & 0 \\
        0 & 0 & P & 0 \\
        0 & 0 & 0 & P \\
    \end{pmatrix}.
\end{align}
In the fluid's own rest frame, the state of the ideal fluid is thus entirely described by its energy density and some isotropic pressure. In a boosted frame, however, all of the off-diagonal terms can potentially be nonzero. The components $T^{0i}$ (where $i \neq 0$) correspond to the energy flux in the $i$-direction or, equivalently, the $i$ component of the momentum density. The off-diagonal spatial components $T^{ij}$ correspond to fluxes of momentum between different spatial directions, which can arise from the bulk motion of the fluid. Fig.~\ref{fig:Tmunu} shows the overall structure of the stress-energy tensor.

\begin{figure}[!t]
  \definecolor{Cenergy}{RGB}{233,50,50}   
  \definecolor{Cflux}{RGB}{255,255,140}    
  \definecolor{Cpress}{RGB}{60,150,165}    
  \definecolor{Cshear}{RGB}{170,150,200}   
  \centering
  \newsavebox{\figTmunu}
  \begin{lrbox}{\figTmunu}%
  \begin{tikzpicture}[font=\small,
          cell/.style={minimum size=11mm, inner sep=1pt, anchor=center,
                       draw=black!55, line width=0.5pt},
          swatch/.style={draw=black!55, line width=0.4pt, minimum width=4.2mm,
                         minimum height=4.2mm, inner sep=0pt}]
        \matrix (T) [matrix of nodes, nodes={cell},
            column sep=-0.4pt, row sep=-0.4pt,]
        {
          $T^{00}$ & $T^{01}$ & $T^{02}$ & $T^{03}$ \\
          $T^{10}$ & $T^{11}$ & $T^{12}$ & $T^{13}$ \\
          $T^{20}$ & $T^{21}$ & $T^{22}$ & $T^{23}$ \\
          $T^{30}$ & $T^{31}$ & $T^{32}$ & $T^{33}$ \\
        };
        \begin{scope}[on background layer]
          \fill[Cenergy] (T-1-1.north west) rectangle (T-1-1.south east);
          \fill[Cflux]   (T-1-2.north west) rectangle (T-1-4.south east);
          \fill[Cflux]   (T-2-1.north west) rectangle (T-4-1.south east);
          \foreach \r/\c in {2/2,3/3,4/4}{
            \fill[Cpress] (T-\r-\c.north west) rectangle (T-\r-\c.south east);}
          \foreach \r/\c in {2/3,2/4,3/2,3/4,4/2,4/3}{
            \fill[Cshear] (T-\r-\c.north west) rectangle (T-\r-\c.south east);}
        \end{scope}
        \foreach \c/\lab in {1/0,2/1,3/2,4/3}{
            \node[font=\scriptsize, black!70] at (T-1-\c |- T.north) [above=1pt] {$\lab$};}
        \node[font=\scriptsize] at (T.north) [above=4.5mm] {$\nu$};
        \foreach \r/\lab in {1/0,2/1,3/2,4/3}{
            \node[font=\scriptsize, black!70] at (T-\r-1 -| T.west) [left=1pt] {$\lab$};}
        \node[font=\scriptsize] at (T.west) [left=4.5mm] {$\mu$};
        \draw[black!75, line width=0.9pt]
            ($(T-1-1.north east)+(0,1pt)$) -- ($(T-4-1.south east)-(0,1pt)$);
        \draw[black!75, line width=0.9pt]
            ($(T-1-1.south west)-(1pt,0)$) -- ($(T-1-4.south east)+(1pt,0)$);
        \coordinate (L) at ($(T.east)+(12mm,15mm)$);
        \node[swatch, fill=Cenergy] (e0) at (L) {};
        \node[anchor=west, font=\scriptsize] at (e0.east) {$T^{00}$: energy density};
        \node[swatch, fill=Cflux, below=2.2mm of e0.south west, anchor=north west] (f0) {};
        \node[anchor=west, font=\scriptsize] at (f0.east) {$T^{0i}\!=\!T^{i0}$: energy flux / momentum\ density};
        \node[swatch, fill=Cpress, below=2.2mm of f0.south west, anchor=north west] (p0) {};
        \node[anchor=west, font=\scriptsize] at (p0.east) {$T^{ii}$: normal stresses / directional pressures};
        \node[swatch, fill=Cshear, below=2.2mm of p0.south west, anchor=north west] (s0) {};
        \node[anchor=west, font=\scriptsize] at (s0.east) {$T^{ij}, i\neq j$: shear stress};
      \end{tikzpicture}%
  \end{lrbox}%
  \resizebox{\textwidth}{!}{\usebox{\figTmunu}}
  \caption[Components of the stress-energy tensor $T^{\mu\nu}$]{Components of the stress-energy tensor $T^{\mu\nu}$ and their physical interpretation.}
  \label{fig:Tmunu}
\end{figure}

In ordinary relativistic hydrodynamics, the stress-energy tensor is taken to be symmetric. More generally, when spin degrees of freedom are treated explicitly, total angular momentum conservation involves both the stress-energy tensor and a spin current, and \(T^{\mu\nu}\) may acquire an antisymmetric part. This is the setting of spin hydrodynamics, which has become a useful tool for studying the effects of vorticity within the QGP and their observable consequences for hadron polarization~\cite{Florkowski:2017ruc}. In this work, however, I will exclusively stick to spinless hydrodynamics and neglect such effects.

We have not yet defined the flow velocity $u^\mu$, and there is some inherent ambiguity in how to do so. This is known as the choice of hydrodynamic ``frame''. I will use the conventional Landau frame, in which flow velocity represents the flow of energy. In other words, $u^\mu$ is the timelike eigenvector of the stress-energy tensor, with eigenvalue $\varepsilon$. This is given by
\begin{align}
    T^{\mu \nu} u_\nu = \varepsilon u^\mu.
\end{align}

The choice of Landau frame was already implicit in the absence of energy flow in the local rest frame; in general, $T^{0i}$ need not vanish in other frames. Another common choice of hydrodynamic frame is the Eckart frame~\cite{Eckart:1940te}, in which $u^\mu$ defines the flow of a conserved charge such as baryon number. In the Eckart frame, the diffusion current of that charge vanishes in the local rest frame, which can be convenient in studies focused on a single conserved charge, such as models of systems at large baryon density. 

The flow velocity $u^\mu$ is a normalized four-vector, and so is also subject to the condition
\begin{align}
    u^\mu u_\mu = 1.
\end{align}
Thus, $u^\mu$ has three independent components. Together with the energy density $\varepsilon$, this gives four independent hydrodynamic fields in ideal hydrodynamics.  With four independent equations given by Eq.~\ref{eq:dmuTmunu}, we have a closed set of differential equations that can be solved.

\subsection{Viscous corrections}

While many features of the medium's expansion are captured by ideal hydrodynamics, viscous effects are necessary for a more realistic description. The full stress-energy tensor as it is typically used in viscous relativistic hydrodynamics becomes 
\begin{align}
    T^{\mu\nu} = \varepsilon u^\mu u^\nu - (P + \Pi) \Delta^{\mu \nu} + \pi^{\mu \nu},
\end{align}
where $\Pi$ is the bulk pressure and $\pi^{\mu \nu}$ is the shear stress tensor. These are additional hydrodynamic fields whose evolution is controlled by transport coefficients of the medium, most prominently the bulk viscosity, $\zeta$, and the shear viscosity, $\eta$. As was discussed in more detail in section~\ref{section14}, the bulk pressure modifies the effective isotropic pressure, with larger $\zeta/s$ generally tending to suppress radial expansion, while larger $\eta/s$ tends to damp the development of anisotropic flow.

Naively generalizing the conventional Navier-Stokes equations of non-relativistic fluid dynamics yields the relations
\begin{align}
    \Pi = - \zeta \theta, \quad \pi^{\mu\nu} = 2 \eta \sigma^{\mu \nu},
\end{align}
where $\theta = \partial_\mu u^\mu$ is the expansion rate of the fluid and $\sigma^{\mu \nu}$ is the velocity shear tensor, given by
\nopagebreak[4]
\begin{align}
    \sigma^{\mu \nu} = \nabla^{\langle \mu} u^{\nu \rangle} = \frac{1}{2}(\nabla^\mu u^\nu + \nabla^\nu u^\mu) - \frac{1}{3}\theta \Delta^{\mu \nu},
\end{align}
with $\nabla^\mu = \Delta^{\mu \nu} \partial_\nu$, and where the angle brackets denote the symmetric, traceless part transverse to $u^\mu$:
\begin{align} 
A^{\langle \mu \nu \rangle} \equiv \left[\frac{1}{2}\left(\Delta^\mu_{\ \alpha}\Delta^\nu_{\ \beta}+\Delta^\mu_{\ \beta}\Delta^\nu_{\ \alpha}\right)-\frac{1}{3}\Delta^{\mu\nu}\Delta_{\alpha\beta}\right]A^{\alpha\beta}. 
\end{align}

However, these relations, corresponding to relativistic first-order viscous hydrodynamics, lead to problems with both stability and causality in their Navier-Stokes form. Stability in the hydrodynamic context means the condition that small perturbations do not grow uncontrollably. Distinct but related, causality is the condition that signal propagation is never faster than light. Both are formal requirements for physical relativistic hydrodynamic evolution, although in numerical simulations, small or localized violations of these criteria may not lead to visible instabilities. While recent work has shown that first-order viscous hydrodynamics can be made causal and stable with appropriate choices of hydrodynamic frame and transport coefficients~\cite{Bemfica:2017wps, Bemfica:2019knx, Kovtun:2019hdm}, the more conventional approach in heavy-ion phenomenology has been to use second-order hydrodynamics by introducing relaxation equations for the shear stress tensor and bulk pressure.

Most modern hydrodynamic simulations use one of a family of Israel-Stewart-type formulations~\cite{Israel:1979wp}, which were originally introduced to address the stability and causality problems of relativistic Navier-Stokes theory. The formulation used in MUSIC, the hydrodynamics code used in this work~\cite{Schenke:2010nt, Schenke:2010rr}, has the equations of motion 
\begin{align}
    \tau_\pi \dot{\pi}^{\langle \mu \nu \rangle} + \pi^{\mu \nu}
    &= 2 \eta \sigma^{\mu \nu}
    + 2 \tau_\pi \pi_\alpha^{\langle\mu} \omega^{\nu \rangle \alpha}
    - \delta_{\pi \pi} \pi^{\mu \nu} \theta
    \notag \\
    &\quad
    + \varphi_7 \pi_\alpha^{\langle\mu} \pi^{\nu \rangle \alpha}
    - \tau_{\pi \pi} \pi_\alpha^{\langle\mu} \sigma^{\nu \rangle \alpha}
    + \lambda_{\pi \Pi} \Pi \sigma^{\mu \nu}
    + \varphi_6 \Pi \pi^{\mu \nu}
    \notag \\
    \tau_{\Pi} \dot{\Pi} + \Pi
    &= -\zeta \theta
    - \delta_{\Pi \Pi} \Pi \theta
    + \varphi_1 \Pi^2
    + \lambda_{\Pi \pi} \pi^{\mu \nu} \sigma_{\mu \nu}
    + \varphi_3 \pi^{\mu \nu} \pi_{\mu \nu},
\end{align}
where $\omega^{\mu\nu} = \tfrac{1}{2}\left(\nabla^{\mu}u^{\nu} - \nabla^{\nu}u^{\mu}\right)$ is the fluid vorticity tensor, the overdot denotes the comoving derivative $\dot{A} \equiv u^{\lambda}\partial_{\lambda}A$, and the eleven new coefficients are second-order transport coefficients. In the implementation used here, most of these coefficients are determined from the first-order transport coefficients, $\eta$ and $\zeta$, together with thermodynamic quantities given by the equation of state, while the remaining coefficients are set to zero.

The essential feature of these equations is that $\Pi$ and $\pi^{\mu\nu}$ no longer instantaneously take their Navier-Stokes values, but instead relax toward them over finite timescales $\tau_\Pi$ and $\tau_\pi$. Modern heavy-ion simulations therefore evolve not only the local energy density and flow velocity, but also the dissipative stresses themselves.

\begin{figure}[!t]
  \centering
  \includegraphics[width=\textwidth]{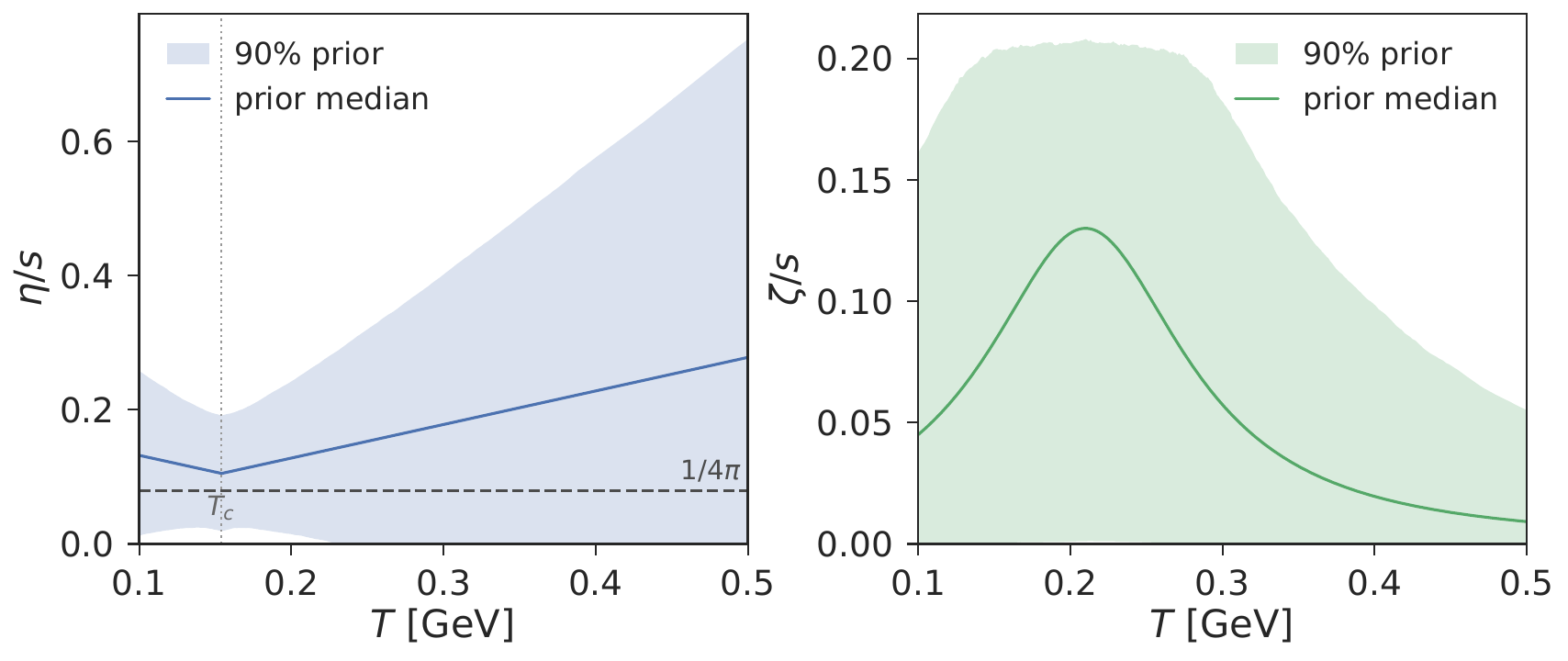}
  \caption[Parameterizations of the specific shear and bulk viscosities versus temperature]{Parameterizations of the specific shear viscosity $\eta/s$ (left) and specific bulk viscosity $\zeta/s$ (right) as functions of temperature. Solid curves use the prior medians; shaded bands span the central $90$\% of the prior.}
  \label{fig:viscosity}
\end{figure}

The shear and bulk viscosities are not precisely known from first principles, and in practice their temperature dependence is parameterized, with selected parameters constrained together with the other model parameters, as is done in chapter~\ref{chapter5}. For the parameterizations we adopt forms based on those introduced by the JETSCAPE collaboration~\cite{JETSCAPE:2020shq, JETSCAPE:2020mzn}. The specific shear viscosity $\eta/s$ is taken to be piecewise linear in temperature, with a kink at the critical temperature $T_c$ and independent slopes above and below it. The specific bulk viscosity $\zeta/s$ is taken to be a peaked, possibly skewed Cauchy distribution centered near the transition. These parameterized forms are shown according to the design ranges used in chapter~\ref{chapter5} in Fig.~\ref{fig:viscosity}.

A frequently quoted benchmark for the shear viscosity comes from the anti--de Sitter/conformal field theory (AdS/CFT) correspondence, a conjectured duality from string theory relating a strongly coupled gauge theory to a weakly coupled gravitational theory in one higher dimension~\cite{Maldacena:1997re}. The duality makes otherwise intractable strong-coupling transport calculations feasible, and for a broad class of strongly interacting field theories possessing such a gravitational dual it yields a lower bound on the specific shear viscosity, $\eta/s = 1/4\pi$~\cite{Kovtun:2004de}. However, QCD is neither conformal nor supersymmetric and has no known gravitational dual, so whether this bound applies to QCD remains unclear. I include $\eta/s = 1/4\pi$ in the figure as a useful heuristic for the characteristic scale of the shear viscosity in a strongly coupled fluid rather than a constraint imposed on the model --- consistent with the prior adopted in this work, which permits smaller values. 

\subsection{Equation of state}
\label{subsection233}

The hydrodynamic equations must be closed by an equation of state (EoS), which relates thermodynamic quantities and encodes the equilibrium properties of matter. In particular, at vanishing chemical potential, the EoS gives the relationship $P(\varepsilon)$ between the local pressure and energy density, and therefore controls how pressure gradients drive the expansion of the fluid.

In the high-temperature regime relevant for the deconfined QGP, the equilibrium thermodynamics of QCD can be calculated nonperturbatively using lattice QCD~\cite{Borsanyi:2013bia, HotQCD:2014kol}. In brief, lattice QCD evaluates the QCD path integral on a discrete Euclidean spacetime grid, allowing thermodynamic quantities to be computed from first principles. Temperature enters through the finite extent of the Euclidean time direction: for a lattice with spacing $a$ and $N_\tau$ sites in the temporal direction, the temperature is given by
\begin{align}
    T = \frac{1}{aN_\tau}.
\end{align}

The spatial volume is similarly set by the spatial lattice size, $V=(aN_s)^3$, for $N_s$ sites in each spatial direction. The calculation is repeated with different lattice spacings and temporal lattice sizes to determine the trend with respect to $a$, which can then be extrapolated to the continuum limit as $a \rightarrow 0$.

A central quantity determined in lattice calculations of the equation of state is the trace of the energy-momentum tensor, typically called the trace anomaly or interaction measure, $\varepsilon - 3P$. This measures the deviation of the equation of state from the conformal limit, in which $\varepsilon = 3P$ and the stress-energy tensor is traceless. A conformal equation of state has no intrinsic scale, so in $3+1$ dimensions the pressure and energy density scale as
\begin{align}
    P \propto T^4, \qquad \varepsilon \propto T^4,
\end{align}
with the additional constraint $\varepsilon=3P$. QCD is not conformal, however, because it contains physical scales associated with quark masses, confinement, and the running of the coupling. The trace anomaly therefore quantifies how strongly the thermodynamics of QCD departs from this scale-invariant limit.

In the high-temperature limit, asymptotic freedom weakens the interactions among quarks and gluons, and the medium approaches a gas of nearly free partons. Its thermodynamics then tend toward the Stefan-Boltzmann limit of an ideal ultrarelativistic gas, in which the temperature-scaled quantities $\varepsilon/T^4$, $P/T^4$, and $s/T^3$ approach constants fixed solely by the number of bosonic and fermionic degrees of freedom. For a gas of gluons and $N_f$ massless quark flavors,
\begin{align}
  \frac{P_\text{SB}}{T^4}
    = \frac{\pi^2}{90}\left[\, 2\,(N_c^2 - 1) + \frac{7}{2}\,N_c N_f \,\right],
  \label{eq:sb}
\end{align}
which for $N_c = 3$ colors and $N_f = 3$ flavors evaluates to $95\pi^2/180$. This limit is itself conformal, so $\varepsilon = 3P$ and the trace anomaly vanishes.

The trace anomaly can be calculated on the lattice as a function of the QCD partition function $Z$, lattice spacing $a$, and lattice volume $V$: 
\begin{align}
    \varepsilon - 3P = -\frac{T}{V}\frac{d \log Z}{d \log a}.
\end{align}

Once the trace anomaly is known, the pressure can be determined using the relation
\begin{align}
    \frac{\varepsilon - 3P}{T^4} = T \frac{d}{dT} \left( \frac{P}{T^4} \right)
\end{align}
and integrating over temperature
\begin{align}
    \frac{P}{T^4} = \frac{P_0}{T_0^4} + \int_{T_0}^T dT' \frac{(\varepsilon - 3P)}{T'^5},
\end{align}
with respect to some reference pressure $P_0$ at $T_0$. The energy density as a function of temperature can then be calculated by simply summing $3P$ with the trace anomaly, and the entropy density $s$ is given by the derivative of the pressure with respect to temperature,
\begin{align}
    s = \frac{dP}{dT}.
\end{align}

The reference pressure is typically taken from a hadron resonance gas (HRG) calculation~\cite{Huovinen:2009yb} at a sufficiently low reference temperature $T_0$, where QCD is in the confined regime and the low-temperature thermodynamics are well described in terms of hadronic degrees of freedom. In the HRG model, the relevant states are hadrons and resonances, where resonances are short-lived unstable hadronic states that appear as peaks in scattering cross sections and eventually decay into lighter hadrons. The thermodynamics is then approximated by a gas of these species, often treated as non-interacting. This provides a natural low-temperature baseline for the lattice equation of state and connects smoothly to the hadronic degrees of freedom that later appear in particlization and hadronic transport.

The hadron resonance gas energy density and pressure are calculated from hadronic distribution functions $f_i(p)$ as:
\begin{equation}
\begin{aligned}
    \varepsilon &= \sum_i g_i \int \frac{d^3 p}{(2\pi)^3} E_p f_i(p), \\
    P &= \sum_i g_i \int \frac{d^3 p}{(2\pi)^3} \frac{p^2}{3E_p} f_i(p),
\end{aligned}
\end{equation}
where $i$ denotes the hadron species, $g_i$ and $f_i$ are the respective degeneracy and distribution function for each species, and $E_p = \sqrt{p^2 + m_i^2}$ is the individual hadron energy given a mass of $m_i$ and momentum $p$. In equilibrium, the chemical potentials vanish and the distributions reduce to the standard quantum-statistical forms, with bosonic mesons following Bose-Einstein statistics and fermionic baryons following Fermi-Dirac statistics:
\begin{align}
    f_i(p) = \frac{1}{\exp{(p \cdot u/T)}\pm 1},
    \label{eq:bose_fermi}
\end{align}
where the $+$ describes fermions and the $-$ describes bosons.

Because QCD undergoes a crossover and not a true phase transition at zero chemical potential~\cite{Aoki:2006we}, the lattice QCD and HRG descriptions match smoothly at intermediate temperatures. Modern hydrodynamic simulations predominantly use such an interpolated EoS. 

One particularly important quantity derived from the EoS is the speed of sound, $c_s^2 = dP/d\varepsilon$. This characterizes the stiffness of the medium and is typically minimized in the transition region. Note that in the conformal limit, $c_s^2 = 1/3$. In hydrodynamic evolution, this matters because pressure gradients provide the acceleration of the fluid. A smaller speed of sound in the crossover region means that pressure responds less strongly to changes in energy density, softening the hydrodynamic expansion relative to the conformal limit.

\begin{figure}[!htbp]
  \centering
  \includegraphics[width=0.9\textwidth]{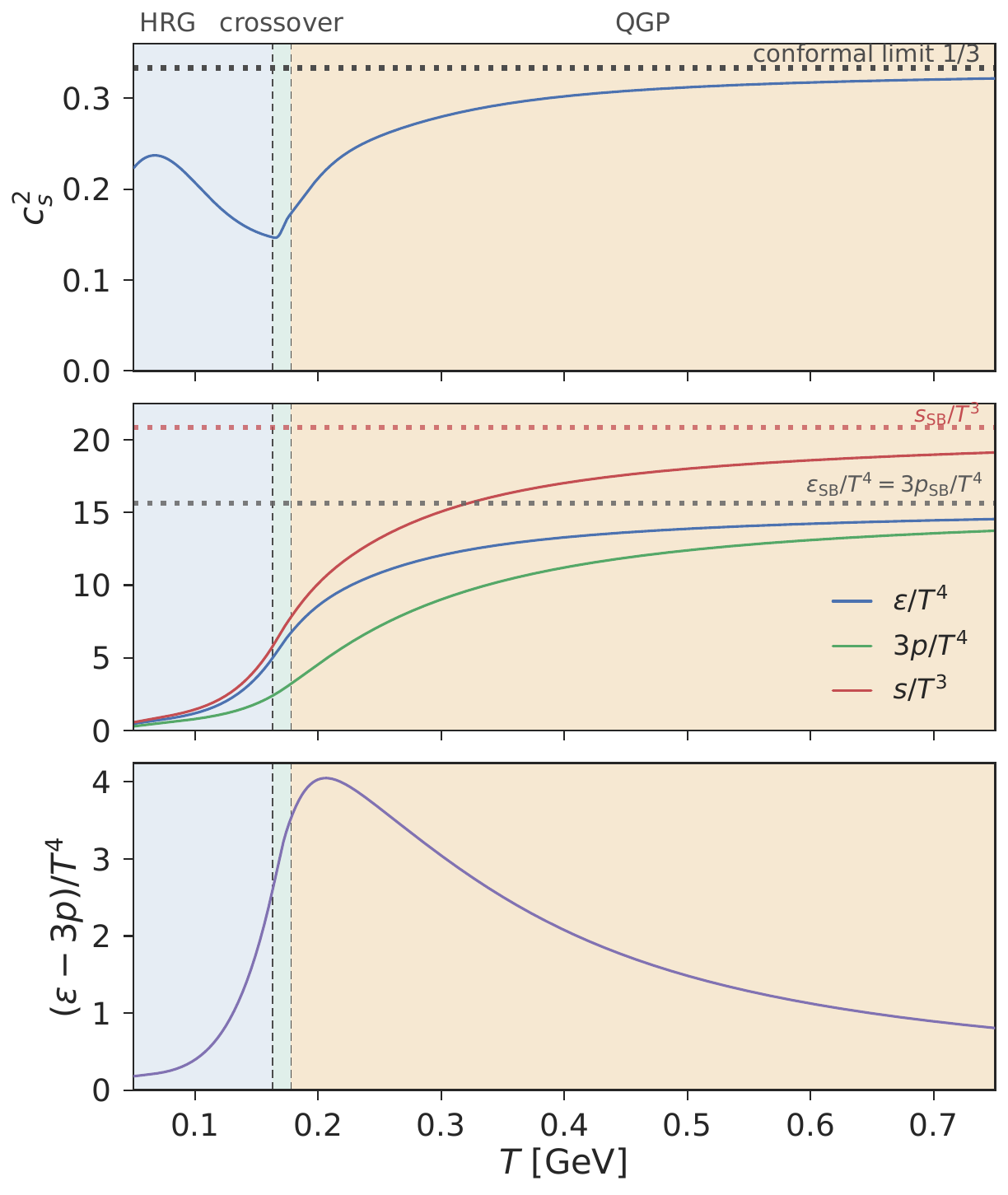}
  \caption[Equilibrium equation of state used as the hydrodynamic baseline]{Equilibrium equation of state ($\mu = 0$, unit fugacities) used as the hydrodynamic baseline in this work. \emph{Top:} speed of sound $c_s^2 = dP/d\varepsilon$, which softens to a minimum in the crossover region and approaches the conformal value $1/3$ at high temperature. \emph{Middle:} the energy density, pressure, and entropy density scaled by temperature, $\varepsilon/T^4$, $3P/T^4$, and $s/T^3$, rising across the crossover toward their Stefan-Boltzmann limits (dotted). \emph{Bottom:} trace anomaly $(\varepsilon - 3P)/T^4$, peaking just above the crossover. The tabulated equation of state connects a hadron resonance gas at low temperature~\cite{Huovinen:2009yb} to the HotQCD lattice equation of state at high temperature~\cite{HotQCD:2014kol} through an interpolating window; the shaded bands mark the hadron resonance gas, crossover, and lattice regions.}
  \label{fig:eos}
\end{figure}

These features are summarized in Fig.~\ref{fig:eos}, which shows the equilibrium equation of state used in this work. The trace anomaly $(\varepsilon - 3P)/T^4$ rises to a peak just above the crossover and falls toward zero at high temperature; the temperature-scaled thermodynamic quantities climb across the crossover toward their Stefan-Boltzmann limits; and the speed of sound develops its characteristic minimum in the transition region before approaching the conformal value $1/3$. The shaded bands indicate the hadron resonance gas, interpolation, and lattice regions of the construction. This equilibrium, zero chemical potential equation of state serves as the baseline that the flavor-dependent construction of chapter~\ref{chapter3} modifies.

The discussion thus far has neglected the effects of chemical potential on the equation of state. This is because lattice QCD has a limited regime of validity. At finite baryon chemical potential, lattice QCD is complicated by the sign problem, so the EoS must be extended using methods such as Taylor expansions, analytic continuation, or phenomenological modeling. Even then, such methods become less controlled as one moves farther from $\mu_B=0$. For example, Taylor-expansion-based approaches encounter convergence limitations at sufficiently large $\mu_B/T$, while newer expansion or resummation schemes can extend the range to roughly $\mu_B/T \sim\!3$--$3.5$~\cite{Borsanyi:2021sxv}.

In general, however, for systems with nonzero chemical potentials, the EoS must relate not only $\varepsilon$, $P$, $s$, and $T$, but also the relevant chemical potentials and conserved charge densities. One of the main thrusts of the present work is to extend the typical EoS to encompass systems with finite flavor chemical potentials, as described in chapter~\ref{chapter3}.

Another assumption here has been that the EoS is uniform throughout the medium for the entire duration of hydrodynamics. This fixed-EoS approach is standard in heavy-ion collision simulations, but it must be understood as a modeling assumption rather than a consequence of the hydrodynamic framework itself. More recent work has begun exploring families of possible equations of state, allowing the EoS to be varied and constrained alongside other model parameters~\cite{Pratt:2015zsa}.

\subsection{Hydrodynamic modeling in practice}

Hydrodynamics is only well justified in regions where gradients are not too large and the matter remains sufficiently dense and strongly interacting. Near the dilute edges of the medium, or at very early or late times, the separation of scales underlying hydrodynamics can begin to break down. This limitation has both physical and numerical aspects. Physically, once a system becomes too dilute or rapidly varying, a description in terms of smooth local fluid fields is no longer well justified. This is one of the reasons hydrodynamics must eventually be replaced by a microscopic hadronic description, as discussed in the following sections. Numerically, hydrodynamic simulations often still evolve cells near the edge of this regime, but overextending the model can lead to instabilities in the numerical solver. These practical limitations are closely related to the earlier discussion of stability and causality: large gradients and large dissipative corrections can push the hydrodynamic equations outside the regime where their formal properties and numerical evolution are well controlled.

As discussed in section~\ref{section22}, hydrodynamic modeling also requires choices for the initial flow field and dissipative stresses at the hydrodynamic starting time. A realistic pre-equilibrium calculation can supply these quantities through the full stress-energy tensor, while simplified frameworks often initialize hydrodynamics directly from an energy or entropy density profile with additional assumptions about $u^\mu$, $\pi^{\mu\nu}$, and $\Pi$. In particular, a common choice, which I also make in this work, is to initialize the transverse flow velocity to zero. These choices can affect the subsequent evolution, particularly because the early-time expansion is characterized by large gradients and potentially large dissipative corrections.

In addition to these physical initialization choices, hydrodynamic simulations also differ in their treatment of the longitudinal direction. In principle, one would like to always evolve the fluid in full $3+1$ dimensions: the two transverse spatial directions, spacetime rapidity, and proper time. There are programs that do so, but it can quickly become computationally expensive due to the number of cells and timesteps needed to cover the full simulation at sufficient resolution. Because of this, along with many observables of interest being measured near midrapidity, it is common to instead use a $(2+1)$-dimensional calculation by assuming longitudinal boost invariance.

Boost invariance means that the hydrodynamic fields are taken to be independent of spacetime rapidity, such that only the transverse structure is evolved explicitly. This approximation is best motivated near midrapidity at sufficiently high collision energies, where the system is approximately invariant under longitudinal boosts over a limited rapidity window. However, the computational and conceptual simplicity of this approximation comes with the loss of genuine longitudinal structure. With $(2+1)$D simulations, it is no longer possible to study rapidity-dependent observables, baryon stopping, or decorrelations along the beam direction.

Several numerical packages have been developed to solve the equations of viscous relativistic hydrodynamics in heavy-ion collisions, including widely used software such as MUSIC, VISHNew~\cite{Shen:2014vra}, and vHLLE~\cite{Karpenko:2013wva}. MUSIC solves the second-order viscous hydrodynamic equations on a spacetime grid using the Kurganov-Tadmor algorithm~\cite{Kurganov:2000ovy}, a high-resolution scheme designed to handle large gradients in conservation laws. In practice, this means that the continuous hydrodynamic fields are discretized on a grid and advanced forward in steps of proper time. At each timestep, MUSIC solves the conservation equations together with the relaxation equations for the dissipative fields, while also applying numerical procedures designed to regulate steep gradients and maintain stability of the discretized evolution. For the purposes of multistage models, the primary output of this stage is the hypersurface on which the hydrodynamic evolution is terminated, which is then passed to particlization.

In this work, the hydrodynamic evolution is performed using MUSIC in a $(2+1)$D boost-invariant setup. Hydrodynamics is initialized at $\tau_0 = 0.6$ fm/$c$ with an initial energy density profile from T\textsubscript{R}ENTo initial conditions and vanishing initial transverse flow. The simulation includes viscous evolution with both shear and bulk degrees of freedom, and uses the modified equation of state described in chapter~\ref{chapter3}. This hydrodynamic stage provides the spacetime evolution of the QGP medium until it reaches the switching condition for particlization.

\section{Particlization}
\label{section24}

As the QGP expands and cools, the medium gradually reforms into hadrons. Although hydrodynamics can describe hadronic matter to some extent --- hence the applicability of a hadron resonance gas equation of state at low temperatures --- the system eventually becomes too dilute for a fluid description to remain valid. This breakdown of hydrodynamics necessitates an interface between fluid and particle descriptions of the medium.

This interface is called particlization, in which continuous hydrodynamic fields are converted into discrete hadrons on a switching hypersurface. Informally, this conversion is often called ``freeze-out'', but that term risks confusion with chemical and kinetic freeze-out in the hadronic phase. Although particlization is closely related to the physical process of hadronization, the two are not identical: hadronization refers to the microscopic conversion of deconfined degrees of freedom into hadrons, while particlization is an effective modeling prescription for switching between descriptions. In modern multistage simulations, this conversion is usually performed using the Cooper-Frye prescription, which converts the local thermodynamic information on the switching hypersurface into hadronic momentum distributions.
\Needspace{4\baselineskip}
\subsection{Cooper-Frye particlization}
\label{subsection241}

The standard prescription for particlization is the Cooper-Frye formula~\cite{Cooper:1974mv}, which computes the momentum distribution of hadrons emitted from a fluid hypersurface. This conversion conserves energy and momentum provided the hadron distribution functions reproduce the stress-energy tensor of the fluid on the surface. The switching hypersurface is usually defined by a fixed temperature or energy density, chosen to lie near the QCD crossover transition between the QGP and HRG descriptions. 

\begin{figure}[!htbp]
  \centering
  \resizebox{0.55\textwidth}{!}{%
  \begin{tikzpicture}[font=\small, >=Stealth,
      normal/.style={-{Stealth[length=2.4mm]}, line width=0.8pt, black!85},
      had/.style={->, line width=0.5pt, black!55}]
    \definecolor{Cfluid}{RGB}{233,150,90}
    \def\surfacepath{(1.8,0.0) .. controls (2.3,0.85) and (2.2,1.75) .. (2.3,2.4)
                     .. controls (2.45,2.9) and (3.0,2.95) .. (2.8,3.5)
                     .. controls (2.6,3.9) and (1.3,4.15) .. (0,4.3)}
    \fill[Cfluid!55] (0,0) -- \surfacepath -- (0,0) -- cycle;
    \draw[->, line width=0.8pt] (0,0) -- (4.0,0) node[below right=-1pt and -2pt]{$r$};
    \draw[->, line width=0.8pt] (0,0) -- (0,5.0) node[left]{$\tau$};
    \node[below left, font=\scriptsize] at (0,0.2) {$\tau_0$};
    \node[below left, font=\scriptsize] at (0.25,0.05) {$0$};
    \draw[line width=1.3pt, black] \surfacepath;
    \coordinate (TL) at (2.09,1.62); \coordinate (TR) at (2.43,1.84);
    \coordinate (BR) at (2.39,1.18); \coordinate (BL) at (2.05,0.96);
    \draw[fill=black!12, draw=black!85, line width=0.6pt] (TL)--(TR)--(BR)--(BL)--cycle;
    \draw[normal] (2.24,1.40) -- (3.12,1.36);
    \node[font=\scriptsize, anchor=west] at (3.16,1.46) {$d\Sigma_\mu$};
    \filldraw[black!65] (3.05,3.25) circle (1.3pt); \draw[had] (3.05,3.25) -- ++(0.25,0.12);
    \filldraw[black!65] (2.85,3.70) circle (1.3pt); \draw[had] (2.85,3.70) -- ++(0.40,0.24);
    \filldraw[black!65] (2.25,4.05) circle (1.3pt); \draw[had] (2.25,4.05) -- ++(0.88,0.26);
    \node[font=\scriptsize, anchor=west] at (2.40,4.34) {$p^{\mu}$};
    \node[font=\scriptsize, anchor=west, align=left] at (3.18,3.48) {hadrons};
    \node[font=\scriptsize, align=center, black!85] at (1.0,1.3)
         {fluid \\[-1pt] $T>T_{\text{part}}$};
    \node[font=\scriptsize, anchor=west] (sig) at (0.5,4.62) {$\Sigma:\ T=T_{\text{part}}$};
    \draw[black, line width=0.4pt] (sig.west) -- (0.15,4.26);
  \end{tikzpicture}
  }
  \caption[Schematic of the particlization hypersurface in the $\tau-r$ plane]{Schematic of the particlization hypersurface in the $\tau$--$r$ plane, with $\tau_0$ the hydrodynamic starting time. The shaded region is fluid above the switching condition $T=T_{\text{part}}$: cooler outer cells cross the surface early (large $r$, near $\tau_0$), while radial expansion carries the maximum radius to intermediate times before the hotter center particlizes last. The shaded tile is a surface element with outward-oriented normal $d\Sigma_\mu$; via the Cooper-Frye prescription each such element emits hadrons with momenta $p^\mu$.}
  \label{fig:hypersurface}
\end{figure}
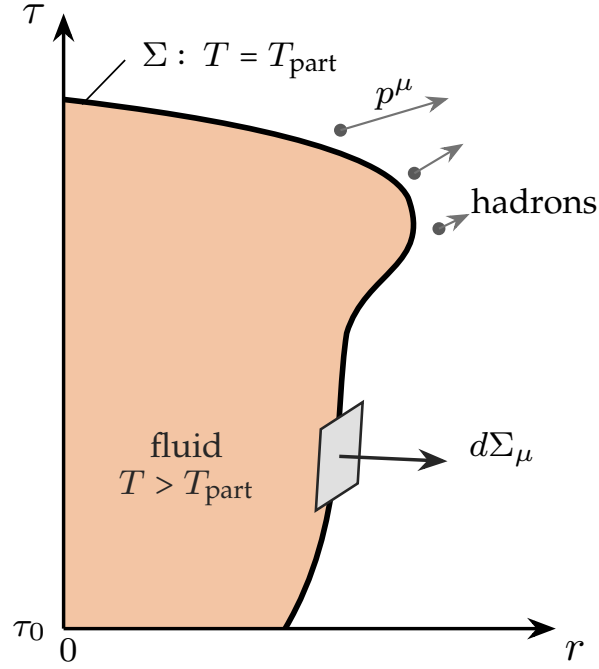

To visualize the hypersurface, imagine seeing the temperature field of the expanding fireball at one point in proper time, as shown in Fig. \ref{fig:hypersurface}. At this point, the switching criterion $T(x) = T_\text{part}$ defines a contour surrounding the hotter region of the medium. As the system expands and cools, this contour evolves through the transverse plane and eventually disappears once the entire system has fallen below the switching temperature. Stacking these contours over time forms a hypersurface in spacetime, and it is precisely along this hypersurface that particlization applies. Thus, particlization should not be pictured as the whole medium converting into particles at a single given time. Instead, different fluid cells cross the switching condition at different times, with cooler outer regions typically particlizing earlier and hotter central regions surviving longer. In a boost-invariant $(2+1)$D calculation, this hypersurface is assumed to be independent of spacetime rapidity, just like the hydrodynamic fields.

In the Cooper-Frye prescription, each surface element contributes to particle production according to its local flow velocity, temperature, chemical potentials, and orientation in spacetime:

\begin{align}
    E \frac{dN_i}{d^3p} = \frac{g_i}{(2\pi)^3} \int_\Sigma p^\mu d\Sigma_\mu \, f_i(x,p),
    \label{eq:cooper_frye}
\end{align}
where $i$ labels the hadron species, $N_i$ is the total multiplicity of that species, $g_i$ is its degeneracy, $p^\mu$ is the on-shell particle four-momentum, $d\Sigma_\mu$ is the outward-directed hypersurface element, and $f_i(x,p)$ is the local distribution function. Importantly, the hadron species sampled here must match those in the HRG equation of state used by the hydrodynamics and those propagated in the hadronic transport, for physical consistency. The left-hand side is the Lorentz-invariant momentum distribution, whose integral gives the total multiplicity, 
\begin{align} 
     N_i = \int \frac{d^3p}{E} \, E \frac{dN_i}{d^3p}. 
\end{align}

The basic equilibrium contribution is determined by thermal distribution functions for each hadron species included in the chosen hadronic resonance list. In thermal and chemical equilibrium, these are given by the same Bose-Einstein and Fermi-Dirac distribution functions introduced in the context of the hadron resonance gas in Eq.~\ref{eq:bose_fermi}.

In practice, the Cooper-Frye formula is typically used as the basis for stochastic sampling. Rather than retaining only continuous momentum distributions, one samples a discrete list of hadrons that can then be passed to a hadronic ``afterburner'', the colloquial term for hadronic transport models at the end of heavy-ion collision simulations. The final observed hadrons will overwhelmingly be dominated by light species such as pions, kaons, and protons. However, even if one is not specifically interested in rare particle yields, the choice of hadron species included in the sampling is important because unstable resonances contribute to the final stable particle yields through their subsequent decays.

In a numerical simulation, the Cooper-Frye integral cannot be evaluated on a perfectly continuous hypersurface. The hydrodynamic fields are known only on a discrete spacetime grid, so the switching hypersurface must itself be constructed from the discrete grid. Cornelius is a widely used surface-finding algorithm~\cite{Huovinen:2012is} which identifies the local isosurface element associated with a fixed temperature or energy density and returns its position, size, and normal vector. In this discretized form, the Cooper-Frye integral becomes a sum over surface elements,
\begin{align}
    E \frac{dN_i}{d^3p} = \frac{g_i}{(2\pi)^3} \sum_a p^\mu \Delta \Sigma_{\mu,a}  f_i(x_a,p),
\end{align}
where $a$ labels the reconstructed surface elements, $x_a$ is the spacetime position of each element, and $\Delta\Sigma_{\mu,a}$ is the corresponding oriented hypersurface element. The expected multiplicity of each species $i$ is then obtained by integrating this distribution over momentum,
\begin{align}
    \langle N_i \rangle = \frac{g_i}{(2\pi)^3} \sum_a \int \frac{d^3p}{E} p^\mu \Delta \Sigma_{\mu,a} f_i(x_a,p).
    \label{eq:cooper_frye_discrete}
\end{align}

In a sampled event, $N_i$ is an integer particle count drawn from this underlying distribution, while Eq.~\ref{eq:cooper_frye_discrete} gives its corresponding sampling expectation value.

\subsection{Viscous corrections}
\label{section242}

For physical consistency, the stress-energy tensor represented by the hydrodynamic fields should be matched by the particle distributions used at particlization. In ideal hydrodynamics, the equilibrium distribution functions in the Cooper-Frye formula reproduce the ideal fluid stress-energy tensor. In viscous hydrodynamics, however, the hydrodynamic stress-energy tensor also contains the shear stress tensor $\pi^{\mu\nu}$ and the bulk viscous pressure $\Pi$. These non-equilibrium contributions are not part of the equilibrium HRG equation of state, and must instead be encoded through corrections to the local distribution functions. 

The local distribution functions are written as
\begin{align}
    f_i(x,p) = f_{i,\text{eq}}(x,p) + \delta f_i(x,p),
\end{align}
where $f_{i,\text{eq}}(x,p)$ are the local equilibrium Bose-Einstein and Fermi-Dirac distributions given in Eq.~\ref{eq:bose_fermi}. The correction $\delta f_i$ encodes the fact that the local momentum distribution of particles emitted from a viscous fluid cell is not exactly thermal.

In the absence of conserved charge diffusion, $\delta f_i$ can be broken down into components associated with the shear stress tensor $\pi^{\mu\nu}$ and the bulk viscous pressure $\Pi$,
\begin{align}
\delta f_i = \delta f_i^{(\pi)} + \delta f_i^{(\Pi)}.
\end{align}
The shear correction is associated with the shear stress tensor $\pi^{\mu\nu}$ and therefore depends on the contraction $p^\mu p^\nu \pi_{\mu\nu}$, which modifies the anisotropic part of the local momentum distribution. The bulk correction is associated with the scalar bulk pressure $\Pi$ and modifies the isotropic momentum dependence of the distribution. These corrections affect particle spectra and flow observables directly at particlization, supplementing the effects that viscosity already had on the hydrodynamic evolution and the particlization hypersurface.

There is inherent model dependence in actually determining these corrections, because hydrodynamics fixes only a limited number of moments of the distribution function. The shear stress tensor and bulk pressure constrain the part of $\delta f_i$ that contributes to the energy-momentum tensor, but they do not uniquely determine the full momentum dependence of $\delta f_i$ for every hadron species. Fixing it requires input beyond the hydrodynamic moments; one approach computes species-dependent corrections directly from linearized kinetic theory~\cite{Molnar:2014fva}. In practice, different particlization prescriptions choose different kinetic-theory-inspired ansatzes for these correction functions.

In kinetic theory, macroscopic quantities are obtained by integrating the microscopic distribution function against powers of the particle momentum. These momentum-weighted integrals are called moments of the distribution function. For each species $i$, define the Lorentz-invariant momentum integral
\nopagebreak[4]
\begin{align}
    \int dP_i \equiv g_i \int \frac{d^3p}{(2\pi)^3 E_i}.
\end{align}

A conserved current is a first moment of the distribution function, while the stress-energy tensor is a second moment:
\begin{align}
    J_Q^\mu = \sum_i q_i \int dP_i p^\mu f_i(x,p), \qquad T^{\mu\nu} = \sum_i \int dP_i p^\mu p^\nu f_i(x,p),
\end{align}
where $q_i$ is the corresponding charge of species $i$. The local equilibrium energy density and pressure are particular projections of the equilibrium stress-energy tensor,
\begin{align}
    \varepsilon = u_\mu u_\nu T_{\text{eq}}^{\mu\nu}, \qquad P = -\frac{1}{3}\Delta_{\mu\nu}T_{\text{eq}}^{\mu\nu}.
\end{align}

In this sense, hydrodynamics does not determine the full momentum dependence of $f_i(x,p)$. It determines only the particular momentum integrals corresponding to the conserved currents and the stress-energy tensor.

A moment expansion uses this relationship in reverse. Instead of trying to determine an arbitrary function $\delta f_i(x,p)$, one expands it in a finite set of momentum structures whose coefficients are chosen to reproduce the hydrodynamic moments. One common choice, and the one used in this work, is Grad's 14-moment approximation~\cite{Grad:1949zz}. It keeps the scalar, vector, and rank-two tensor structures needed to match the dissipative hydrodynamic quantities, while discarding higher moments that are not fixed by hydrodynamics.

The non-equilibrium correction $\delta f_i$ is constrained by requiring its moments to reproduce the dissipative hydrodynamic fields. In the Landau frame, the correction should not change the local energy density, while its scalar and traceless tensor projections reproduce the bulk pressure and shear stress tensor:
\begin{align}
    u_\mu u_\nu \delta T^{\mu\nu}=0, \qquad \Pi = -\frac{1}{3}\Delta_{\mu\nu}\delta T^{\mu\nu}, \qquad \pi^{\mu\nu} = \delta T^{\langle\mu\nu\rangle},
\end{align}
where
\begin{align}
    \delta T^{\mu\nu} = \sum_i \int dP_i p^\mu p^\nu \delta f_i.
\end{align}

These equations determine only a finite number of momentum integrals of $\delta f_i$, not the full function $\delta f_i(x,p)$. The remaining freedom is the origin of the model dependence in viscous corrections at particlization.

The allowed form of the correction is constrained by tensor structure. In the local rest frame of the fluid, the equilibrium distribution depends only on the energy of the particle, and is therefore isotropic. Non-equilibrium corrections can then be organized according to how they transform under spatial rotations in this local rest frame. A scalar correction changes the isotropic part of the distribution and is associated with the bulk pressure $\Pi$. A vector correction introduces a dipole-like distortion and is associated with conserved charge diffusion currents. A rank-two traceless tensor correction introduces a quadrupole-like distortion and is associated with the shear stress tensor $\pi^{\mu\nu}$.

Covariantly, these structures are built using projections transverse to the fluid velocity. The scalar correction can depend on scalar combinations such as $u\cdot p$ and $m_i^2$. If there is a diffusion current, denoted $V^\mu$, then the vector correction is proportional to $p^{\langle\mu\rangle}V_\mu$. The shear correction is proportional to $p^{\langle\mu}p^{\nu\rangle}\pi_{\mu\nu}$, matching the symmetric, traceless, transverse structure of the shear stress tensor. Thus, before choosing a specific ansatz for the momentum dependence, the viscous correction can be decomposed schematically as
\begin{align}
    \delta f_i = \delta f_i^{(\Pi)} + \delta f_i^{(V)} + \delta f_i^{(\pi)}.
\end{align}

In the calculations considered here, charge diffusion is not included, so the vector correction is omitted. The remaining task is to specify the scalar bulk correction and the rank-two shear correction in a way that satisfies the matching conditions above. Grad's 14-moment approximation specifies the momentum dependence of $\delta f_i$ as a polynomial expansion in the particle four-momentum, truncated at quadratic order. The correction is then written as the polynomial ansatz
\begin{align}
    \delta f_i = f_{i,\text{eq}}\tilde f_{i,\text{eq}}\, c_{\mu\nu}p^\mu p^\nu,
\end{align}
where the coefficients $c_{\mu\nu}$ are chosen to reproduce the hydrodynamic dissipative fields, and $\tilde f_{i,\text{eq}} \equiv 1 + f_{i,\text{eq}}$ for bosons (Bose enhancement) or $1 - f_{i,\text{eq}}$ for fermions (Pauli blocking), consistent with the quantum statistics of Eq. \ref{eq:bose_fermi}. The prefactor $f_{i,\text{eq}}\tilde f_{i,\text{eq}}$ is not an independent choice: it is the derivative of the equilibrium distribution with respect to its argument,
\begin{align}
    \frac{\partial f_{i,\text{eq}}}{\partial(u\cdot p)} = -\frac{1}{T}f_{i,\text{eq}}\tilde f_{i,\text{eq}},
\end{align}
so that $\delta f_i$ takes the form of a linear response of $f_{i,\text{eq}}$ to a shift of its argument quadratic in the particle momentum. In the Boltzmann limit $f_{i,\text{eq}}\ll 1$, one has $\tilde f_{i,\text{eq}}\to 1$, and the correction reduces to $f_{i,\text{eq}}$ multiplying a polynomial in momentum. This expression should be understood as an ansatz for the momentum dependence of $\delta f_i$, not as a unique consequence of hydrodynamics.

The next step is to decompose the coefficients $c_{\mu\nu}$ into their scalar and rank-two tensor pieces. Written in terms of irreducible structures in the local rest frame, the correction takes the schematic form
\begin{align}
    \delta f_i = f_{i,\text{eq}}\tilde f_{i,\text{eq}}\left[c_T m_i^2 + c_E(u\cdot p)^2 + c_\pi^{\langle\mu\nu\rangle}p_{\langle\mu}p_{\nu\rangle}\right].
\end{align}

This expression makes the tensor content explicit. The terms proportional to $c_T$ and $c_E$ are scalar structures and contribute to the bulk correction. The term proportional to $c_\pi^{\langle\mu\nu\rangle}$ is the rank-two traceless structure associated with the shear stress tensor. The coefficients are then fixed by matching to the hydrodynamic fields. In the 14-moment approximation, this matching gives
\begin{align}
    c_T = A_T\Pi, \qquad c_E = A_E\Pi, \qquad c_\pi^{\langle\mu\nu\rangle} = A_\pi\pi^{\mu\nu}.
\end{align}

The coefficients $A_T$, $A_E$, and $A_\pi$ are common to all species and are determined by momentum integrals summed over the full hadron resonance gas, since $\Pi$ and $\pi^{\mu\nu}$ represent the total dissipative response of all species. Their role is to ensure that $\delta f_i$ satisfies the Landau matching conditions and reproduces the hydrodynamic values of the bulk pressure and shear stress tensor.

Grad's 14-moment approximation is only one way of resolving the undetermined momentum dependence of $\delta f_i$. Because the hydrodynamic fields constrain only a finite set of moments, the same dissipative currents can be reproduced by correction functions that differ away from those moments, and the resulting spread in particle spectra is itself a source of theoretical uncertainty at particlization. The choice of viscous correction method has been found to have a significant effect on observables, with the Grad model being one of the better-performing options \cite{JETSCAPE:2020mzn}.

One alternative prescription in common use is the Chapman-Enskog expansion~\cite{Jaiswal:2013npa}, in which $\delta f_i$ is obtained by solving the Boltzmann equation perturbatively around local equilibrium, in practice within the relaxation time approximation with a common relaxation time for all hadron species. At first order the correction is again linear in the dissipative fields, but its momentum dependence follows from the gradient term of the Boltzmann equation, which introduces inverse powers of the particle energy, rather than from a fixed polynomial ansatz. As a result, the Chapman-Enskog and Grad corrections can be matched to the same hydrodynamic moments while differing in their predicted particle spectra.

Another widely used alternative is the class of modified-equilibrium prescriptions, which treat the viscous correction as a deformation of the local equilibrium distribution rather than as a purely additive linear term. The most well-known of these are the Pratt-Torrieri-Bernhard (PTB) and Pratt-Torrieri-McNelis (PTM) distributions, both refinements of the original prescription of Pratt and Torrieri~\cite{Bernhard:2018hnz, McNelis:2021acu, Pratt:2010jt}. Instead of adding $\delta f_i$ to $f_{i,\text{eq}}$, the particle momenta are rescaled so that the resulting distribution carries the anisotropy and bulk modification required by the hydrodynamic fields, while remaining positive-definite by construction, a property the linearized forms do not share. At small dissipative corrections they reduce to a linear viscous correction of the same general form as the additive $\delta f$. Their positivity comes at a price, however: unlike the other models, the modified-equilibrium distributions do not reproduce the input shear stress and bulk pressure exactly, and this mismatch grows as the viscous corrections become larger.

The above correction schemes apply to any system whose momentum distributions are distorted by shear and bulk viscosity (and, more generally, by the diffusion of conserved charges), but they are all built on a local equilibrium baseline $f_{i,\text{eq}}$ that assumes the quark abundances have reached chemical equilibrium. The central concern of this work is that this assumption need not hold. In that case, the equilibrium baseline must be modified to reflect the out-of-equilibrium quark abundances, and the viscous corrections built on top of it accordingly; the construction is given in section \ref{section31}.

\subsection{Particlization in practice}

As mentioned at the end of section \ref{subsection241}, particlization is, in practice, used to generate an ensemble of sampled particles rather than continuous distributions. This introduces a number of practical issues. The most general is that conservation laws now hold only in a statistical sense: an individual sampled event need not exactly conserve energy, momentum, or net charge, and these quantities are recovered only on average over the ensemble.

Another issue is the geometric nature of applying Cooper-Frye to the particlization hypersurface. The intuitive picture is a hot fireball at the center, with fluid cells flowing out through the hypersurface to form particles. However, depending on the orientation of the hypersurface element, particles with momenta directed back toward the fluid can carry a negative Cooper-Frye flux factor $p^\mu d\Sigma_\mu$, corresponding to backflow into the hydrodynamic medium. In practice, these negative contributions are usually discarded, and the transition from hydrodynamics to hadronic transport is treated as one-way, even though it is not truly so --- hadrons can and do fly into the fluid. The interactions between individual particles and the medium constitute a rich subfield of their own in the context of jet physics, but a stray pion entering the QGP is insignificant in practical effect. 

A further caveat from the previous subsection is that viscous corrections can become large enough that the underlying linear expansion around equilibrium breaks down. For the linearized 14-moment and Chapman-Enskog schemes, $\delta f_i$ can grow comparable to $f_{i,\text{eq}}$ in magnitude, breaking the assumption $|\delta f_i| \ll f_{i,\text{eq}}$; where the correction is negative, the total distribution $f_{i,\text{eq}} + \delta f_i$ can itself become negative. A negative phase space density cannot be sampled, so practical implementations regulate this behavior, often by setting negative values of the distribution to zero. Thankfully, this occurs mainly at very high particle momenta or where the bulk pressure is large, so for most soft-sector observables the effect is modest.

In this work, particlization is performed with iS3D~\cite{McNelis:2019auj} using the 14-moment method, on a hypersurface found within MUSIC via the Cornelius algorithm. Unusually, though, the particlization hypersurface is not defined by a single universal switching temperature. Instead, the particlization temperature depends on the local fugacities, as described in chapter \ref{chapter3}. To reduce sampling fluctuations, the hypersurface is oversampled up to a hundred times per event, stopping sooner only if a threshold multiplicity is reached. This produces an ensemble of oversampled events from every single hydrodynamic event, each of which is then propagated independently through the final phase of hadronic transport.

\section{Hadronic transport}
\label{section25}

After particlization, the outgoing hadrons from the collision continue to interact for some time. At this stage, the system is too dilute for hydrodynamics to be the most appropriate description, but still dense enough that hadrons can rescatter before reaching their final observed state. The late-stage evolution is therefore modeled microscopically, by propagating individual hadrons and allowing them to scatter or decay according to hadronic cross sections and resonance properties.

The underlying framework is relativistic Boltzmann transport. For each hadron species $i$, the phase-space distribution $f_i(x,p)$ evolves according to the Boltzmann equation,
\begin{align}
    p^\mu \partial_\mu f_i(x,p) = C_i[f],
\end{align}
where $C_i[f]$ is the collision term. The collision term accounts for all of the ways in which particle interactions can modify the phase space densities. In the special case of $C_i[f] = 0$, we recover free streaming. In practical simulations, the collision term is implemented stochastically by propagating particles along classical trajectories and checking which pairs scatter, form resonances, or undergo other allowed reactions.

The most common interactions are two-body processes. In elastic scattering, the outgoing particles are identical to the incoming ones, so the scattering changes their momenta but not their particle identities. In inelastic scattering, the particle content can change while conserving the total energy, momentum, and relevant quantum numbers. Important examples of inelastic reactions include resonance formation, as well as baryon–antibaryon annihilation and its inverse, regeneration. Resonance decays are also part of the hadronic transport stage, but are one-body processes rather than scatterings. These reactions become less frequent as the hadron gas expands, leading gradually to chemical and kinetic freeze-out. After their last interaction or decay, particles stream freely to the detector.

Many of the hadrons sampled at particlization are unstable resonances that do not survive to detection, but instead decay into lighter and longer-lived hadrons. Their decays are governed by lifetimes and branching ratios, while their masses are often sampled from a spectral function rather than fixed to a single pole mass. A common schematic form is the Breit-Wigner distribution,
\begin{align}
    A_R(m) = \frac{1}{2\pi}\frac{\Gamma_R}{(m-M_R)^2+\Gamma_R^2/4},
\end{align}
where $M_R$ is the resonance pole mass and $\Gamma_R$ is its width. More detailed implementations can use relativistic spectral functions and mass-dependent widths, but the basic point is that short-lived resonances appear as distributed mass states that subsequently decay into stable or longer-lived particles.

Hadronic transport can significantly modify final observables. Resonance decays feed down into stable hadron yields, especially pions, and can change the observed particle composition relative to the initially sampled hadron list. Rescattering changes the momentum spectra by redistributing momentum among hadrons after particlization. This is especially important for identified particles and mean transverse momenta, because different species continue to interact for different lengths of time and have different scattering cross sections. The hadronic stage can also modify anisotropic flow: some flow is inherited from the hydrodynamic stage, but additional rescattering can further build or damp the final momentum anisotropies.

The main limitation of hadronic transport is that the collision term must be specified phenomenologically. In practice, transport codes rely on experimentally constrained hadronic cross sections, resonance properties, decay branching ratios, and symmetry arguments when direct data are unavailable. Most standard afterburner calculations include binary collisions, resonance formation, selected inelastic channels, and decays, while neglecting many-particle scatterings. This is physically reasonable in the dilute late hadronic stage, where the probability of many particles interacting simultaneously is small, but it remains an approximation. Uncertainties in poorly known cross sections or missing reaction channels can therefore propagate into the final hadronic observables.

The two most widely used Boltzmann transport models in heavy-ion phenomenology are UrQMD (Ultrarelativistic Quantum Molecular Dynamics)~\cite{Bass:1998ca, Bleicher:1999xi} and SMASH (Simulating Many Accelerated Strongly-interacting Hadrons)~\cite{SMASH:2016zqf}. Both implement the same broad physical picture: they are microscopic Boltzmann transport models that propagate hadrons and simulate scattering, resonance formation, and decay. UrQMD is older and well-established, while SMASH is a newer refinement of the same approach that has held up equally well in comparisons to data \cite{JETSCAPE:2020mzn}. The differences between the two mostly lie in details such as resonance treatments, numerical implementation, and SMASH's support for multi-particle reactions. Both models have applicability well beyond their role as afterburners. In lower energy collisions where a hydrodynamic stage is not included, hadronic transport can also be used to model the full microscopic evolution.

In this work, the hadronic transport stage is performed using SMASH. The sampled hadrons from iS3D are passed to SMASH, which evolves them through hadronic rescattering and resonance decays until the system has effectively frozen out. 

\section{Comparing to experimental data}
\label{section26}

The end product of the model chain described thus far is an ensemble of final-state particles. To compare with experiments, these particles must be transformed into the observables described in section \ref{section13}. This requires more than simply counting all particles produced by the simulation. Experiments have particular criteria for particle selection, detector acceptance, and event classification, and these choices must be mirrored as closely as possible for a model-to-data comparison to be meaningful.

Some observables are comparatively direct to compute from the final particle list. Multiplicities are obtained by counting particles of a given species within the relevant kinematic cuts. Mean transverse momenta and transverse momentum spectra are obtained by averaging or binning the transverse momenta of the selected particles. The transverse energy can similarly be computed as a sum over final-state particles, for example through
\begin{align}
    E_T = \sum_i E_i \sin\theta_i,
\end{align}
where $\theta_i$ is the polar angle of particle $i$. In practice, however, the precise definition of $E_T$ can be experiment-dependent, especially in how particle masses, weak decays, and acceptance cuts are handled.

Anisotropic flow coefficients require more care. They are not simple single-particle averages, but are extracted from azimuthal correlations among particles, making them particularly sensitive to event statistics, centrality definitions, and the details of the particle selection cuts. In this work, I calculate them using the cumulant method of Ref.~\cite{Bilandzic:2010jr}. The relevant correlations are built from the per-event flow vector $Q_n = \sum_{k=1}^{M} e^{i n \varphi_k}$, where $M$ is the event multiplicity and $\varphi_k$ the azimuthal angle of particle $k$. The two-particle azimuthal correlation in a single event follows from $Q_n$ once the self-pairs are removed,
\begin{align}
    \langle 2 \rangle = \frac{|Q_n|^2 - M}{M(M-1)},
\end{align}
with the denominator counting the distinct pairs. Averaging this single-event quantity over all events in a centrality class, each weighted by its number of pairs, defines the two-particle cumulant $c_n\{2\} = \langle\langle 2 \rangle\rangle$, from which the flow coefficient is recovered as $v_n\{2\} = \sqrt{c_n\{2\}}$. I take the cumulant $c_n\{2\}$, rather than $v_n\{2\}$, as the primary quantity: the two carry the same information, but $c_n\{2\}$ is what the correlations measure directly, whereas $v_n\{2\}$ is real only when $c_n\{2\}$ is non-negative. Under finite event statistics a small flow signal can drive the measured $c_n\{2\}$ below zero, where the square root ceases to be meaningful, so working directly in $c_n\{2\}$ avoids a pathology that becomes important when the signal is carried through the statistical inference of chapter~\ref{chapter4}.

A two-particle cumulant by itself retains short-range non-flow correlations, such as those from resonance decays, which are not part of the collective flow signal. To suppress these, I correlate particles across a pseudorapidity gap, dividing the acceptance into forward and backward subevents and pairing each particle in one window only with particles in the other. Built from the two subevent flow vectors $Q_n^A$ and $Q_n^B$, the gapped correlation
\begin{align}
    \langle 2 \rangle_{\Delta\eta} = \frac{\mathrm{Re}\!\left[Q_n^A \, Q_n^{B*}\right]}{M_A M_B}
\end{align}
requires no self-pair subtraction, since the two windows share no particles. This mirrors standard experimental practice and is what allows the model observable to correspond to the measured one: an ungapped correlation would retain short-range non-flow that experiments attempt to suppress and that the present soft-sector model is not designed to reproduce in full. Because the hydrodynamic evolution here is boost-invariant, the collective flow is independent of pseudorapidity, so the gap suppresses short-range non-flow without diminishing the genuine flow signal. Where flow is reported for identified species, it is computed differentially, correlating each particle of interest against the charged particle reference rather than against the same species in isolation~\cite{Bilandzic:2013kga}. With the particles of interest and the reference particles placed on opposite sides of the gap, the two sets remain disjoint, so the differential cumulant $d_n\{2\}$ (the analogous pair-weighted average of the particle-of-interest--reference correlation) carries no overlap correction and the differential flow follows as
\begin{align}
    v_n'\{2\} = \frac{d_n\{2\}}{\sqrt{c_n\{2\}}},
\end{align}
normalized by the charged particle reference cumulant $c_n\{2\}$.

Experiments set particular kinematic cuts based on both their geometry and the detectors' capabilities. These typically include cuts on particle transverse momentum and either rapidity or pseudorapidity, which can vary on an observable-by-observable basis. Pseudorapidity is used for unidentified charged particles, since computing rapidity requires the particle mass, whereas identified particle analyses can use true rapidity. 

For example, STAR reports the charged particle multiplicity density within $|\eta| < 0.5$ and various strange hadron yields within $|y| < 0.5$, with the identified spectra measured over species-dependent transverse momentum ranges and extrapolated to obtain $p_T$-integrated yields. Such cuts must be reproduced in the analysis of the model output, rather than applying a single universal acceptance to all observables. This is especially important when comparing identified hadrons, since the experimental cuts for pions, kaons, protons, and strange hadrons are not always the same.

The definition of centrality should also be consistent between model and experiment. In an initial condition model such as T\textsubscript{R}ENTo, it is straightforward to inspect the impact parameter or initial entropy of each simulated event. Experiments, however, do not have direct access to these quantities. Instead, centrality is usually inferred from final-state particle production, most commonly through charged particle multiplicity or detector signals correlated with multiplicity. I therefore sort simulated events into centrality classes using final charged particle multiplicity, rather than directly using impact parameter or initial entropy.

More elaborate treatments can include detector response, limited tracking efficiency, particle misidentification, and other experimental effects. I do not include a full detector simulation here. Instead, the model-to-data comparison is performed at the particle level, with the experimental kinematic cuts and centrality definitions applied as closely as possible.

All model results come with statistical uncertainties. Particle yields for abundant light hadrons and $\langle p_T \rangle$ typically converge relatively quickly and can be determined with modest event samples. Flow coefficients, on the other hand, require many events because they are extracted from correlations. Rare hadron species such as the $\Omega$ are also statistically challenging because their multiplicities are small. Flow coefficients for rare identified species are particularly difficult, since they combine both challenges: low yields and correlation-based observables. For this reason, rare-species flow observables are not included in this work.

Ultimately, all of the models described thus far rely heavily on freely chosen parameters. Some choices are fixed directly by the experimental system, such as the collision energy and species of colliding nuclei. Others are much less direct. The viscosities of the QGP, for example, can be described with a variety of functional forms, and changing them affects many different observables simultaneously. At the same time, many observables are affected by multiple parameters at once, producing degeneracies that are difficult to disentangle by manual inspection.

Historically, many studies tuned model parameters manually until a reasonable agreement with data was obtained. This can be useful for exploratory work, but it is not sufficient for a high-dimensional model with many correlated observables and uncertain parameters. A more systematic approach is Bayesian parameter estimation, detailed in chapter \ref{chapter4}. Bayesian analyses make it possible to quantify parameter constraints, propagate model uncertainties, and account for correlations among observables.

As detailed in chapter \ref{chapter4}, Bayesian analyses require robust model statistics. The multistage model chain must be run repeatedly for many different parameter sets, and each parameter point must include enough minimum-bias events to populate all centrality classes. Here, minimum-bias means that events are generated over the full relevant range of collision geometries before centrality selections are applied. In this work, I typically use $10^4$ hydrodynamic events for each parameter point, with up to $100$ particlization oversamples per event to reduce the statistical uncertainty associated with Cooper-Frye sampling.

As the field of heavy-ion collisions develops into more of a precision science, specialized computing infrastructure is increasingly essential to make such production campaigns viable. In my case, the results shown here required $\sim\!10^7$ CPU-hours to produce. I have primarily relied on the National Energy Research Scientific Computing Center (NERSC) and the Open Science Grid (OSG), which provide high-performance computing and high-throughput computing resources, respectively. 
\chapter{Quark flavor equilibration}
\label{chapter3}

As discussed in sections~\ref{section16} and~\ref{section21}, the phenomenological success of gluon-saturated models of the initial state motivates treating the early QCD medium as a gluon-dominated system, in which quarks and antiquarks are initially scarce. Perturbative estimates of the relevant equilibration timescales suggest that gluons equilibrate more rapidly than quarks and antiquarks, owing to the significantly larger gluon-gluon cross sections relative to those involving quarks~\cite{Shuryak:1992wc, Biro:1993qt}. As a result, while the gluonic medium hydrodynamizes on a timescale of $\approx\!1$~fm/$c$ (section~\ref{section15}), chemical equilibration of quarks and antiquarks may take considerably longer, in some estimates exceeding $3$~fm/$c$~\cite{Xu:2004mz, Kurkela:2018oqw, Kurkela:2018xxd}. The QGP may therefore become well described by hydrodynamics while still substantially undersaturated in quarks, spending a significant portion of its evolution out of chemical equilibrium, with potentially observable effects on its properties and the particles it ultimately produces.

Previous studies have examined hydrodynamic evolution in partial chemical equilibrium, in which the medium forms in a gluon-saturated initial state and subsequently equilibrates according to a time-dependent equation of state~\cite{Vovchenko:2015yia, Vovchenko:2016ijt}. These studies found sensitivity to chemical equilibration in photonic and dileptonic observables, suggesting that electromagnetic probes can reveal information about the early stages of the collision. To date, however, this scheme has not been extended to a precision study of hadronic observables, which are measured extensively at the Relativistic Heavy-Ion Collider (RHIC) and the Large Hadron Collider (LHC) and may provide complementary signals of the chemical composition of the QGP.

This chapter develops such a study in two stages. We first construct a model of the QGP in partial chemical equilibrium in which all quark flavors equilibrate together, governed by a single fugacity and a single equilibration timescale (section~\ref{section31}), and examine the resulting effects on hadronic and electromagnetic observables in complete heavy-ion collision events (section~\ref{section32}). Strangeness has long been regarded as a signature of QGP formation~\cite{Rafelski:1982pu}, and the larger strange quark mass suggests that strange and light flavors need not equilibrate at the same rate. We therefore extend the model to treat light and strange quarks with independent fugacities and equilibration timescales (section~\ref{section33}), and study the distinct effects of light and strange flavor equilibration on final-state observables (section~\ref{section34}).

The partial chemical equilibrium model of section~\ref{section31} and the flavor-independent results of section~\ref{section32} are adapted from my first publication of this work~\cite{Gordeev:2025vog}. The remaining sections extend that work to differential flavor equilibration and are presented here for the first time.

\section{Modeling partial chemical equilibrium}
\label{section31}

Our model evolves the QCD medium in partial chemical equilibrium: with fully thermalized gluons but zero initial (anti)quark content, relaxing toward a chemically equilibrated quark-gluon plasma over the course of the hydrodynamic phase. This requires a particular construction of the equation of state, but it does not modify the fundamental equations of motion of relativistic viscous hydrodynamics introduced in chapter~\ref{chapter2},
\begin{align}
    \partial_\mu T^{\mu \nu} = 0,
    \label{eq:eom}
\end{align}
where the energy-momentum tensor is
\begin{align}
    T^{\mu \nu} = \varepsilon u^\mu u^\nu - (P + \Pi)(g^{\mu \nu} - u^\mu u^\nu) + \pi^{\mu \nu},
    \label{eq:Tmunu}
\end{align}
with $\varepsilon$ the energy density, $u^\mu$ the local flow velocity, $P$ the pressure, $\Pi$ the bulk pressure, $g^{\mu \nu}$ the metric tensor in the mostly-minus convention $g^{\mu \nu} = \mathrm{diag}(+1,-1,-1,-1)$, and $\pi^{\mu \nu}$ the shear stress tensor. We use the Israel-Stewart-type second-order viscous hydrodynamics described in section \ref{section23}.

The distinction in partial chemical equilibrium is chemical rather than dynamical: under the usual assumption of chemical equilibrium, $P$ and $\varepsilon$ are related by a fixed equation of state while $\Pi$ and $\pi^{\mu \nu}$ encode deviations from local thermodynamic equilibrium, whereas in partial chemical equilibrium all of these quantities additionally depend on the local quark fugacity.
\Needspace{4\baselineskip}
\subsection{Equation of state}
\label{subsec:eos_pce}

The equation of state encodes the chemistry of the QGP by relating the primary state variables that define the medium. Defining the non-equilibrium EoS is therefore essential to evolving the medium out of chemical equilibrium. We do so by calculating the pressure $P$ and energy density $\varepsilon$ as functions of the temperature $T$ and a quark fugacity $\gamma_q$. This is sufficient to define the EoS. We additionally derive the entropy density $s(T,\gamma_q)$ in appendix~\ref{app:entropy}: although the equation of state itself requires only $P(T,\gamma_q)$ and $\varepsilon(T,\gamma_q)$, the entropy density enters the dissipative sector through the viscosity normalization described in section \ref{subsec:implementation_pce}.

\subsubsection{High temperature}

In equilibrium, the EoS at high temperatures is derived from lattice QCD calculations. Ideally, we would calculate the non-equilibrium EoS the same way. However, due to the numerical sign problem, such a calculation is thus far impractical, and lattice studies have largely been limited to equilibrium calculations at finite but small chemical potentials~\cite{Bazavov:2017dus}. Instead, we adopt an approach based on that proposed in Ref.~\cite{Vovchenko:2016ijt}, which interpolates between two equilibrium lattice calculations: one in ($2+1$)-flavor QCD~\cite{HotQCD:2014kol}, and one in pure SU($3$) gauge theory~\cite{Borsanyi:2012ve}. One should note that pure SU($3$) gauge theory is only an approximation for a gluon-dominated QCD system. Even in a chemical non-equilibrium environment where gluons predominate and quarks are scarce, virtual quark contributions still impact gluon interactions and the running of the strong coupling constant.

We assume that quarks and antiquarks have equal densities characterized by the quark fugacity $\gamma_q$, which linearly interpolates the pressure and energy density of the two lattice equations of state as
\begin{equation}
\begin{aligned}
\frac{P}{T^4}(T,\gamma_q) &= \gamma_q \frac{P_3}{T^4}
   \left(T \frac{T_3}{T_{\text{c}}(\gamma_q)}\right) + (1-\gamma_q) \frac{P_0}{T^4}
   \left(T \frac{T_0}{T_{\text{c}}(\gamma_q)}\right), \\
\frac{\varepsilon}{T^4}(T,\gamma_q) &= \gamma_q \frac{\varepsilon_3}{T^4}
   \left(T \frac{T_3}{T_{\text{c}}(\gamma_q)}\right) + (1-\gamma_q) \frac{\varepsilon_0}{T^4}
   \left(T \frac{T_0}{T_{\text{c}}(\gamma_q)}\right),
\end{aligned}
\label{eq:e_T4}
\end{equation}
where $P_3(T)$, $\varepsilon_3(T)$, and $T_3$ are respectively the pressure, energy density, and critical temperature of the full QCD EoS, and $P_0(T)$, $\varepsilon_0(T)$, and $T_0$ are the corresponding quantities for the pure glue EoS. At $\gamma_q = 1$ we recover exactly the ($2+1$)-flavor QCD EoS, and at $\gamma_q = 0$ we recover exactly the gluonic EoS. This quark fugacity can be generalized to flavor-specific fugacities, which is the central extension developed in section~\ref{section33}; for the flavor-independent model of this section and its application in section~\ref{section32}, we consider only a single $\gamma_q$ that encompasses all quark flavors.

As there is a first-order phase transition in $N_f = 0$ at $T_0 \approx 260$~MeV, as opposed to a crossover in $N_f = 2+1$ at $T_3 \approx 158$~MeV, simply interpolating between $\frac{P_3}{T^4}(T)$ and $\frac{P_0}{T^4}(T)$ without additional modifications would cause the system to experience separate transitions at each critical temperature for any $0 < \gamma_q < 1$. To avoid this, we rescale all temperatures according to a fugacity-dependent critical temperature $T_\text{c}(\gamma_q)$ so that there is only ever one transition at a given fugacity~\cite{Moreau:unpublished}. We define
\begin{align}
    T_\text{c}(\gamma_q) = \sqrt{\gamma_q}\, T_3 + (1-\sqrt{\gamma_q})\, T_0.
    \label{eq:Tc}
\end{align}

\begin{figure}[!htbp]
    \centering
    \includegraphics[width=0.7\linewidth]{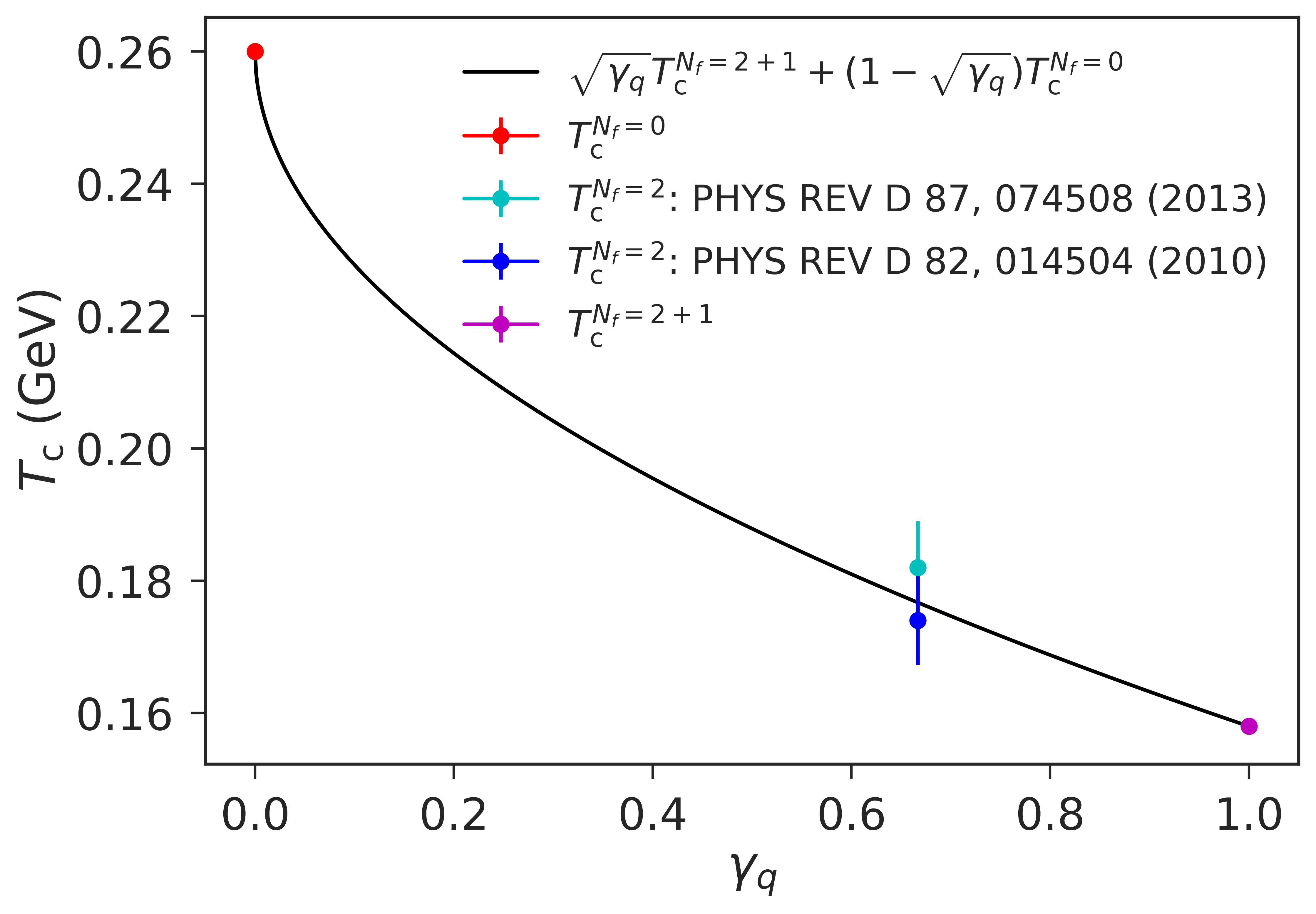}
    \caption[$T_\text{c}(\gamma_q)$ compared to lattice critical temperatures]{$T_\text{c}(\gamma_q)$ compared to the critical temperatures for $N_f = 0$~\cite{Borsanyi:2012ve}, $N_f = 2$~\cite{Burger:2011zc, Bornyakov:2009qh}, and $N_f = 2+1$~\cite{HotQCD:2014kol} results.}
    \label{fig:Tc}
\end{figure}

The behavior of the phase transition at intermediate quark fugacities is largely unknown. As a guiding approximation, we construct the fugacity-dependent critical temperature such that it smoothly interpolates between the lattice QCD results for $T_0$ and $T_3$ while passing through the $N_f = 2$ critical temperature~\cite{Burger:2011zc, Bornyakov:2009qh} at $\gamma_q = 2/3$, although it should be noted that these $N_f = 2$ results are derived with heavier-than-physical quark masses. Fig.~\ref{fig:Tc} plots the result.

With this temperature rescaling, the actual pressure $P(T,\gamma_q)$ and energy density $\varepsilon(T,\gamma_q)$ take the forms
\begin{equation}
\begin{aligned}
P(T,\gamma_q) &= \gamma_q
   \left(\frac{T_{\text{c}}(\gamma_q)}{T_3}\right)^4
   P_3\!\left(T \frac{T_3}{T_{\text{c}}(\gamma_q)}\right) + (1-\gamma_q)
   \left(\frac{T_{\text{c}}(\gamma_q)}{T_0}\right)^4
   P_0\!\left(T \frac{T_0}{T_{\text{c}}(\gamma_q)}\right), \\
\varepsilon(T,\gamma_q) &= \gamma_q
   \left(\frac{T_{\text{c}}(\gamma_q)}{T_3}\right)^4
   \varepsilon_3\!\left(T \frac{T_3}{T_{\text{c}}(\gamma_q)}\right) + (1-\gamma_q)
   \left(\frac{T_{\text{c}}(\gamma_q)}{T_0}\right)^4
   \varepsilon_0\!\left(T \frac{T_0}{T_{\text{c}}(\gamma_q)}\right).
\end{aligned}
\label{eq:e}
\end{equation}

\subsubsection{Low temperature}

Conventionally, the lattice equation of state at high temperature is matched to a hadron resonance gas equation of state at low temperature~\cite{Huovinen:2009yb}. We do the same here, only with modifications to the hadron distribution functions to account for nonzero quark chemical potential. In principle, the low-temperature pure glue EoS corresponds to a gas of glueballs~\cite{Stoecker:2015zea} rather than a gas of hadrons. With the addition of a Hagedorn spectrum~\cite{Meyer:2009tq}, the glueball gas model agrees well with lattice results while additionally explaining the very low pressure and energy density at $T < T_\text{c}$ through the large masses of the constituent glueballs~\cite{Borsanyi:2012ve}. However, in this work we use the hadron resonance gas equation of state for all $0 \leq \gamma_q \leq 1$, and instead introduce hadronic fugacities that suppress the hadron distributions at low $\gamma_q$.

The hadron resonance gas energy density and pressure are calculated as
\nopagebreak[4]
\begin{equation}
\begin{aligned}
    \varepsilon &= \sum_i g_i \int \frac{d^3 p}{(2\pi)^3} E_p f_i(p), \\
    P &= \sum_i g_i \int \frac{d^3 p}{(2\pi)^3} \frac{p^2}{3E_p} f_i(p),
\end{aligned}
\label{eq:ep_hrg}
\end{equation}
where $i$ denotes the hadron species, $g_i$ and $f_i$ are the respective degeneracy and distribution function for each species, and $E_p = \sqrt{p^2 + m_i^2}$ is the individual hadron energy given a mass $m_i$. In equilibrium, mesons are described by a Bose-Einstein distribution and baryons by a Fermi-Dirac distribution. In chemical non-equilibrium, the distribution functions are modified to the form
\begin{align}
    f_i(p, \lambda_i) = \frac{1}{\lambda_i^{-1} e^{E_p/T} \pm 1},
    \label{eq:f_i}
\end{align}
where $\lambda_i$ is a species-specific fugacity factor distinct from the $\gamma_q$ used in the high-temperature EoS.

There is some ambiguity in how the hadronic fugacities should map to the quark fugacities. One constraint is that, as baryons have three (anti)quarks and mesons have two, baryons should be further suppressed by a power of $3/2$. Another is that we seek to construct an EoS where all thermodynamic variables are continuous and differentiable over the transition region, for all quark fugacities between $0$ and $1$. These constraints lead us to define
\begin{equation}
\begin{aligned}
    \lambda_{\text{meson}} &= 0.85\, \gamma_q + 0.15, \\
    \lambda_{\text{baryon}} &= \lambda_{\text{meson}}^{3/2},
    \label{eq:lambda_i}
\end{aligned}
\end{equation}
where the numerical constants are fit to ensure smooth matching with the high-temperature EoS. Using these hadronic fugacities, the distribution functions in Eq.~\ref{eq:f_i} can be used in Eq.~\ref{eq:ep_hrg} to construct the low-temperature EoS.

\begin{figure*}[!t]
    \centering
    \includegraphics[width=0.49\textwidth]{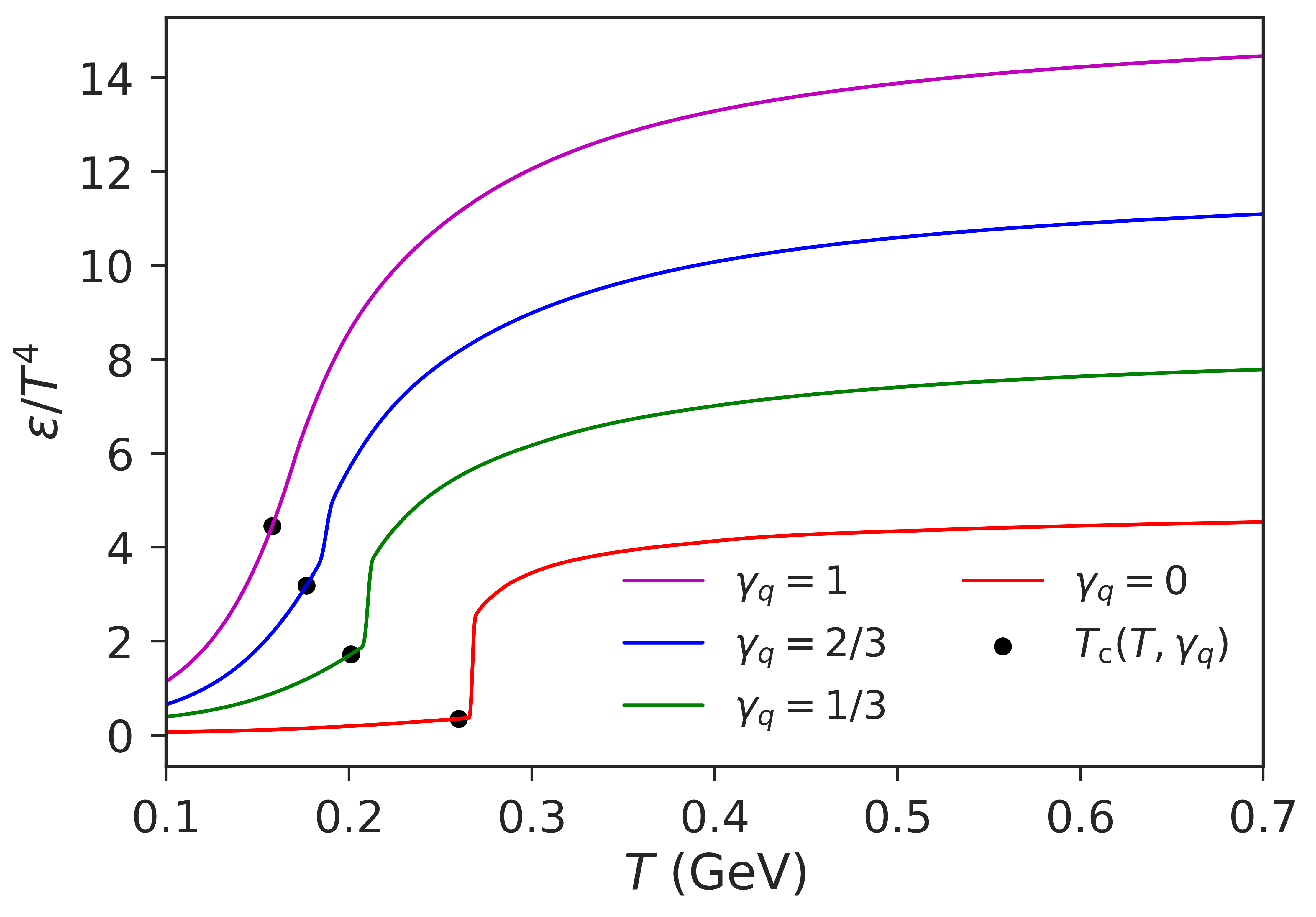}%
    \hfill
    \includegraphics[width=0.49\textwidth]{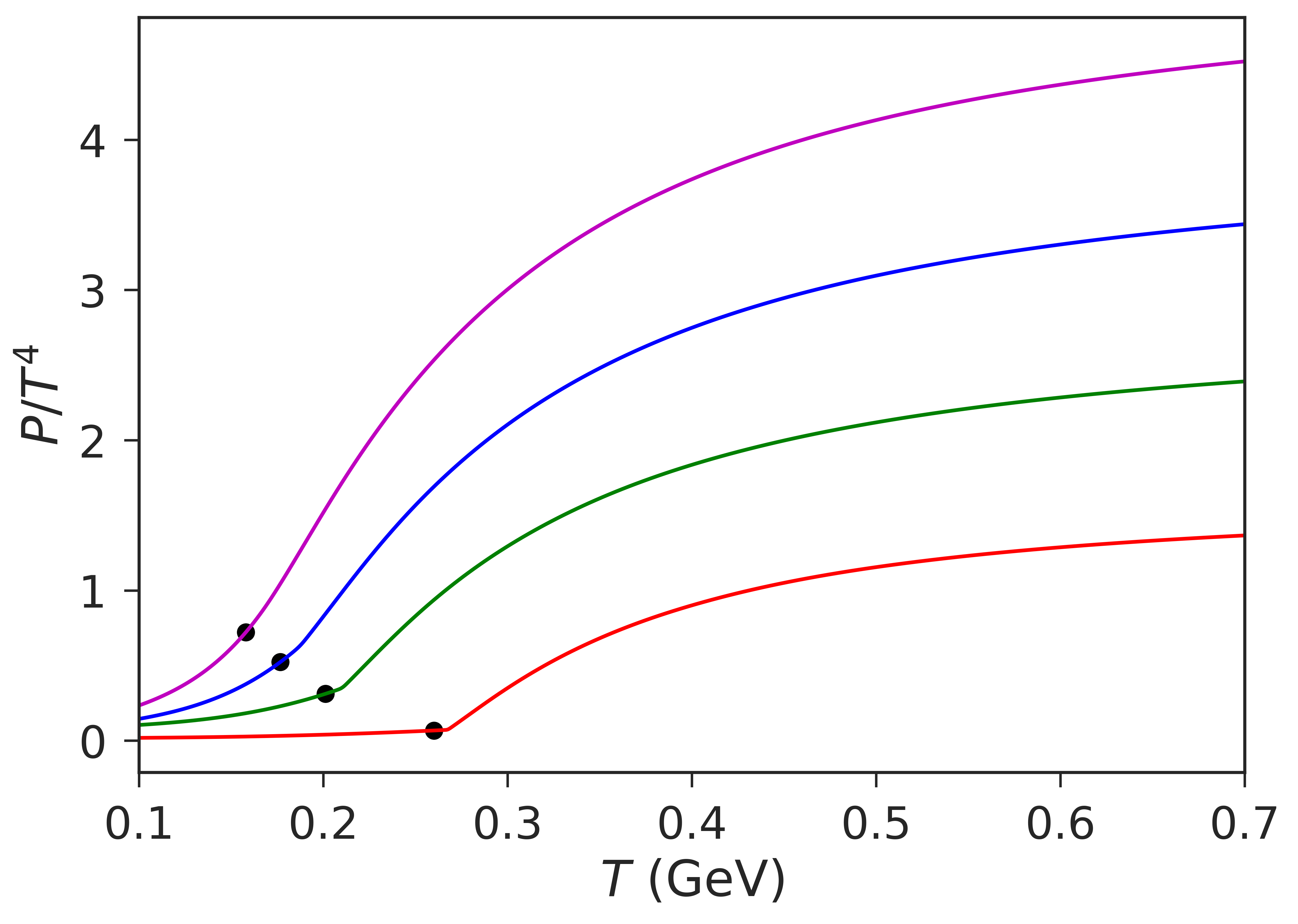}
    \caption[Energy density and pressure of the partial chemical equilibrium EoS]{Energy density (left) and pressure (right) of the partial chemical equilibrium EoS, constructed by matching a linear interpolation in $\gamma_q$ of lattice data at high $T$ to a non-equilibrium hadron resonance gas at low $T$. Note the first-order phase transition in the pure glue EoS at $T_\text{c} = 260$~MeV. The black points indicate the respective critical temperature $T_\text{c}(\gamma_q)$ for each value of $\gamma_q$.}
    \label{fig:EoS}
\end{figure*}

This low-temperature result for $\varepsilon(T,\gamma_q)$ and $P(T,\gamma_q)$ is conventionally matched to the high-temperature values over a crossover range of $T \approx 160$--$200$~MeV. We do the same, using Krogh interpolation, only for a fugacity-dependent temperature range defined as $(T_\text{c}(\gamma_q) + 5~\text{MeV},\ T_\text{c}(\gamma_q) + (10 + 100\gamma_q)~\text{MeV})$. Fig.~\ref{fig:EoS} shows the resulting $\varepsilon(T,\gamma_q)$ and $P(T,\gamma_q)$ for several choices of $\gamma_q$.

Because we interpolate with the pure glue equation of state for all intermediate fugacities, there is a first-order phase transition for all $\gamma_q < 1$ with a magnitude $(1-\gamma_q)$. As discussed in section~\ref{subsec:particlization_pce}, we do not observe a significant impact of this transition on the QGP evolution, as we particlize at $T_\text{c}(\gamma_q)$.

\subsection{Quark fugacity}
\label{subsec:fugacity}

\begin{figure}[!b]
    \vspace{12pt}
    \centering
    \includegraphics[width=0.7\linewidth]{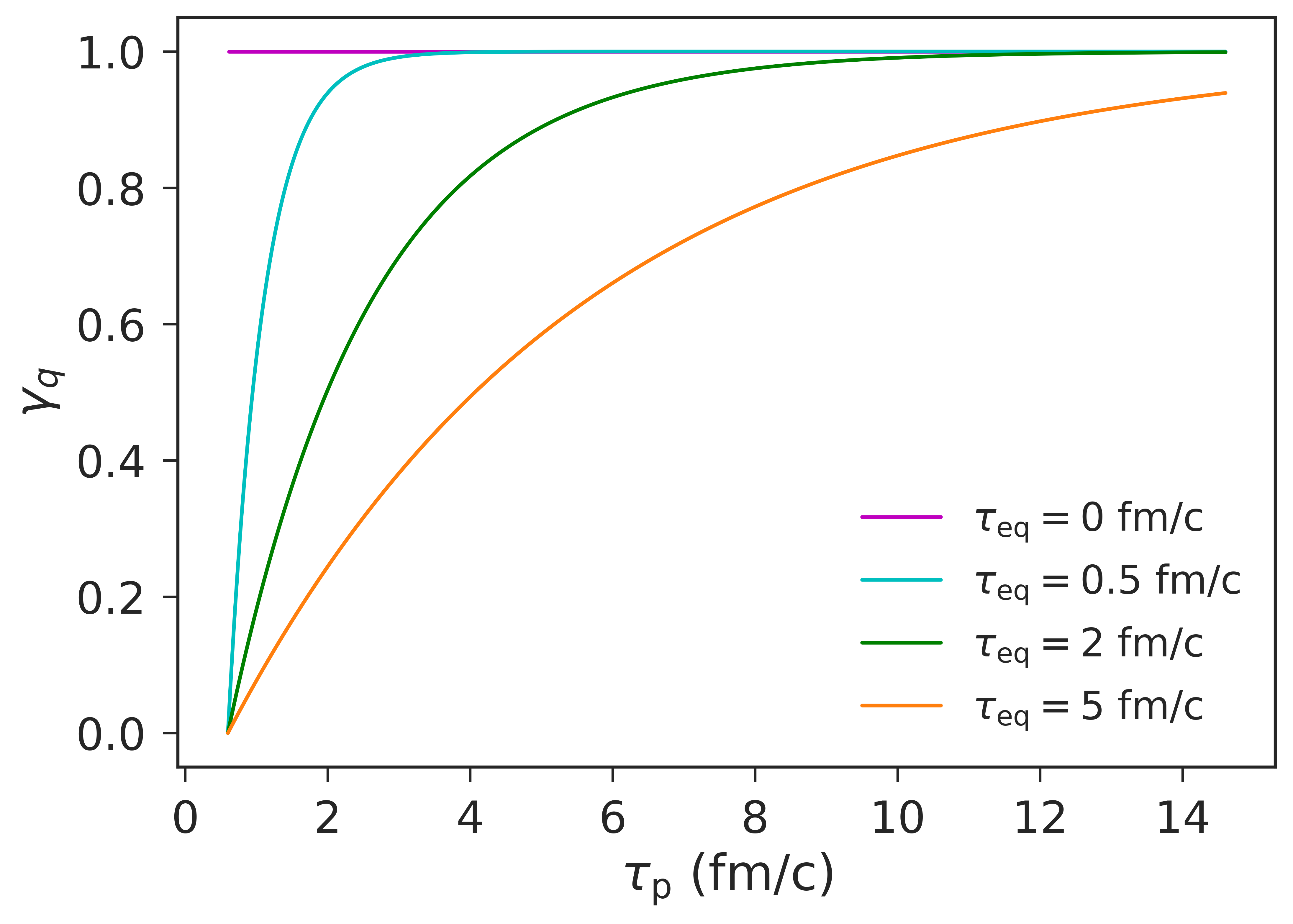}
    \caption[Quark fugacity $\gamma_q$ versus proper time for several $\tau_\text{eq}$]{$\gamma_q$ as a function of proper time for several equilibration timescales $\tau_\text{eq}$, for $\tau_0 = 0.6$~fm/$c$.}
    \label{fig:gamma_q}
\end{figure}

As defined above, the equation of state in partial chemical equilibrium depends on two variables: the temperature $T$ and the quark fugacity $\gamma_q$. In practice, we invert $\varepsilon(T,\gamma_q)$ to treat $\varepsilon$ as an independent field evolved hydrodynamically. One should, in principle, introduce additional rate equations that govern the time evolution of the quark fugacity $\gamma_q$~\cite{Biro:1993qt}. To allow for more direct control of the equilibration timescale, we instead parameterize $\gamma_q$ as a simple function of the local proper time $\tau_\text{p}$,
\begin{align}
    \gamma_q(\tau_\text{p}) = 1 - \exp\left(\frac{\tau_0 - \tau_\text{p}}{\tau_\text{eq}}\right),
    \label{eq:fugacity}
\end{align}
where $\tau_0$ is the global initial time of the system and $\tau_\text{eq}$ is a free parameter corresponding to the effective chemical equilibration time. The limit $\tau_\text{eq}\rightarrow 0$ is treated as instantaneous chemical equilibration, with $\gamma_q = 1$ throughout the hydrodynamic evolution. Fig.~\ref{fig:gamma_q} shows the form of this function.

Note that $\tau_\text{p}$ is distinct from the time coordinate $\tau$; due to transverse flow and time dilation, the local proper time of each fluid cell evolves at a different rate. During the hydrodynamic evolution we therefore solve the additional equation
\begin{align}
    u^\mu \partial_\mu \tau_\text{p} = 1,
    \label{eq:proper_time}
\end{align}
with the initial condition $\tau_\text{p}(\tau_0) = \tau_0$, to correctly determine the fugacity, and thus the EoS, everywhere in the fluid. We treat $\tau_\text{p}$ as an auxiliary field and evolve it alongside $\varepsilon$ and $u^\mu$. At each timestep, while $T^{\mu \nu}$ and its constituent fields are updated according to the viscous hydrodynamic equations of motion, $\tau_\text{p}$ is advected along the fluid velocity $u^\mu$. As a result, regions of the fluid with greater flow velocities evolve in proper time more gradually, and the periphery of the system typically chemically equilibrates more slowly than the center in the lab frame.

\subsection{Particlization}
\label{subsec:particlization_pce}

\begin{figure}[!htbp]
    \centering
    \includegraphics[width=0.7\linewidth]{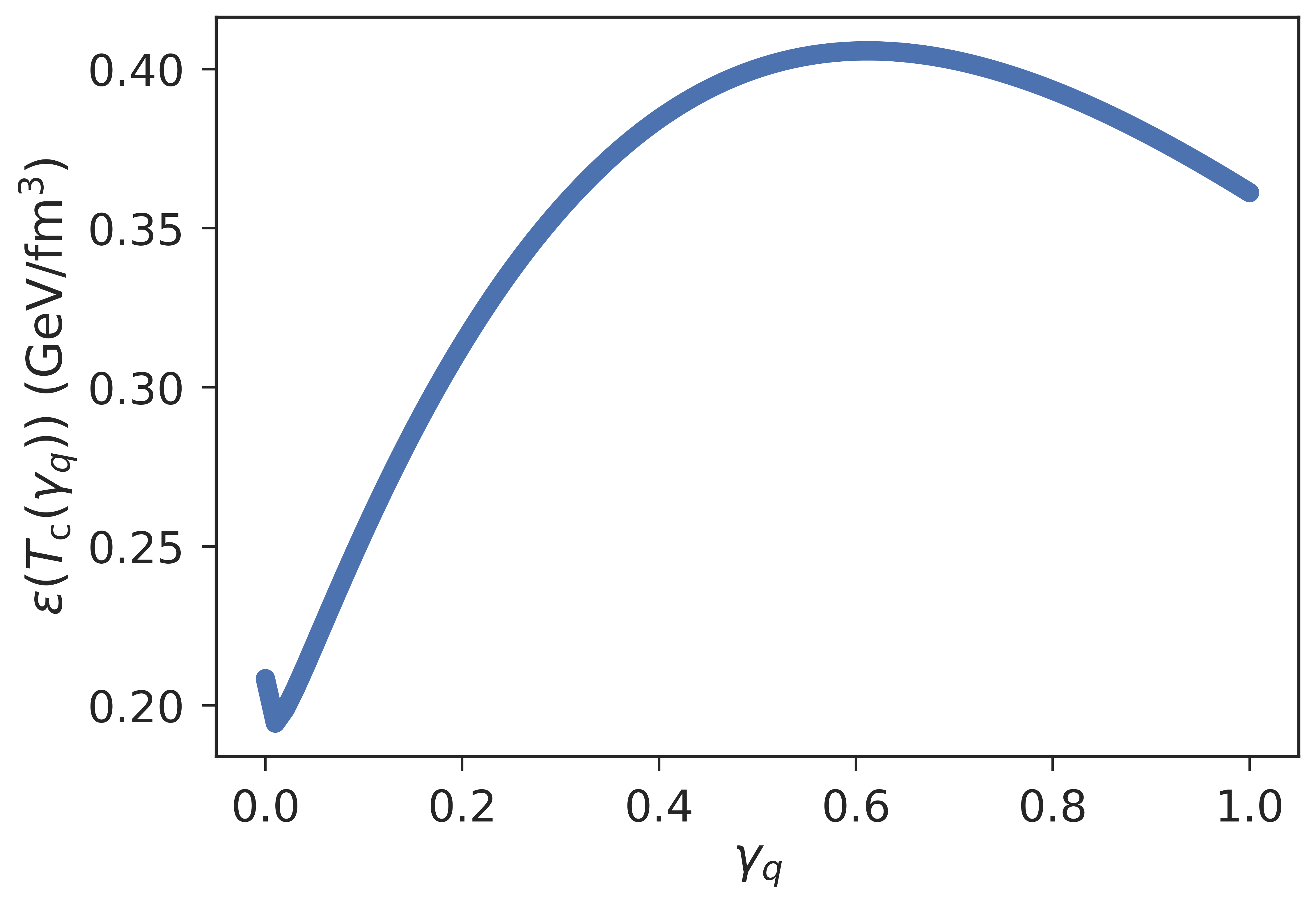}
    \caption[Energy density along the $T_\text{c}(\gamma_q)$ hypersurface]{Energy density along the $T_\text{c}(\gamma_q)$ hypersurface plotted against $\gamma_q$.}
    \label{fig:e_Tc}
\end{figure}

After hydrodynamics, we use the Cooper-Frye prescription~\cite{Cooper:1974mv} introduced in chapter~\ref{chapter2} to convert from a continuous fluid to a set of discrete particles,
\begin{align}
    E\frac{d^3N_i}{dp^3} = g_i \int_\Sigma \frac{d\Sigma_\mu p^\mu}{(2\pi)^3} f_i(x,p),
\end{align}
where $i$ denotes each hadron species, $g_i$ their respective degeneracies, and $\Sigma$ the particlization hypersurface with normal surface elements $d\Sigma_\mu$. The hypersurface is typically defined by a condition of constant temperature or energy density~\cite{Huovinen:2012is}. To account for the fugacity-dependent nature of the transition in our equation of state, we instead define $\Sigma$ according to the critical temperature $T_\text{c}(\gamma_q)$ of Eq.~\ref{eq:Tc}. The primary effect is that fluid cells with low fugacities both hadronize and particlize at higher temperatures. Fig.~\ref{fig:e_Tc} shows the corresponding energy densities at $T_\text{c}(\gamma_q)$. $\varepsilon(T_\text{c}(\gamma_q))$ changes non-monotonically with $\gamma_q$ as a result of our interpolation scheme, and can vary by as much as $\approx\!50$\%. As shown in section~\ref{section32}, however, few fluid cells reach fugacities near zero for all but the largest equilibration timescales, and the variation in practice is typically $\approx\!20$\%.

In ideal hydrodynamics, the only further modification required is that the hadron distribution functions $f_i(x,p)$ be modified with fugacity factors, as in Eq.~\ref{eq:f_i}. This is equivalent to requiring that the equation of state remain consistent across the transition from hydrodynamics to particles.

In viscous hydrodynamics, we additionally account for viscous corrections to the distribution functions,
\begin{align}
    f_i(x,p) = f_{i,\text{eq}}(x,p) + \delta f_i(x,p),
    \label{eq:df}
\end{align}
using the Grad 14-moment expansion~\cite{Grad:1949zz} developed in section~\ref{section24}.

Partial chemical equilibrium requires only a single modification to that treatment, in the identification of the local equilibrium distribution. Although the medium is out of chemical equilibrium, we assume it remains close enough to local thermal equilibrium for hydrodynamics to apply, and identify $f_{i,\text{eq}}$ with the fugacity-modified distributions of Eq.~\ref{eq:f_i} rather than the standard Bose-Einstein and Fermi-Dirac forms.

The matching coefficients $A_T$, $A_E$, and $A_\pi$ of section~\ref{section242} are built from thermal integrals over the hadron resonance gas,
\begin{align}
    J_{kq,i} = \int dP_i\, \frac{(u\cdot p)^{k-2q}\,(-p\cdot\Delta\cdot p)^q}
    {(2q+1)!!}\, f_{i,\text{eq}}\,\tilde f_{i,\text{eq}},
    \label{eq:J}
\end{align}
with the invariant momentum measure $\int dP_i$ defined there. In partial chemical equilibrium these integrals inherit a dependence on the quark fugacity through the equilibrium distributions: when particlizing a fluid cell of fugacity $\gamma_q$, the $J_{kq,i}$ --- and hence $A_T$, $A_E$, and $A_\pi$ --- are evaluated with the fugacity-modified distributions $f_{i,\text{eq}}(\gamma_q)$ of Eq.~\ref{eq:f_i}, so that $\delta f_i$ is matched to the dissipative fields of the fugacity-modified medium. As in that treatment, the charge diffusion correction is omitted; here this follows from working exclusively at $\mu_B = 0$, where quarks and antiquarks are produced in equal number and quark number is not conserved, so the entire viscous correction is carried by the shear and bulk terms. In principle, the shear and bulk viscosities, and thus $\Pi$ and $\pi^{\mu\nu}$ themselves, should also depend on the quark fugacity; due to the difficulty of determining this effect we neglect any such dependence here, and return to its consequences in section~\ref{section32}.

\subsection{Implementation}
\label{subsec:implementation_pce}

We realize this model within the multistage heavy-ion framework of chapter~\ref{chapter2}, with modifications confined to the hydrodynamic and particlization stages. Initial conditions are generated with T\textsubscript{R}ENTo~\cite{Moreland:2014oya} and hadronic transport is handled by SMASH~\cite{SMASH:2016zqf} with no modifications, exactly as described there. The partial chemical equilibrium effects enter through modifications to hydrodynamics and particlization. First, the hydrodynamic evolution, carried out with MUSIC~\cite{Schenke:2010nt, Schenke:2010rr, Paquet:2015lta}, uses the non-equilibrium equation of state of section~\ref{subsec:eos_pce} together with the auxiliary proper time field of Eq.~\ref{eq:proper_time}, so that the local fugacity follows Eq.~\ref{eq:fugacity} cell by cell. The dimensionful shear and bulk viscosities supplied to MUSIC are obtained from the specific viscosities via $\eta = (\eta/s)\,s$ and $\zeta = (\zeta/s)\,s$, evaluated with the non-equilibrium entropy density $s(T,\gamma_q)$ of appendix~\ref{app:entropy}. Second, particlization with iS3D~\cite{McNelis:2019auj} incorporates the fugacity factors of Eq.~\ref{eq:f_i} and the surface condition $T_\text{c}(\gamma_q)$. Although the hadron distributions at particlization are modified by the quark fugacity, their subsequent dynamics in SMASH carry no explicit fugacity dependence. 

\newpage
\section{Effects of flavor-independent chemical equilibration}
\label{section32}

We now apply the partial chemical equilibrium model of section~\ref{section31} to complete Pb+Pb collision events, simulated with the pipeline described in section~\ref{subsec:implementation_pce}, to examine the effects of quark chemical equilibration on the QGP evolution and on final-state observables. The T\textsubscript{R}ENTo parameters were set without specific calibration to experimental multiplicities at a particular center-of-mass energy, as the aim here is to demonstrate the effects of varying the chemical equilibration timescale rather than empirically constrain the model parameters. As a result, the simulated charged particle multiplicities are intermediate between those measured at $\sqrt{s_{NN}} = 2.76$~TeV and $\sqrt{s_{NN}} = 5.02$~TeV. We carry out ($2+1$)-dimensional boost-invariant hydrodynamic evolution for the same set of initial conditions, but with different equilibration timescales $\tau_\text{eq}$ ranging between $0$ and $10$~fm/$c$. The shear viscosities are set to match the Bayesian model-averaged posterior in Ref.~\cite{JETSCAPE:2020shq}, and the bulk viscosity is set to zero.

\subsection{Hydrodynamic evolution}
\label{subsec:evolution_pce}

Before examining the effects of quark chemical equilibration on final particle observables, we should understand how the hydrodynamic evolution itself is impacted. We consider a single event-averaged energy density profile corresponding to a central ($b = 0$) event, and hydrodynamically evolve this same initial condition with varying $\tau_\text{eq}$.

\begin{figure}[!t]
    \centering
    \includegraphics[width=0.7\linewidth]{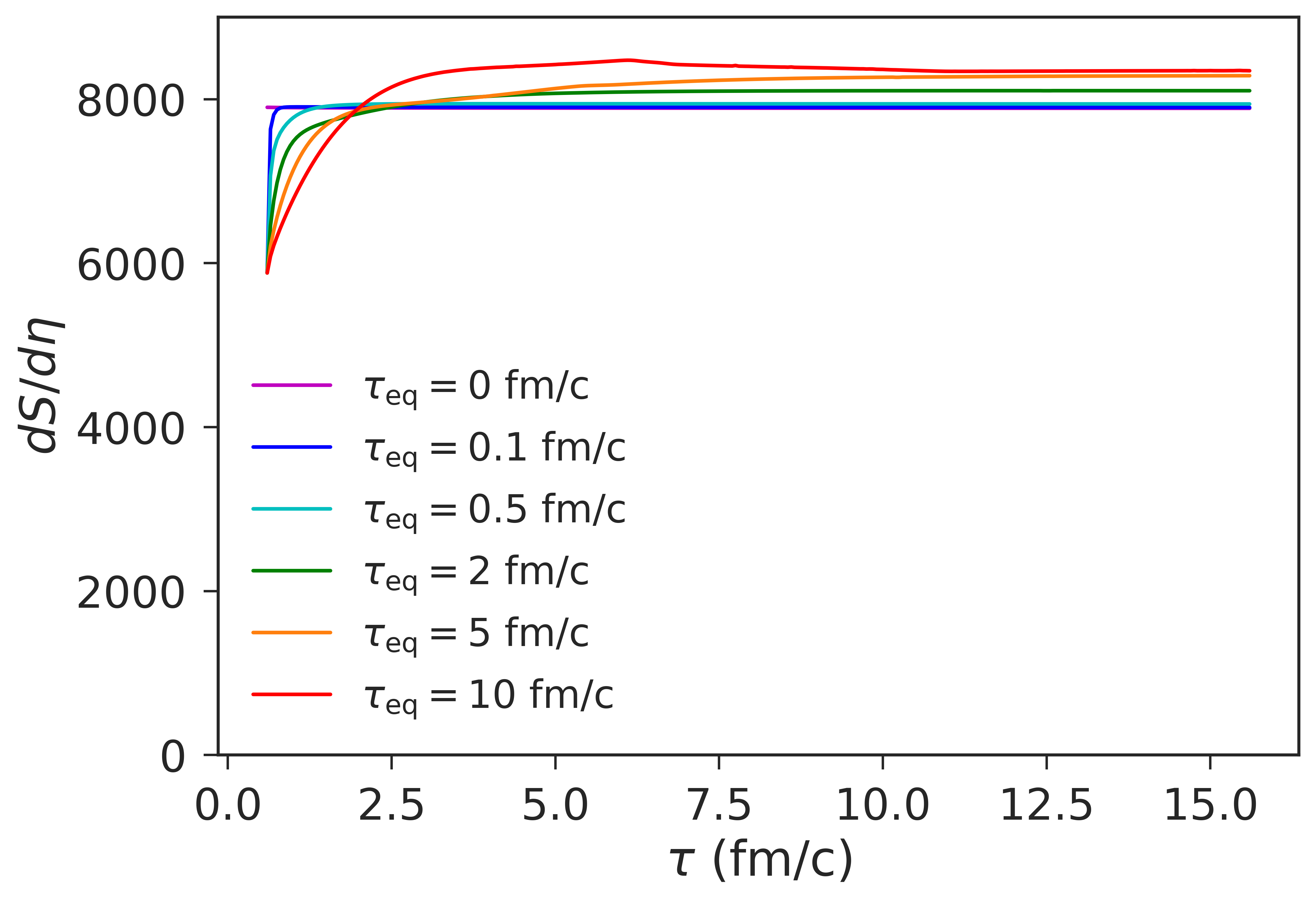}
    \caption[Entropy per unit spacetime rapidity for varying $\tau_\text{eq}$]{Total entropy per unit spacetime rapidity over time for an averaged central event evolved with varying $\tau_\text{eq}$ using ideal hydrodynamics.}
    \label{fig:entropy}
\end{figure}

Entropy production is one signature of quark chemical equilibration that was demonstrated in Refs.~\cite{Vovchenko:2015yia, Vovchenko:2016ijt, Kurkela:2018xxd}. Entropy is conserved in ideal hydrodynamics with a fixed equation of state, but has been shown to increase when the system chemically equilibrates. We reproduce this result in Fig.~\ref{fig:entropy}. Due to the initial energy density being kept constant, the total entropy starts lower for all $\tau_\text{eq} > 0$~fm/$c$ as the medium is initialized with the pure glue EoS in those cases. Depending on $\tau_\text{eq}$, as much as one third of the final entropy may be produced by chemical equilibration during the hydrodynamic evolution. The entropy does not strictly increase for each timestep in Fig.~\ref{fig:entropy} because it shows the entropy per unit spacetime rapidity within the midrapidity slice of the ($2+1$)-dimensional computational grid, and not the total entropy of the ($3+1$)-dimensional physical system. The observed decrease at certain times is due to longitudinal expansion transporting matter outside the grid boundaries.

\begin{figure}[!htbp]
    \centering
    \includegraphics[width=0.65\linewidth]{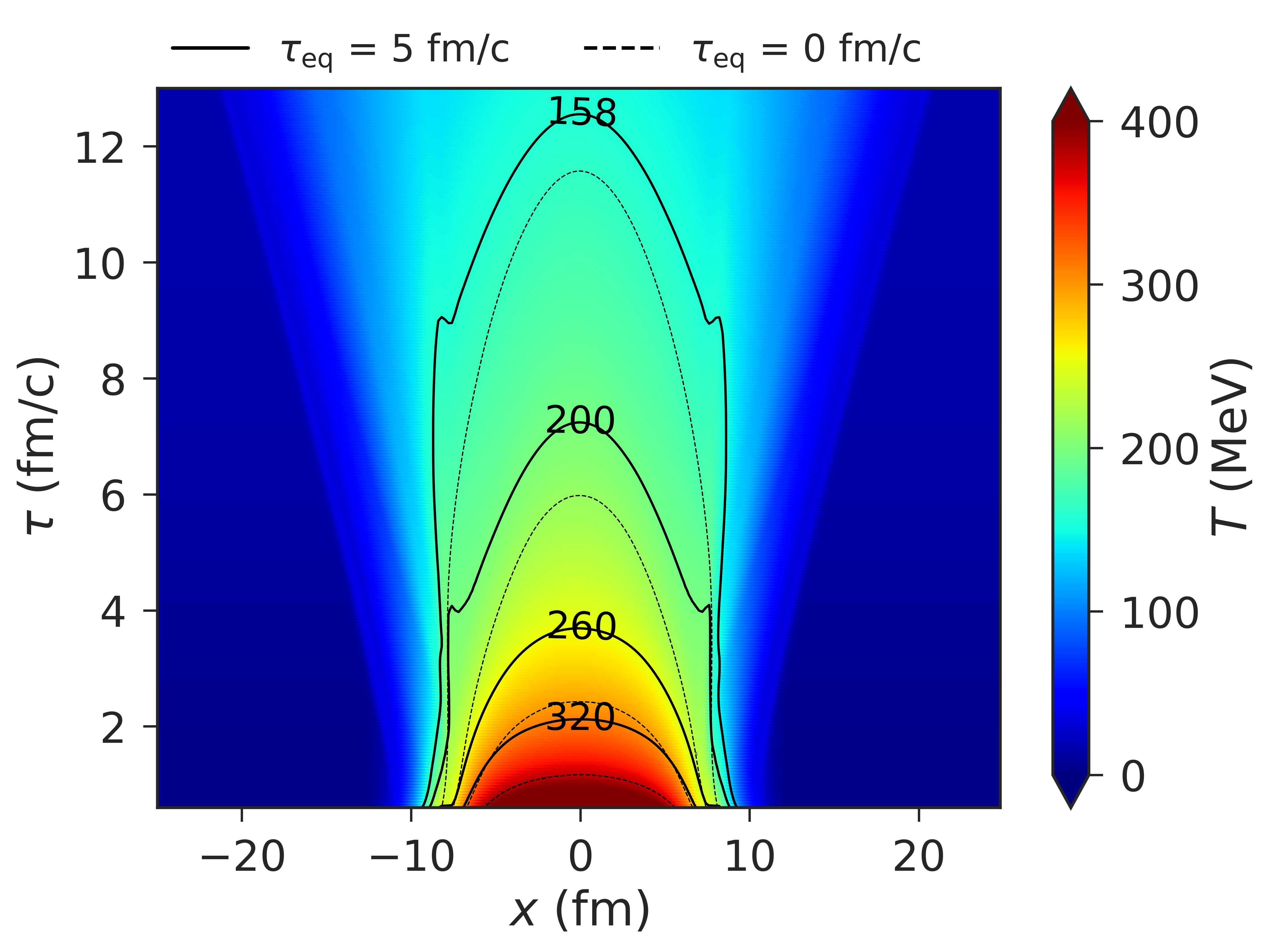}
    \caption[Temperature in the $x-\tau$ plane at $\tau_\text{eq} = 5$~fm/$c$]{Contour plot of temperature in the $x-\tau$ plane for an averaged central event evolved with $\tau_\text{eq} = 5$~fm/$c$. Solid lines show contours of $T$ in MeV, while the dashed lines show the respective isotherms when $\tau_\text{eq} = 0$~fm/$c$.}
    \label{fig:temp_profile}
\end{figure}

Fig.~\ref{fig:temp_profile} shows the temperature of the medium in the $x-\tau$ plane for $\tau_\text{eq} = 5$~fm/$c$. In general, increasing $\tau_\text{eq}$ makes the medium hotter and pushes the isotherms farther out. This is to be expected, as reducing the number of quark degrees of freedom increases the temperature at a given energy density.

\begin{figure}[!htbp]
    \centering
    \includegraphics[width=0.65\linewidth]{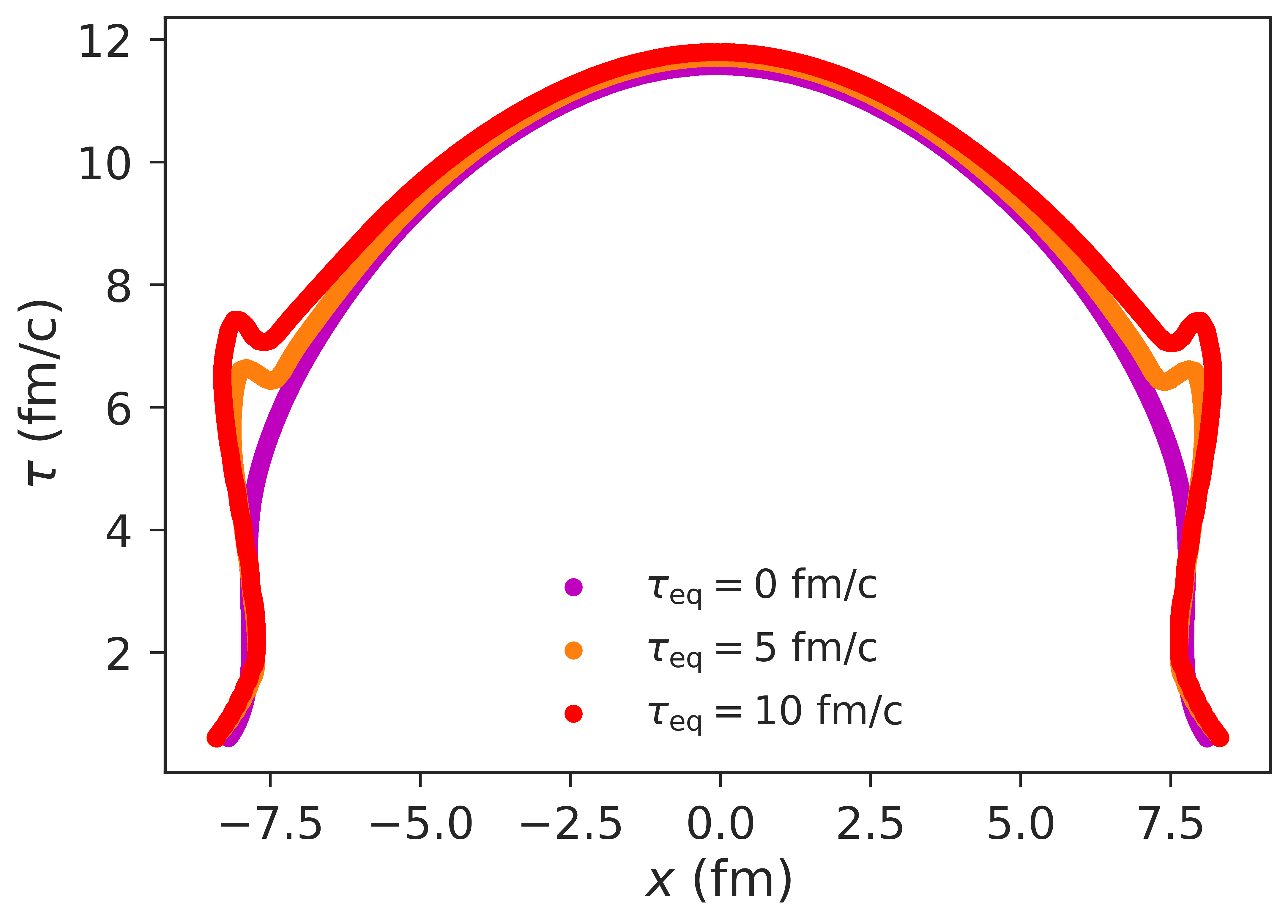}
    \caption[Particlization hypersurfaces for varying $\tau_\text{eq}$]{Particlization hypersurfaces at $T_\text{c}(\gamma_q)$ in the $x-\tau$ plane for an averaged central event evolved with varying $\tau_\text{eq}$.}
    \label{fig:surface_comparison}
\end{figure}

While the isotherms are notably farther from the origin with larger equilibration times, the effect is almost negligible for the particlization hypersurfaces. This is due to the choice to particlize at $T_\text{c}(\gamma_q)$, which is higher for lower fugacities. Thus, when $\tau_\text{eq}$ is larger, we particlize across a surface with higher temperature and reduced quark content, but the volume and energy density change little. Fig.~\ref{fig:surface_comparison} demonstrates this by comparing the hypersurfaces for several values of $\tau_\text{eq}$.

\FloatBarrier

\begin{figure}[!t]
    \centering
    \includegraphics[width=0.7\linewidth]{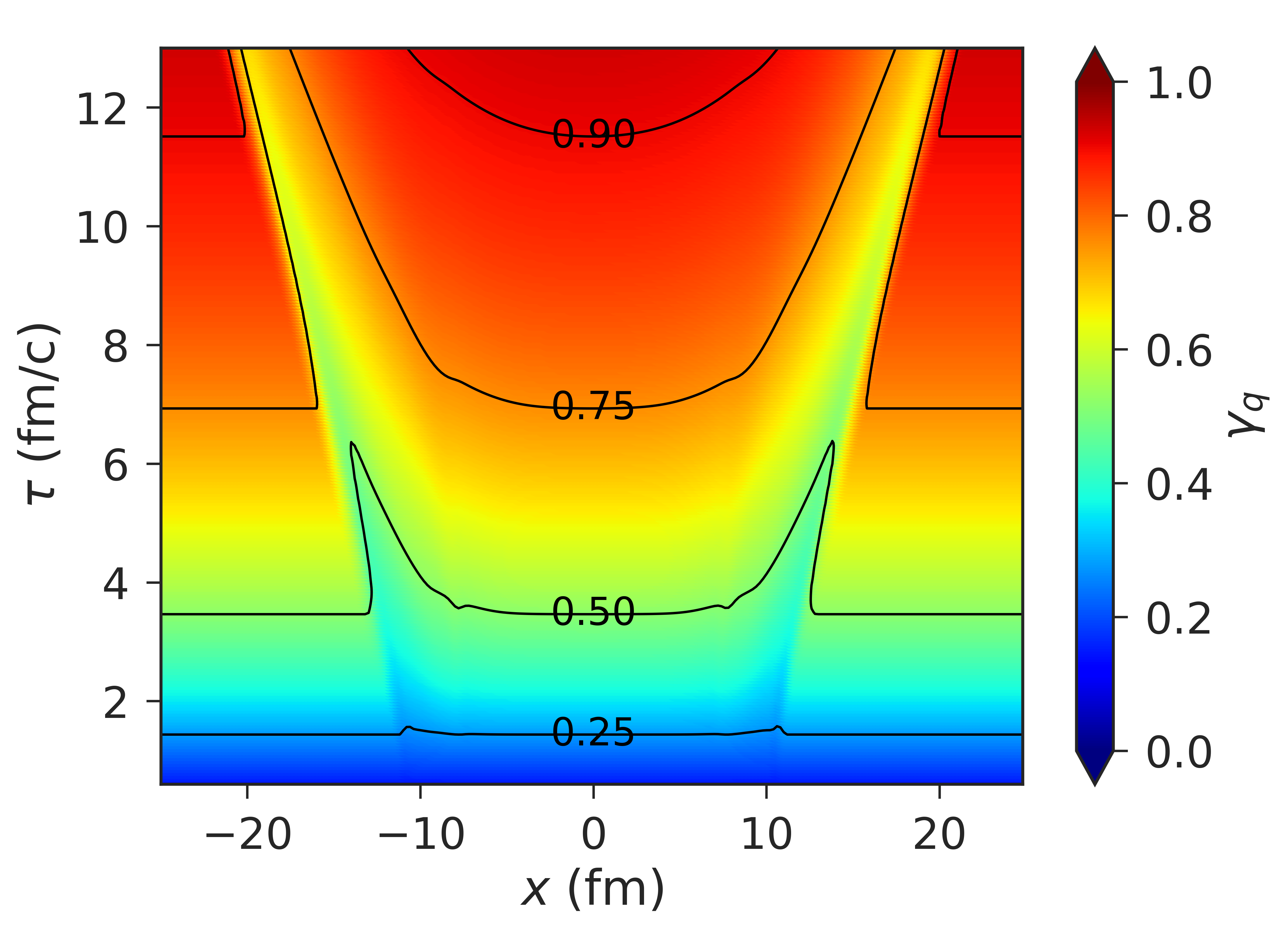}
    \caption[Fugacity in the $x-\tau$ plane at $\tau_\text{eq} = 5$~fm/$c$]{Contour plot of fugacity in the $x-\tau$ plane for an averaged central event evolved with $\tau_\text{eq} = 5$~fm/$c$.}
    \label{fig:fugacity_profile}
\end{figure}

Fig.~\ref{fig:fugacity_profile} shows the fugacity in the $x-\tau$ plane for the same event at $\tau_\text{eq} = 5$~fm/$c$. As expected from the proper time dependence of $\gamma_q$ in Eq.~\ref{eq:fugacity}, cells nearer the periphery of the fluid equilibrate more slowly due to their greater velocities. Note that for large enough $x$, where there is effectively zero energy density, the local proper time $\tau_\text{p}$ is equivalent to the global simulation time $\tau$. This is irrelevant in practice, as this dilute region is well outside the particlization hypersurface.

\begin{figure}[!htbp]
    \centering
    \includegraphics[width=0.7\linewidth]{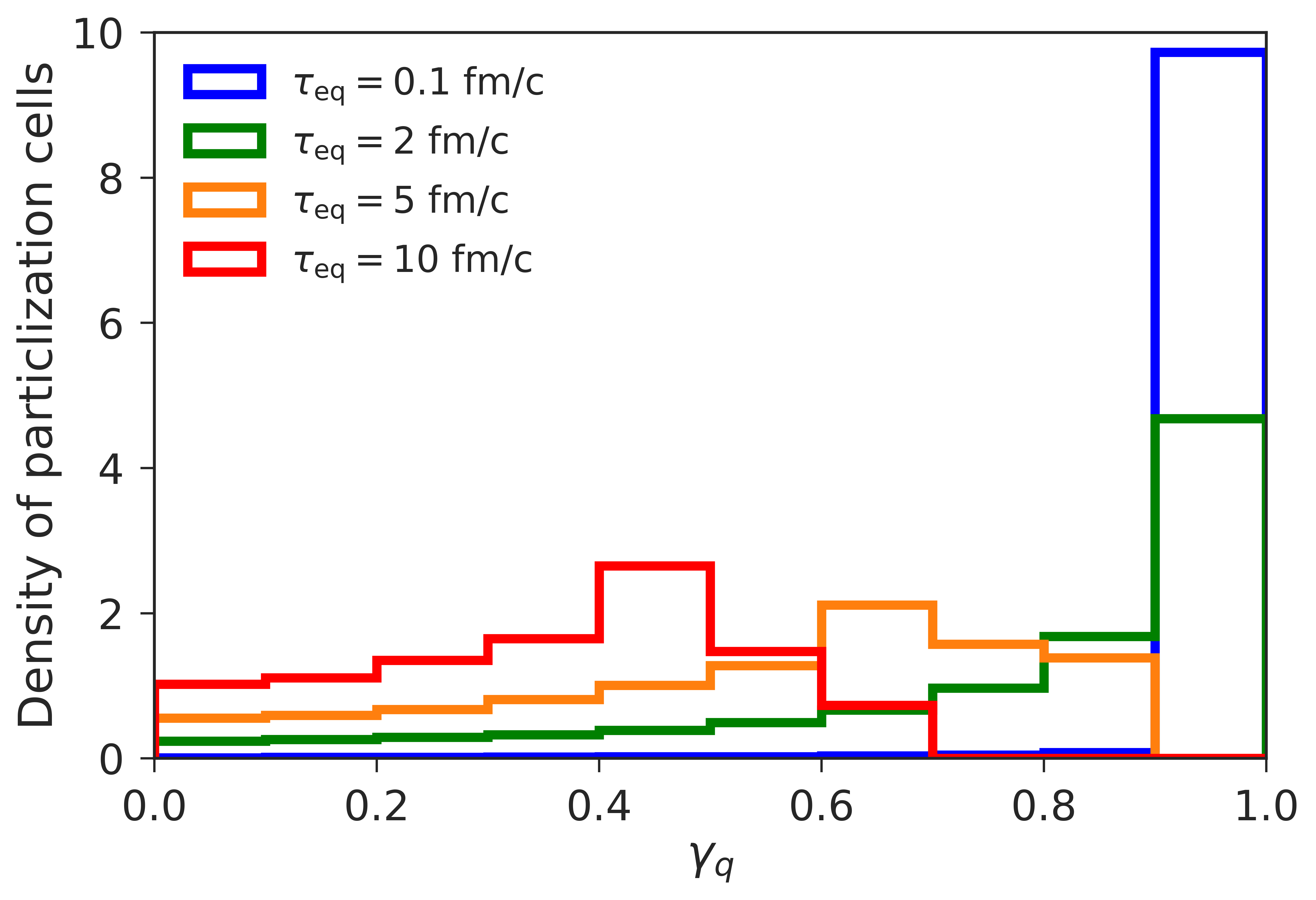}
    \caption[Fugacity distribution of hypersurface cells for varying $\tau_\text{eq}$]{Fugacity distribution of cells in the particlization hypersurface for an averaged central event evolved with varying $\tau_\text{eq}$.}
    \label{fig:fugacity_density}
\end{figure}

It is also worth noting that the entire surface at $\tau_\text{eq} = 5$~fm/$c$ is within the region where $\gamma_q < 0.9$, and much of it falls within $\gamma_q < 0.5$. To show this more clearly, Fig.~\ref{fig:fugacity_density} shows the fugacity distribution of cells in the surface for various $\tau_\text{eq}$. Unsurprisingly, selecting a small enough equilibration timescale produces a surface that is almost entirely equilibrated. For large enough $\tau_\text{eq}$, on the other hand, the entire medium can particlize far from equilibrium --- as low as $\gamma_q < 0.7$ for the largest timescale we consider, $\tau_\text{eq} = 10$~fm/$c$.

\begin{figure}[!htbp]
    \centering
    \includegraphics[width=0.7\linewidth]{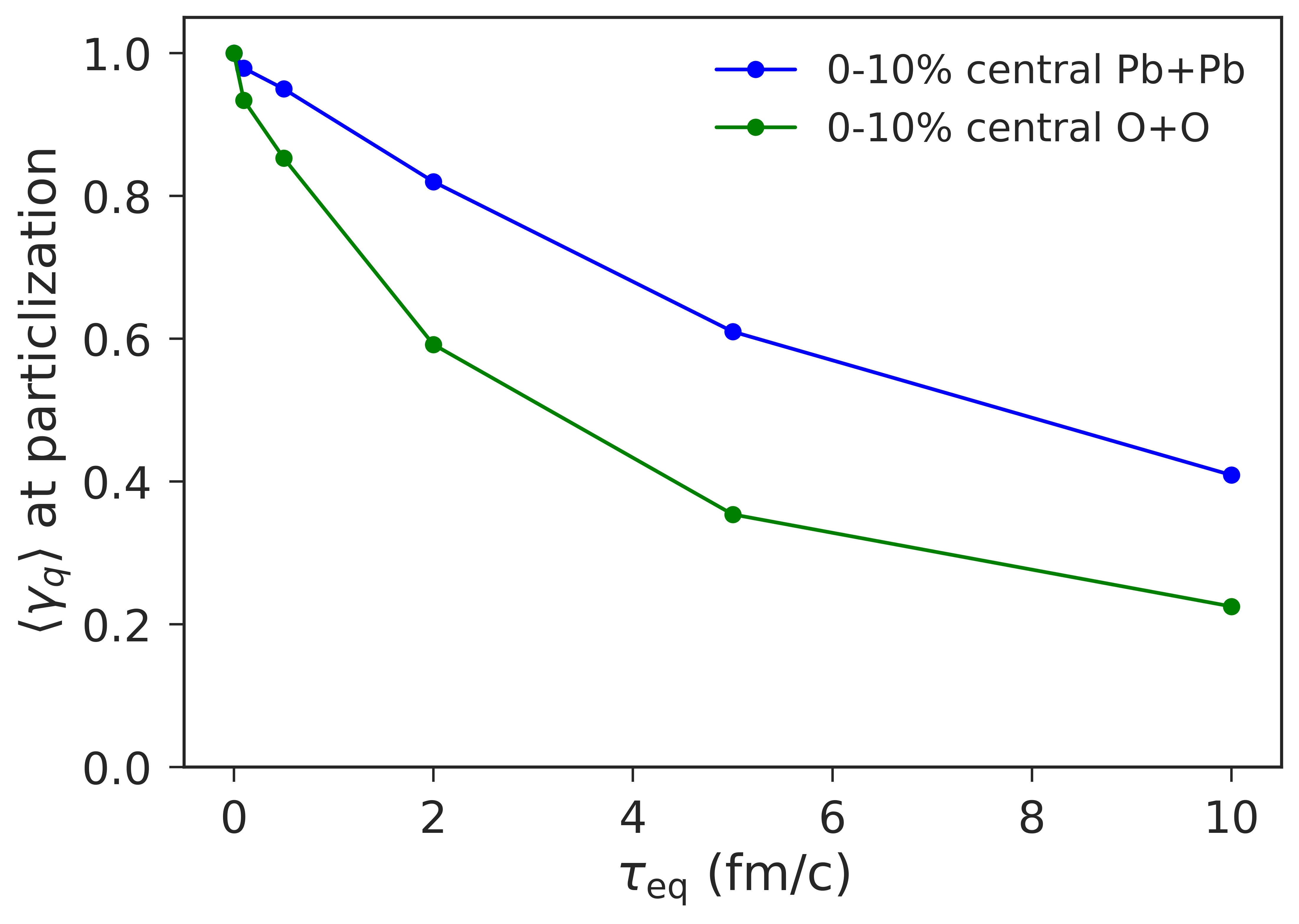}
    \caption[Mean hypersurface fugacity for Pb+Pb and O+O collisions]{Mean fugacity of the particlization hypersurface for averaged central Pb+Pb and O+O events evolved with varying $\tau_\text{eq}$.}
    \label{fig:fugacity_means}
\end{figure}

\begin{figure}[!htbp]
    \centering
    \includegraphics[width=0.7\linewidth]{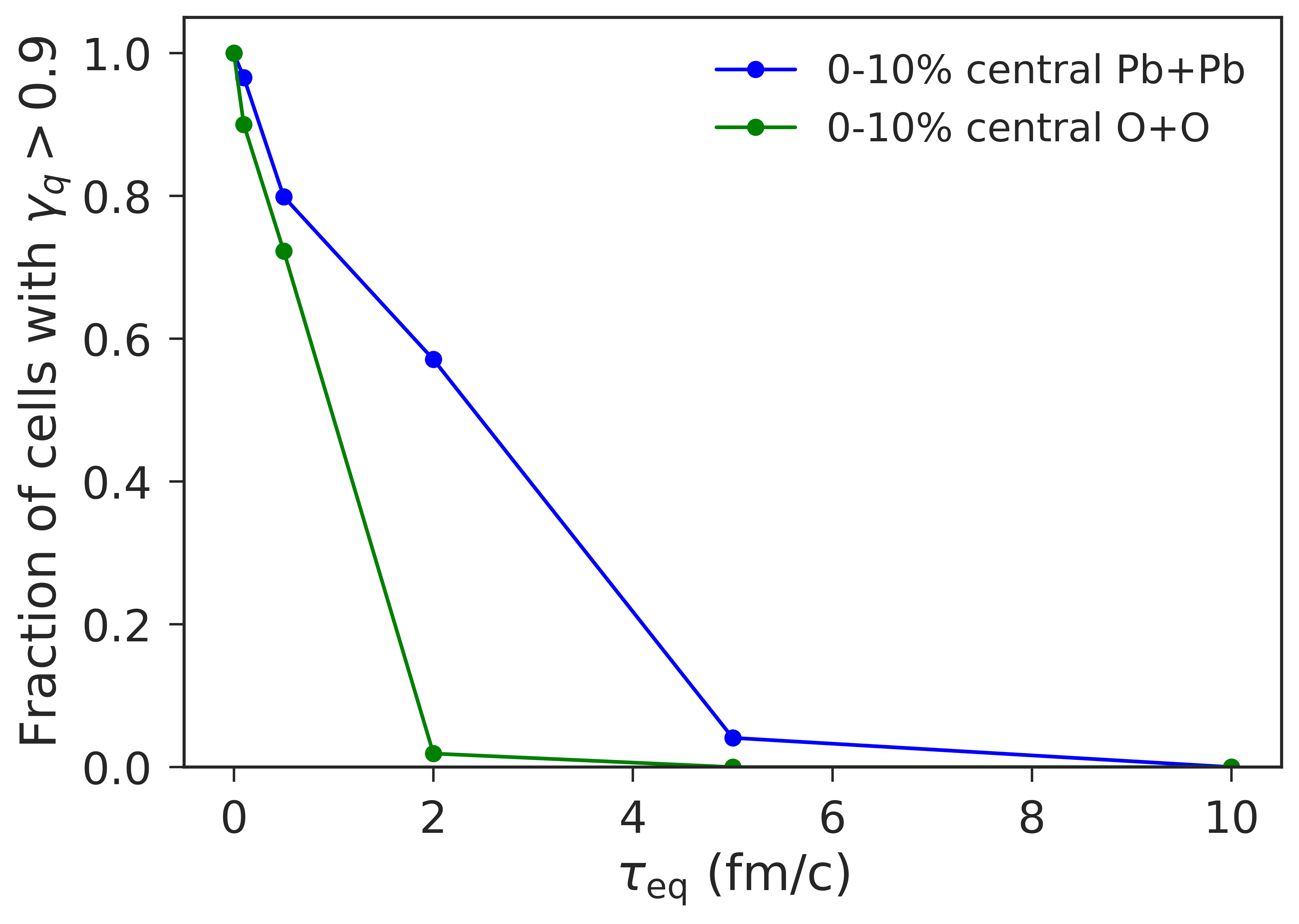}
    \caption[Fraction of near-equilibrium hypersurface cells for Pb+Pb and O+O]{Proportion of particlization hypersurface cells near chemical equilibrium for averaged central Pb+Pb and O+O events evolved with varying $\tau_\text{eq}$.}
    \label{fig:fugacity_0p9}
\end{figure}

In this work, we primarily consider Pb+Pb events, which are among the largest nuclei collided experimentally. However, the model being presented is general enough to apply to any collision sufficiently large to produce a QGP droplet. Smaller systems will tend to have less time to chemically equilibrate before hadronization, and we expect this to be reflected in the quark fugacity at particlization. To demonstrate this effect, we compare the particlization hypersurfaces obtained from event-averaged central Pb+Pb and O+O initial conditions. Fig.~\ref{fig:fugacity_means} shows the resulting mean fugacities as a function of $\tau_\text{eq}$. One can see that for any given $\tau_\text{eq} > 0$, the mean fugacity is consistently lower for O+O, as expected. Additionally, Fig.~\ref{fig:fugacity_0p9} shows the proportion of fluid cells in the hypersurface with a fugacity of $\gamma_q > 0.9$, which one can interpret as roughly the fraction of cells at or near chemical equilibrium. In this light, it is apparent that even for a moderate timescale of $\tau_\text{eq} = 2$~fm/$c$, almost none of the system is near chemical equilibrium in the case of an O+O collision.

\FloatBarrier

\subsection{Hadron production}
\label{subsec:hadrons_pce}

It is now apparent that for larger equilibration timescales, particlization occurs at a higher temperature and at lower quark fugacities. This leads to two competing effects on hadron production: Higher temperatures at particlization generally increase the yields of hadrons, as can be seen from their distribution functions in Eq.~\ref{eq:f_i}. Conversely, lower fugacities correspond to larger suppression factors in the hadron distribution functions. The net effect on hadron production is thus determined by the interplay between these two factors.

\begin{figure}[!b]
    \vspace{12pt}
    \centering
    \includegraphics[width=0.7\linewidth]{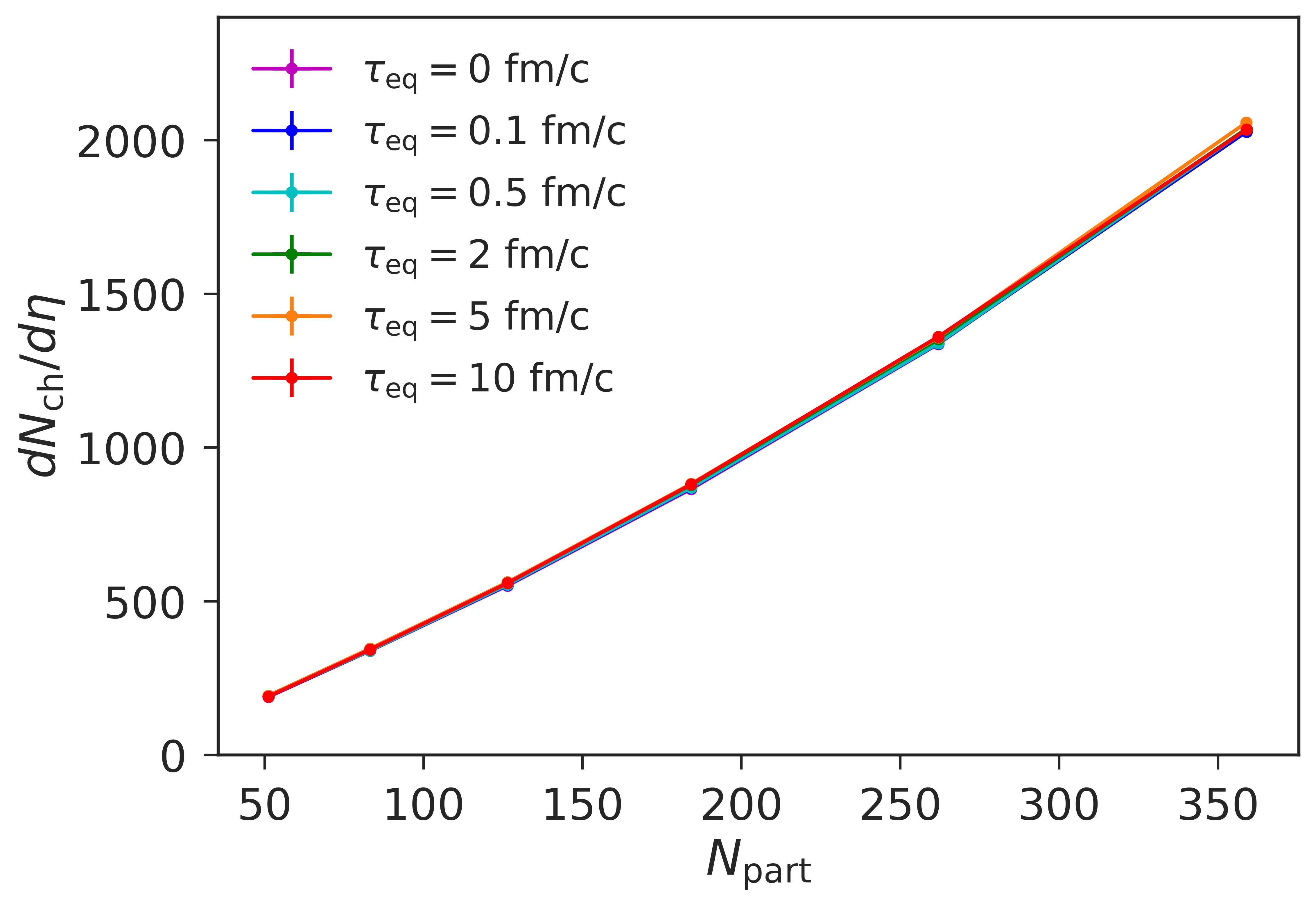}
    \caption[Charged particle multiplicity for varying $\tau_\text{eq}$]{Charged particle multiplicity for $0$--$60$\% centrality events evolved with varying $\tau_\text{eq}$. Each point corresponds to a $10$\% centrality bin.}
    \label{fig:dNch}
\end{figure}

For the following results, we consider an ensemble of $10,000$ minimum-bias events, each evolved at several values of $\tau_\text{eq}$. All observables are calculated at midrapidity with the cut $|\eta| < 0.5$, and the events are binned by centrality in increments of $10$\% according to their total initial entropy. Fig.~\ref{fig:dNch} shows the resulting charged particle multiplicity for the six most central bins ($0$--$60$\%), plotted against the number of participant nucleons in the initial collision, $N_\text{part}$. Surprisingly, when varying $\tau_\text{eq}$ from $0$ to $10$~fm/$c$, the multiplicities consistently agree within $\approx\!2$\%. This suggests that there is a close cancellation between the effects due to higher temperatures and lower fugacities at particlization.
\enlargethispage{2\baselineskip}

\begin{figure*}[!t]
    \centering
    \includegraphics[width=0.49\textwidth]{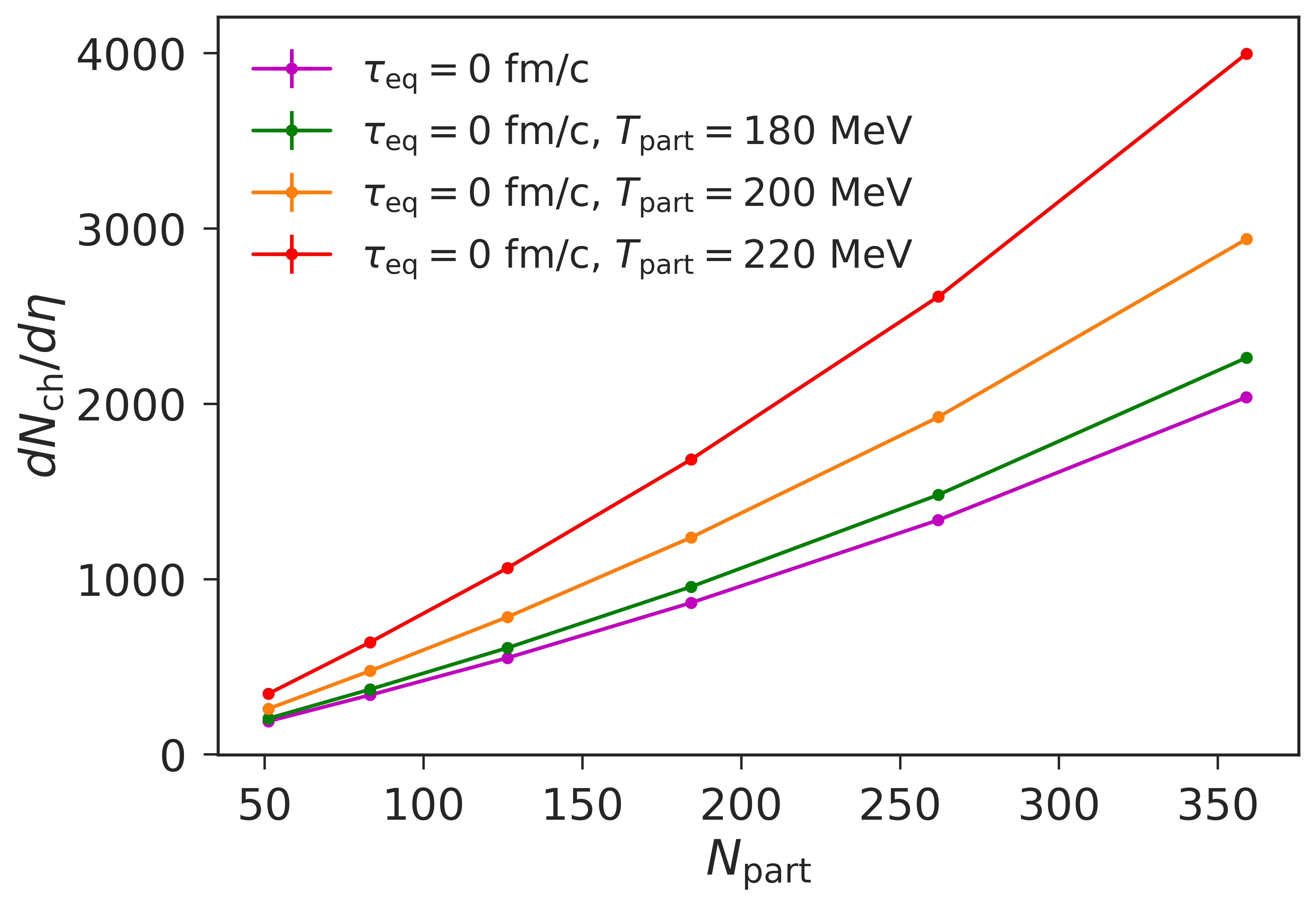}%
    \hfill
    \includegraphics[width=0.49\textwidth]{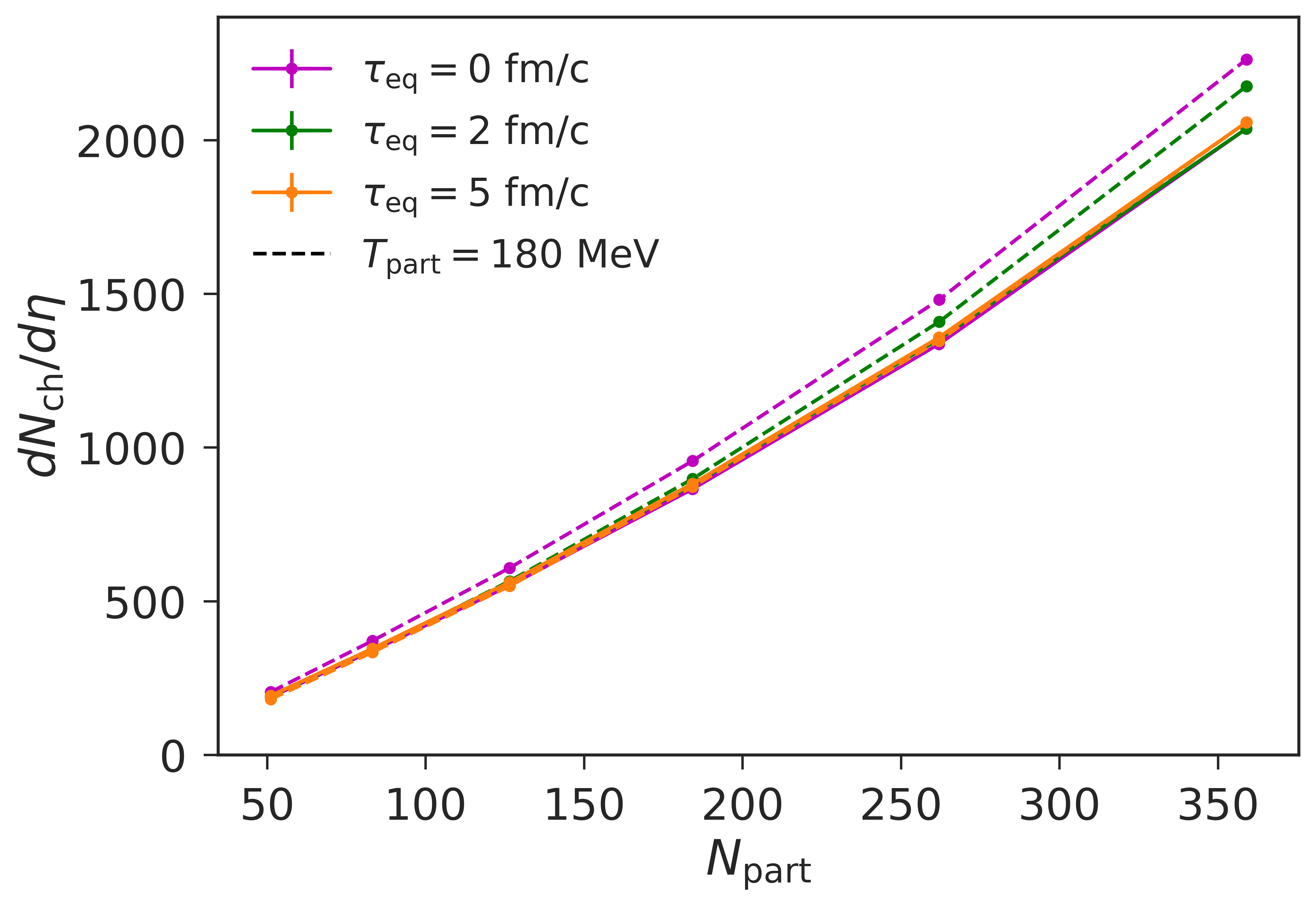}
    \caption[Charged particle multiplicity: separated temperature and fugacity effects]{Charged particle multiplicity for $0$--$60$\% centrality events evolved with varying particlization temperature $T_\text{part}$ (left) or varying $\tau_\text{eq}$ at a fixed $T_\text{part}$ (right). The curves are color-coded such that the value of $T_\text{part}$ for each curve on the left is approximately the mean particlization temperature for the corresponding colored solid curve on the right (e.g., $\langle T_\text{part} \rangle \approx 200$~MeV for $\tau_\text{eq} = 5$~fm/$c$). Each point corresponds to a $10$\% centrality bin.}
    \label{fig:dNch_temperature}
\end{figure*}

Fig.~\ref{fig:dNch_temperature} shows the effects due to temperature and fugacity individually on the charged particle multiplicities. It is clear from the left plot that increasing the particlization temperature substantially increases the multiplicity, by as much as a factor of two. At the same time, the right plot demonstrates that at a fixed particlization temperature, increasing the equilibration timescale reduces the multiplicity, albeit not as drastically.

\begin{figure*}[!t]
    \centering
    \includegraphics[width=0.49\textwidth]{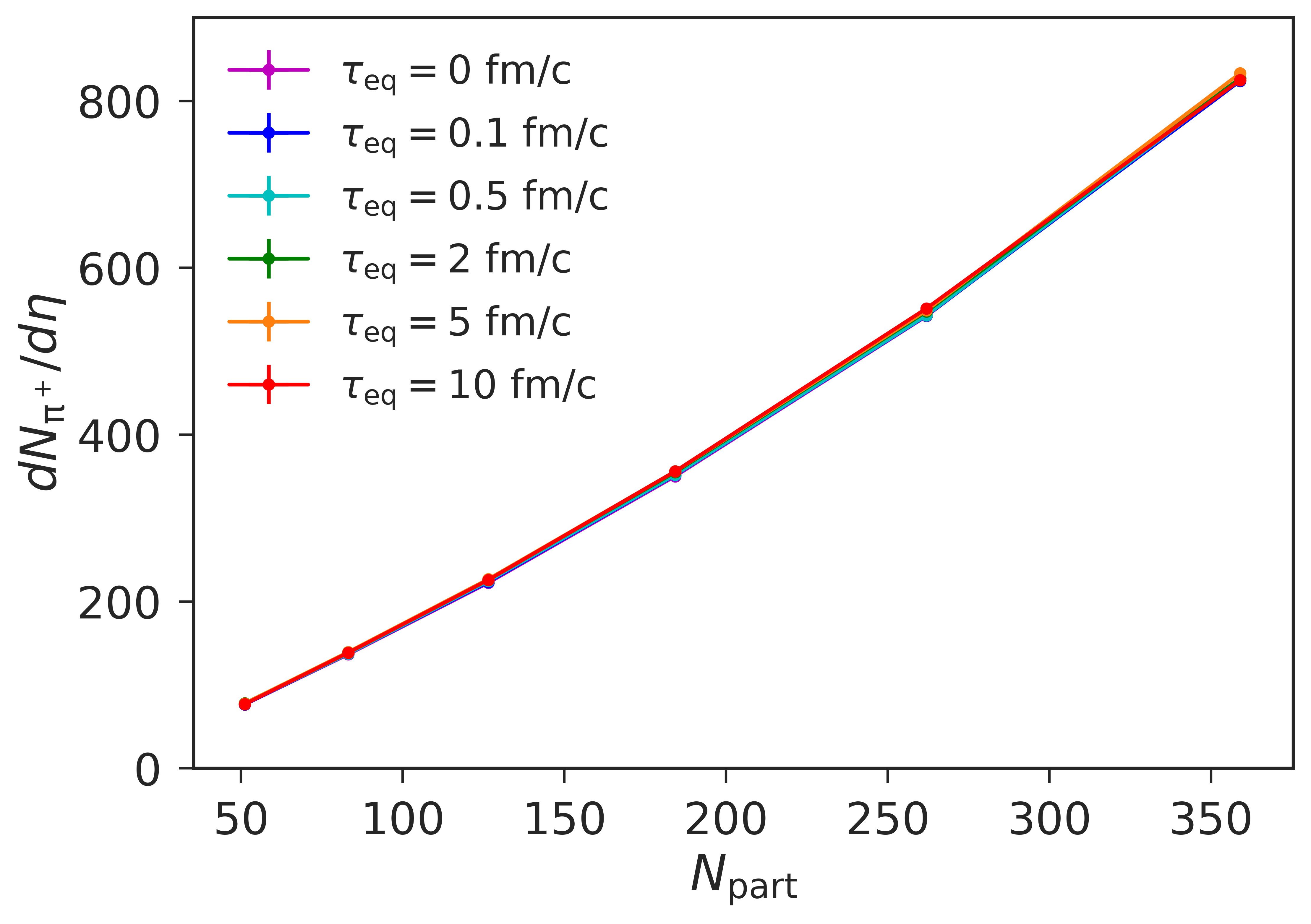}%
    \hfill
    \includegraphics[width=0.49\textwidth]{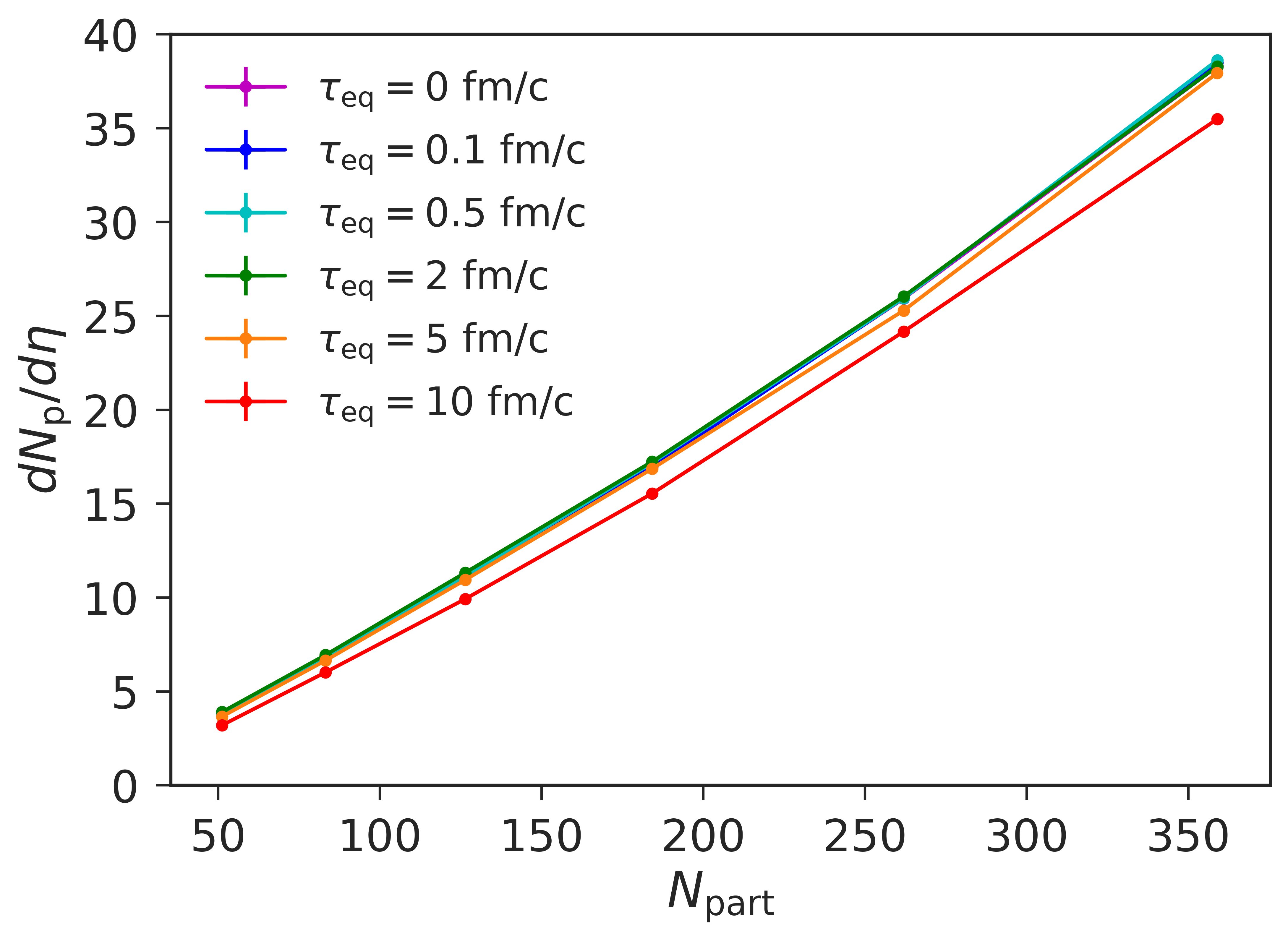}
    \caption[Pion and proton multiplicities for varying $\tau_\text{eq}$]{Pion (left) and proton (right) multiplicities for $0$--$60$\% centrality events evolved with varying $\tau_\text{eq}$. Each point corresponds to a $10$\% centrality bin.}
    \label{fig:dNppi}
\end{figure*}

\begin{figure}[!t]
    \centering
    \includegraphics[width=0.7\linewidth]{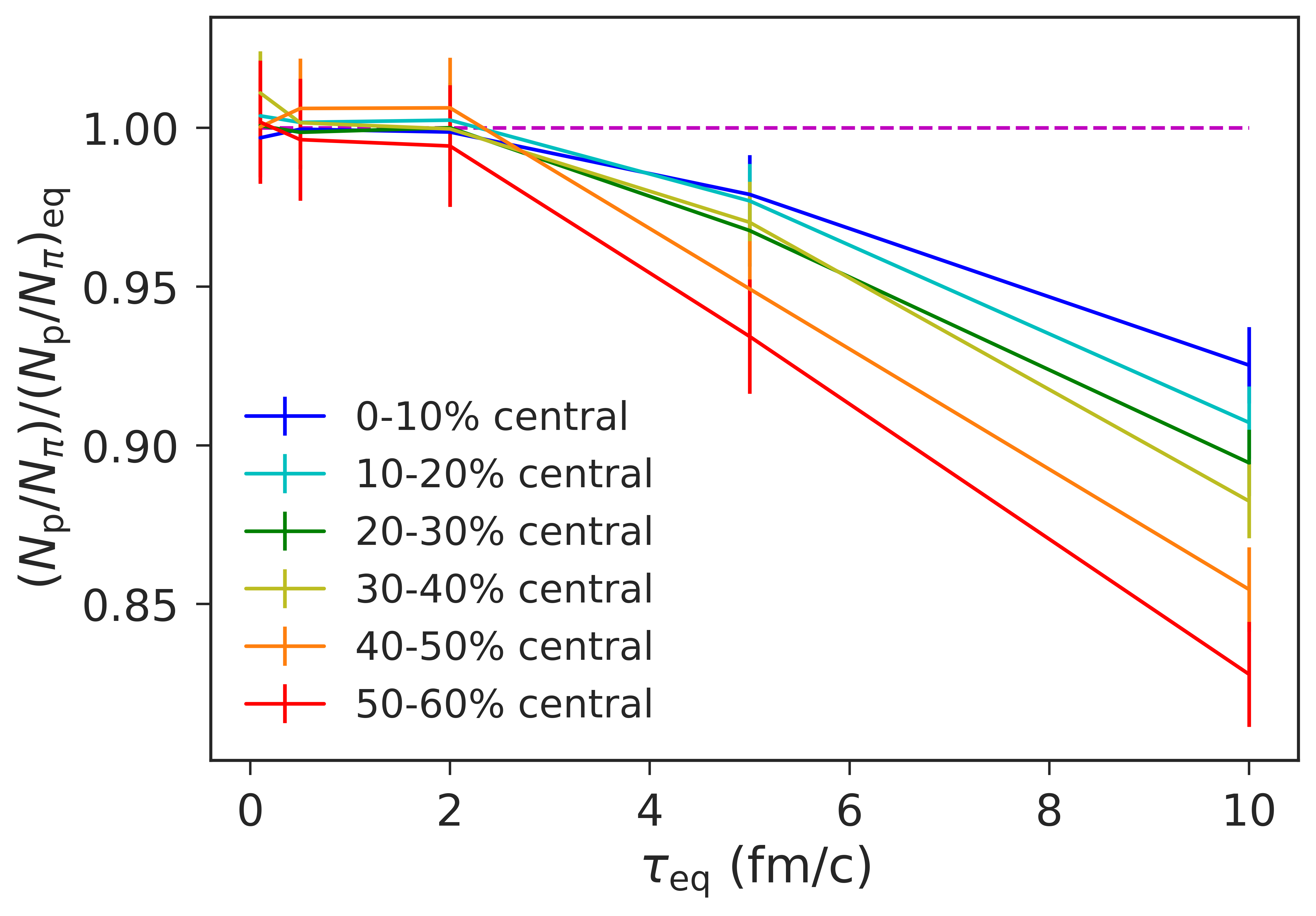}
    \caption[Proton-to-pion ratio for varying $\tau_\text{eq}$]{Modification to proton-to-pion ratio for $0$--$60$\% centrality events evolved with varying $\tau_\text{eq}$, as compared to this ratio with $\tau_\text{eq} = 0$~fm/$c$.}
    \label{fig:p_pi}
\end{figure}

Baryon suppression by a factor of $\approx\!1.5$ relative to expectations from thermal models has been observed experimentally in $\sqrt{s_{NN}} = 2.76$~TeV Pb+Pb collisions at the LHC~\cite{ALICE:2013mez}. This has been understood as a consequence of baryon--antibaryon annihilation into pions~\cite{Steinheimer:2012rd, Karpenko:2012yf, Becattini:2012xb}, but it has also been hypothesized that chemical undersaturation of the QGP may contribute, particularly for peripheral collisions~\cite{Vovchenko:2015yia}. This is evident in our model, as the fugacity factors assigned to the hadron distribution functions as defined in Eq.~\ref{eq:lambda_i} naturally suppress baryons more than mesons. Fig.~\ref{fig:dNppi} shows the pion and proton multiplicities produced by our model, while Fig.~\ref{fig:p_pi} shows the modification to the proton-to-pion ratio as a function of $\tau_\text{eq}$. For equilibration times up to $\tau_\text{eq} = 2$~fm/$c$, the variation in both the pion and proton multiplicities is negligible. Only for $\tau_\text{eq} \geq 5$~fm/$c$ do we observe the proton multiplicity and proton-to-pion ratio decrease, providing a weak signal of baryon suppression. However, this is a $\lesssim\!20$\% effect that does not explain the much larger suppression of the $p/\pi$ ratio seen in experiments, and instead should be considered as a correction in conjunction with other models of baryon suppression.

\Needspace{4\baselineskip}
\subsection{Transverse flow}
\label{subsec:flow_pce}

\begin{figure}[!b]
    \vspace{12pt}
    \centering
    \includegraphics[width=0.7\linewidth]{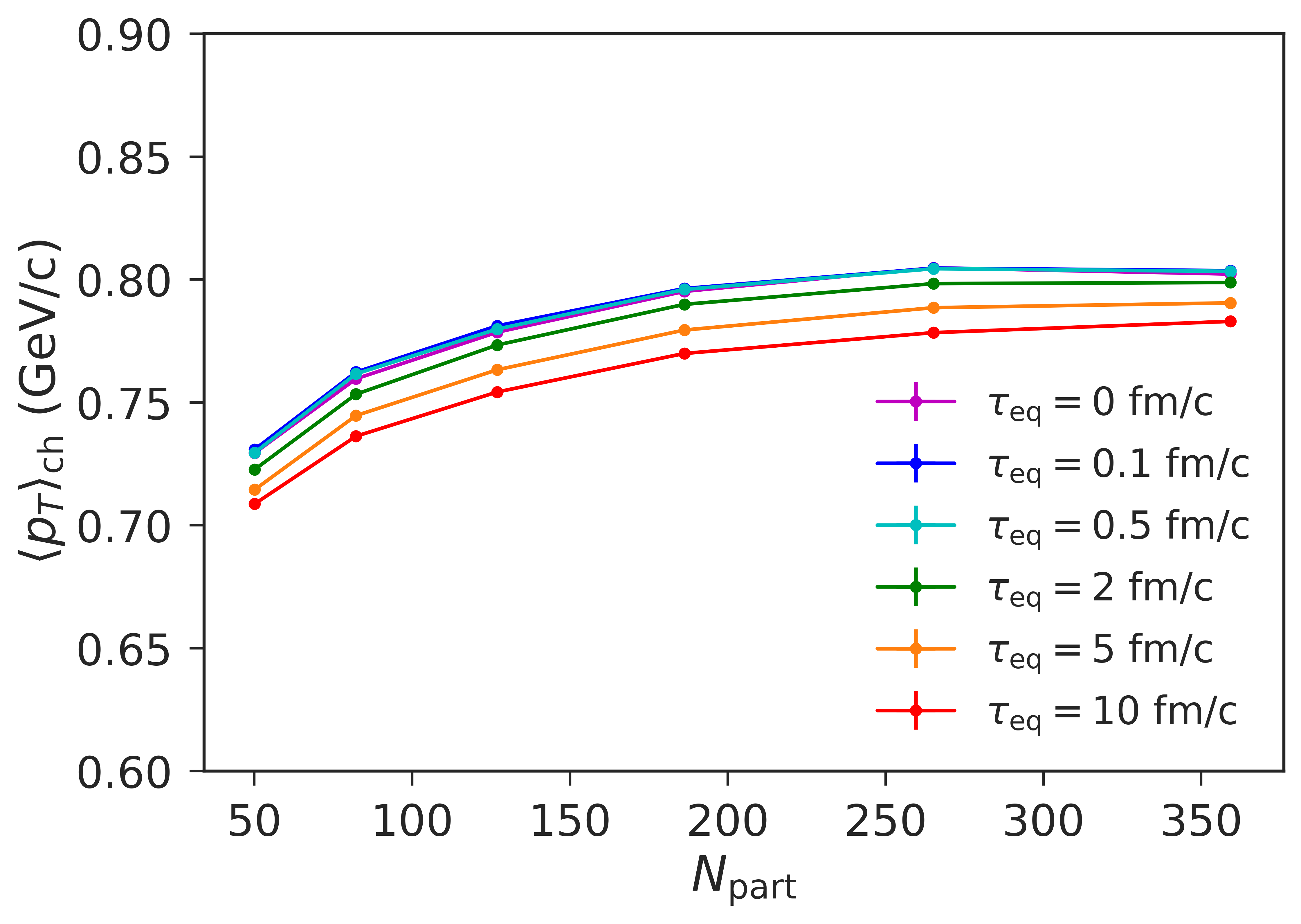}
    \caption[Charged particle $\langle p_T \rangle$ for varying $\tau_\text{eq}$]{Mean transverse momentum of charged particles for $0$--$60$\% centrality events evolved with varying $\tau_\text{eq}$. Each point corresponds to a $10$\% centrality bin.}
    \label{fig:pTch}
\end{figure}

\begin{figure*}[!htbp]
    \centering
    \includegraphics[width=0.49\textwidth]{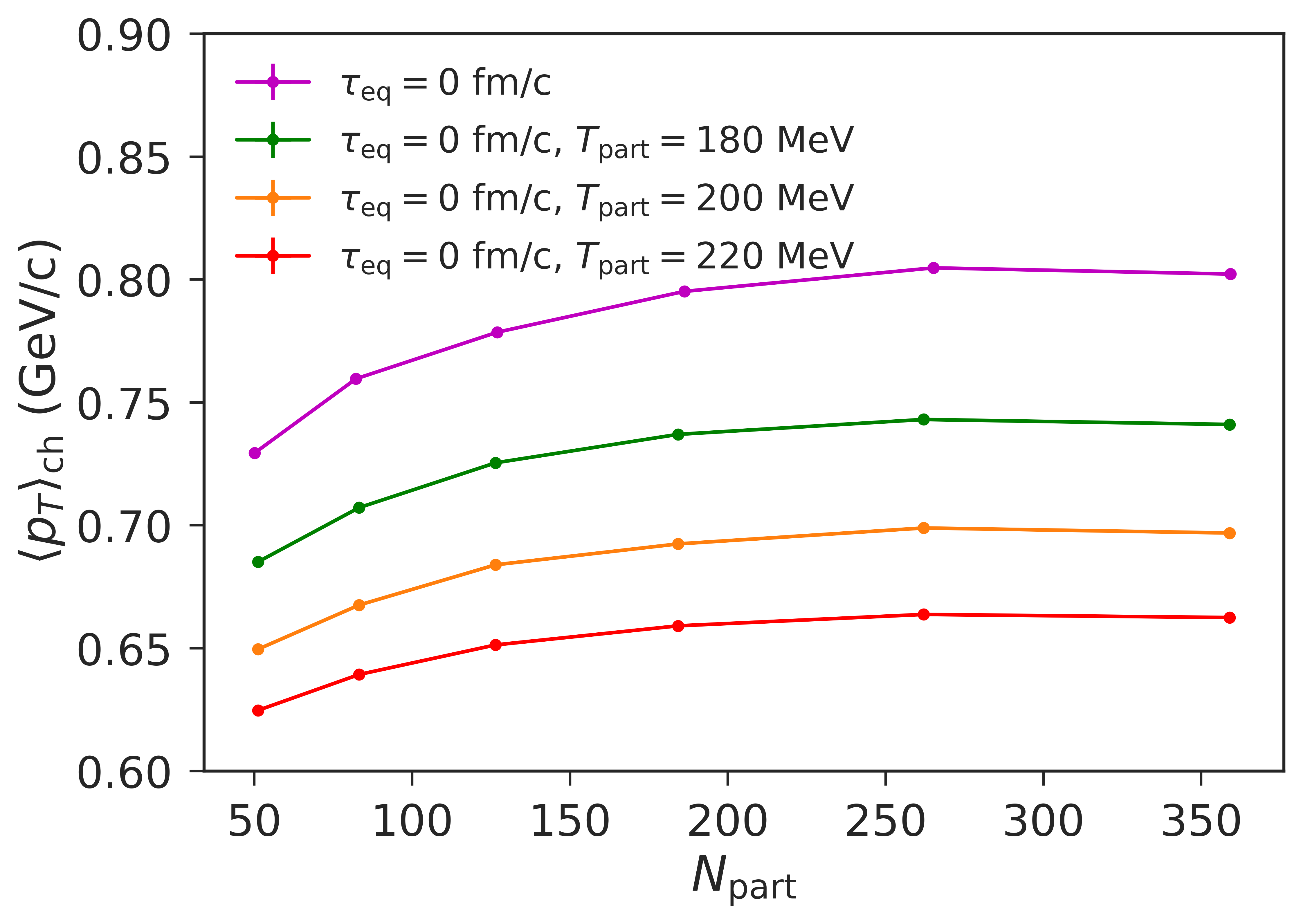}%
    \hfill
    \includegraphics[width=0.49\textwidth]{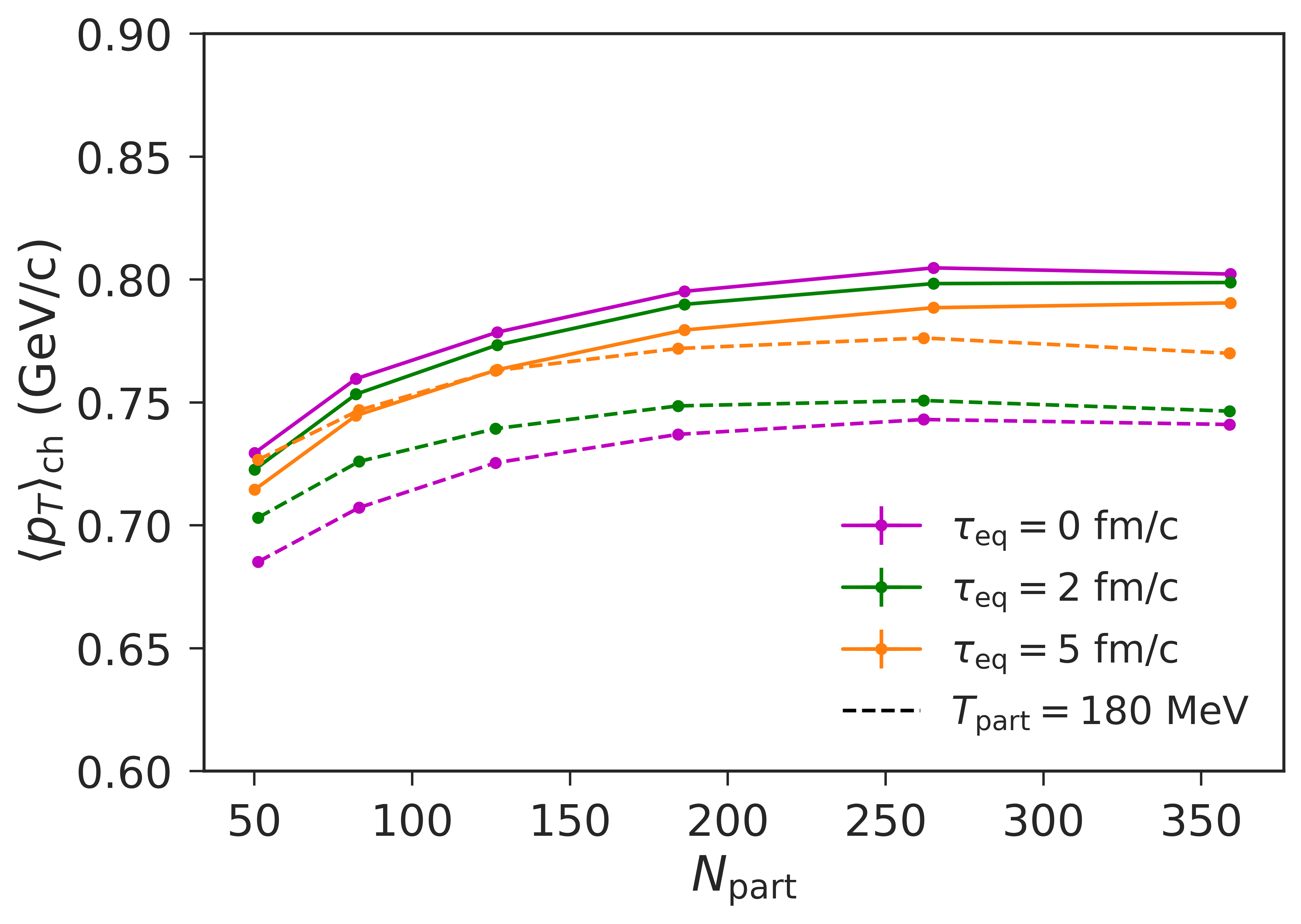}
    \caption[Charged particle $\langle p_T \rangle$: separated temperature and fugacity effects]{Mean transverse momentum of charged particles for $0$--$60$\% centrality events evolved with varying particlization temperature $T_\text{part}$ (left) or varying $\tau_\text{eq}$ at a fixed $T_\text{part}$ (right). The curves are color-coded such that the value of $T_\text{part}$ for each curve on the left is approximately the mean particlization temperature for the corresponding colored solid curve on the right (e.g., $\langle T_\text{part} \rangle \approx 200$~MeV for $\tau_\text{eq} = 5$~fm/$c$). Each point corresponds to a $10$\% centrality bin.}
    \label{fig:pTch_temperature}
\end{figure*}

Quark chemical equilibration can affect not only the production of hadrons at particlization, but also the development of flow during the evolution of the QGP. This is ultimately reflected in the momenta of the final-state hadrons. Fig.~\ref{fig:pTch} shows the mean transverse momentum $\langle p_T \rangle$ of charged particles, while Fig.~\ref{fig:pTch_temperature} shows the independent effects of the particlization temperature and fugacity.

As $\tau_\text{eq}$ increases, the mean transverse momentum systematically decreases. In chemical equilibrium, increasing the particlization temperature reduces the evolution time and thus allows less flow to develop, as in Fig.~\ref{fig:pTch_temperature}. However, as shown in Fig.~\ref{fig:surface_comparison}, the evolution time does not decrease with $\tau_\text{eq}$. In fact, when $\tau_\text{eq}$ is increased and the particlization temperature is fixed, as in Fig.~\ref{fig:pTch_temperature}, the evolution time and thus mean $\langle p_T \rangle$ increase. The decrease in $\langle p_T \rangle$ should instead be attributed to the reduced pressure out of equilibrium, as lower quark fugacities correspond to lower $P$ throughout the evolution, and it is pressure gradients that drive the development of transverse flow.

\begin{figure}[!htbp]
    \centering
    \includegraphics[width=0.7\linewidth]{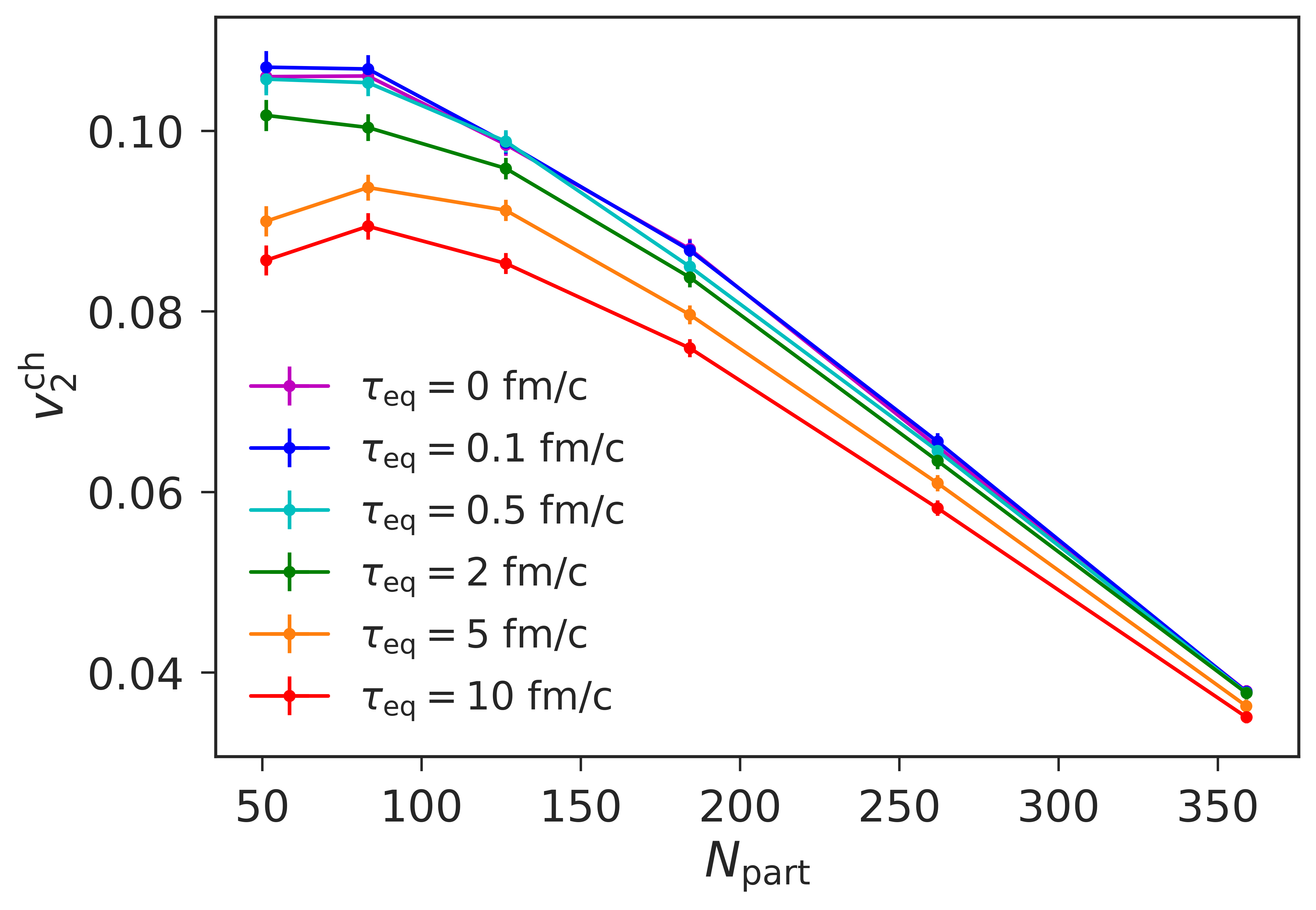}
    \caption[Charged particle elliptic flow for varying $\tau_\text{eq}$]{Elliptic flow for $0$--$60$\% centrality events evolved with varying $\tau_\text{eq}$. Each point corresponds to a $10$\% centrality bin.}
    \label{fig:v2ch}
\end{figure}

A similar effect can be observed with the flow anisotropy. Transverse momentum anisotropy is commonly expressed in terms of the coefficients $v_n$ of the Fourier expansion
\begin{align}
    \frac{dN}{d\phi} \propto 1 + 2 \sum_{n=1}^\infty v_n \cos{n(\phi - \Phi_n)},
\end{align}
where $\phi = \mathrm{atan2}(p_y,p_x)$ is the azimuthal angle and $\Phi_n$ is the event-plane angle. The elliptic flow $v_2$ is particularly useful as a measure of the conversion from elliptical anisotropy in the initial condition to momentum anisotropy. For the ensemble of simulated events, the $v_2$ of charged particles is shown in Fig.~\ref{fig:v2ch}. The suppression of $v_2$ due to quark chemical equilibration is greater than that of $\langle p_T \rangle$, and this is particularly noticeable for less central events. This is to be expected, as the smaller size and shorter lifetime of less central events amplify the effects of weaker pressure gradients.

It is known that $v_2$ tends to decrease with greater shear and bulk viscosities~\cite{Song:2007ux, Shen:2011kn, Dusling:2011fd, Ryu:2017qzn}, so one can interpret quark chemical equilibration as effectively increasing these viscosities. This mimics the chemical potential dependence of the specific viscosities anticipated in section~\ref{section14}, here arising from the finite quark chemical potential of the undersaturated medium; it is best read at the level of observables, as the suppression of $v_2$ and $\langle p_T \rangle$ originates in the reduced pressure of the medium and not in any change to the input $\eta/s$. In this work, we did not vary $\Pi$ and $\pi^{\mu \nu}$ with the quark fugacity, but this result highlights the necessity of doing so in future studies. When modeling the QCD medium in partial chemical equilibrium, smaller viscosities may yield better agreement with experimental data due to the effective viscosity increase implied by quark chemical equilibration. The interplay between chemical equilibration and the QGP transport coefficients is examined directly in the Bayesian calibration of chapter~\ref{chapter5}.

\subsection{Thermal photon production}
\label{subsec:photons_pce}

Photons are emitted from various sources throughout the stages of a heavy-ion collision: prompt photons are produced from nucleon interactions in the initial stages, thermal photons are emitted by the thermalized medium as it expands, and additional photons are produced by hadronic decays. We focus on thermal photons here as the most direct probe of the deconfined phase.

To leading order, there are four processes by which thermal photons are produced in the QGP: gluon-photon Compton scattering, quark-antiquark annihilation, bremsstrahlung, and inelastic pair annihilation~\cite{Arnold:2001ms}. With a medium in partial chemical equilibrium, the photon production rate for each process should be suppressed by a factor of $\gamma_q$ for each (anti)quark. Gluon-photon Compton scattering ($q/\bar{q} + g \rightarrow q/\bar{q} + \gamma$) is then suppressed linearly by $\gamma_q$, and quark-antiquark annihilation ($q + \bar{q} \rightarrow g + \gamma$) is suppressed quadratically by $\gamma_q^2$. It is more difficult to determine the correct suppression for the two inelastic processes. Bremsstrahlung may occur with either two (anti)quarks or one (anti)quark and one gluon, and inelastic pair annihilation involves scattering of a (anti)quark on another parton that may be either a (anti)quark or gluon.

Adopting the scheme of Ref.~\cite{Vovchenko:2016ijt}, we compare two approximations in which the inelastic processes are suppressed either linearly or quadratically with respect to $\gamma_q$. Defining $\Gamma(k,T,\gamma_q)$ as the production rate for photons with energy $k$ in a fluid cell with temperature $T$ and fugacity $\gamma_q$, we have:
\begin{equation}
\begin{aligned}
\Gamma(k,T,\gamma_q) = \gamma_q\, \Gamma_{\text{Compton}}(k,T)
 + \gamma_q^2\, \Gamma_{\text{annihilation}}(k,T) + \gamma_q^n\, \Gamma_{\text{inelastic}}(k,T),
\end{aligned}
\end{equation}
where $n \in \{1,2\}$, and $\Gamma_{\text{Compton}}$, $\Gamma_{\text{annihilation}}$, and $\Gamma_{\text{inelastic}}$ are the respective photon production rates of gluon-photon Compton scattering, quark-antiquark annihilation, and inelastic processes. $n = 1$ corresponds to a scheme that overestimates the total thermal photon production rate, and $n = 2$ corresponds to a scheme that underestimates it. Using the expressions for these three rates given in Ref.~\cite{Arnold:2001ms}, we calculate thermal photon emission for the single averaged central event evolved at varying $\tau_\text{eq}$ referred to in section~\ref{subsec:evolution_pce}. We also neglect viscous corrections here, and only consider the ideal photon production rate.

\begin{figure}[!htbp]
    \centering
    \includegraphics[width=0.75\linewidth]{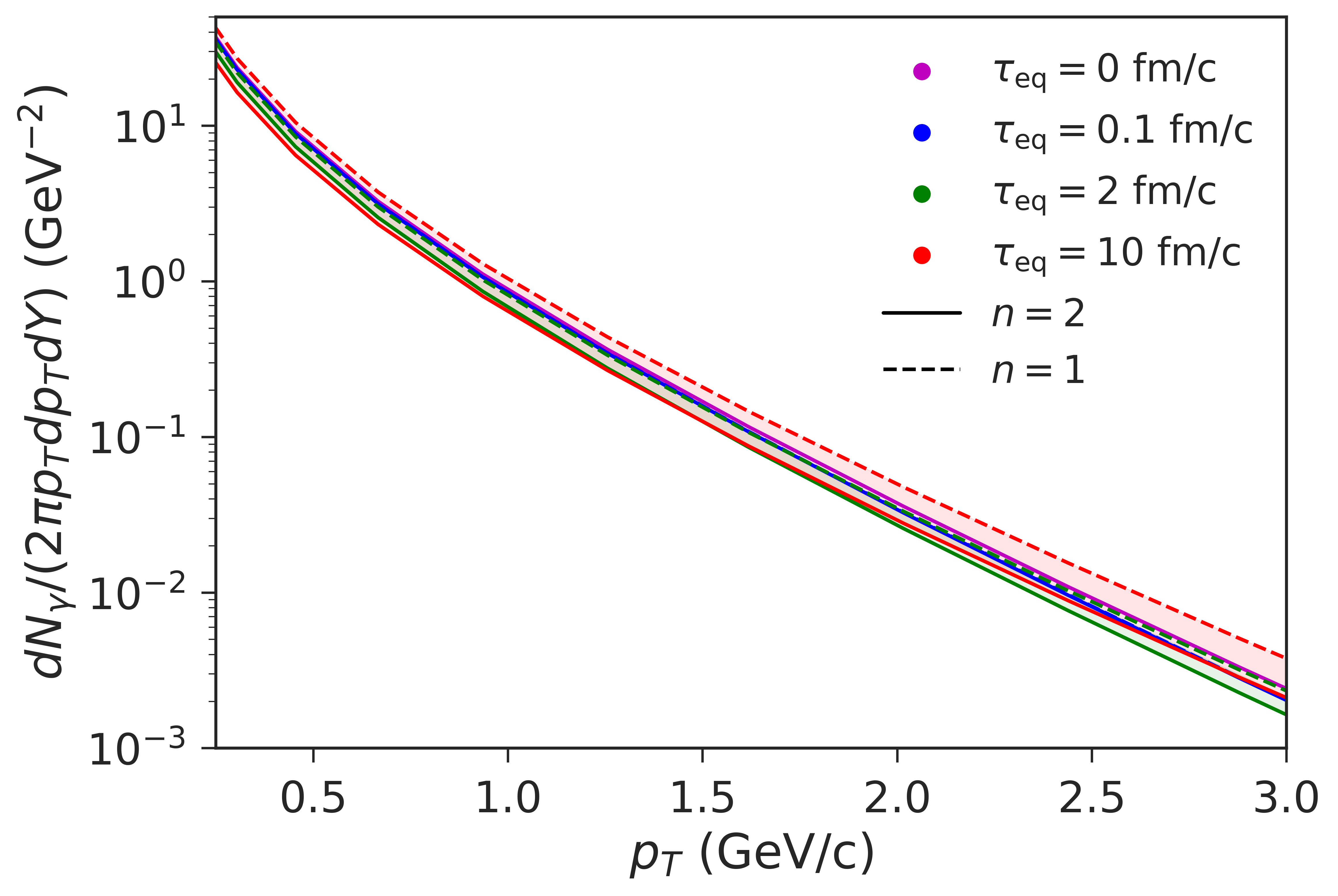}
    \includegraphics[width=0.75\linewidth]{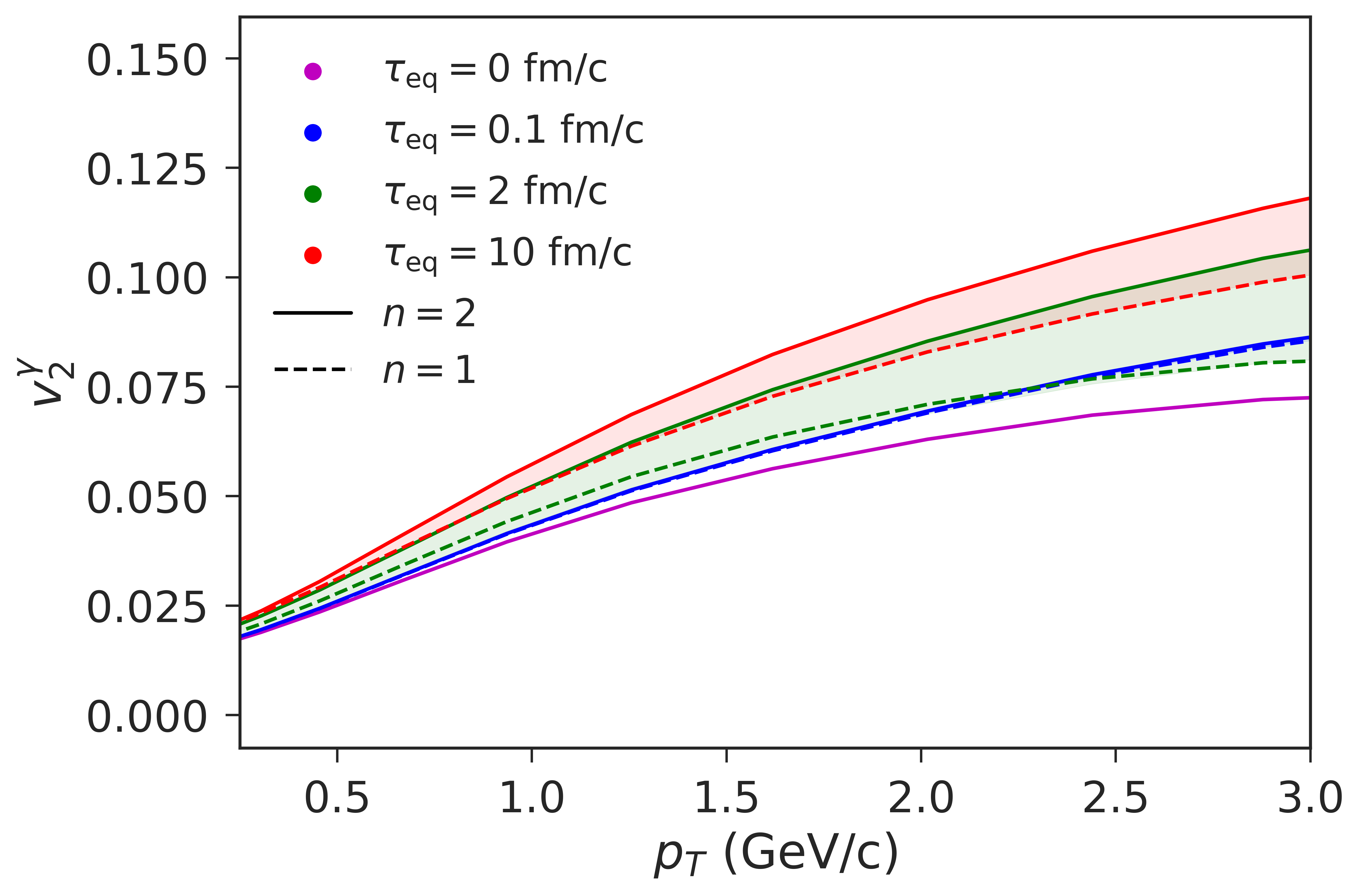}
    \caption[Thermal photon spectrum and elliptic flow for varying $\tau_\text{eq}$]{Thermal photon spectrum (top) and elliptic flow (bottom) for an averaged central event evolved with varying $\tau_\text{eq}$, using a cutoff temperature of $150$~MeV. Solid lines correspond to quadratic scaling of $\Gamma_{\text{inelastic}}$ with respect to $\gamma_q$ and dashed lines correspond to linear scaling.}
    \label{fig:photons}
\end{figure}

For ($2+1$)-dimensional boost-invariant hydrodynamics, the $p_T$ spectrum of thermal photons is given by
\begin{equation}
\frac{dN_\gamma}{d^2p_T\,dy}
= \int d^2x_T d\tau d\eta_s \tau
  \Gamma(k,T,\gamma_q)\,
  \theta\!\left(T - T_{\text{min}}\right),
\end{equation}
where $\theta(x)$ is the Heaviside step function and $T_{\text{min}}$ is the minimum temperature for emission, here set to $150$~MeV. In other words, we neglect late-stage photon emission from fluid cells with $T < 150$~MeV. Fig.~\ref{fig:photons} shows the resulting spectrum together with the elliptic flow of thermal photons, which is given by
\begin{align}
    v_2^\gamma (p_T) = \frac{\int_0^{2\pi} d\phi \frac{dN_\gamma}{d^2p_Tdy} \cos(2\phi)}{\int_0^{2\pi} d\phi \frac{dN_\gamma}{d^2p_Tdy}}.
\end{align}

The results in Fig.~\ref{fig:photons} show that both the spectra and elliptic flow of thermal photons are sensitive to the quark chemical equilibration timescale. Most notably, $v_2^\gamma$ tends to be enhanced at larger $\tau_\text{eq}$, similar to what was observed in Ref.~\cite{Vovchenko:2016ijt}. However, the theoretical uncertainty due to the choice of linear or quadratic scaling for $\Gamma_{\text{inelastic}}$ is large enough that it is not possible to meaningfully quantify this effect. A more precise determination of how inelastic processes that produce thermal photons scale with the quark fugacity will be essential to use experimental data as a way to determine $\tau_\text{eq}$.

\FloatBarrier

\section{Separating light and strange flavor equilibration}
\label{section33}

The flavor-independent model of section~\ref{section31} assumes that all quark flavors equilibrate together through a single quark fugacity $\gamma_q$. This is a useful first approximation, but it cannot distinguish the chemical equilibration of light quarks from that of strange quarks. We therefore extend the model by introducing independent light and strange fugacities, $\gamma_l$ and $\gamma_s$, with corresponding effective equilibration timescales $\tau_{\text{eq},l}$ and $\tau_{\text{eq},s}$ and initial values $\gamma_l^0$ and $\gamma_s^0$. Here $\gamma_l$ describes the chemical occupancy of light (up and down) quarks and antiquarks, while $\gamma_s$ describes that of strange quarks and antiquarks. 

These parameters are not intended to model the microscopic production channels directly. Rather, they provide a phenomenological way to vary the chemical occupancy of light and strange quarks independently and determine how final-state observables respond. For this reason, the model does not impose a prior hierarchy between the two equilibration timescales, even though one might expect strange quarks to equilibrate more slowly on microscopic grounds. The hydrodynamic equations of motion remain unchanged; the extension enters through the equation of state, the local fugacity evolution, and the particlization prescription.

\subsection{Equation of state}
\label{subsec:eos_strange}

The flavor-dependent equation of state generalizes the construction of section~\ref{subsec:eos_pce}. Instead of interpolating between the pure glue and $(2+1)$-flavor equations of state, we interpolate among three limiting systems: pure glue QCD, two-flavor QCD, and $(2+1)$-flavor QCD. The pure glue limit corresponds to $\gamma_l = \gamma_s = 0$, the two-flavor limit corresponds to equilibrated light quarks with absent strange quarks, and the $(2+1)$-flavor limit corresponds to $\gamma_l = \gamma_s = 1$.

\begin{figure}[!b]
    \vspace{12pt}
    \centering
    \includegraphics[width=0.7\textwidth]{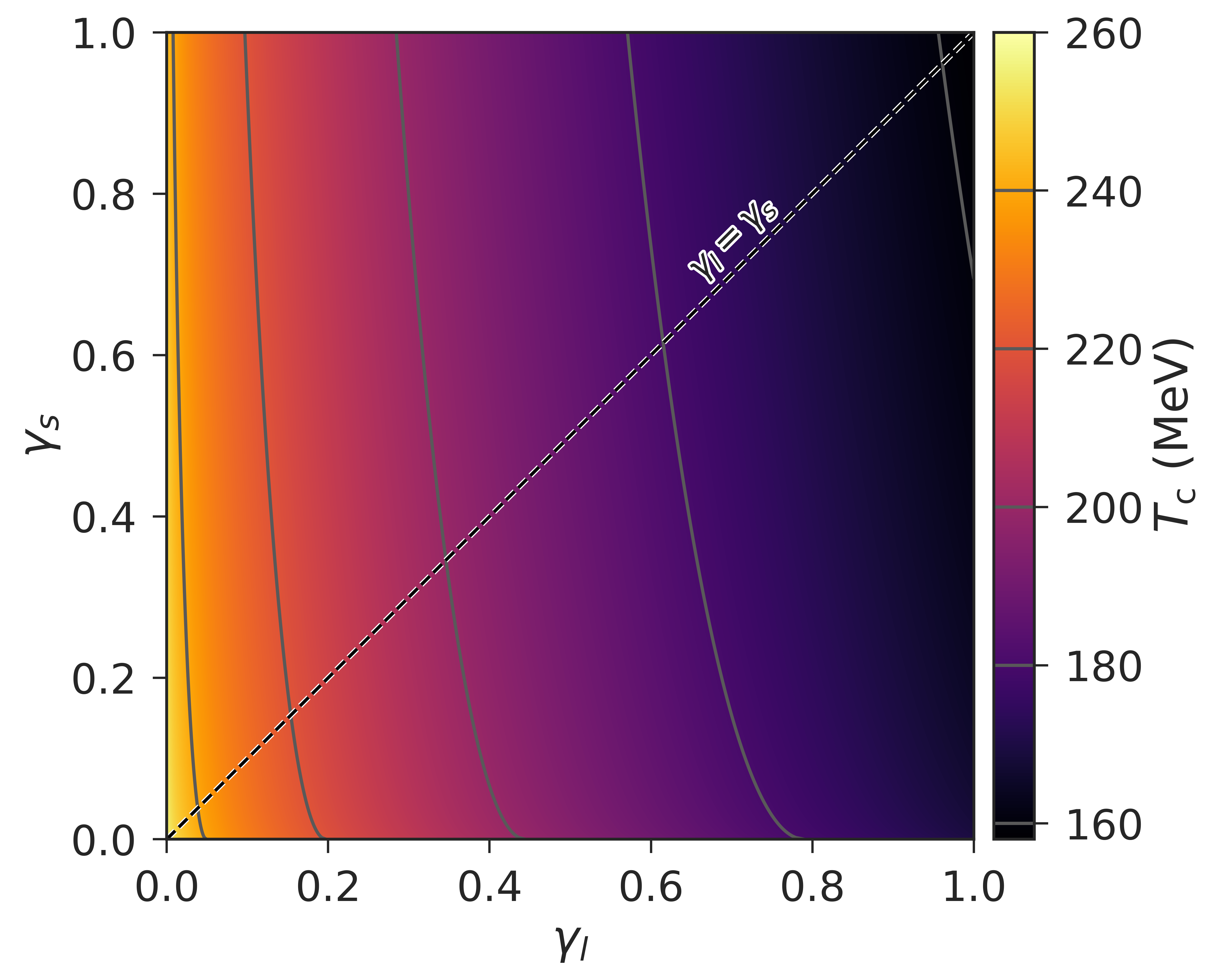}
    \caption[Fugacity dependence of the critical temperature]{Critical
    temperature $T_\text{c}(\gamma_l,\gamma_s)$ from Eq.~\ref{eq:Tc_strange},
    shown across the light and strange fugacity plane with $T_0=260$, $T_2=170$,
    and $T_3=158$~MeV and gray contours every $20$~MeV. The three corners are
    the pure glue ($\gamma_l=\gamma_s=0$), two-flavor ($\gamma_l=1$,
    $\gamma_s=0$), and $(2+1)$-flavor ($\gamma_l=\gamma_s=1$) limits, at which
    $T_\text{c}$ equals $T_0$, $T_2$, and $T_3$ respectively. Along the dashed
    diagonal $\gamma_l=\gamma_s$ the construction reduces to the single-flavor
    critical temperature of Fig.~\ref{fig:Tc}.}
    \label{fig:Tc_strange}
\end{figure}

As in the flavor-independent case, the temperature dependence of each branch is rescaled so that the transition occurs at a single fugacity-dependent critical temperature. We define
\begin{align}
    T_\text{c}(\gamma_l,\gamma_s)
    =
    \sqrt{\gamma_s} T_3
    +
    (\sqrt{\gamma_l} - \sqrt{\gamma_s}) T_2
    +
    (1-\sqrt{\gamma_l}) T_0,
    \label{eq:Tc_strange}
\end{align}
where $T_0$, $T_2$, and $T_3$ denote the characteristic transition temperatures of the pure glue, two-flavor, and $(2+1)$-flavor equations of state, respectively. This construction reduces to Eq.~\ref{eq:Tc} when $\gamma_l = \gamma_s = \gamma_q$. The value $T_2 = 170$~MeV is introduced as an ansatz: the two-flavor transition temperature is not sharply defined, depending on both the quark mass and the observable used to identify the transition, and heavier-than-physical lattice studies~\cite{Burger:2011zc, Bornyakov:2009qh} place it above the physical $(2+1)$-flavor value. Because $T_2$ lies only $\approx\!12$~MeV from $T_3$, its precise value has little effect on $T_\text{c}$ away from the pure glue-dominated region. This asymmetry is apparent in Fig.~\ref{fig:Tc_strange}: $T_\text{c}$ varies by $\approx\!90$~MeV across the light fugacity direction but by only $\approx\!12$~MeV across the strange fugacity direction, so the light quark occupancy sets the transition temperature far more strongly than the strange quark occupancy.

\enlargethispage{\baselineskip}

The high-temperature pressure is then constructed as
\begin{equation}
    \begin{aligned}
    \frac{P}{T^4}(T,\gamma_l,\gamma_s)
    =&
    \gamma_s \frac{P_3}{T^4}
    \left(T \frac{T_3}{T_{\text{c}}(\gamma_l,\gamma_s)}\right)
    +
    (\gamma_l-\gamma_s) \frac{P_2}{T^4}
    \left(T \frac{T_2}{T_{\text{c}}(\gamma_l,\gamma_s)}\right) \\
    &\quad+
    (1-\gamma_l) \frac{P_0}{T^4}
    \left(T \frac{T_0}{T_{\text{c}}(\gamma_l,\gamma_s)}\right),
    \end{aligned}
    \label{eq:P_T4_strange}
\end{equation}
and the corresponding energy density is constructed as
\begin{equation}
    \begin{aligned}
    \frac{\varepsilon}{T^4}(T,\gamma_l,\gamma_s)
    &=
    \gamma_s \frac{\varepsilon_3}{T^4}
    \left(T \frac{T_3}{T_{\text{c}}(\gamma_l,\gamma_s)}\right)
    +
    (\gamma_l-\gamma_s) \frac{\varepsilon_2}{T^4}
    \left(T \frac{T_2}{T_{\text{c}}(\gamma_l,\gamma_s)}\right) \\
    &\quad+
    (1-\gamma_l) \frac{\varepsilon_0}{T^4}
    \left(T \frac{T_0}{T_{\text{c}}(\gamma_l,\gamma_s)}\right).
    \end{aligned}
    \label{eq:e_T4_strange}
\end{equation}
Equivalently, the dimensionful pressure and energy density are
\begin{equation}
    \begin{aligned}
    P(T,\gamma_l,\gamma_s)
    &=
    \gamma_s
    \left(\frac{T_{\text{c}}(\gamma_l,\gamma_s)}{T_3}\right)^4
    P_3\left(T \frac{T_3}{T_{\text{c}}(\gamma_l,\gamma_s)}\right) \\
    &\quad+
    (\gamma_l-\gamma_s)
    \left(\frac{T_{\text{c}}(\gamma_l,\gamma_s)}{T_2}\right)^4
    P_2\left(T \frac{T_2}{T_{\text{c}}(\gamma_l,\gamma_s)}\right) \\
    &\quad+
    (1-\gamma_l)
    \left(\frac{T_{\text{c}}(\gamma_l,\gamma_s)}{T_0}\right)^4
    P_0\left(T \frac{T_0}{T_{\text{c}}(\gamma_l,\gamma_s)}\right),
    \end{aligned}
    \label{eq:P_strange}
\end{equation}
and
\begin{equation}
    \begin{aligned}
    \varepsilon(T,\gamma_l,\gamma_s)
    &=
    \gamma_s
    \left(\frac{T_{\text{c}}(\gamma_l,\gamma_s)}{T_3}\right)^4
    \varepsilon_3\left(T \frac{T_3}{T_{\text{c}}(\gamma_l,\gamma_s)}\right) \\
    &\quad+
    (\gamma_l-\gamma_s)
    \left(\frac{T_{\text{c}}(\gamma_l,\gamma_s)}{T_2}\right)^4
    \varepsilon_2\left(T \frac{T_2}{T_{\text{c}}(\gamma_l,\gamma_s)}\right) \\
    &\quad+
    (1-\gamma_l)
    \left(\frac{T_{\text{c}}(\gamma_l,\gamma_s)}{T_0}\right)^4
    \varepsilon_0\left(T \frac{T_0}{T_{\text{c}}(\gamma_l,\gamma_s)}\right).
    \end{aligned}
    \label{eq:e_strange}
\end{equation}

When $\gamma_s \leq \gamma_l$, this construction can be interpreted as an interpolation from pure glue QCD, to light quark QCD, to full $(2+1)$-flavor QCD. Because the purpose of the model is to explore independent light and strange equilibration timescales while remaining agnostic to the microscopic physics, we do not enforce this hierarchy in general. When $\gamma_s > \gamma_l$, the coefficient of the two-flavor branch becomes negative, so the construction should be understood as a controlled extrapolation rather than a strict interpolation. Because the two-flavor branch is itself a positive rescaling of the $(2+1)$-like equation of state, this extrapolation remains numerically well behaved; we verify that the pressure, energy density, and squared speed of sound remain positive across the full fugacity grid.

The pure glue and $(2+1)$-flavor limits are taken from the same equations of state used in section~\ref{subsec:eos_pce}. The intermediate $N_f = 2$ branch is less direct. Ideally, this branch would be supplied by a two-flavor lattice QCD equation of state with physical quark masses. Such an input is not available in a form directly compatible with the present construction, and attempts to use existing two-flavor parameterizations with heavier-than-physical quark masses~\cite{Burger:2011zc} led to numerical difficulties in the interpolation. We therefore approximate the two-flavor branch by rescaling the $(2+1)$-flavor HotQCD equation of state according to the ideal-gas ratio of effective degrees of freedom,
\begin{align}
    P_2(T) &\approx \frac{37}{47.5} P_3(T), \qquad 
    \varepsilon_2(T) \approx \frac{37}{47.5} \varepsilon_3(T).
    \label{eq:nf2_rescaling}
\end{align}

This approximation preserves a distinct light quark interpolation direction while avoiding numerical artifacts from an incompatible two-flavor input. It should therefore be viewed as a pragmatic component of the phenomenological model, not as a first-principles determination of the physical two-flavor equation of state.

At low temperature, the high-temperature equation of state is matched to a hadron resonance gas equation of state, as in section~\ref{subsec:eos_pce}. The hadron resonance gas calculation is performed using \texttt{frzout}~\cite{frzout} for the purpose of constructing the equation of state. The distribution functions are modified by species-dependent hadronic fugacity factors determined by the light and strange valence quark content of each hadron. Since all calculations are performed at $\mu_B = 0$, quarks and antiquarks of a given flavor are assigned the same fugacity.

We define light and strange hadronic fugacity factors
\begin{align}
    \lambda_l = 0.85\gamma_l + 0.15, \qquad
    \lambda_s = 0.85\gamma_s + 0.15.
    \label{eq:lambda_ls}
\end{align}

For a hadron species $i$, the corresponding fugacity factor is assigned according to
\begin{align}
    \lambda_i = \lambda_l^{(n_{l,i}+\bar{n}_{l,i})/2} \lambda_s^{(n_{s,i}+\bar{n}_{s,i})/2},
    \label{eq:lambda_i_strange}
\end{align}
where $n_{l,i} (\bar{n}_{l,i})$ and $n_{s,i} (\bar{n}_{s,i})$ are the total light and strange valence quark (antiquark) contents of species $i$, respectively. This reduces to the baryon and meson prescriptions of Eq.~\ref{eq:lambda_i} when $\gamma_l=\gamma_s=\gamma_q$: mesons carry one power of the common fugacity factor, while baryons carry a power of $3/2$.

The low-temperature hadron resonance gas and high-temperature lattice-based equations of state are matched using the same Krogh interpolation procedure as in section~\ref{subsec:eos_pce}, but with a matching range that depends on both fugacities. We define the interpolation interval as
\begin{align}
T_a = T_\text{c}(\gamma_l,\gamma_s) + 5~\text{MeV}, \qquad 
T_b = T_\text{c}(\gamma_l,\gamma_s) + (10 + 5\gamma_l + 5\gamma_s)~\text{MeV}.
\label{eq:matching_range_strange}
\end{align}

\begin{figure}[!htbp]
    \centering
    \includegraphics[width=\textwidth]{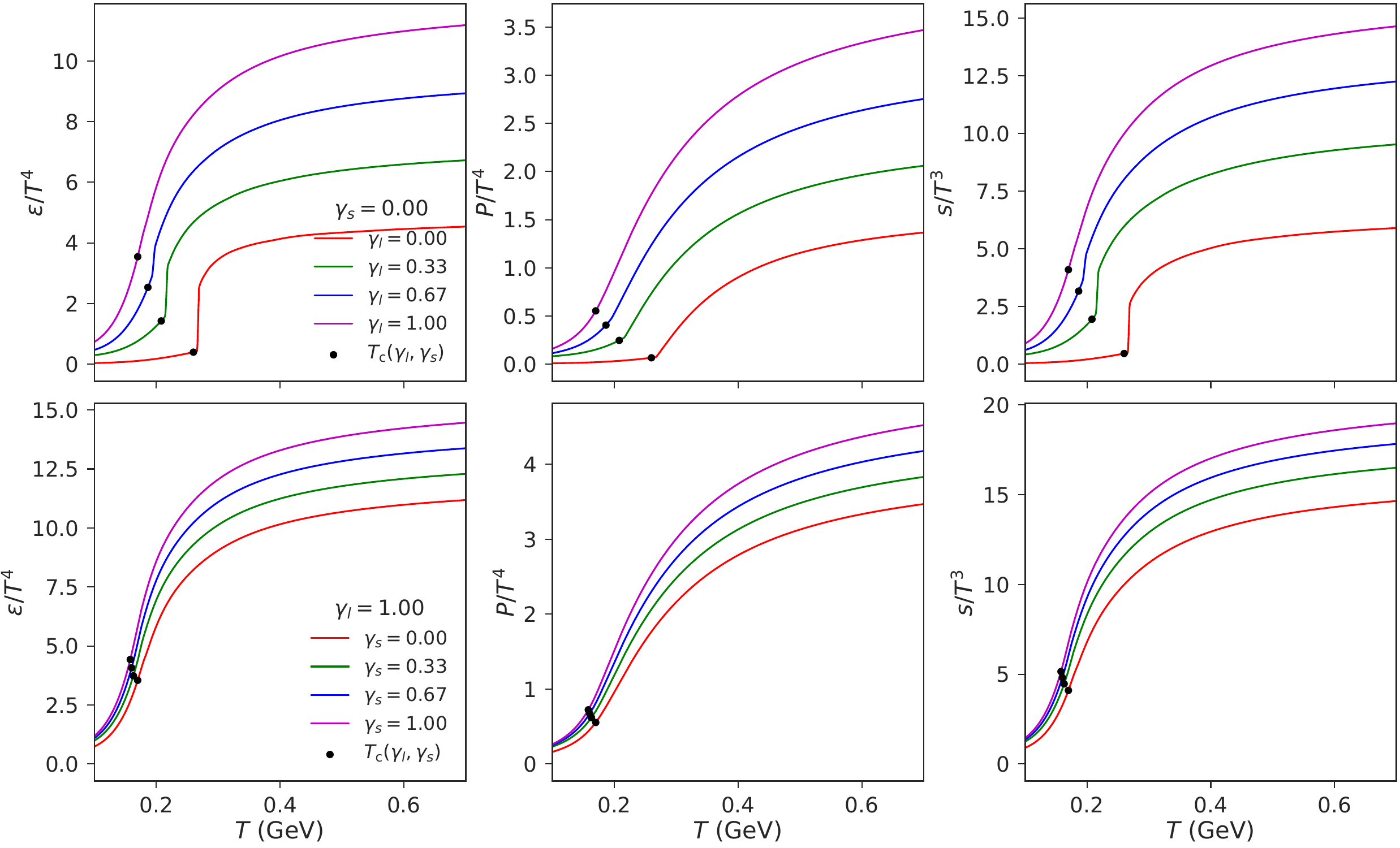}
    \caption[Thermodynamics of the two-flavor equation of state along the light and strange directions]{Dimensionless energy density $\varepsilon/T^4$, pressure $P/T^4$, and entropy density $s/T^3$ (left to right) of the flavor-dependent equation of state as functions of temperature. Top row: the light fugacity $\gamma_l$ is varied at fixed $\gamma_s = 0$, spanning the pure glue and two-flavor limits. Bottom row: the strange fugacity $\gamma_s$ is varied at fixed $\gamma_l = 1$, spanning the two-flavor and $(2+1)$-flavor limits. Both fixed values lie on the boundary of the physical wedge $\gamma_s \leq \gamma_l$. Black points mark the particlization temperature $T_\text{c}(\gamma_l,\gamma_s)$ from Eq.~\ref{eq:Tc_strange}. The pure glue first-order transition is visible in the low-$\gamma_l$ curves of the top row and is absent along the $\gamma_l = 1$ edge.}
    \label{fig:EoS_strange}
\end{figure}

The particular constants of this interpolation are chosen for numerical smoothness, and should not be taken to have deeper physical significance. The resulting equation of state provides $P(T,\gamma_l,\gamma_s)$ and $\varepsilon(T,\gamma_l,\gamma_s)$ across the full temperature and fugacity range used in the hydrodynamic evolution. Fig.~\ref{fig:EoS_strange} shows this equation of state along two edges of the fugacity grid. Increasing $\gamma_l$ at $\gamma_s = 0$ carries the thermodynamics from the pure glue limit to the two-flavor limit, and increasing $\gamma_s$ at $\gamma_l = 1$ carries it from the two-flavor limit to full equilibrium. The first-order transition of the pure glue sector appears in the low-$\gamma_l$ curves and weakens as $(1-\gamma_l)$, softening into a crossover as the light quarks approach equilibrium; because this coefficient vanishes along the $\gamma_l = 1$ edge, the strange direction is a smooth crossover throughout, with a smaller effect on the thermodynamics that mirrors the weak dependence of $T_\text{c}$ on $\gamma_s$ seen in Fig.~\ref{fig:Tc_strange}.

\Needspace{4\baselineskip}
\subsection{Fugacity}
\label{subsec:fugacity_strange}

The proper time prescription of section~\ref{subsec:fugacity} is generalized by assigning separate equilibration timescales to light and strange quarks. We also introduce initial fugacity factors $\gamma_l^0$ and $\gamma_s^0$, with $0 \leq \gamma_l^0,\gamma_s^0 \leq 1$, corresponding to the quark content already present in the medium at the onset of hydrodynamics. These can be interpreted as effective measures of the quark content produced during pre-equilibrium, which was assumed to vanish in the flavor-independent model. The flavor-dependent fugacities thus take the form
\begin{equation}
\begin{aligned}
\gamma_l(\tau_\text{p}) = 1 - (1-\gamma_l^0)\exp\left(\frac{\tau_0 - \tau_\text{p}}{\tau_{\text{eq},l}}\right), \\
\gamma_s(\tau_\text{p}) = 1 - (1-\gamma_s^0)\exp\left(\frac{\tau_0 - \tau_\text{p}}{\tau_{\text{eq},s}}\right),
\end{aligned}
\label{eq:flavor_fugacities}
\end{equation}
where $\tau_\text{p}$ is the local proper time evolved according to Eq.~\ref{eq:proper_time}. The limits $\tau_{\text{eq},l}\rightarrow0$ and $\tau_{\text{eq},s}\rightarrow0$ are treated as instantaneous equilibration for the corresponding flavor, with that fugacity set to unity throughout the hydrodynamic evolution. When $\tau_{\text{eq},l}=\tau_{\text{eq},s}$ and $\gamma_l^0 = \gamma_s^0 = 0$, the model reduces to the flavor-independent fugacity evolution of section~\ref{subsec:fugacity}.

\subsection{Particlization}
\label{subsec:particlization_strange}

Particlization is modified analogously to section~\ref{subsec:particlization_pce}. The hypersurface is defined by the local flavor-dependent critical temperature,
\begin{align}
T(x) = T_\text{c}(\gamma_l(x),\gamma_s(x)),
\label{eq:surface_condition_strange}
\end{align}
so that fluid cells with different light and strange occupancies particlize according to the transition temperature appropriate to their local equation of state.

The equilibrium distribution functions used in iS3D are also modified by the species-dependent fugacity factors of Eq.~\ref{eq:lambda_i_strange}. Thus, for species $i$,
\begin{align}
f_{i,\text{eq}}(p,\lambda_i)
=
\frac{1}{\lambda_i^{-1} e^{E_p/T} \pm 1}.
\label{eq:fi_strange}
\end{align}

The viscous correction is treated with the same Grad 14-moment expansion described in section~\ref{section24}, but the thermal integrals entering the matching coefficients are evaluated with the flavor-dependent equilibrium distributions. In this way, the coefficients $A_T$, $A_E$, and $A_\pi$ inherit a dependence on the local values of $\gamma_l$ and $\gamma_s$. As before, the charge diffusion correction is omitted because the calculation is performed at $\mu_B = 0$ and quarks and antiquarks of each flavor are produced symmetrically.

After particlization, the particles are passed to SMASH for hadronic transport. The subsequent hadronic evolution carries no explicit memory of the fugacity fields except through the particle species and momenta generated at particlization.

\subsection{Implementation}
\label{subsec:implementation_strange}

The flavor-dependent equation of state depends on three independent variables: the energy density, the light quark fugacity, and the strange quark fugacity. We therefore tabulate the thermodynamic quantities on a three-dimensional grid in $(\varepsilon,\gamma_l,\gamma_s)$. The table stores $P$, $s$, $T$, and $c_s^2$ for each grid point, with $30$ points in each fugacity direction and $10,000$ points in energy density.

The entropy density is computed numerically from the final tabulated equation of state, 
\begin{align}
s(T, \gamma_l, \gamma_s) = \frac{\varepsilon + P}{T} - \frac{1}{T} \sum_{f=l,s} \gamma_f \ln\gamma_f \left( \frac{\partial P}{\partial \gamma_f} \right)_{V,T},
\label{eq:entropy_numerical_strange}
\end{align}
with the fugacity derivatives evaluated by finite differences of the tabulated pressure at fixed temperature (appendix~\ref{app:entropy}). This avoids extending the cumbersome closed-form one-fugacity expression to the two-fugacity case and ensures that the entropy used by the hydrodynamic evolution is consistent with the final pressure table.

MUSIC was modified to read and interpolate the three-dimensional equation of state table during the hydrodynamic evolution. At each spacetime point, the local pressure, temperature, entropy density, and speed of sound are evaluated from the local values of $\varepsilon$, $\gamma_l$, and $\gamma_s$. The dimensionful shear and bulk viscosities are then obtained from the specific viscosities using this flavor-dependent entropy density, as in section~\ref{subsec:implementation_pce}.

iS3D was also modified to read the local values of $\gamma_l$ and $\gamma_s$ from the particlization hypersurface. These fugacities enter both the surface condition of Eq.~\ref{eq:surface_condition_strange} and the species-dependent distribution functions of Eq.~\ref{eq:fi_strange}. The initial condition model and hadronic afterburner are otherwise unchanged from the framework described in chapter~\ref{chapter2}.
\Needspace{4\baselineskip}
\section{Effects of light and strange flavor equilibration}
\label{section34}

We now apply the flavor-dependent model of section~\ref{section33} to complete Pb+Pb collision events, extending the flavor-independent sensitivity study of section~\ref{section32} to independent light and strange equilibration. As before, the aim is to demonstrate the effects of chemical non-equilibrium on final-state observables rather than to constrain the model against experimental data, which is the subject of chapter~\ref{chapter5}.

We use the same setup as in section~\ref{section32}, with the following exceptions. In place of the single timescale $\tau_\text{eq}$, we independently vary the light and strange equilibration timescales $\tau_{\text{eq},l}$ and $\tau_{\text{eq},s}$, isolating the separate effects of light and strange chemical equilibration.

The flavor-dependent equation of state is numerically more demanding than its flavor-independent counterpart, and unlike the flavor-independent runs, these simulations do not begin from a purely gluonic state. Instead, we fix the initial fugacities to a small nonzero value, $\gamma_l^0 = \gamma_s^0 = 0.1$, to maintain numerical stability. The medium is therefore never fully devoid of quarks: because the fugacities relax upward from their initial values, $0.1$ acts as an effective lower bound on $\gamma_l$ and $\gamma_s$ throughout the hydrodynamic evolution. The origin of the instability that motivates this choice, the additional safeguards adopted alongside it, and the limitations they reflect are discussed in section~\ref{subsec:numerics_strange}.

The hydrodynamic evolution otherwise proceeds as described in section~\ref{subsec:evolution_pce}, with the light and strange sectors equilibrating to different extents according to their respective timescales. The qualitative features are unchanged: reducing the quark content raises the temperature at a given energy density and shifts the particlization hypersurface, now with independent dependence on the light and strange fugacities. The central distinction from the flavor-independent case appears not in the bulk evolution but in the flavor composition of the final state, which we examine below.

\subsection{Hadron production}
\label{subsec:hadrons_strange}

As in the flavor-independent model of section~\ref{subsec:hadrons_pce}, larger equilibration timescales cause particlization to occur at a higher temperature and at lower quark fugacities, with two competing effects on hadron production. With independent light and strange fugacities, the particlization temperature remains a common property of each fluid cell, while the hadronic fugacity factors become flavor dependent: increasing $\tau_{\text{eq},l}$ or $\tau_{\text{eq},s}$ suppresses the light or strange fugacity contribution to the distribution functions, respectively. We again consider an ensemble of $10,000$ minimum-bias events, with all observables calculated at midrapidity ($|\eta| < 0.5$).

\begin{figure}[!t]
    \centering
    \includegraphics[width=0.7\linewidth]{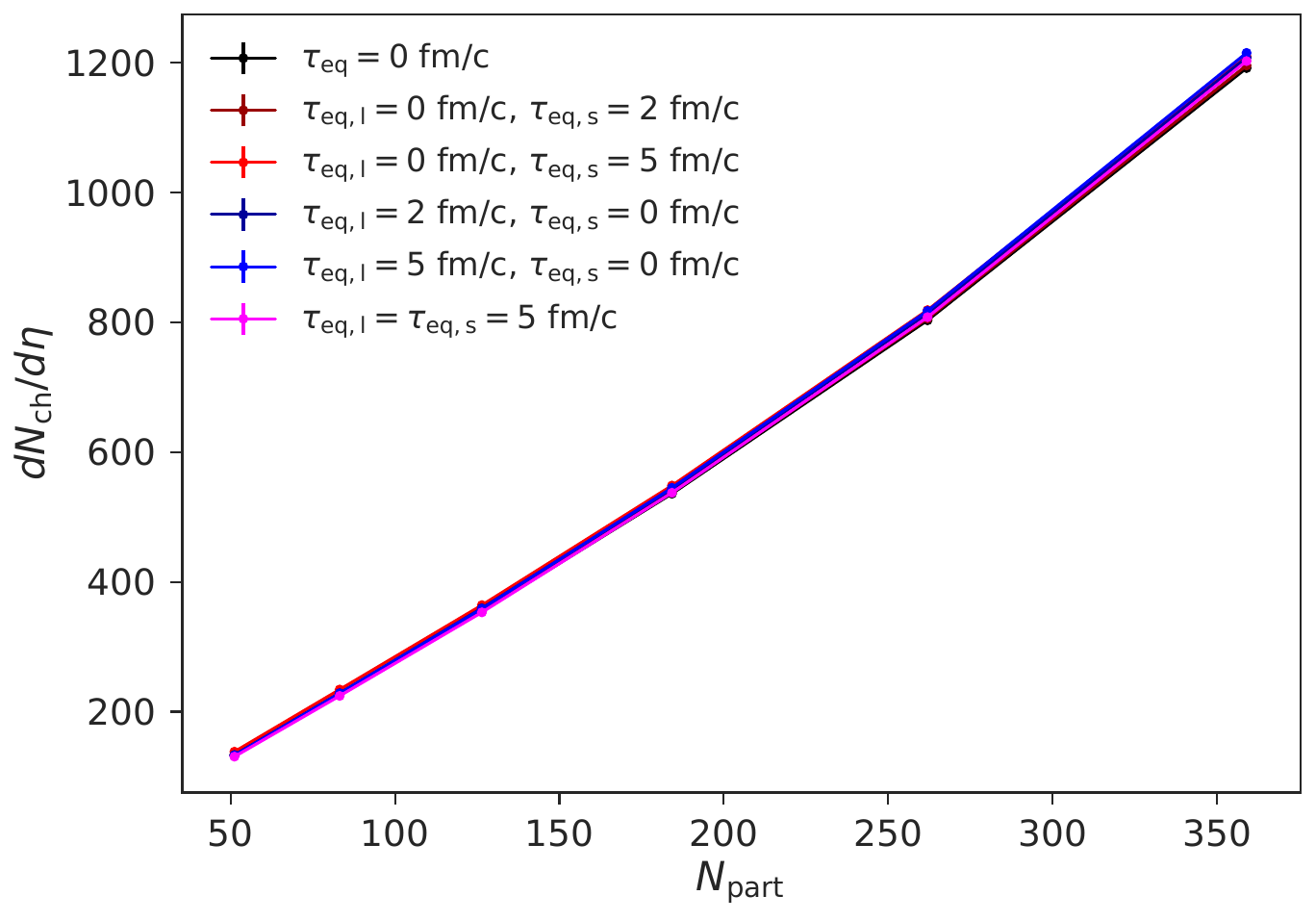}
    \caption[Charged particle multiplicity for events evolved with varying $\tau_{\text{eq},l}$ and $\tau_{\text{eq},s}$]{Charged particle multiplicity for $0$--$60$\% centrality events evolved with varying light and strange equilibration timescales $\tau_{\text{eq},l}$ and $\tau_{\text{eq},s}$. Each point corresponds to a $10$\% centrality bin.}
    \label{fig:dNch_strange}
\end{figure}

Fig.~\ref{fig:dNch_strange} shows the resulting charged particle multiplicity. As in the flavor-independent case, it is remarkably insensitive to chemical equilibration: varying either $\tau_{\text{eq},l}$ or $\tau_{\text{eq},s}$ leaves the charged multiplicity invariant to within a few percent. The close cancellation between the effects of higher particlization temperature and lower fugacity identified in section~\ref{subsec:hadrons_pce} therefore persists when the two flavors are separated, and holds under the independent variation of each.

\begin{figure}[!htbp]
    \centering

    \begin{subfigure}{0.48\linewidth}
        \centering
        \includegraphics[width=\linewidth]{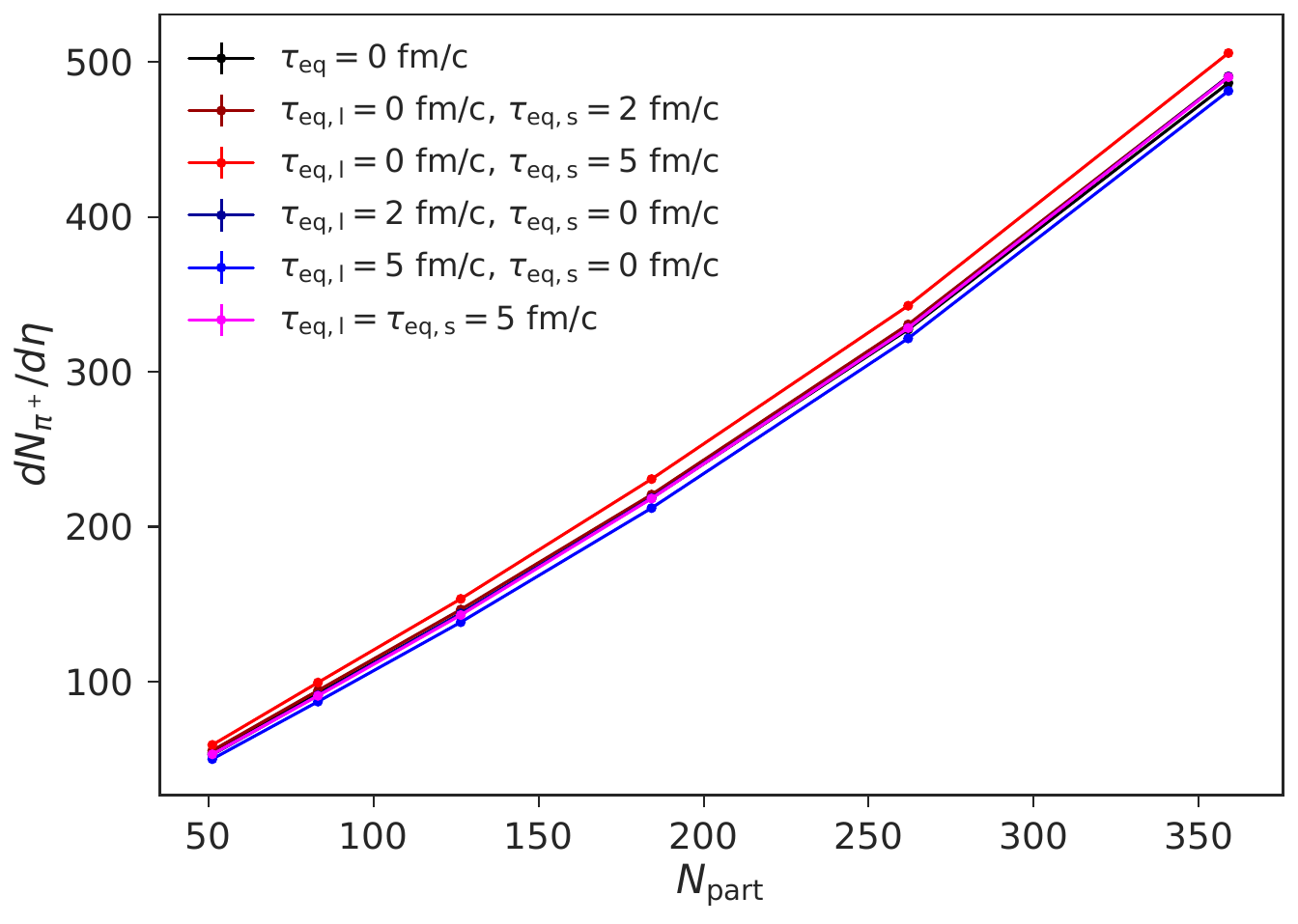}
        \caption{$\pi^+$}
        \label{fig:yields_pions}
    \end{subfigure}
    \hfill
    \begin{subfigure}{0.48\linewidth}
        \centering
        \includegraphics[width=\linewidth]{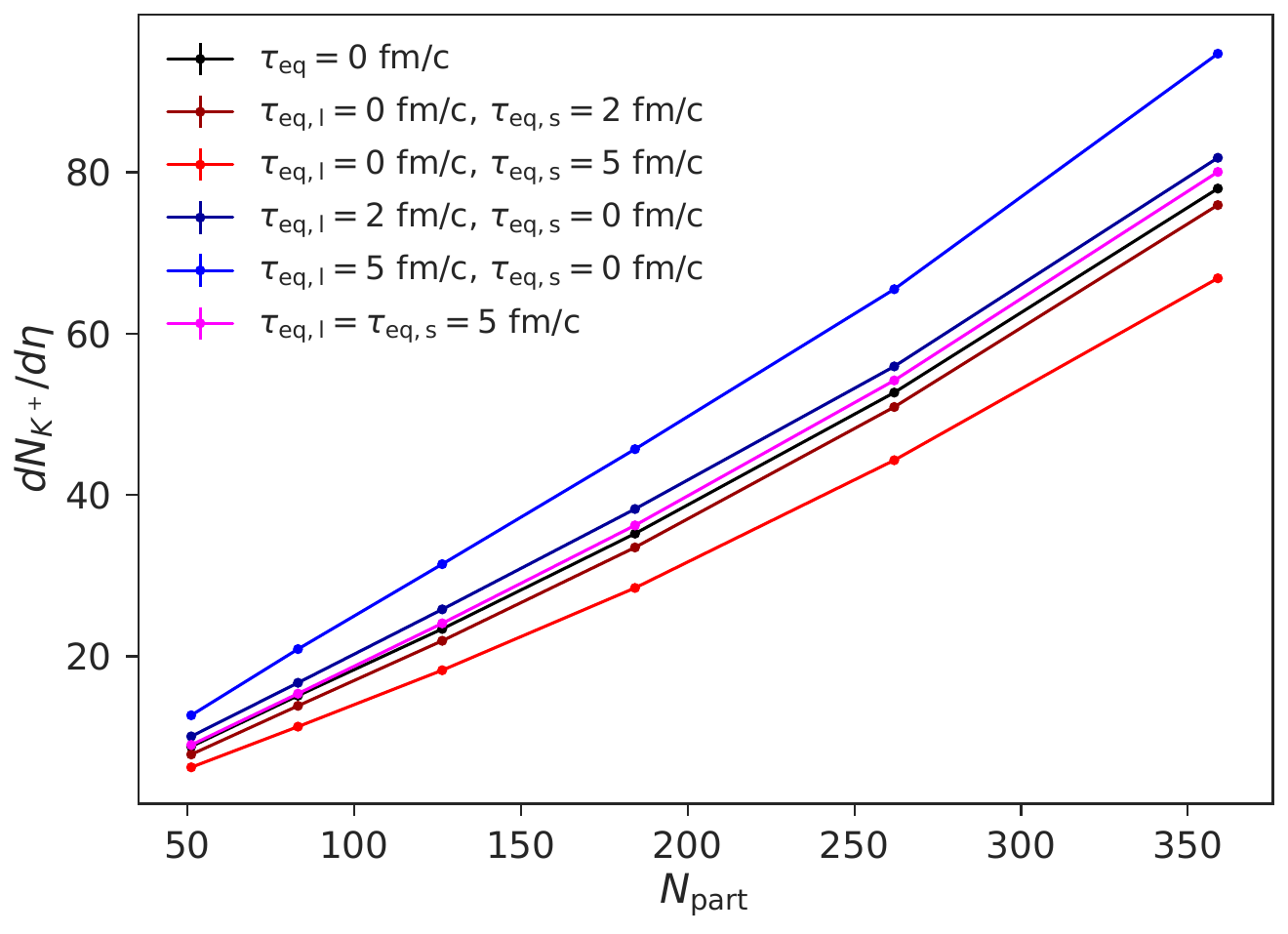}
        \caption{$K^+$}
        \label{fig:yields_kaons}
    \end{subfigure}

    \vspace{0.4cm}

    \begin{subfigure}{0.48\linewidth}
        \centering
        \includegraphics[width=\linewidth]{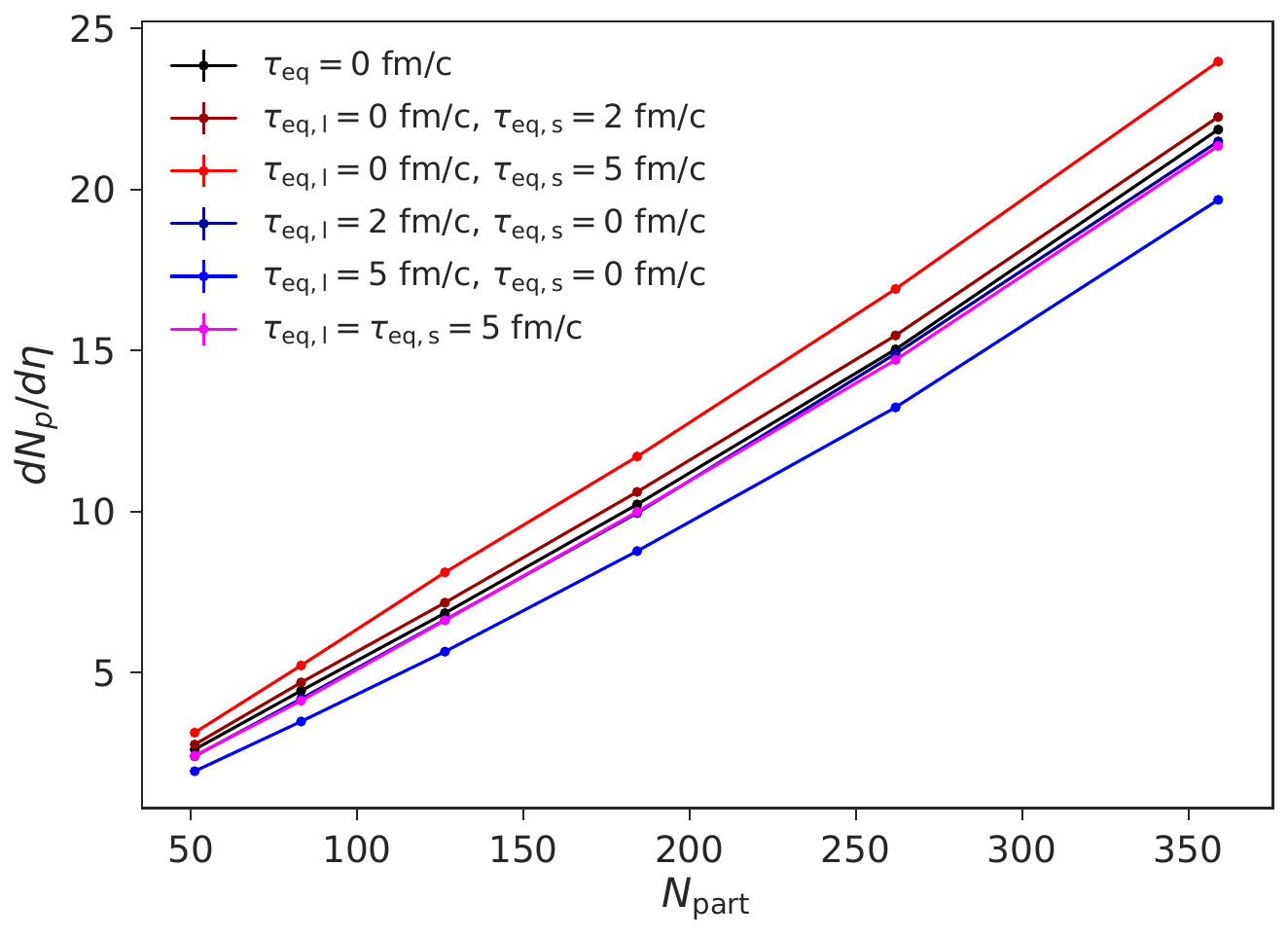}
        \caption{$p$}
        \label{fig:yields_protons}
    \end{subfigure}
    \hfill
    \begin{subfigure}{0.48\linewidth}
        \centering
        \includegraphics[width=\linewidth]{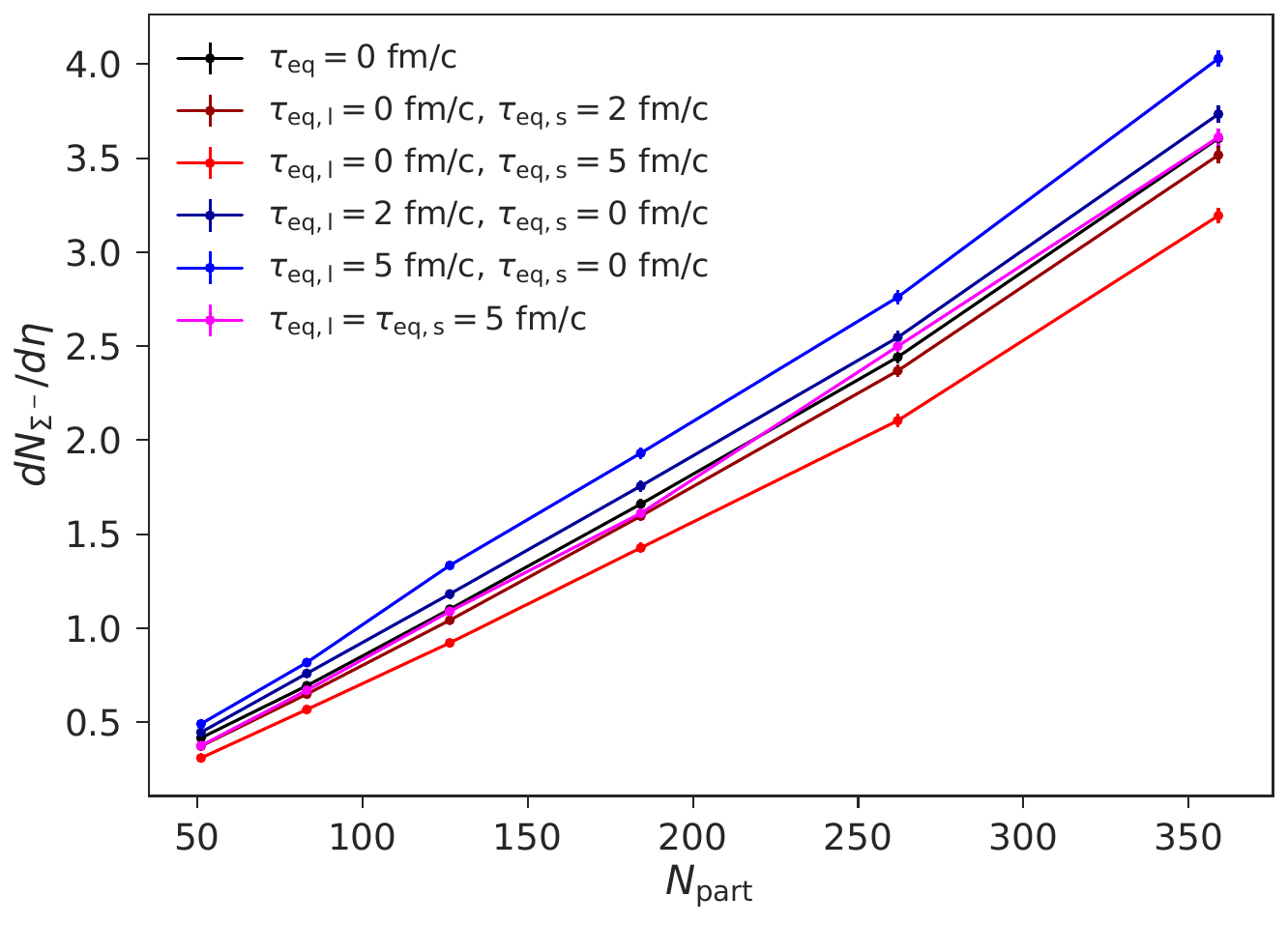}
        \caption{$\Sigma^-$}
        \label{fig:yields_sigma_minus}
    \end{subfigure}

    \vspace{0.4cm}

    \begin{subfigure}{0.48\linewidth}
        \centering
        \includegraphics[width=\linewidth]{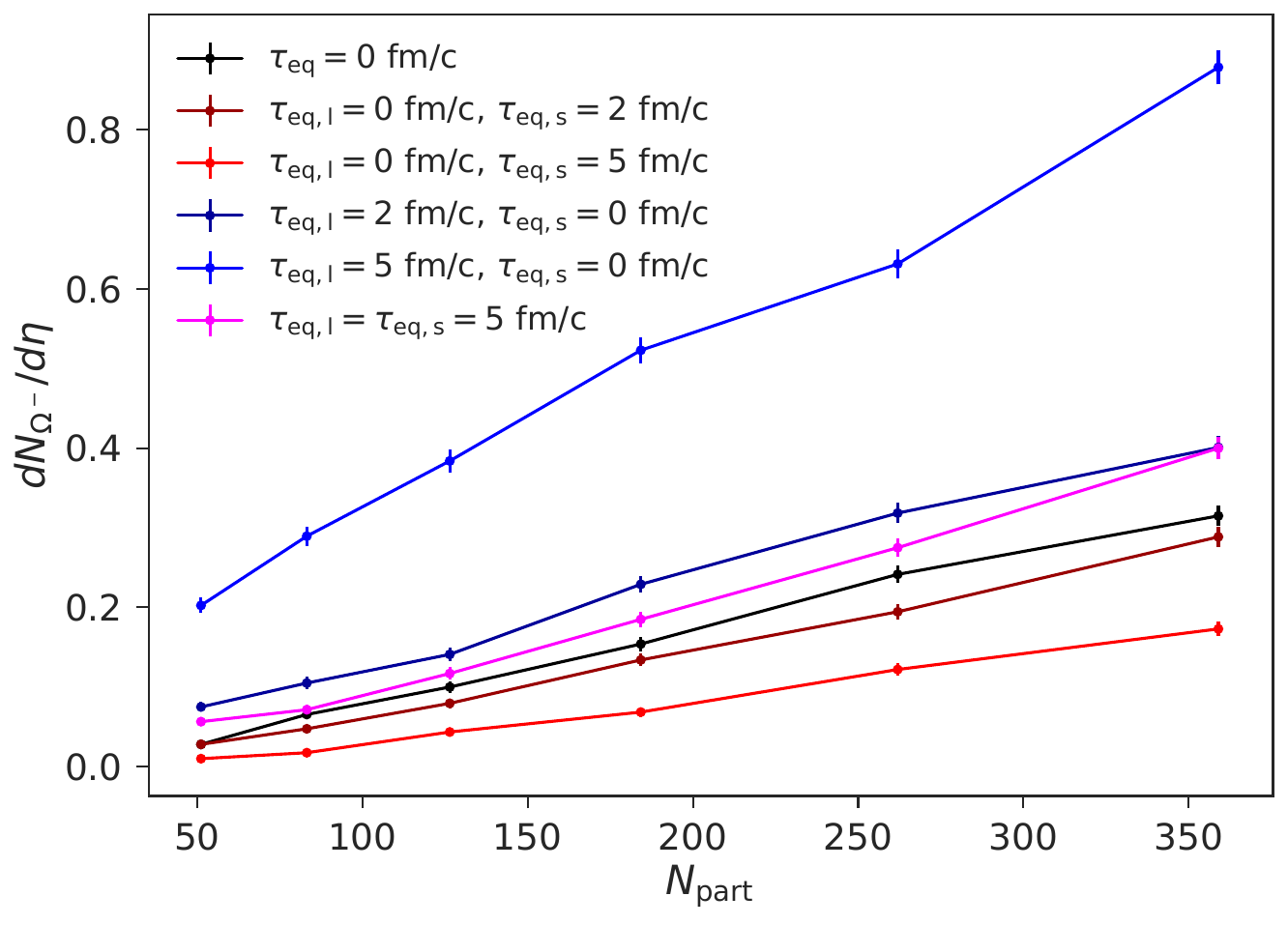}
        \caption{$\Omega^-$}
        \label{fig:yields_omega_minus}
    \end{subfigure}

    \caption[Identified particle yields of $\pi^+$, $K^+$, $p$, $\Sigma^-$, and $\Omega^-$ for events evolved with varying $\tau_{\text{eq},l}$ and $\tau_{\text{eq},s}$]{Identified particle yields of $\pi^+$, $K^+$, $p$, $\Sigma^-$, and $\Omega^-$ for $0$--$60$\% centrality events evolved with varying $\tau_{\text{eq},l}$ and $\tau_{\text{eq},s}$. Each point corresponds to a $10$\% centrality bin.}
    \label{fig:yields_strange}
\end{figure}

The identified particle yields, by contrast, depend strongly on flavor equilibration. Fig.~\ref{fig:yields_strange} shows the yields of $\pi^+$, $K^+$, $p$, $\Sigma^-$, and $\Omega^-$, whose valence content spans a range of light and strange quark numbers: pions and protons are purely light, kaons and $\Sigma^-$ each carry a single strange quark, and $\Omega^-$ is composed entirely of strange quarks. Suppressing the light sector through a larger $\tau_{\text{eq},l}$ suppresses the light-dominated pion and proton yields while enhancing the strange-carrying kaon, $\Sigma^-$, and $\Omega^-$ yields; suppressing the strange sector through a larger $\tau_{\text{eq},s}$ reverses the effect. The magnitude of the response increases with strange content, so among the species shown, the triply strange $\Omega^-$ is the most strongly enhanced when the light sector is suppressed.

This anti-correlation follows directly from the construction of the model. Because every event is evolved from the same initial energy density profile irrespective of the equilibration timescales, and energy is conserved throughout the evolution, the total energy carried into the final state is fixed; the near-invariance of the charged multiplicity in Fig.~\ref{fig:dNch_strange} shows that the overall scale of production is likewise stable. Within this fixed energy budget, the species-dependent hadronic fugacity factors of Eq.~\ref{eq:lambda_i_strange} redistribute production among hadron species according to their light and strange content. Suppressing $\gamma_l$ shifts production away from light-dominated hadrons, which must be compensated by an enhancement of strange-dominated hadrons, and vice versa.

\begin{figure}[!t]
    \centering
    \includegraphics[width=0.7\linewidth]{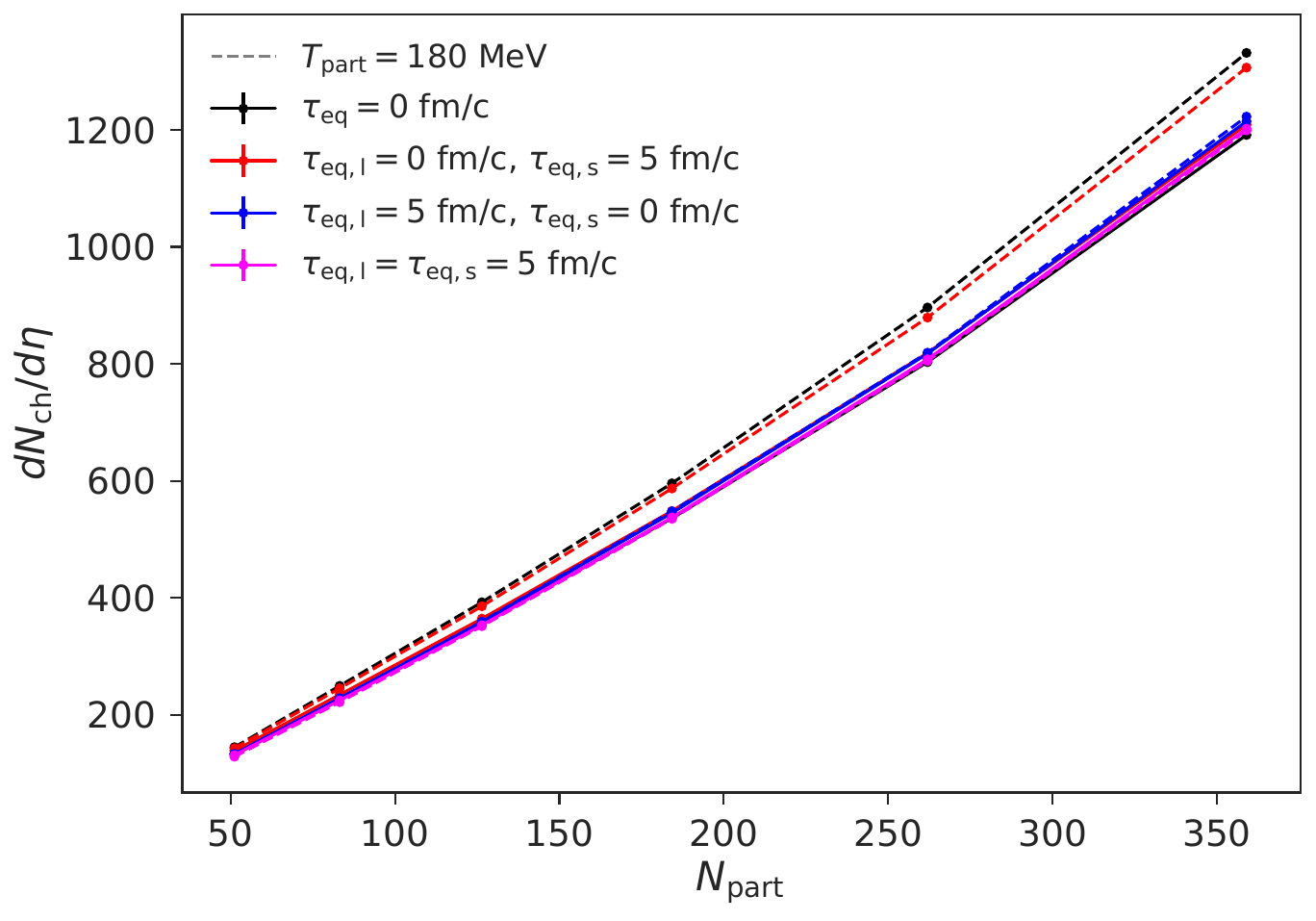}
    \caption[Charged particle multiplicity for events evolved with varying $\tau_{\text{eq},l}$ and $\tau_{\text{eq},s}$, comparing the fugacity-dependent particlization surface with a fixed $T_\text{part}$]{Charged particle multiplicity for $0$--$60$\% centrality events evolved with varying $\tau_{\text{eq},l}$ and $\tau_{\text{eq},s}$, comparing the two particlization prescriptions: solid curves particlize on the fugacity-dependent surface $T = T_\text{c}(\gamma_l, \gamma_s)$, while dashed curves particlize at a fixed $T_\text{part} = 180~\text{MeV}$. Each point corresponds to a $10$\% centrality bin.}
    \label{fig:dNch_temperature_strange}
\end{figure}

As in section~\ref{subsec:hadrons_pce}, we can separate the effect of the particlization temperature from that of the fugacities by holding the particlization temperature fixed. Fig.~\ref{fig:dNch_temperature_strange} shows the charged multiplicity in this case. In contrast to the full result, it is no longer invariant: the charged multiplicity is suppressed when the light sector is suppressed, but remains close to invariant when the strange sector is suppressed. This supports the interpretation that the invariance in Fig.~\ref{fig:dNch_strange} arises from the same cancellation between particlization temperature and fugacity as in the flavor-independent model, and that this cancellation is carried predominantly by the light sector, consistent with strange hadrons making up only a small fraction of the charged multiplicity.

\FloatBarrier
\subsection{Transverse flow}
\label{subsec:flow_strange}

\begin{figure}[!t]
    \centering
    \includegraphics[width=0.7\linewidth]{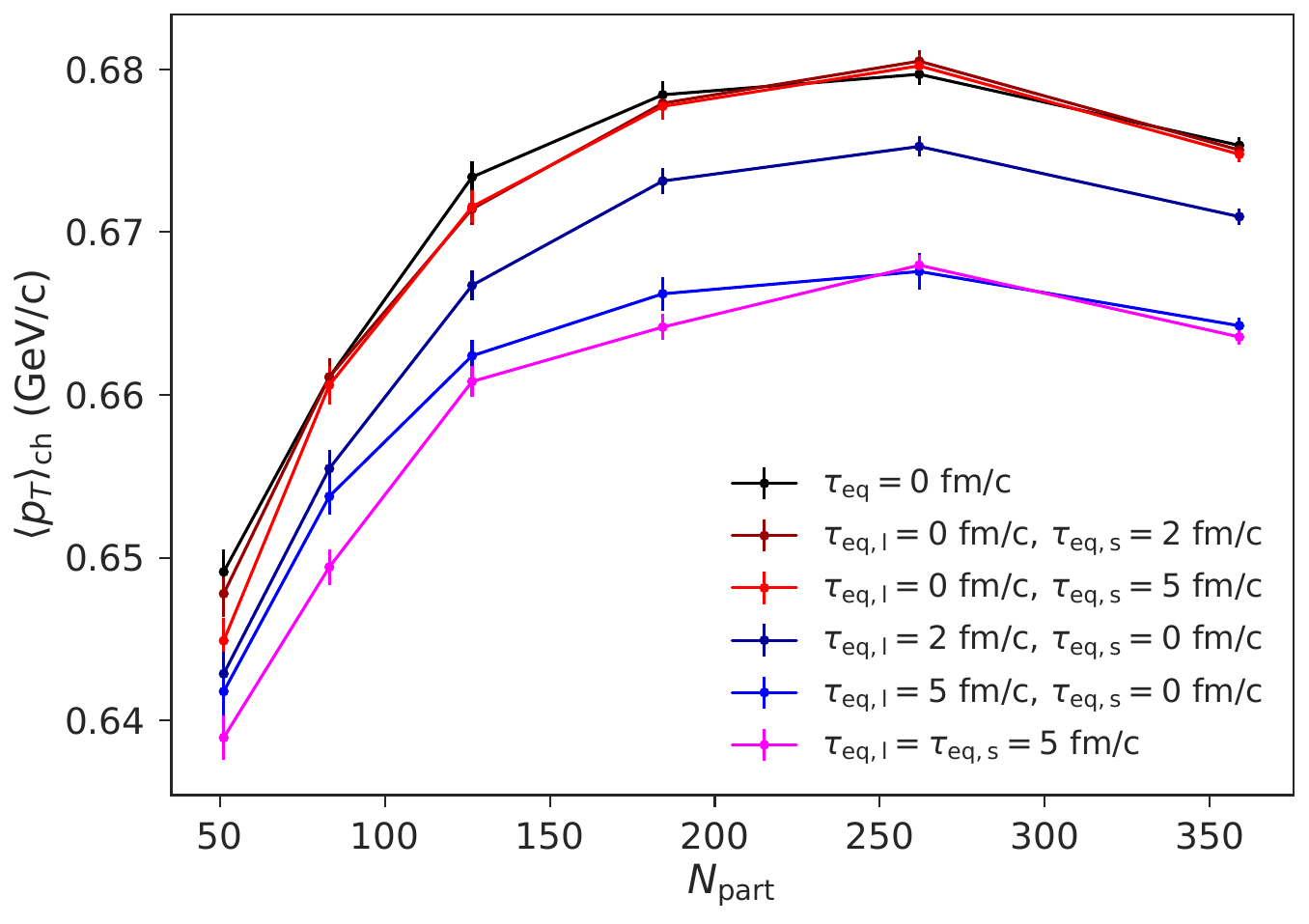}
    \caption[Mean transverse momentum of charged particles for events evolved with varying $\tau_{\text{eq},l}$ and $\tau_{\text{eq},s}$]{Mean transverse momentum of charged particles for $0$--$60$\% centrality events evolved with varying light and strange equilibration timescales $\tau_{\text{eq},l}$ and $\tau_{\text{eq},s}$. Each point corresponds to a $10$\% centrality bin.}
    \label{fig:pTch_strange}
\end{figure}

As in the flavor-independent model of section~\ref{subsec:flow_pce}, chemical equilibration affects not only the production of hadrons at particlization but also the development of transverse flow during the hydrodynamic evolution, which is ultimately reflected in the momenta of the final-state hadrons. Fig.~\ref{fig:pTch_strange} shows the mean transverse momentum $\langle p_T \rangle$ of charged particles.

The charged particle $\langle p_T \rangle$ depends modestly on chemical equilibration, and this dependence is driven predominantly by the light sector: increasing $\tau_{\text{eq},l}$ systematically reduces $\langle p_T \rangle$, while varying $\tau_{\text{eq},s}$ has comparatively little effect. This follows the flavor-independent result of section~\ref{subsec:flow_pce}, where the suppression of $\langle p_T \rangle$ out of equilibrium was attributed to the reduced pressure of the undersaturated medium. Because the light sector controls the larger change in the thermodynamics and dominates the final charged particle content, it dominates this pressure, while the strange sector, a small fraction of the total, has little influence on the bulk transverse flow.

\begin{figure}[!t]
    \centering

    \begin{subfigure}{0.48\linewidth}
        \centering
        \includegraphics[width=\linewidth]{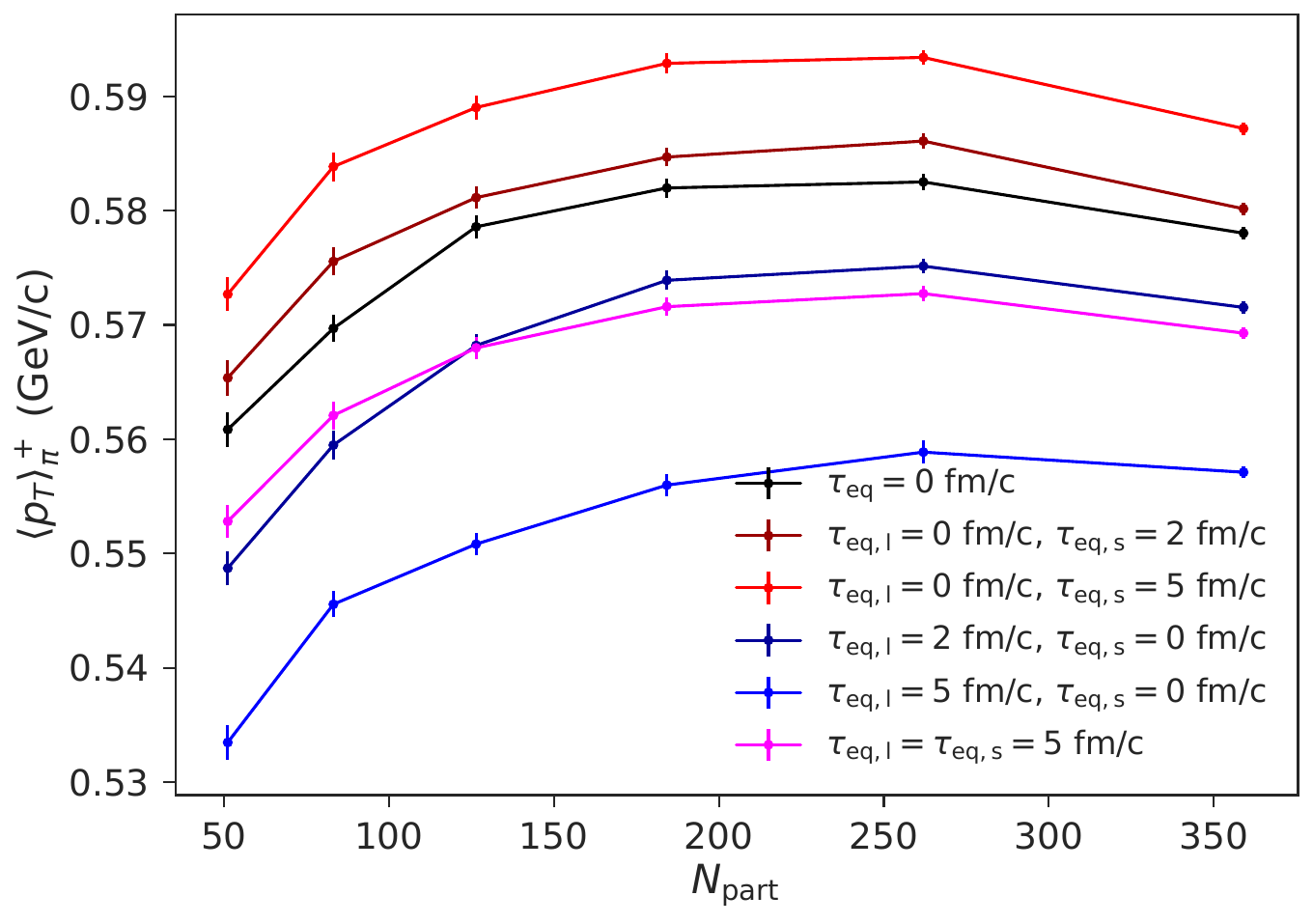}
        \caption{$\pi^+$}
        \label{fig:pT_identified_pions}
    \end{subfigure}
    \hfill
    \begin{subfigure}{0.48\linewidth}
        \centering
        \includegraphics[width=\linewidth]{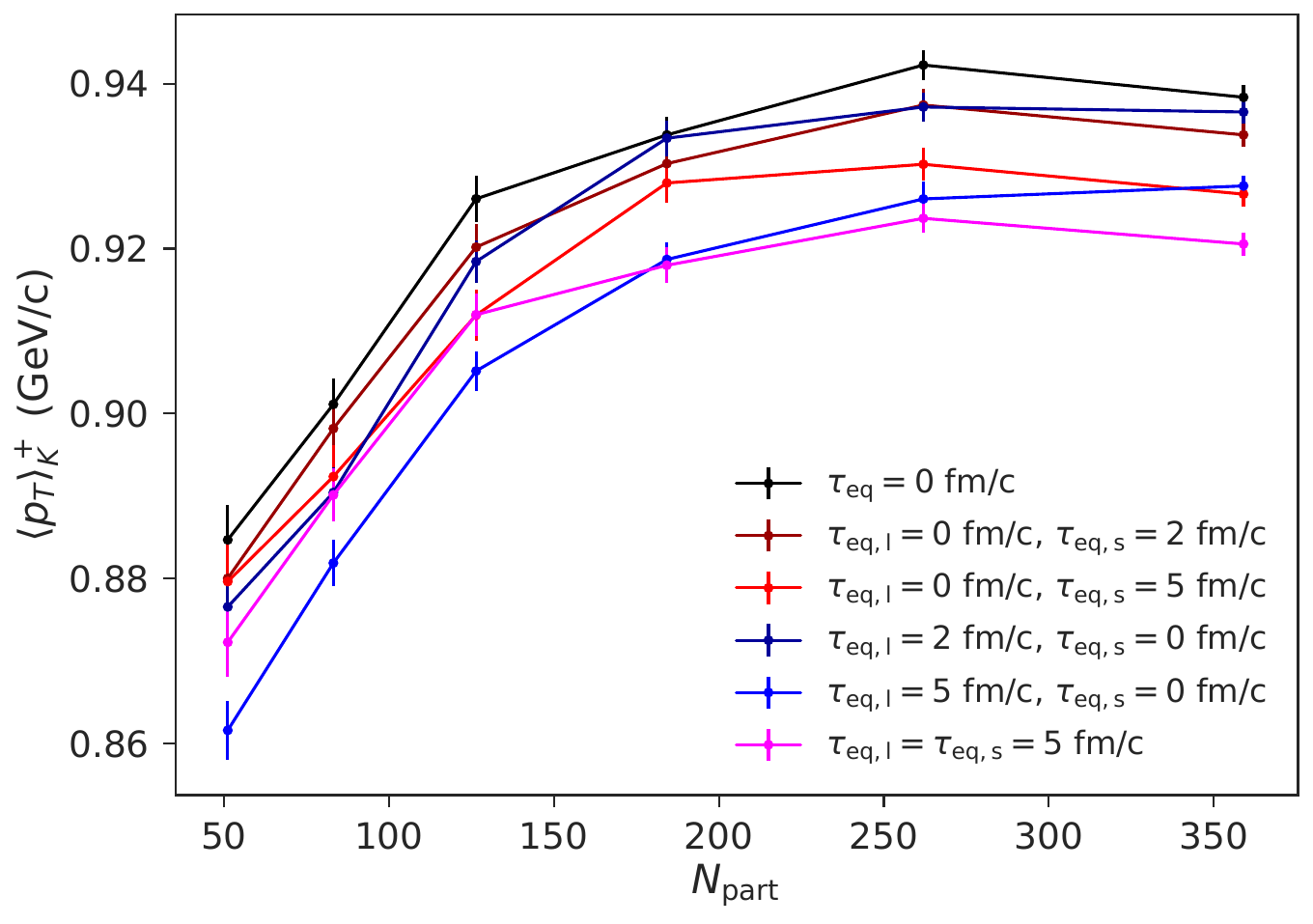}
        \caption{$K^+$}
        \label{fig:pT_identified_kaons}
    \end{subfigure}

    \vspace{0.4cm}

    \begin{subfigure}{0.48\linewidth}
        \centering
        \includegraphics[width=\linewidth]{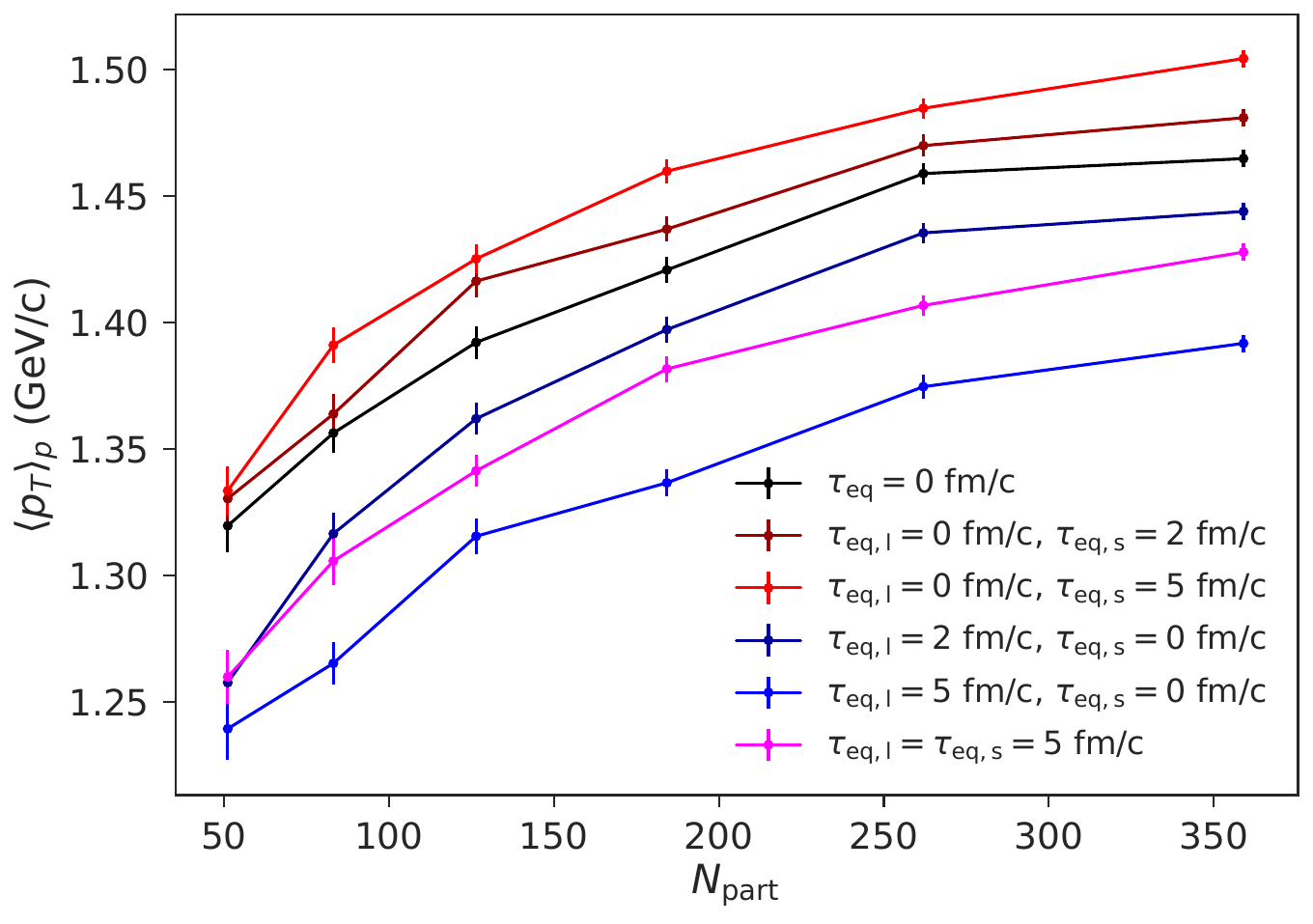}
        \caption{$p$}
        \label{fig:pT_identified_protons}
    \end{subfigure}

    \caption[Mean transverse momentum of pions, kaons, and protons for events evolved with varying $\tau_{\text{eq},l}$ and $\tau_{\text{eq},s}$]{Mean transverse momentum of pions, kaons, and protons for $0$--$60$\% centrality events evolved with varying $\tau_{\text{eq},l}$ and $\tau_{\text{eq},s}$. Each point corresponds to a $10$\% centrality bin.}
    \label{fig:pT_identified_strange}
\end{figure}

The identified particle $\langle p_T \rangle$, shown in Fig.~\ref{fig:pT_identified_strange} for pions, kaons, and protons, reflects a superposition of the common hydrodynamic response and species-dependent particlization effects. Each species predominantly follows the same light-driven trend as the charged $\langle p_T \rangle$, but is additionally shifted in correlation with its own yield: a species tends to be enhanced in $\langle p_T \rangle$ when its yield is enhanced, and suppressed when its yield is suppressed, following the flavor redistribution of section~\ref{subsec:hadrons_strange}. The remaining identified species are too noisy to resolve at the available statistics and are not shown.

\FloatBarrier

\begin{figure}[!htbp]
    \centering
    \includegraphics[width=0.7\linewidth]{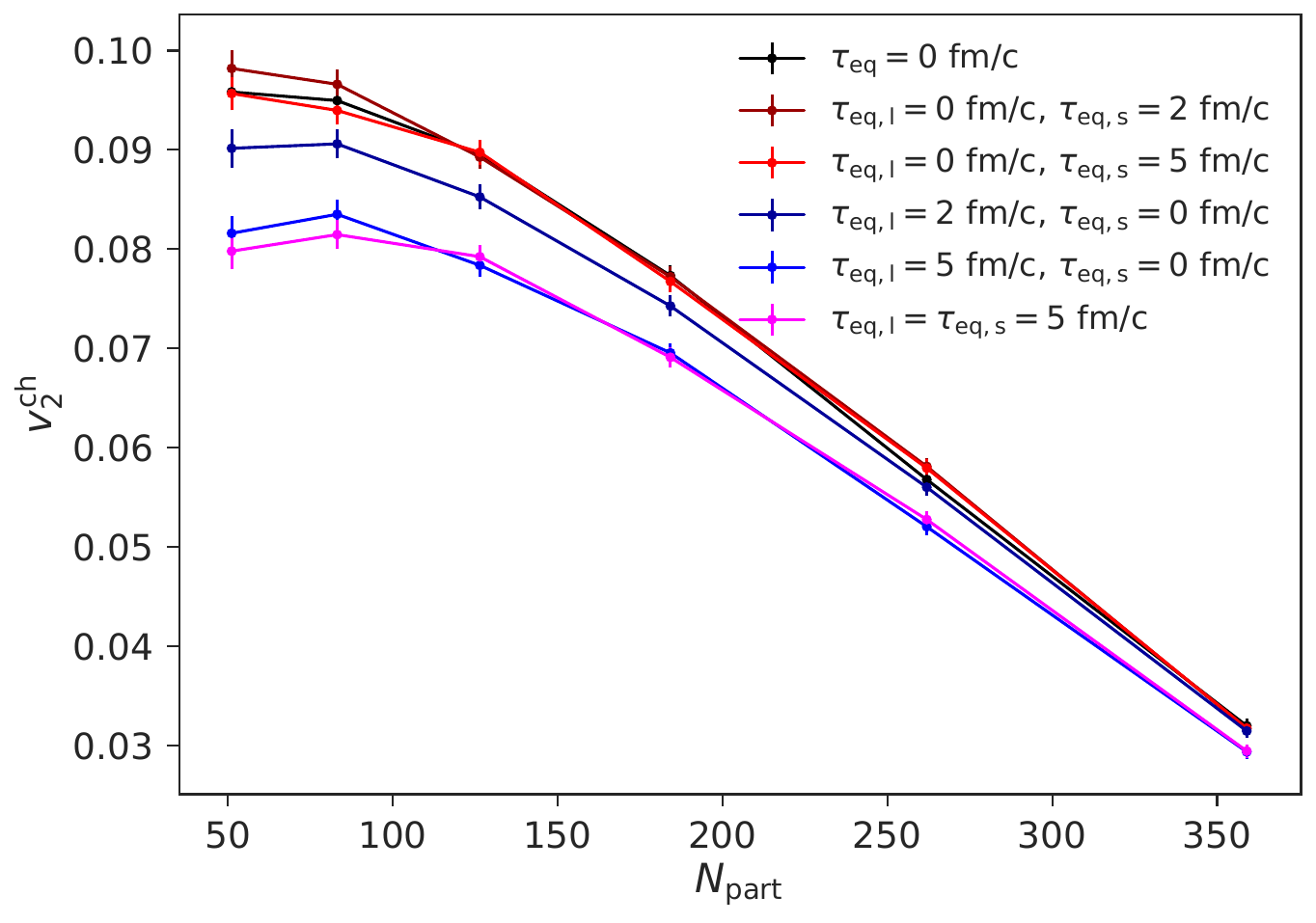}
    \caption[Elliptic flow of charged particles for events evolved with varying $\tau_{\text{eq},l}$ and $\tau_{\text{eq},s}$]{Elliptic flow of charged particles for $0$--$60$\% centrality events evolved with varying $\tau_{\text{eq},l}$ and $\tau_{\text{eq},s}$. Each point corresponds to a $10$\% centrality bin.}
    \label{fig:v2ch_strange}
\end{figure}

The flow anisotropy behaves much as in the flavor-independent case. Fig.~\ref{fig:v2ch_strange} shows the elliptic flow $v_2$ of charged particles. Its dependence on chemical equilibration closely follows the flavor-independent result: the suppression of $v_2$ tracks $\tau_{\text{eq},l}$, while variations in $\tau_{\text{eq},s}$ have little effect. As discussed in section~\ref{subsec:flow_pce}, this suppression mimics at the level of final observables the effects of larger shear or bulk viscosity.

\FloatBarrier

\subsection{Computational considerations}
\label{subsec:numerics_strange}

The partial chemical equilibrium model introduces a feature absent from standard hydrodynamic simulations with a fixed equation of state: the thermodynamic relation between pressure and energy density changes during the evolution. This has consequences for the numerical stability of the solver. Although the flavor-independent model of section~\ref{section31} is affected as well, these considerations are most acute for the flavor-dependent model, and I collect them here. I document them both to record the controls required for stable evolution and because they bear on the interpretation of the model.

In a conventional simulation the equation of state is fixed, and a fluid cell evolves adiabatically in ideal hydrodynamics, with entropy increasing only through dissipation. Here the quark fugacities are instead prescribed as explicit functions of proper time, Eq.~\ref{eq:flavor_fugacities}, so the equation of state relating $P$ and $\varepsilon$ drifts as a cell evolves, and entropy is produced as the medium chemically equilibrates (section~\ref{subsec:evolution_pce}) without an explicit microscopic source term. The evolution still enforces the conservation law $\partial_\mu T^{\mu\nu} = 0$ and treats the medium as a relativistic fluid; what departs from a standard treatment is that the relation between $P$ and $\varepsilon$ is no longer fixed. It is in this sense that the prescription strictly speaking violates the assumptions of hydrodynamics, and this has practical consequences.

The most visible of these appears in the recovery of the fluid velocity. At each timestep the primitive variables --- the local energy density and flow velocity, with the pressure then following from the equation of state --- must be reconstructed from the conserved components of $T^{\mu\nu}$, which requires solving a nonlinear equation in which the equation of state enters. As discussed in section~\ref{subsec:eos_strange}, interpolating with the pure glue equation of state imprints a first-order transition of magnitude $(1-\gamma_l)$ on the medium for every light fugacity $\gamma_l < 1$, and by construction (Eq.~\ref{eq:Tc_strange}) this transition coincides with the particlization temperature $T_\text{c}(\gamma_l,\gamma_s)$. As a fluid cell at low fugacity cools toward $T_\text{c}(\gamma_l,\gamma_s)$, the speed of sound $c_s^2 = \partial P / \partial \varepsilon$ becomes small near this transition. The flattening of $P(\varepsilon)$ is a natural source of difficulty, since the velocity reconstruction becomes more ill-conditioned when the pressure changes only weakly with energy density. I do not attempt to quantify the separate contribution of the small speed of sound, but in practice the reconstruction failures occur in this region; without adequate controls, the velocity solution breaks down and produces highly nonphysical evolution. I mitigate this using the hybrid Newton-based root-finding routine within MUSIC~\cite{Schenke:2010nt, Schenke:2010rr, Paquet:2015lta}, with internal parameters tuned so that cells failing to yield a physical solution are reverted more aggressively to a safe state. This is a configuration of the existing solver, not a modification of the reconstruction itself.

This solver tuning is the first of the controls used to stabilize the evolution, and for the flavor-independent results of section~\ref{section32} it is sufficient on its own. The flavor-dependent equation of state is numerically more demanding, however, and two further safeguards are used. The first addresses the cells most strongly affected by the ill-conditioning, which are those at low temperature and low fugacity, near the problematic region of the equation of state. These cells typically lie at the dilute edge of the medium and contribute little to final-state observables, so cells that persistently fail to yield a physical solution are forced to vacuum, removing them from the active evolution with minimal impact on the results.

The second safeguard is a floor on the quark fugacities. The flavor-independent runs of section~\ref{section32} begin from a purely gluonic state and require no such floor; the flavor-dependent runs instead fix the initial fugacities to $\gamma_l^0 = \gamma_s^0 = 0.1$. Because the fugacities relax upward from their initial values, this holds $\gamma_l, \gamma_s \geq 0.1$ throughout the evolution, keeping every cell away from the immediate vicinity of zero fugacity, where the reconstruction is most poorly conditioned. The same lower bound is used in the broader calibration of chapter~\ref{chapter5}, which samples the equilibration parameters across a wide prior and can therefore drive more of the medium toward the difficult regime.

These safeguards greatly reduce, but do not eliminate, failures of the velocity solver, so their extent must be monitored. I track the fraction of cells flagged with nonphysical velocities over the course of each evolution, and treat a growing fraction as a signal that a given parameter point is straining the model.

These numerical difficulties are not merely technical; they reflect a genuine limitation of applying the present hydrodynamic treatment to a chemically undersaturated medium. Chemical non-equilibrium and kinetic non-equilibrium are distinct: a medium with $\gamma_l < 1$ or $\gamma_s < 1$ need not have large gradients, so the validity of the gradient expansion is not in itself the issue. The limitation is more specific to the closure used here. Our construction places a first-order transition of magnitude $(1-\gamma_l)$ into the equation of state, and the region near this transition --- where the speed of sound softens and the equation of state responds most sharply to changes in fugacity --- is where both the solver and the underlying hydrodynamic description are under the most strain. Moreover, the transport coefficients used throughout were extracted under the assumption of a chemically equilibrated QGP, so applying them at low fugacity is an extrapolation rather than a controlled limit. The fugacity floor restricts access to this region rather than resolving it. The numerical breakdown and the regime where the physics is least defensible therefore largely coincide, and the floor serves a dual purpose: it keeps the evolution away from the most ill-conditioned fugacity range while restricting the model to the range of fugacities where the hydrodynamic treatment is most justified.

Together, the results of this chapter show that chemical equilibration can leave observable imprints on both the hydrodynamic response and the final hadron chemistry. In the flavor-independent model, delayed equilibration primarily modifies the pressure gradients and suppresses transverse flow while leaving the charged multiplicity nearly unchanged. Separating light and strange equilibration reveals a more differential signal: bulk observables remain controlled mainly by the light sector, while identified hadron yields redistribute according to valence flavor content. These sensitivities motivate the Bayesian calibration of chapter~\ref{chapter5}, where the equilibration parameters are varied simultaneously with the remaining model parameters and confronted with experimental data.

\chapter{Bayesian inference for heavy-ion collisions}
\label{chapter4}

We have seen that our paradigm for heavy-ion collisions relies on complex, multistage models. The initial conditions, hydrodynamics, and particlization all depend on a number of model parameters, and allowing for quark chemical non-equilibrium only adds to these. Given a particular choice of parameters, it is at least conceptually straightforward (even if often nontrivial in practice) to run this multistage model setup and obtain observables as outputs. Ultimately, however, this alone tells us too little about the underlying theory; we already know from experiments the yields and other properties of particles produced in heavy-ion collisions. What we really seek to solve here is a much trickier inverse problem: figuring out which model parameters produce outputs that agree with the data. That is where Bayesian inference plays a critical role.

This chapter therefore shifts focus from discussing the physics model itself to the statistical framework used to connect that model to experimental measurements. I will introduce the framework of Bayesian inference in a general form, agnostic to our specific use case. Then, in chapter \ref{chapter5}, I apply these techniques to the model introduced in chapter \ref{chapter3} in order to empirically constrain the quark flavor equilibration timescales alongside all the other model parameters.

Since learning by doing is often more effective than solely reading about techniques in the abstract, I have also made available an \href{https://github.com/andrewgordeev/bayesian_inference_tutorial}{interactive Bayesian inference tutorial} illustrating most of the concepts discussed below.

\section{Motivation for Bayesian inference}
\label{section41}

If we want to explore the range of possibilities for a given physics model, perhaps the most natural thing to try is to simply choose different values of our parameters and run the model with each. This was essentially the strategy for all of the results in chapter \ref{chapter3}. Such direct parameter scans are useful for building intuition, since they allow us to see quantitatively how changes in individual inputs affect the final observables.

However, this approach becomes increasingly inadequate once the model contains many parameters and many observables. Varying one or two parameters at a time may be informative in a low-dimensional setting, but realistic heavy-ion collision models generally contain a large number of coupled parameters whose effects on observables can overlap or partially compensate one another. In such cases, it becomes difficult to determine by inspection which regions of parameter space are genuinely favored by the data. A brute-force exploration of the full parameter space is also computationally prohibitive, especially in higher dimensions, as will be discussed further in section \ref{section43}.

Bayesian inference provides a robust framework for addressing this problem by systematically connecting model parameters to measured observables. Rather than relying on ad hoc parameter choices or limited scans, the Bayesian approach uses probability theory to quantify how strongly different regions of parameter space are supported by the experimental data. In this way, it becomes possible to calibrate a complex model against many observables simultaneously in a statistically consistent manner.

It is important to stress that what comes out of a Bayesian analysis is not merely a single best-fit value for each parameter, but rather, a posterior distribution over the parameter space — a full probability distribution quantifying which parameter values are supported by the data. This is particularly valuable because there may be broadly favored regions, strong correlations between parameters, or approximate degeneracies in which different combinations of parameters produce similar physical predictions. The posterior distribution captures this structure directly in a way that a simple fit generally cannot.

In addition to constraining our model parameters, Bayesian inference also provides a rigorous framework for quantifying uncertainties. Experimental uncertainties enter explicitly into the statistical analysis, and additional sources of uncertainty, such as those associated with model emulation, can likewise be incorporated. The result is therefore a probabilistic characterization of not only what the data imply about the model, but also a measure of how comprehensive or limited this knowledge may be.

Applying these techniques to heavy-ion collisions is a modern but not entirely novel idea; a number of major analyses in the field have already been done with Bayesian inference \cite{Bernhard:2015hxa, Bernhard:2016tnd, Bernhard:2018hnz, Moreland:2019szz, Bernhard:2019bmu, JETSCAPE:2020mzn, Nijs:2020roc}. These prior studies sought to constrain the properties of the initial state and the transport coefficients of the quark-gluon plasma, alongside quantities such as the particlization temperature, free-streaming time, jet-quenching parameters, and many others. The present work follows this general program, while extending it to the question of quark flavor equilibration.

\section{Bayesian formalism}
\label{section42}

The entire Bayesian framework rests on Bayes' theorem, which is a straightforward consequence of the definition of conditional probability:
\begin{equation}p(\boldsymbol{\theta}|\mathbf{y}) = \frac{p(\mathbf{y}|\boldsymbol{\theta})p(\boldsymbol{\theta})}{p(\mathbf{y})}.\end{equation}

This equation describes how our knowledge of the model parameters is updated after taking the experimental data into account. Note that since the model parameters are continuous, the quantities here are understood as probability densities, which I denote by lowercase $p$. The parameter vector $\boldsymbol{\theta}$ contains all model parameters involved in the inference, while the observable vector $\mathbf{y}$ contains all experimental data used to constrain them. For any given parameter choice, the model produces predictions in the form of a function $\mathbf{y}_\text{model}(\boldsymbol{\theta})$. Bayesian inference uses the comparison between $\mathbf{y}_\text{model}(\boldsymbol{\theta})$ and $\mathbf{y}$ to determine which regions of parameter space are favored by the data.

$p(\boldsymbol{\theta})$ is the prior distribution, which encodes any pre-existing knowledge or assumptions we have regarding the true values of the model parameters. In many applications, including the present one, the prior is chosen to be uniform within some physically reasonable range for each parameter, but this can, in principle, take any form.

The likelihood, $p(\mathbf{y}|\boldsymbol{\theta})$, is a function that measures how probable the observed data are under any given parameter choice. In practice, it quantifies the degree of agreement between the experimental measurements and the model predictions at particular parameter values. Assuming Gaussian uncertainties, the likelihood can be written as

\begin{equation} p(\mathbf{y}|\boldsymbol{\theta}) = \frac{1}{\sqrt{(2\pi)^N |\Sigma|}} \exp\left[-\frac{1}{2}\left(\mathbf{y} - \mathbf{y}_\text{model}(\boldsymbol{\theta})\right)^T
\Sigma^{-1}
\left(\mathbf{y} - \mathbf{y}_\text{model}(\boldsymbol{\theta})\right)\right], \end{equation}
where $\Sigma$ is the covariance matrix describing the uncertainties entering the comparison and $N$ is the number of observables. This form makes it clear that parameter choices leading to better agreement with the data are assigned larger likelihood, modulated by the scale of the uncertainties.

The posterior distribution $p(\boldsymbol{\theta}|\mathbf{y})$ represents the updated probability distribution of the parameters after incorporating information from experimental data. This posterior is the main object of interest in Bayesian inference, since it contains the full statistical information about which regions of parameter space are favored or disfavored. These statements are necessarily conditional on the assumed model, likelihood, and prior; Bayesian inference quantifies uncertainty within that specified statistical framework.

The evidence \(p(\mathbf{y})\) serves as a normalization constant ensuring that the posterior integrates to unity. Although it is often not evaluated explicitly in parameter estimation, it becomes important when comparing different models. If the model choice itself is treated as a discrete variable, then each model \(M\) has an associated evidence
\begin{equation}p(\mathbf{y}|M) = \int p(\mathbf{y}|\boldsymbol{\theta},M)\,p(\boldsymbol{\theta}|M)\,d\boldsymbol{\theta},\end{equation}
obtained by integrating the likelihood over the prior distribution of that model's parameters. The ratio of evidences for two competing models, \(M_A\) and \(M_B\), is known as the Bayes factor \cite{Kass:1995},
\begin{equation}\frac{p(\mathbf{y}|M_A)}{p(\mathbf{y}|M_B)}.\end{equation}

If this ratio is large, $M_A$ is favored by the evidence relative to the prior odds; if small, $M_B$ is favored. This can be seen by comparing the ratios of the prior and posterior odds assigned to the two candidate models:
\begin{equation}\frac{p(M_A|\mathbf{y})}{p(M_B|\mathbf{y})} = \frac{p(\mathbf{y}|M_A)}{p(\mathbf{y}|M_B)} \frac{p(M_A)}{p(M_B)}.\end{equation}

In this way, the evidence provides a natural framework for Bayesian model comparison, even though it plays only a passive role in the inference considered in the present work.

Once the posterior distribution has been determined, it can be analyzed in a number of useful ways. Marginal posterior distributions may be constructed for individual parameters or pairs of parameters, making it possible to visualize constraints, correlations, and degeneracies. Credible intervals may likewise be extracted in order to quantify the range of parameter values favored by the data at a given probability level. The specific procedure to determine the posterior and the resulting understanding that can be gleaned from it will be explained in more depth in section \ref{section45}, but we first must think more precisely about the model and parameters going into the analysis.

\section{Parameter space}
\label{section43}

Bayesian calibration requires defining a parameter space consisting of the model parameters to be constrained and the ranges over which they are allowed to vary. These parameters are not necessarily identical to all inputs taken by the model, since some inputs may be fixed externally rather than freely tunable. One of the first steps in any Bayesian analysis is therefore to identify which quantities are to be treated as genuine model parameters and which are not.

For each such parameter, we must specify a range of values worth considering. This is one of the main places where our existing physics intuition is essential, as we can exclude regions that are not physically meaningful. At the same time, these ranges should not be chosen too narrowly, since doing so may artificially restrict the final posterior distribution. It is generally better to err on the side of being inclusive, since values that fit the data poorly will naturally be disfavored. However, the range should not be made so broad that the limited set of feasible model evaluations is spread too thinly across physically implausible or difficult-to-emulate regions.

\subsection{Choice of priors}

Priors are then assigned over this parameter space. This is a nontrivial choice: the resulting posterior distribution will always be sensitive, to some degree, to the prior. A common convention is a uniform prior if we have no reason to favor one particular region of the space over another. Note that this is not equivalent to a truly unbiased prior: uniformity is parameterization-dependent, since a prior that is flat in a parameter is not flat in a nonlinear transformation of it, so even this choice encodes a nontrivial assumption. Also, a sharp cutoff at the boundary is often problematic: it is hard to imagine that a normalization of $19.99$ would be as likely as any other choice, but a normalization of $20.01$ would be outright forbidden.

\subsection{Design point selection}

With a fast enough model, we could immediately proceed to evaluate the model all across the parameter space once we have selected our priors. However, in practice, heavy-ion collision simulations are far too computationally intensive for this. The precise computational cost depends on the particulars of the models being used and the observables under consideration, which typically necessitate many repeated runs at one point in parameter space for adequate statistics. Typically, though, it is on the order of thousands of CPU-hours per design point. This limitation is the primary motivation for introducing a finite set of design points at which the model is evaluated. An emulator, discussed in more detail in section \ref{section44}, is trained on the model inputs and outputs at these points, and then acts as a surrogate model to produce model-like outputs across the remainder of the parameter space.

How many design points to use depends on the model and, in particular, its smoothness across the parameter space. There is always a tradeoff to consider: adding design points generally will improve the emulator performance, but also necessitates higher computational costs in the form of additional model evaluations. Higher dimensionality parameter spaces naturally necessitate more design points for better coverage, but one must be cautious with selecting these points. 

An intuitive choice would be to define a grid of uniformly spaced design points for even coverage. This works well enough in one- or two-dimensional spaces, but scales very poorly to higher dimensions. If we suppose we want $m$ samples of each parameter for each of $n$ parameters, then we would need $m^n$ design points. With modest choices of $m=5$ and $n=10$, we would need $5^{10} \approx 9.8 \times 10^6$ design points. And in practice, this still likely would not be enough to provide good enough coverage. Such a strategy is therefore computationally impossible for realistic heavy-ion physics.

One might suppose that purely random sampling could be more reasonable in high dimensions, since we circumvent the rigid exponential scaling of a lattice. However, this comes with a high risk of leaving large regions of our parameter space inadequately sampled in higher dimensional spaces. If the model has any noteworthy behavior in such a region, then the emulator trained on that design will fail to reproduce that structure, thus hindering the final Bayesian analysis. Efficient sampling of parameter space therefore requires a method that avoids both the exponential cost of a grid and the uneven coverage of naive random sampling.

The standard method of choice for this is Latin hypercube sampling \cite{McKay:1979}. In this approach, each dimension is divided into subintervals, and the design is constructed so that each subinterval in each dimension contains exactly one sample point. This is analogous to filling out a Sudoku grid, which is itself a special type of Latin square. The scaling to many dimensions then becomes manageable, as we can choose an arbitrary number $k$ of design points. This avoids the rigid $m^n$ scaling of the grid, although the number of points required for adequate emulator accuracy depends on the dimensionality and complexity of the model response. As a rough heuristic, though, $\sim\!10$ design points per dimension is needed for acceptable emulator accuracy with a smooth model \cite{Loeppky:2009}, and more is typically better.

Due to their simplicity and ubiquity in heavy-ion collision analyses, I will exclusively adopt Latin hypercube sampling in this work. However, it should be noted that there are alternatives. For example, Sobol sequences \cite{Sobol:1967} are a type of quasi-random low-discrepancy sequence that can also be used to generate well-distributed samples in high-dimensional parameter spaces.

There are likewise many algorithms for Latin hypercube sampling, including more complex adaptive methods that weight the distribution of points differently depending on the choice of prior, but I will restrict myself to a simple and widely used maximin criterion \cite{Morris:1995}, which maximizes the minimum distance between points. This helps avoid clustering of design points in order to improve coverage of the parameter space.

\section{Model emulation}
\label{section44}

Recall that the motivation for choosing a limited number of design points was that directly evaluating the model across a vast range of parameter values becomes entirely impractical for expensive models, as in the case of heavy-ion collision simulations. This is particularly true for Markov chain Monte Carlo methods that rely on extensive sampling through repeated runs of the model to probe statistical distributions. 

\begin{figure}[!htbp]
    \centering
    \includegraphics[width=\textwidth]{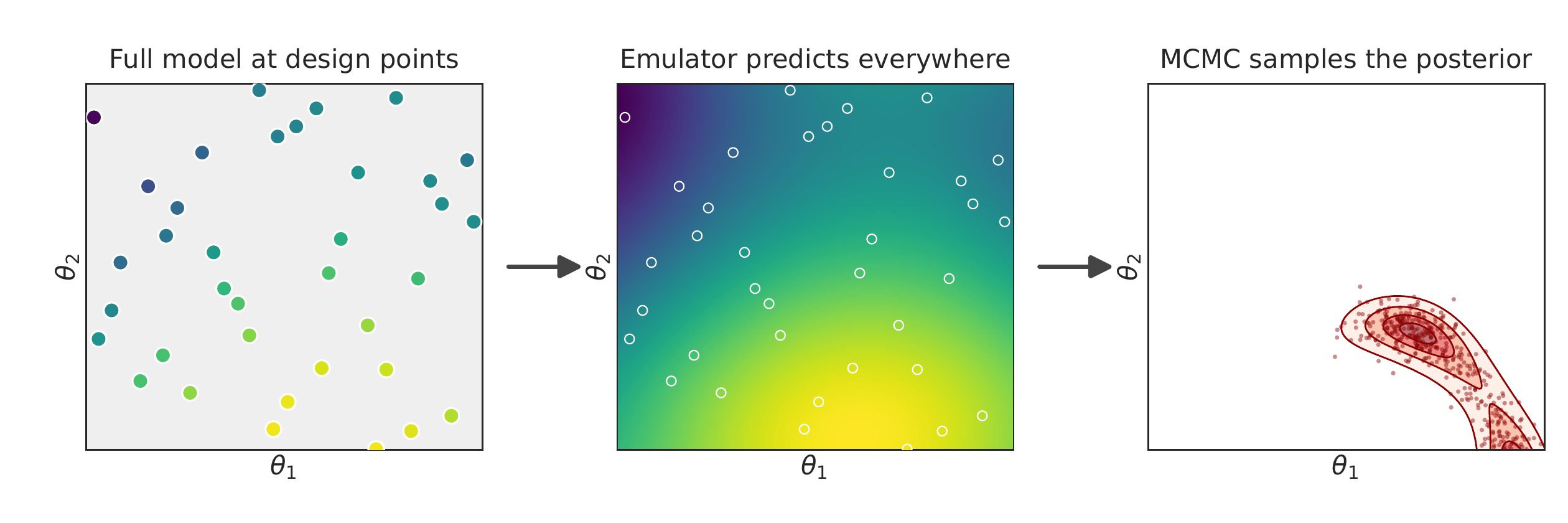}
    \caption[Schematic of the Bayesian calibration workflow in a
    two-dimensional parameter space]{Schematic of the Bayesian calibration workflow in a
    two-dimensional parameter space. The full model is evaluated at a
    finite set of design points, and an emulator is trained to predict
    the model response throughout the parameter space. The emulator is
    then used within MCMC to sample the posterior distribution. The
    colors in the first two panels represent the value of a model
    observable.}
    \label{fig:bayesian_workflow}
\end{figure}

An emulator is a surrogate model trained to map model inputs to model outputs \cite{Kennedy:2001}. This is typically done without detailed knowledge of the internal calculations performed by the model itself, instead relying on statistical relationships inferred from the outputs at the design points. If constructed correctly, the emulator can accurately predict model outputs even for parameter values at which the full model was not actually run, thereby reducing the computational cost by many orders of magnitude. The central challenge, naturally, is ensuring that these emulator predictions remain faithful to the original model. The overall calibration workflow is illustrated schematically in Fig.~\ref{fig:bayesian_workflow}.

\Needspace{4\baselineskip}
\subsection{Gaussian process emulators}

In heavy-ion physics, most widely used emulators are based on Gaussian processes. These are chosen because they make relatively few assumptions about the model, can flexibly interpolate many-dimensional functions, and naturally quantify their own uncertainty. A Gaussian process is a stochastic process, or a collection of random variables \cite{Rasmussen:2006}. It is defined by the property that for any finite set of input points $\{\boldsymbol{\theta}_1, \dots, \boldsymbol{\theta}_n\}$, the corresponding function values
\begin{equation}\mathbf{f} = \big(f(\boldsymbol{\theta}_1), \dots, f(\boldsymbol{\theta}_n)\big)\end{equation}
have a multivariate normal distribution,
\begin{equation}\mathbf{f} \sim \mathcal{N}(\boldsymbol{\mu}, \mathbf{K}).\end{equation}

The Gaussian process is fully specified by a mean function $\mu(\boldsymbol{\theta})$ and a covariance function, or kernel, $k(\boldsymbol{\theta}, \boldsymbol{\theta'})$, such that
\begin{equation}\mu_i = \mu(\boldsymbol{\theta}_i), \qquad K_{ij} = k(\boldsymbol{\theta}_i, \boldsymbol{\theta}_j).\end{equation}

In other words, a Gaussian process defines a probability distribution over functions. It is thus a nonparametric regression method, in which we do not assume any particular functional form for the mapping between parameters and observables. Given training data consisting of the model inputs at the design points $\boldsymbol{\theta}_\text{t}$ and the corresponding model outputs $\mathbf{y}_\text{t} = \mathbf{f}(\boldsymbol{\theta}_\text{t})$, we place a Gaussian process prior over the unknown function $f(\boldsymbol{\theta})$. Conditioning this prior on the observed training data yields a posterior distribution over functions. From this posterior, we can obtain for any new input $\boldsymbol{\theta_{*}}$ a predictive mean and variance by averaging over all functions in the posterior weighted by their probability.

\subsection{Kernel functions and hyperparameter optimization}

Conventionally, one often sets $\mu(\boldsymbol{\theta}) = 0$ and encodes the structure in the kernel. The kernel $k(\boldsymbol{\theta}_i, \boldsymbol{\theta}_j)$ encodes assumptions about smoothness and correlation structure in parameter space. The emulator does not learn the underlying physics; rather, it learns the statistical structure of how the model outputs vary across parameter space.

There are many possible choices for the kernel. One standard choice is the squared exponential, or radial basis function, kernel, which tends to work well for models that vary relatively smoothly with their parameters:
\begin{equation}k(\boldsymbol{\theta}_i, \boldsymbol{\theta}_j) = \sigma_f^2 \exp{\left(-\frac{1}{2} \sum_k \frac{|\theta_{ik}-\theta_{jk}|^2}{\ell_k^2}\right)},\end{equation}
for some overall variance scale $\sigma_f$ and characteristic length scales $\ell_k$. $\ell_k$ controls how rapidly the function varies with each $k$th parameter. Small values of $\ell_k$ correspond to rapid variation along the \(k\)th parameter direction, while large values correspond to smoother dependence.

In many applications, including heavy-ion collisions, the model outputs are not fully deterministic. Initial-state fluctuations and particle sampling introduce statistical noise, so that the model outputs are more accurately written as
\begin{equation}
    y_{\text{model},i}(\theta) = f_i(\theta) + \epsilon_i
\end{equation}
for some noise term $\epsilon_i$. This can be accounted for in the kernel by adding a noise term:
\begin{equation}k(\boldsymbol{\theta}_i, \boldsymbol{\theta}_j) = \sigma_f^2 \exp{\left(-\frac{1}{2} \sum_k \frac{|\theta_{ik}-\theta_{jk}|^2}{\ell_k^2}\right)} + \sigma_n^2 \delta_{ij},
\end{equation}
where $\sigma_n$ is an additional hyperparameter, $\delta_{ij}$ is the Kronecker delta, and we rely on an assumption similar to earlier that the noise is Gaussian and uncorrelated between training points. This is the simplest homoscedastic noise model; if point-dependent statistical variances are known, they can instead be supplied as a heteroscedastic diagonal contribution.

The quantities \(\ell_k\), \(\sigma_f\), and \(\sigma_n\) are examples of hyperparameters: tunable parameters that control the behavior of the emulator rather than the underlying physical model. In practice, these hyperparameters are usually chosen by maximizing the marginal likelihood of the training data under the Gaussian process prior. This quantity measures how probable the observed training outputs are after integrating over all functions allowed by the Gaussian process. For training outputs \(\mathbf{y}\) at inputs \(X\), the marginal likelihood is
\begin{equation}p(\mathbf{y}|X,\phi) = \frac{1}{\sqrt{(2\pi)^{n_\text{t}} |K|}} \exp\left(-\frac{1}{2}\mathbf{y}^T K^{-1} \mathbf{y}\right),\end{equation}
where \(\phi\) denotes the full set of hyperparameters, \(K\) is the covariance matrix evaluated at the training points, and \(n_\text{t}\) is the number of training points. Equivalently, one may maximize the log marginal likelihood,
\begin{equation}\log p(\mathbf{y}|X,\phi)=-\frac{1}{2}\mathbf{y}^T K^{-1} \mathbf{y}-\frac{1}{2}\log |K|-\frac{n_\text{t}}{2}\log(2\pi). \end{equation}

The first term favors hyperparameter choices that reproduce the training data well, while the second disfavors overly flexible covariance structures. In this way, hyperparameter optimization balances overfitting against underfitting in a principled manner. An overfit model will treat even small amounts of noise in the model outputs as true variation with respect to the parameters; such a model spreads its predictive probability over a wider range of possible data and therefore assigns lower probability to the data actually observed, which is reflected in a lower marginal likelihood. On the other hand, an underfit model will wash out real parameter dependence, and thus will poorly fit the data.

\subsection{Principal component analysis}

A standard Gaussian process emulator predicts a scalar output as a function of a vector input. Naively, one might then attempt to construct a separate Gaussian process for every observable under consideration. However, this would ignore the often substantial correlations between different observables. For example, particle multiplicities across all centrality bins will typically increase with the energy deposition of the initial condition, while flow coefficients will tend to decrease with the magnitude of the shear viscosity. At the same time, these observables are not redundant; they still retain distinct sensitivity to other parameters. To reduce the dimensionality of the problem without discarding the dominant correlated structure in the data, it is standard to apply dimensionality reduction, replacing the full suite of model observables with a smaller set of related variables that encapsulate their main features.

The most widely used technique for dimensionality reduction is principal component analysis (PCA), which transforms the original observables into a new orthonormal basis of uncorrelated linear combinations known as principal components (PCs). These PCs are ordered such that the first components capture the largest possible fraction of the variance in the training data. In practice, only the leading principal components are retained, typically enough to explain some large fraction of the total variance, such as $95$\% or $99$\%. The emulator is then trained on these principal components rather than on the original observables directly, and predictions for the original observables are reconstructed from the predicted PCs afterward \cite{Higdon:2008}. Before applying PCA, the outputs are centered and often standardized when the observables have different units or characteristic scales; otherwise, observables with the largest numerical variance would dominate the decomposition. 
\Needspace{4\baselineskip}
\subsection{Emulator uncertainty and validation}

One of the main advantages of Gaussian process emulators is that they output a well-defined predictive uncertainty. This uncertainty reflects the variance among the plausible functions allowed by the posterior Gaussian process and generally grows in regions of parameter space that are farther from the training points. In the Bayesian calibration procedure, this emulator uncertainty must be propagated into the likelihood, so that poorly constrained emulator predictions do not lead to artificially overconfident posterior constraints.

Because the final Bayesian analysis depends crucially on the emulator being faithful to the original model, emulator validation is essential. While plots comparing emulator predictions to model outputs can provide useful intuition, cross-validation offers a more systematic assessment. The core idea is to retrain the emulator on only part of the design and then test its predictions on held-out design points that were not used during training. Two common approaches are:

\begin{enumerate}[
    label=\arabic*.,
    leftmargin=2.2em,
    labelsep=0.6em,
    itemsep=0.8em,
    topsep=0.8em,
    parsep=0pt
]

    \item \textbf{Leave-$p$-out cross-validation:}
    Use $p$ samples as the validation set and the remaining samples for training, then repeat for all possible choices of the $p$-sample subset.

    \item \textbf{$k$-fold cross-validation:}
    Use one of $k$ equal-sized subsamples, or folds, as the validation set and the remaining samples for training, then repeat $k$ times. In the limit that $k$ equals the number of samples, this method becomes equivalent to leave-one-out cross-validation.

\end{enumerate}

Both are generally reasonable methods for measuring the emulator's performance at predicting model outputs for new parameter values.

Several quantitative diagnostics are useful for this purpose. One is the root-mean-square error (RMSE), which measures the typical size of the difference between emulator predictions and the true model outputs at the validation points:
\begin{equation}\text{RMSE} = \sqrt{\frac{1}{N}\sum_i (y_{\text{true},i} - \mu_{\text{emu},i})^2},\end{equation}
where the sum runs over the $N$ validation points. Another is the standardized residual,
\begin{equation}z = \frac{y_{\text{true}} - \mu_{\text{emu}}}{\sigma_{\text{emu}}},\end{equation}
which tests whether the emulator uncertainty is properly calibrated. If the predictive uncertainty is accurate, these residuals should be distributed approximately as $\sim \mathcal{N}(0,1)$. Finally, one may examine the empirical coverage of nominal $68$\% and $95$\% predictive intervals. If significantly fewer validation points fall within these intervals than expected, the emulator is overconfident; if significantly more do, it is underconfident.

\section{Posterior determination}
\label{section45}

Everything so far has been setting the pieces in place for a Bayesian calibration. With an emulator in place, we can now combine its predictions with experimental data in order to determine the posterior distribution over the model parameters. The first step is to specify the likelihood. As introduced in section~\ref{section42}, assuming that the residuals between the data and the model predictions are Gaussian-distributed, the likelihood takes the multivariate form
\begin{equation}
p(\mathbf{y}|\boldsymbol{\theta}) = \frac{1}{\sqrt{(2\pi)^N |\Sigma|}} \exp\left[ -\frac{1}{2} \left(\mathbf{y} - \mathbf{y}_{\text{model}}(\boldsymbol{\theta})\right)^T \Sigma^{-1} \left(\mathbf{y} - \mathbf{y}_{\text{model}}(\boldsymbol{\theta})\right) \right],
\end{equation}
where $N$ is the number of observables and $\Sigma$ is the covariance matrix describing the uncertainties entering the comparison. This form allows one to account not only for the size of the uncertainties on individual observables, but also for correlations between them. Here $\mathbf{y}_\text{model}(\boldsymbol{\theta})$ is in practice the emulator's predictive mean, and, assuming the experimental and emulator uncertainties are independent, $\Sigma$ combines their covariances as $\Sigma = \Sigma\text{exp} + \Sigma_\text{emu}$, so that regions where the emulator is poorly constrained are appropriately down-weighted. In the special case that the uncertainties between observables are uncorrelated, $\Sigma$ becomes diagonal, $\Sigma_{ij} = \sigma_i^2 \delta_{ij}$, and the likelihood reduces to a product of independent Gaussians:
\begin{equation}p(\mathbf{y}|\boldsymbol{\theta}) = \prod_{i=1}^{N} \frac{1}{\sqrt{2\pi \sigma_i^2}} \exp\left(-\frac{\left(y_i - y_{\text{model},i}(\boldsymbol{\theta})\right)^2}{2\sigma_i^2}\right).\end{equation}

In practice, the likelihood often evaluates to very small numbers. For numerical reasons, it is easier to work with the logarithm of the likelihood:
\begin{equation}
\log p(\mathbf{y}|\boldsymbol{\theta}) = -\frac{1}{2} \left(\mathbf{y} - \mathbf{y}_{\text{model}}(\boldsymbol{\theta})\right)^T \Sigma^{-1} \left(\mathbf{y} - \mathbf{y}_{\text{model}}(\boldsymbol{\theta})\right) -\frac{1}{2}\log |\Sigma| -\frac{N}{2}\log(2\pi).
\end{equation}

This avoids numerical underflow and turns products into sums, which is advantageous when the likelihood is evaluated many times. Additionally, using the logarithm puts Bayes' theorem in the convenient form:
\begin{equation}\log{p(\boldsymbol{\theta}|\mathbf{y})} = \log p(\mathbf{y}|\boldsymbol{\theta}) + \log p(\boldsymbol{\theta}) - \log{p(\mathbf{y})}\end{equation}
Because $p(\mathbf{y})$ is a constant with respect to our parameters, one can ignore the evidence term for now and calculate an unnormalized posterior.

\subsection{Markov chain Monte Carlo sampling}

Even with an emulator in place, it is still not practical to evaluate the posterior directly over a high-dimensional parameter space. As with the design points, a direct grid-based evaluation scales exponentially with dimension. A grid with $m$ points in each of $n$ dimensions requires $m^n$ evaluations. Even with a modest $m=50$, a $10$-dimensional problem would already require about $10^{17}$ evaluations. In realistic applications, this becomes impractical very quickly.

Instead of evaluating the posterior everywhere, we would like to focus computational effort where the posterior density is largest. That is the primary motivation for using Markov chain Monte Carlo (MCMC) methods. The core principle of MCMC sampling is to construct a Markov chain whose stationary distribution is the probability distribution of interest, in this case the posterior. Starting from an initial point in parameter space, the chain proposes new parameter values, accepts or rejects them according to the posterior density, and repeats this process many times. After an initial burn-in period, the resulting sequence of samples is distributed according to the target posterior,
\nopagebreak[4]
\begin{equation}
\{\boldsymbol{\theta}_1,\boldsymbol{\theta}_2,\dots\}\sim p(\boldsymbol{\theta}|\mathbf{y}).
\end{equation}

Importantly, MCMC does not require explicit evaluation of the evidence; it requires the posterior density only up to a constant of proportionality. This is perfectly fine for Bayesian parameter estimation, where the evidence is often intractable.

One of the simplest and most widely used MCMC methods is the Metropolis-Hastings algorithm \cite{Metropolis:1953am, Hastings:1970}. Starting from some initial parameter vector $\boldsymbol{\theta_0}$, the algorithm proceeds as follows:

\begin{enumerate}[
    label=\arabic*.,
    leftmargin=2.2em,
    labelsep=0.6em,
    itemsep=0.8em,
    topsep=0.8em,
    parsep=0pt
]

    \item \textbf{Propose a new point}
    $\boldsymbol{\theta'}$ from a proposal distribution $q(\boldsymbol{\theta'}|\boldsymbol{\theta})$, often taken to be a multivariate normal centered on the current point $\boldsymbol{\theta}$.

    \item \textbf{Compute the acceptance ratio}
    \begin{equation}
        r = \frac{p(\mathbf{y}|\boldsymbol{\theta'})p(\boldsymbol{\theta'})}
        {p(\mathbf{y}|\boldsymbol{\theta})p(\boldsymbol{\theta})}.
    \end{equation}

    \item \textbf{Accept or reject the proposed point}
    with probability $\min{(1,r)}$. Otherwise, remain at the current point.

    \item \textbf{Repeat}
     many times.

\end{enumerate}

Note that this form of the acceptance ratio holds for a symmetric proposal distribution, $q(\boldsymbol{\theta}'|\boldsymbol{\theta}) = q(\boldsymbol{\theta}|\boldsymbol{\theta}')$, as is the case for the Gaussian proposal above; this special case is the original Metropolis algorithm, and the Hastings generalization to asymmetric proposals multiplies $r$ by $q(\boldsymbol{\theta}|\boldsymbol{\theta}')/q(\boldsymbol{\theta}'|\boldsymbol{\theta})$.

While Metropolis-Hastings is conceptually simple, and perhaps the best-known MCMC algorithm, other MCMC methods are often more efficient in practice. In particular, the widely used \texttt{emcee} Python package \cite{Foreman-Mackey:2012any} uses affine-invariant ensemble sampling \cite{Goodman:2010}, in which an ensemble of walkers explores the posterior simultaneously. Using an ensemble rather than a single walker often improves exploration of the parameter space, especially when the posterior exhibits strong correlations or different characteristic scales in different directions. The affine-invariant property makes the sampler insensitive to linear rescalings and distortions of parameter space, which is especially useful with high-dimensional parameter spaces.

It generally takes hundreds of steps, if not many more (depending on the specific problem and algorithm), for the Markov chain to no longer depend on the starting position. Because of this, we almost always discard the first part of the chain as a ``burn-in'' phase. Furthermore, convergence of an MCMC chain is not determined solely by whether it has moved away from its starting point. Successive samples in the chain are generally correlated, so the number of effectively independent samples is smaller than the total chain length. For this reason, one must consider both burn-in and autocorrelation when assessing the quality of the posterior sample. In practice, this is typically monitored through trace plots, acceptance fractions, and estimates of the autocorrelation time.

\subsection{Marginalization and credible intervals}

The full posterior distribution in many dimensions is impossible to directly visualize, so we tend to focus on marginalized distributions. A marginal distribution is simply the posterior distribution for a subset of the parameters, usually one or two, after integrating out the rest. In general, the marginal distribution for a subset of the first $i$ parameters is given by:
\begin{equation}p(\theta_1, ...,\theta_i|\mathbf{y}) = \int d\theta_{i+1}...d\theta_n p(\boldsymbol{\theta}|\mathbf{y}). \end{equation}

From marginal distributions, we can compute credible intervals. These quantify the uncertainties on our parameters by designating a fraction and determining an interval containing that fraction of the posterior density; common conventions are the narrowest such interval (the highest posterior density interval) or the central interval with equal probability in each tail. For example, a $68$\% credible interval contains $68$\% of the posterior density, and indicates a $68$\% chance that the true value of the parameter falls within that interval. The most visible end products of a Bayesian analysis, which one will typically expect to see in talks and papers, are the $1$D and $2$D marginal distributions and corresponding credible intervals. These are often displayed together in a single corner plot, with the $1$D distributions along the diagonal of a right triangle and all $2$D correlations filling out the rest of the triangle.

A closely related quantity is the maximum a posteriori (MAP) estimate, given by the mode of the posterior distribution. This can be viewed as a Bayesian analogue of a best-fit point. However, one should treat it with caution, since a single point estimate can be unrepresentative of a broad or highly non-Gaussian posterior and contains no information about the associated uncertainty. The full posterior structure and the corresponding credible intervals are more informative than the MAP alone.

\subsection{Validation}

Once the posterior has been sampled, a useful diagnostic is the posterior predictive check \cite{Gelman:2013}. In this procedure, one draws parameter samples from the posterior, passes them back through the emulator, and thereby constructs a posterior predictive distribution for the observables. If the calibration is successful, the experimental data should appear statistically consistent with this posterior predictive distribution. This provides a direct check that the inferred parameter regions genuinely reproduce the observed measurements.

Closure tests provide a more robust validation of the Bayesian inference procedure by verifying that the method can recover known parameter values from synthetic data. In a closure test, pseudodata are generated using the model at a known parameter point and the full Bayesian calibration procedure is repeated. Similarly to emulator validation, the most effective way to do this is often cross-validation: taking one design point to represent the ``true'' parameter values, using the model output of that point in place of experimental data, and redoing the calibration using all the other design points. If everything works properly, the calibration should be able to produce a reasonable posterior distribution relative to the true parameter values. In general, one should do this for many different points and verify that the $68$\% credible interval contains the true value $68$\% of the time, and likewise for the $95$\% interval.

The discussion in this chapter has so far been deliberately general, with the aim of introducing the statistical machinery independently of any one specific heavy-ion model. In the next chapter, I apply this full Bayesian framework to the chemically non-equilibrated heavy-ion collision model developed in chapter \ref{chapter3}. There, the abstract ingredients introduced here --- parameter ranges, priors, emulator construction, likelihood evaluation, posterior sampling, and validation through predictive checks and closure tests --- are assembled into a concrete calibration procedure for constraining the quark flavor equilibration timescales alongside the remaining model parameters.
\chapter{Bayesian constraints on flavor equilibration and transport}
\label{chapter5}

The preceding chapters supplied both a forward model of heavy-ion collisions and the statistical machinery required to calibrate it. Chapter~\ref{chapter2} developed a multistage description of the collision: T\textsubscript{R}ENTo initial conditions, viscous hydrodynamic evolution with MUSIC, particlization through iS3D, and hadronic rescattering in SMASH. In chapter~\ref{chapter3}, this framework was extended by a non-equilibrium fugacity sector in which the light and strange quark abundances relax toward their equilibrium values over finite timescales. Chapter~\ref{chapter4} developed the Bayesian framework, combining Gaussian process emulation of the model response with Markov chain Monte Carlo sampling to make inference over the resulting high-dimensional parameter space tractable. This chapter brings these elements together by calibrating the chemical equilibration parameters and the medium's transport coefficients simultaneously against Au+Au collisions at $\sqrt{s_{NN}} = 200~\text{GeV}$. The fugacity parameters are therefore not fit in isolation; they are inferred jointly with the initial-state and transport parameters, so that interplay between chemical non-equilibrium and the bulk sector is reflected in the resulting constraints rather than hidden by fixed choices.

Existing global analyses of heavy-ion data generally assume that the quark-gluon plasma attains chemical equilibrium by the onset of the hydrodynamic stage. Although standard hydrodynamic calculations do not explicitly evolve quark fugacities, this assumption corresponds in the present framework to holding the light and strange fugacities fixed at unity throughout the hydrodynamic evolution. The use of hydrodynamics already presupposes rapid hydrodynamization, the onset of an effective hydrodynamic description of the medium's evolution, but this is distinct from chemical equilibration, the approach of each quark flavor to its equilibrium abundance. It is the latter that the fugacity sector of chapter~\ref{chapter3} renders dynamical. Within this framework, flavor equilibration becomes part of the inferred model structure rather than an imposed initial condition. With the equilibrium assumption lifted, three questions become empirical. First, do the data favor an initially undersaturated plasma, with light or strange fugacities below unity, as expected for an early gluon-dominated state? Second, are the corresponding equilibration timescales resolvable by the data, and does the strange sector equilibrate more slowly than the light sector, as the larger strange quark mass and correspondingly smaller production rates would suggest? Third, and most consequential for transport extraction more broadly, does opening the chemical non-equilibrium sector shift the inferred shear and bulk viscosities, given that bulk viscosity and quark fugacities both affect transverse momentum spectra and can therefore compete in describing the same measurements?

Au+Au collisions at $\sqrt{s_{NN}} = 200~\text{GeV}$ serve throughout as the calibration anchor, with the calibration data taken from the STAR collaboration. The calibration presented here is deliberately confined to this single collision system and beam energy. The phenomenological relevance of the fugacity sector beyond this system was explored in chapter~\ref{chapter3}, where chemical non-equilibrium was shown to leave observable imprints on Pb+Pb and O+O predictions. Extending the present calibration into a joint fit across these systems is a natural next step, and is discussed as future work in section~\ref{section55}.

\section{Parameters and priors}
\label{section51}

\begin{table}[!htbp]
\centering
\caption[Calibration parameters and prior ranges]{Calibration parameters, design ranges, and inference priors or fixed values. All parameters listed here vary across the Latin hypercube design. In the final column, ranges denote parameters sampled in the present calibration with independent uniform priors over the stated intervals, while single numerical entries denote parameters held fixed at that value during inference. The three normalizations in the lower block are varied across the multi-system design but do not enter the present Au+Au likelihood.}
\label{tab:priors}

\renewcommand{\arraystretch}{1.12}
\setlength{\tabcolsep}{0.25pt}

\begin{tabularx}{\textwidth}{
@{}
>{\raggedright\arraybackslash}p{0.80in}
>{\raggedright\arraybackslash}X
>{\centering\arraybackslash}p{1.00in}
>{\centering\arraybackslash}p{1.50in}
@{}
}

\toprule
\textbf{Parameter}
& \textbf{Description}
& \textbf{Design range}
& \textbf{Inference prior/value} \\
\midrule

\multicolumn{4}{l}{\color{linkred} Initial state}\\
\addlinespace[2pt]
$\text{Norm}_{200}$      & Au+Au $200$~GeV normalization               & $[5.0,\,20.0]$ & $[5.0,\,20.0]$ \\
$w$                        & Nucleon width (fm)                        & $[0.5,\,1.5]$ & $[0.5,\,1.5]$ \\
$k$                        & Multiplicity fluctuation shape            & $[0.8,\,2.2]$ & $1.5$ \\

\addlinespace[6pt]
\multicolumn{4}{l}{\color{linkred} Chemical equilibration}\\
\addlinespace[2pt]
$\tau_{\text{eq},l}$     & Light quark equilibration time (fm/$c$)   & $[0.0,\,10.0]$ & $[0.0,\,10.0]$ \\
$\tau_{\text{eq},s}$     & Strange quark equilibration time (fm/$c$) & $[0.0,\,10.0]$ & $[0.0,\,10.0]$ \\
$\gamma_l^{0}$             & Initial light quark fugacity              & $[0.1,\,1.0]$ & $[0.1,\,1.0]$ \\
$\gamma_s^{0}$             & Initial strange quark fugacity            & $[0.1,\,1.0]$ & $[0.1,\,1.0]$ \\

\addlinespace[6pt]
\multicolumn{4}{l}{\color{linkred} Shear viscosity $\eta/s(T)$}\\
\addlinespace[2pt]
$(\eta/s)_{\min}$          & Minimum, at $T_\text{c}$                         & $[0.01,\,0.2]$ & $[0.01,\,0.2]$ \\
$a_{\eta}^{\text{low}}$    & Slope, $T<T_\text{c}$ (GeV$^{-1}$)               & $[-2.0,\,1.0]$ & $-0.5$ \\
$a_{\eta}^{\text{high}}$   & Slope, $T>T_\text{c}$ (GeV$^{-1}$)               & $[-1.0,\,2.0]$ & $0.5$ \\

\addlinespace[6pt]
\multicolumn{4}{l}{\color{linkred} Bulk viscosity $\zeta/s(T)$}\\
\addlinespace[2pt]
$(\zeta/s)_{\max}$         & Peak value                                & $[0.01,\,0.25]$ & $[0.01,\,0.25]$ \\
$T_{\zeta}$                & Peak temperature (GeV)                    & $[0.12,\,0.3]$ & $0.16$ \\
$w_{\zeta}$                & Peak width (GeV)                          & $[0.01,\,0.15]$ & $0.05$ \\
$\lambda_{\zeta}$          & Skewness                                  & $[-1.0,\,1.0]$ & $0.0$ \\

\addlinespace[4pt]
\midrule
\addlinespace[2pt]

\multicolumn{4}{l}{%
  \color{linkred}
  Multi-system design parameters (not calibrated in this analysis)}\\
\addlinespace[2pt]
$\text{Norm}_{2760}$     & Pb+Pb $2.76$~TeV normalization              & $[15.0,\,40.0]$ & \text{not used} \\
$\text{Norm}_{5020}$     & Pb+Pb $5.02$~TeV normalization              & $[20.0,\,50.0]$ & \text{not used} \\
$\text{Norm}_{5360}$     & O+O $5.36$~TeV normalization                & $[25.0,\,60.0]$ & \text{not used} \\

\bottomrule
\end{tabularx}
\end{table}

The design for the forward model of chapters~\ref{chapter2} and~\ref{chapter3} contains seventeen varied model parameters, of which fourteen affect the Au+Au observables analyzed here. These parameters are collected in Table~\ref{tab:priors} and organized into four groups: the initial-state normalization and geometry, the chemical equilibration sector, and the shear and bulk viscosities. The design ranges jointly define the Latin hypercube from which the emulator training points are drawn, each parameter varied independently over its range. The ranges are deliberately broad: rather than encoding a strong prior expectation, they are chosen wide enough to let the data locate the posterior with minimal prior influence, at the cost of a design that must span a large parameter volume. Because the calibration presented here uses only Au+Au data, the three normalizations governing the Pb+Pb and O+O systems (lower block of Table~\ref{tab:priors}) do not enter the likelihood; they vary across the design in anticipation of the multi-system calibration of section~\ref{section55} but are unconstrained by the observables used here. 

Although all fourteen Au+Au-relevant parameters vary across the design and the emulator is trained on the full space, the calibration itself samples only eight and holds the remaining six fixed. The sampled set comprises the normalization $\text{Norm}_{200}$, the nucleon width $w$, the four chemical equilibration parameters, and the scale of each viscous coefficient, $(\eta/s)_{\min}$ and $(\zeta/s)_{\max}$; each is assigned an independent uniform prior over its design range. Fixed are the six shape parameters governing the temperature dependence of the transport coefficients and the multiplicity fluctuations: the multiplicity fluctuation shape $k$, the shear viscosity slopes $a_{\eta}^{\text{low}}$ and $a_{\eta}^{\text{high}}$, and the bulk viscosity shape parameters $T_{\zeta}$, $w_{\zeta}$, and $\lambda_{\zeta}$, at the values given in Table~\ref{tab:priors}. These six are fixed rather than sampled because the realized design of approximately $250$ points proved too sparse to constrain all fourteen parameters jointly. Since the present calibration is primarily targeted at the chemical equilibration sector, the transport shape and multiplicity fluctuation parameters are the least direct targets of the analysis and therefore the natural ones to hold fixed. Because the emulator is trained over the full fourteen-dimensional Au+Au design rather than refit on a reduced eight-dimensional subspace, these six parameters still enter the forward prediction. The emulator is evaluated with them held at the fixed values above while the remaining eight parameters are sampled, so their effect on the observables is retained even though they are not calibrated. A robustness study described in section~\ref{section54} varies the viscosity shape parameters and recovers the chemical equilibration constraints essentially unchanged, indicating that fixing them does not substantially bias the parameters of interest.

The chemical equilibration sector is the element that distinguishes this calibration from standard equilibrium analyses, so its four parameters warrant separate comment. Following chapter~\ref{chapter3}, the light and strange quark fugacities take the initial values $\gamma_l^{0}$ and $\gamma_s^{0}$ at hydrodynamic initialization and relax toward their equilibrium value of unity over the timescales $\tau_{\text{eq},l}$ and $\tau_{\text{eq},s}$. The fugacity priors span $[0.1,\,1.0]$. The upper edge corresponds to a plasma that is chemically equilibrated from the outset, while values below unity describe the undersaturation expected of an early, gluon-dominated state. Oversaturation ($\gamma^{0} > 1$) is excluded by construction, having no clear physical motivation at early times, and the floor at $0.1$ rather than zero retains a small but nonzero initial quark abundance, avoiding the numerical issues of the pure glue limit described in section~\ref{subsec:numerics_strange}. The timescale priors span $[0,\,10]$~fm/$c$, from effectively instantaneous equilibration to timescales comparable to or longer than the mean fireball lifetime at this energy, such that both the strongly non-equilibrium and near-equilibrium regimes are accessible. The conventional assumption of instantaneous chemical equilibrium is itself contained within this prior: in the $\tau \to 0$ limit, the fugacities are driven to unity throughout the evolution regardless of their initial values. The calibration could thus recover the standard equilibrium picture if the data prefer it, rather than imposing non-equilibrium behavior by assumption. Together the ranges admit essentially the full space of possible early-time chemistry.

The transport sector follows the temperature parameterizations of the global Bayesian analysis of Ref.~\cite{JETSCAPE:2020mzn}. The specific shear viscosity is a piecewise-linear function of temperature, taking its minimum value $(\eta/s)_{\min}$ at the pseudocritical temperature $T_\text{c}$ and varying with independent slopes $a_{\eta}^{\text{low}}$ and $a_{\eta}^{\text{high}}$ below and above it. The specific bulk viscosity is a skewed-Cauchy peak of maximum $(\zeta/s)_{\max}$ centered at $T_{\zeta}$, with width $w_{\zeta}$ and skewness $\lambda_{\zeta}$. We adopt these functional forms, together with prior ranges taken from the same reference, so that the transport priors are grounded in an established extraction while remaining broad enough to accommodate the influence of the chemical sector.

A number of further model parameters are held fixed across the entire design and are not included in Table~\ref{tab:priors}. Within the T\textsubscript{R}ENTo initial-state model the reduced-thickness parameter is set to $p = 0$, the geometric-mean limit in which the deposited entropy scales as $\sqrt{T_A T_B}$, with $T_A$ and $T_B$ the two participant thickness functions. This choice approximates the entropy deposition of more microscopically motivated initial-state models and is the value favored by earlier Bayesian analyses of the initial state~\cite{Bernhard:2019bmu}. The minimum inter-nucleon distance is fixed at $d_{\min} = 1.0$~fm, as previous Bayesian studies have found it to have very little constraining power~\cite{Bernhard:2019bmu,JETSCAPE:2020mzn}. The hydrodynamic stage begins at a fixed initialization time of $\tau_0 = 0.6$~fm/$c$, with no free-streaming stage preceding it, and the particlization temperature is constructed according to the prescription in chapter~\ref{chapter3}.
\Needspace{4\baselineskip}
\section{Calibration observables}
\label{section52}

The calibration is performed against Au+Au collisions at $\sqrt{s_{NN}} = 200$~GeV. All observables are taken at midrapidity, and each measurement enters the likelihood at the centralities at which it is reported, spanning $0$--$80$\% across the full set. The observable set is chosen to constrain the four chemical equilibration parameters together with the shear and bulk transport coefficients; the discussion below is organized by the aspect of the model to which each class of observable is most directly sensitive. Table~\ref{tab:observables} collects the full set together with its experimental provenance.

\begin{table}[htbp]
\vspace{16pt}
\centering
\caption[Calibration observables and sources]{Experimental observables for the Au+Au $\sqrt{s_{NN}} = 200$~GeV calibration, with their sources. All are taken at midrapidity. The upper block lists the observables that enter the primary likelihood; the lower block lists the $\Lambda$ and proton observables, which are measured and shown for comparison but excluded from the likelihood (section~\ref{section54}). Charged flow is calibrated in signed-cumulant space using Eq.~\ref{eq:cumulant_transform} over the centrality ranges indicated. Proton data are the feeddown-corrected PHENIX measurement; all other data are from STAR.}
\label{tab:observables}

\renewcommand{\arraystretch}{1.12}
\setlength{\tabcolsep}{0.25pt}

\begin{tabularx}{\textwidth}{
@{}
>{\raggedright\arraybackslash}p{1.35in}
>{\raggedright\arraybackslash}X
>{\centering\arraybackslash}p{1.35in}
@{}
}

\toprule
\textbf{Observable}
& \textbf{Species / channel}
& \textbf{Source} \\
\midrule

\multicolumn{3}{l}{\color{linkred} Primary likelihood}\\
\addlinespace[2pt]

$dN_{\text{ch}}/d\eta$
& charged
& \cite{STAR:2008med} \\

$dE_T/d\eta$
& total (hadronic + electromagnetic)
& \cite{STAR:2004moz} \\

\addlinespace[6pt]

$dN/dy$
& $\pi^\pm,\ K^\pm$
& \cite{STAR:2008med} \\

$dN/dy$
& $\Xi^-+\bar{\Xi}^+,\ \Omega^-+\bar{\Omega}^+$
& \cite{STAR:2006egk} \\

\addlinespace[6pt]

$\langle p_T \rangle$
& $\pi^\pm,\ K^\pm$
& \cite{STAR:2008med} \\

\addlinespace[6pt]

$v_2\{2\}$ ($0$--$60$\%)
& charged
& \cite{STAR:2004jwm} \\

$v_3\{2\}$ ($0$--$50$\%)
& charged
& \cite{STAR:2013qio} \\

\addlinespace[4pt]
\midrule
\addlinespace[2pt]

\multicolumn{3}{l}{%
  \color{linkred}
  Shown for comparison (excluded from likelihood)}\\
\addlinespace[2pt]

$dN/dy$
& $\Lambda+\bar{\Lambda}$
& \cite{STAR:2006egk} \\

$dN/dy$
& $p+\bar{p}$ (PHENIX)
& \cite{PHENIX:2003iij} \\

\bottomrule
\end{tabularx}
\end{table}

The charged particle multiplicity $dN_{\text{ch}}/d\eta$ and transverse energy $dE_T/d\eta$ constrain the overall entropy of the initial state and thereby anchor the normalization $\text{Norm}_{200}$. Because pions carry the bulk of the produced light quarks, the integrated multiplicity also retains sensitivity to the degree of light quark saturation: at fixed initial entropy a suppressed $\gamma_l$ reduces the total particle yield, so these bulk observables contribute to constraining the light sector as well as setting the scale.

The identified particle yields provide the primary constraint on the chemistry. Pion and kaon yields constrain the light and strange quark abundances at particlization, with the kaon, the most abundant strange carrier, especially sensitive to $\gamma_s$. The multi-strange hyperons sharpen the constraint considerably. The $\Xi$ and $\Omega$ carry two and three strange quarks to the kaon's one, so that yields along this sequence scale with increasing powers of the strange fugacity, and the $\Omega$ ($sss$) is in effect a clean probe of $\gamma_s$ alone. This strangeness hierarchy is the central lever on the strange sector, and it is sensitive not only to the strange fugacity but to its equilibration timescale. Because the strange abundance builds up over $\tau_{\text{eq},s}$ relative to the fireball lifetime, an incomplete approach to saturation leaves a characteristic imprint on the relative yields along the strangeness ladder, and it is this imprint that carries information on $\tau_{\text{eq},s}$ and not on the initial fugacity $\gamma_s^{0}$ alone.

The mean transverse momenta of pions and kaons probe the radial expansion of the fireball. The mass ordering $\langle p_T \rangle_\pi < \langle p_T \rangle_K$, part of the general increase of mean transverse momentum with hadron mass, reflects the collective radial flow imparted by the pressure gradients of the hydrodynamic stage. These observables are sensitive to the bulk viscosity, which modifies the radial expansion through the bulk pressure and through bulk viscous corrections at particlization. They also couple to the chemical sector, since the fugacities alter the equation of state and the particle composition, and hence the final spectra. This coupling is the origin of the degeneracy between $\zeta/s$ and the fugacity parameters noted in the introduction: both act on the transverse momentum spectra, and separating them is one of the tasks the joint calibration must accomplish.

The anisotropic flow is represented by the charged particle elliptic and triangular flow coefficients $v_2\{2\}$ and $v_3\{2\}$, measured with the two-particle cumulant method. These observables provide the primary constraint on the shear viscosity: $v_2\{2\}$ responds to $\eta/s$ and the average elliptic geometry of the initial state, while $v_3\{2\}$, driven by event-by-event initial-state fluctuations, adds sensitivity to $\eta/s$ that is largely independent of the average geometry. Although the $v_n\{2\}$ are the quoted experimental observables, the calibration is carried out in terms of the signed two-particle cumulants $c_n\{2\} = (v_n\{2\})^2$. The reason is numerical: the finite-statistics cumulant estimated in a given centrality bin can fluctuate to small negative values wherever the true signal approaches zero, rendering $v_n\{2\} = \sqrt{c_n\{2\}}$ undefined. The signed cumulant is the directly estimated quantity, remains well defined through zero, and is more nearly Gaussian in this regime, avoiding a missing-value pathology that would otherwise corrupt the emulator input. The measured coefficients are transformed to cumulant space accordingly, using first-order error propagation:
\begin{equation}
    c_n^{\text{data}}\{2\} = \left(v_n^{\text{data}}\{2\}\right)^2, \qquad
    \sigma_{c_n\{2\}} = 2 v_n^{\text{data}}\{2\} \sigma_{v_n\{2\}}.
    \label{eq:cumulant_transform}
\end{equation}
Each coefficient is compared over the centrality range within which it is resolved, $0$--$60$\% for $c_2\{2\}$ and $0$--$50$\% for $c_3\{2\}$. The model calculation is matched to the measurement in pseudorapidity gap, transverse momentum range, and $|\eta| < 1$ acceptance.

\subsection{Excluded observables}

Several measured observables are excluded from the primary likelihood. The $\Lambda$ and proton yields are shown against the model for comparison but are not fit. Both are in tension with the data in a way the chemical parameters cannot absorb; the two tensions are of opposite sign and different origin, and neither is removed by a feeddown correction.

The $\Lambda$ is underpredicted in central collisions and converges to the data toward the periphery. The deficit arises in the hadronic phase: baryon--antibaryon annihilation in SMASH, whose rate grows with the local particle density, depletes the $\Lambda$ yield most strongly in the dense central collisions~\cite{Rapp:2000gy,Steinheimer:2012rd}. It acts downstream of particlization, so changing the fugacities at hydrodynamic initialization cannot cleanly compensate for it without also altering the strange hadron yields fixed at particlization. The deficit is not a feeddown artifact --- the model $\Lambda$ already carries the $\Sigma^0$ feeddown correction, a factor of $1.35$ that restores the electromagnetic $\Sigma^0 \to \Lambda\gamma$ contribution present in the STAR $\Lambda$ but left undecayed by SMASH --- and the central deficit remains once that contribution is included.

The proton runs the other way, overpredicted at every centrality, and this cannot be the same effect. Annihilation only depletes, and by the scaling of its cross section in the additive quark model it removes the proton, with its larger light quark content, more strongly than the $\Lambda$~\cite{Mohs:2019iee, Bass:1998ca, Pan:2014caa}. Nor is the surplus a feeddown artifact: the comparison uses the feeddown-corrected PHENIX proton, for which weak-decay feeddown has been removed, matched to the model proton, which likewise contains no weak-decay feeddown, so the excess is present in a like-for-like comparison of primary protons. It reflects instead an overproduction of primordial protons that annihilation cannot repair, since the stronger annihilation needed to bring the proton down would drive the already-deficient $\Lambda$ further below the data. Both baryons are therefore set aside, and are examined quantitatively in section~\ref{section54}.

The multi-strange $\Xi$ and $\Omega$ are expected to be less affected because their smaller light quark content suppresses the annihilation cross sections, so the strange sector is anchored by the clean subset of the kaon, $\Xi$, and $\Omega$. Higher harmonic and identified particle flow are likewise excluded. The quadrangular coefficient $v_4\{2\}$ carries too low a signal-to-noise ratio at the present design size and statistics to be informative, and the identified species differential flow coefficients --- including those of the strange hadrons, whose chemistry sensitivity would otherwise make them attractive calibration targets --- are similarly statistics-limited and are deferred to future work.

The experimental data are drawn primarily from STAR. The charged multiplicity, the pion and kaon yields, and the pion and kaon mean transverse momenta are taken from the STAR identified spectra measurement~\cite{STAR:2008med}; the $\Xi$ and $\Omega$ yields from the STAR strangeness measurement~\cite{STAR:2006egk}, which also provides the $\Lambda$ comparison; and the charged elliptic and triangular flow from Refs.~\cite{STAR:2004jwm} and~\cite{STAR:2013qio} respectively. The transverse energy is taken from Ref.~\cite{STAR:2004moz}. For the proton comparison, feeddown-corrected yields from PHENIX~\cite{PHENIX:2003iij} are used in preference to the STAR values, as the correction for weak-decay feeddown removes a substantial and model-dependent contamination of the primary-proton yield. 

\section{Emulator training and validation}
\label{section53}

Direct evaluation of the forward model is far too costly to embed in a Markov chain: each design point requires the full T\textsubscript{R}ENTo--MUSIC--iS3D--SMASH pipeline over thousands of events, so sampling the posterior by running the model at every proposed parameter set is computationally infeasible. The calibration therefore proceeds through a Gaussian process emulator as introduced in chapter~\ref{chapter4}, a fast surrogate that predicts the model observables, together with an estimate of its own predictive uncertainty, throughout the fourteen-dimensional design space after being trained on a finite set of model evaluations. This section describes the construction of the emulator for the present calibration and the tests used to establish that it reproduces the model faithfully enough to be used in its place.

\begin{figure}[htbp]
\centering
\includegraphics[width=0.7\textwidth]{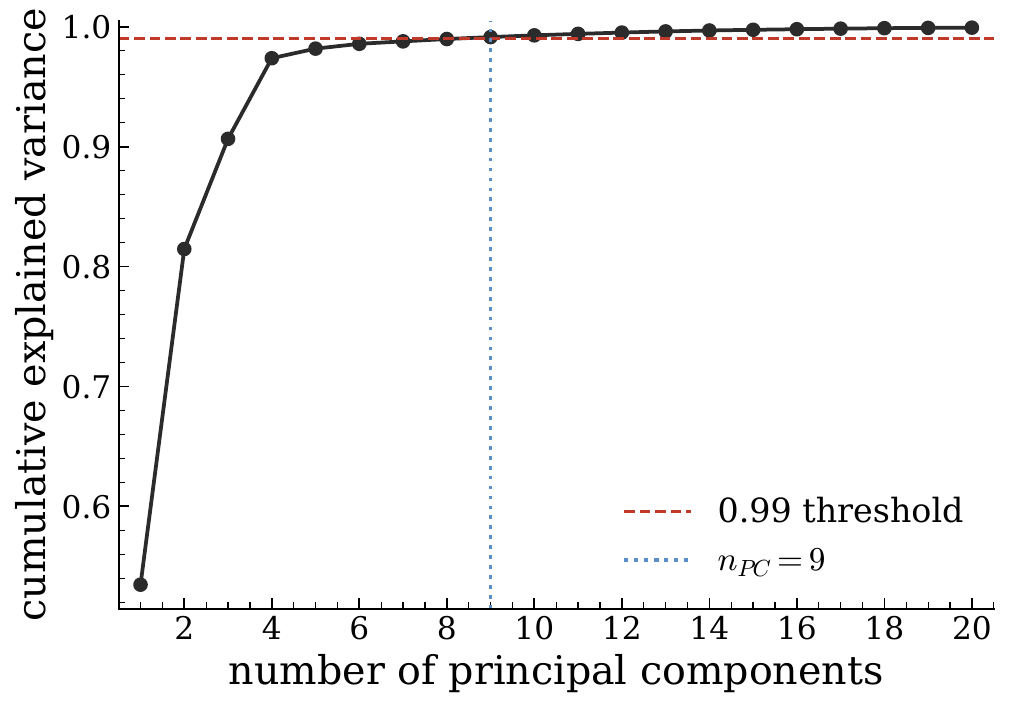}
\caption[Emulator PCA spectrum]{Cumulative fraction of the design variance captured by the leading principal components of the standardized observable vector. Nine components account for $99$\% of the variance (dashed line); the emulator is trained on these, and the remaining components, which are largely associated with statistical noise, are discarded.}
\label{fig:pca_spectrum}
\end{figure}

The training data are generated on a Latin hypercube design of $500$ points spanning the design ranges of Table~\ref{tab:priors}, of which approximately $250$ have completed the full simulation pipeline and constitute the training set used here. At each design point, the model observables retained for calibration are assembled into an output vector. Because these observables span several orders of magnitude --- charged multiplicities of order $10^{2}$ alongside two-particle cumulants of order $10^{-4}$ --- each is first standardized to zero mean and unit variance across the design, so that no single observable dominates the subsequent decomposition by virtue of its scale alone. The standardized outputs are compressed by principal component analysis; nine components suffice to account for $99$\% of the variance across the design, as shown in Fig.~\ref{fig:pca_spectrum}. This reduces the high-dimensional and strongly correlated observable vector to a small number of orthogonal components while discarding only the sub-percent residual, which is largely consistent with statistical noise. 

An independent Gaussian process is then trained to emulate each retained component as a function of the fourteen parameters. Each process uses a squared-exponential, or radial basis function, kernel with an independent length scale in each parameter direction, together with a Gaussian white-noise term that accounts for the finite statistical precision of the training observables and residual small-scale variation not resolved by the emulator. The kernel hyperparameters are obtained by maximizing the marginal likelihood of the training data.

\subsection{Cross-validation}

\begin{figure}[!t]
\centering
\includegraphics[width=\textwidth]{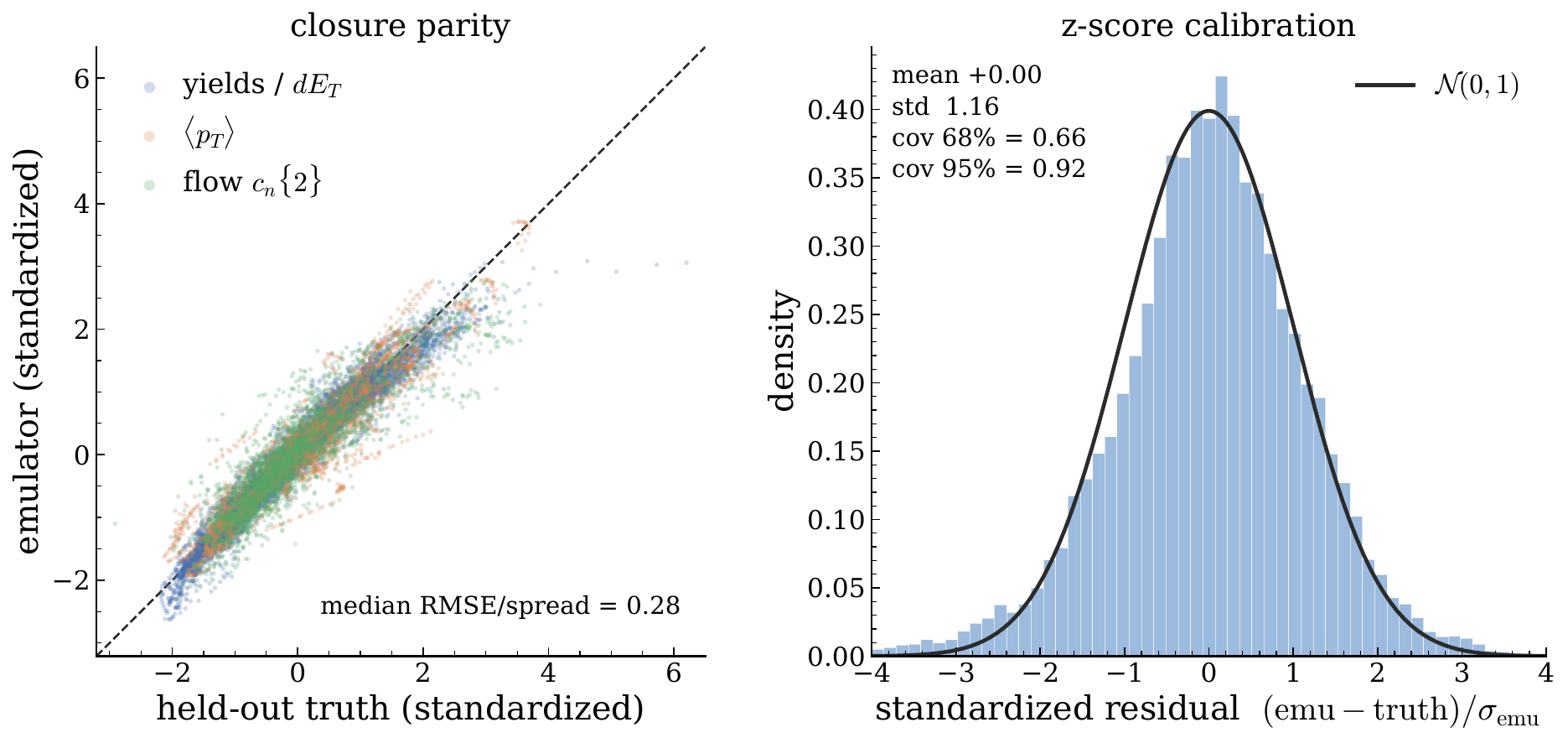}
\caption[Emulator parity and residual calibration]{Cross-validation performance of the emulator. \emph{Left:} standardized held-out predictions against the true model values across all observables and folds; the points track the diagonal with a median error of $0.28$ in units of the design spread. \emph{Right:} distribution of the standardized residuals of Eq.~\ref{eq:zscore}, which is approximately normal but of width $1.16$ rather than unity, indicating mild overconfidence that is corrected by a global uncertainty inflation before inference.}
\label{fig:emulator_parity}
\end{figure}

The accuracy of the trained emulator is assessed by cross-validation. The design is partitioned into eight folds; each fold is held out in turn, the emulator is retrained on the remaining points, and its predictions on the held-out fold are compared with the true model output there. Two aspects matter here. The first is that the central predictions track the model. Fig.~\ref{fig:emulator_parity} shows the standardized held-out predictions against the corresponding true values; the points scatter tightly about the diagonal, with a median error, across all observables and folds, of $0.28$ in units of each observable's standard deviation across the design. An emulator error well below the natural variation of the observables is a necessary condition for the surrogate to stand in for the model, since it is that variation the data must resolve. The residual emulator uncertainty is nonetheless propagated into the likelihood, so that imperfect surrogate accuracy does not appear as artificial constraining power.

The second aspect, equally important for a Bayesian analysis, is that the emulator's uncertainty estimates are honest. For each held-out prediction, the standardized residual
\begin{equation}
    z = \frac{y_{\text{pred}} - y_{\text{true}}}{\sigma_{\text{pred}}}
    \label{eq:zscore}
\end{equation}
measures the prediction error in units of the emulator's own predicted standard deviation, and should follow a unit normal distribution if that uncertainty is neither over- nor underestimated. The distribution of these residuals over the held-out set (Fig.~\ref{fig:emulator_parity}, right) is approximately Gaussian but slightly too wide, with a standard deviation of $1.16$ rather than unity, and the empirical coverage of the nominal $68$\% and $95$\% intervals is correspondingly low, at $0.66$ and $0.92$. The emulator is thus mildly overconfident: its predicted uncertainties are somewhat smaller than the held-out errors warrant. To keep this from propagating into overconfident posteriors, the emulator uncertainty is inflated by a single global factor of approximately $1.16$, the observed width of the standardized residuals, before it enters the likelihood, so that the calibration runs on uncertainties consistent with the observed cross-validation errors.

\begin{figure}[!htbp]
\vspace{12pt}
\centering
\includegraphics[width=0.9\textwidth]{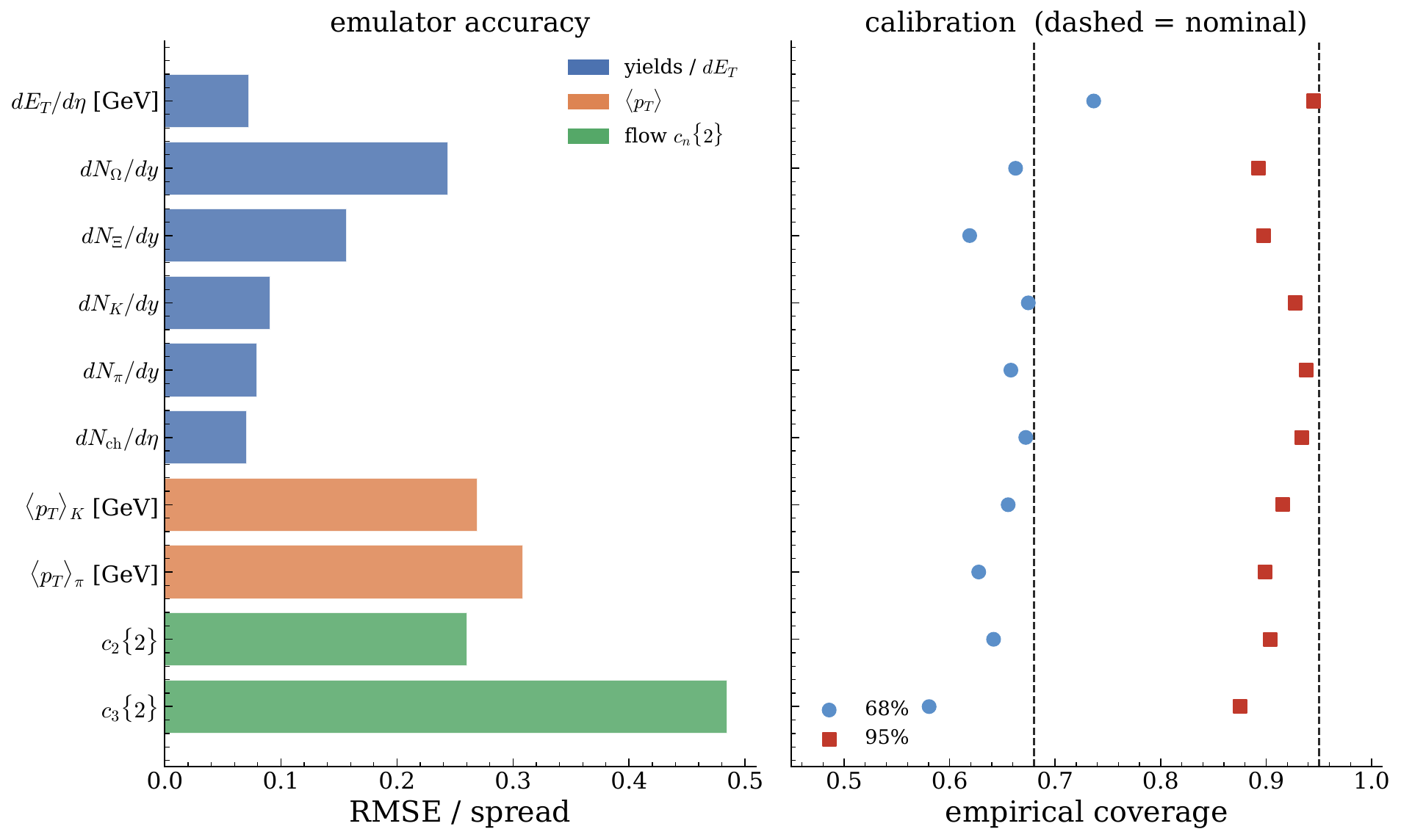}
\caption[Per-observable emulator performance]{Emulator cross-validation error, as a fraction of each observable's design spread, and interval coverage, by observable class. The yields are emulated most accurately and the flow cumulants least, with $c_3\{2\}$ the noisiest; the coverage tracks the same ordering. This pattern is consistent with the statistical character of the observables.}
\label{fig:emulator_metrics}
\end{figure}

The cross-validation errors are not uniform across observables, and their pattern is consistent with the statistical character of the corresponding measurements, as shown in Fig.~\ref{fig:emulator_metrics}. The light hadron yields and transverse energy are emulated most accurately, with errors of $0.07$--$0.09$ of the design spread. The multi-strange hyperons follow, with errors of $0.16$ for $\Xi$ and $0.24$ for $\Omega$, while the mean transverse momenta and flow cumulants are least accurate, with errors of $0.27$--$0.49$ and the triangular cumulant $c_3\{2\}$ the noisiest at $0.49$. 
\enlargethispage{2\baselineskip}

This ordering follows the statistical character of the observables: the yields are first moments of large samples, whereas the flow cumulants are fluctuation-sensitive quantities estimated from the same events, and $c_3\{2\}$ in particular carries the smallest signal and the lowest signal-to-noise of the retained observables. The interval coverage follows the same ordering, with $c_3\{2\}$ the most under-covered observable and the transverse energy slightly over-covered. This non-uniformity is why a single global inflation is a compromise rather than an exact correction: the per-observable residual widths range from $0.96$ to $1.35$ about the pooled value. The inflation nevertheless corrects the overall calibration of the emulator uncertainties and avoids carrying the raw overconfidence of the Gaussian processes into the likelihood.

\begin{figure}[!htbp]
\centering
\includegraphics[width=\textwidth]{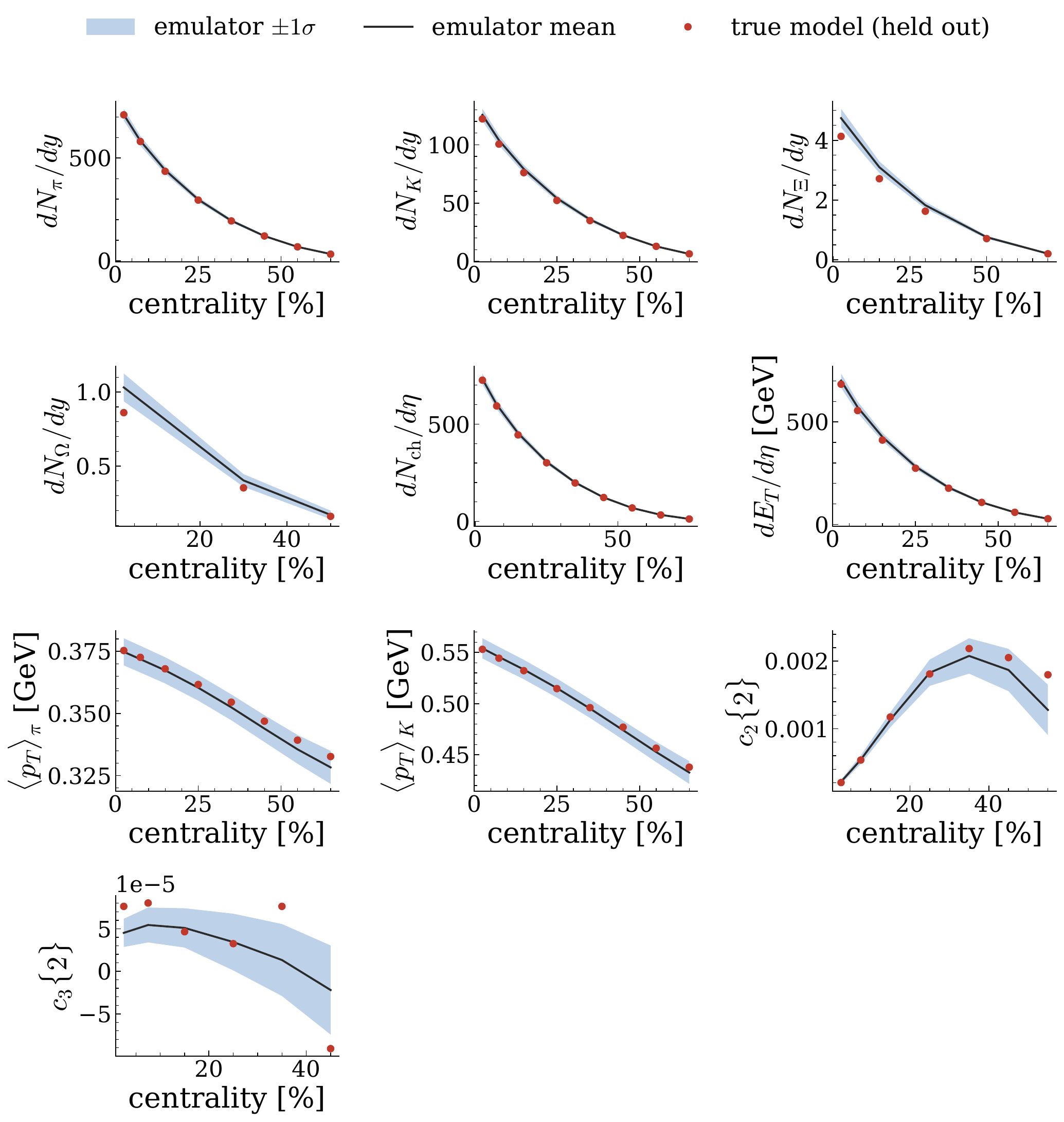}
\caption[Emulator holdout closure]{Emulator prediction (band, $\pm 1\sigma$) against the true model output (points) across centrality for a single held-out design, for a representative set of observables. Flow is shown in signed-cumulant space, the emulator's native output.}
\label{fig:closure_holdout}
\end{figure}

As a direct visual check, Fig.~\ref{fig:closure_holdout} shows the emulator prediction, with its $\pm 1\sigma$ band, against the true model output across centrality for a single design held out of the training. The emulator reproduces the centrality dependence of every observable class, with deviations consistent with its quoted uncertainty. The flow panels are shown in the signed-cumulant space $c_n\{2\}$ in which the emulator is trained and the calibration is performed, rather than the derived $v_n\{2\}$.

\subsection{Closure testing}

The tests so far validate the emulator in isolation. A final and more stringent check closes the entire inference chain through a leave-one-out coverage test~\cite{Cook:2006zir}. For each of one hundred held-out design points, whose true parameters are known, the emulator is retrained on the remaining designs, likelihood-consistent pseudodata are synthesized from the held-out point's model output, and the full sampler is run against them. Recording how often the posterior credible intervals contain the known truth, across many such held-out points, measures whether those intervals are honestly calibrated. In each closure fit, the six parameters not sampled in the calibration were fixed at the held-out point's true values, so the test isolates the inference chain for the eight sampled parameters under a correctly specified model. The sensitivity of the calibration to the fixed values themselves is assessed separately by the robustness study in section~\ref{subsec:robustness}.

\begin{figure}[!b]
\vspace{12pt}
\centering
\includegraphics[width=0.8\textwidth]{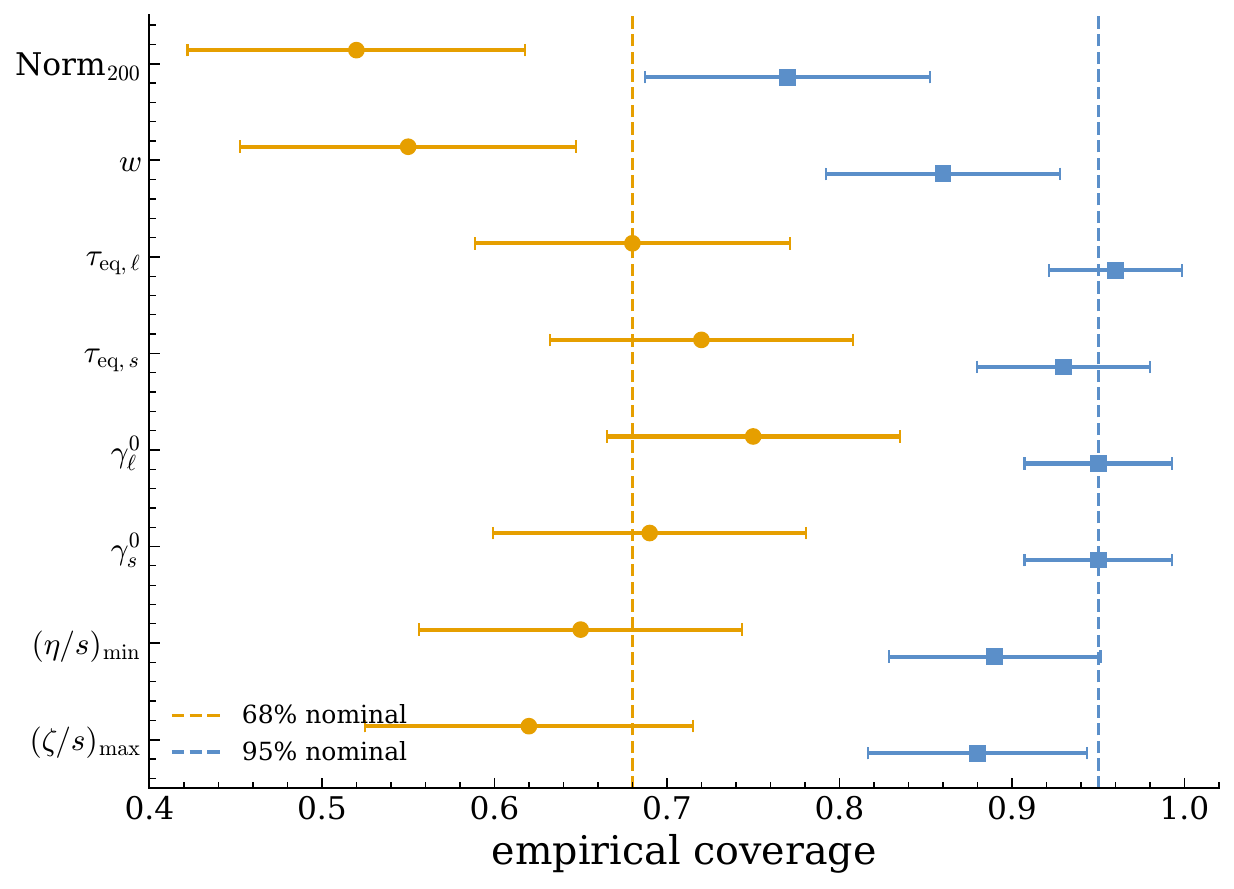}
\caption[Closure test: leave-one-out coverage]{Empirical coverage of the posterior credible intervals against the nominal credible level, from a leave-one-out closure test over one hundred held-out design points that exercises the full inference chain (emulator, likelihood, and Markov chain sampling). For each parameter, the points give the fraction of held-out designs whose $68$\% (amber circles) and $95$\% (blue squares) credible intervals contain the known true value, and the horizontal bars give the statistical uncertainty on that fraction from the hundred-point sample; the vertical dashed lines mark the nominal $68$\% and $95$\% levels. A parameter is well calibrated when its markers coincide with the nominal lines, and markers lying to the left indicate credible intervals that are too narrow.}
\label{fig:closure_coverage}
\end{figure}

Fig.~\ref{fig:closure_coverage} shows the empirical coverage of the nominal $68$\% and $95$\% marginal credible intervals for the eight sampled parameters. The coverage is broadly consistent with expectation, especially for the chemical equilibration parameters. The $95$\% intervals cover the true values at close to the nominal rate for $\tau_{\text{eq},l}$, $\tau_{\text{eq},s}$, $\gamma_l^0$, and $\gamma_s^0$, while the $68$\% intervals are within statistical uncertainty of nominal coverage for the equilibration timescales and $\gamma_s^0$, though slightly over-covering for $\gamma_l^0$. 

The strongest under-coverage appears for the more sharply constrained initial-state parameters, $\text{Norm}_{200}$ and $w$, indicating that their marginal posteriors are narrower than ideal in the closure ensemble. This residual under-coverage reflects the emulator's interpolation error, which is no longer negligible against these unusually narrow posteriors, rather than a flaw in the inference machinery. This motivates interpreting the sharpest constraints with some caution. Importantly, the chemical equilibration sector that is the focus of this analysis does not show a comparable failure of coverage, supporting the use of the posterior distributions below as quantitative, though not exact, constraints.

Two qualifications frame these validation and closure tests. First, the cross-validation establishes the emulator's fidelity at held-out design points, which populate the full fourteen-dimensional design volume. The calibration queries the emulator on the eight-dimensional slice through this volume defined by the six parameters fixed in section~\ref{section51}, a subset of the validated domain. Second, the adequacy of a design of this size in fourteen dimensions is assessed empirically by the cross-validation above rather than assumed from any a priori points-per-dimension rule. With the emulator validated, its uncertainty recalibrated, and the full inference pipeline checked through closure, the calibration proceeds to the Markov chain sampling, whose posterior distributions are presented in section~\ref{section54}.

\section{Posterior distributions}
\label{section54}

\begin{figure}[!htbp]
\centering
\includegraphics[width=\textwidth]{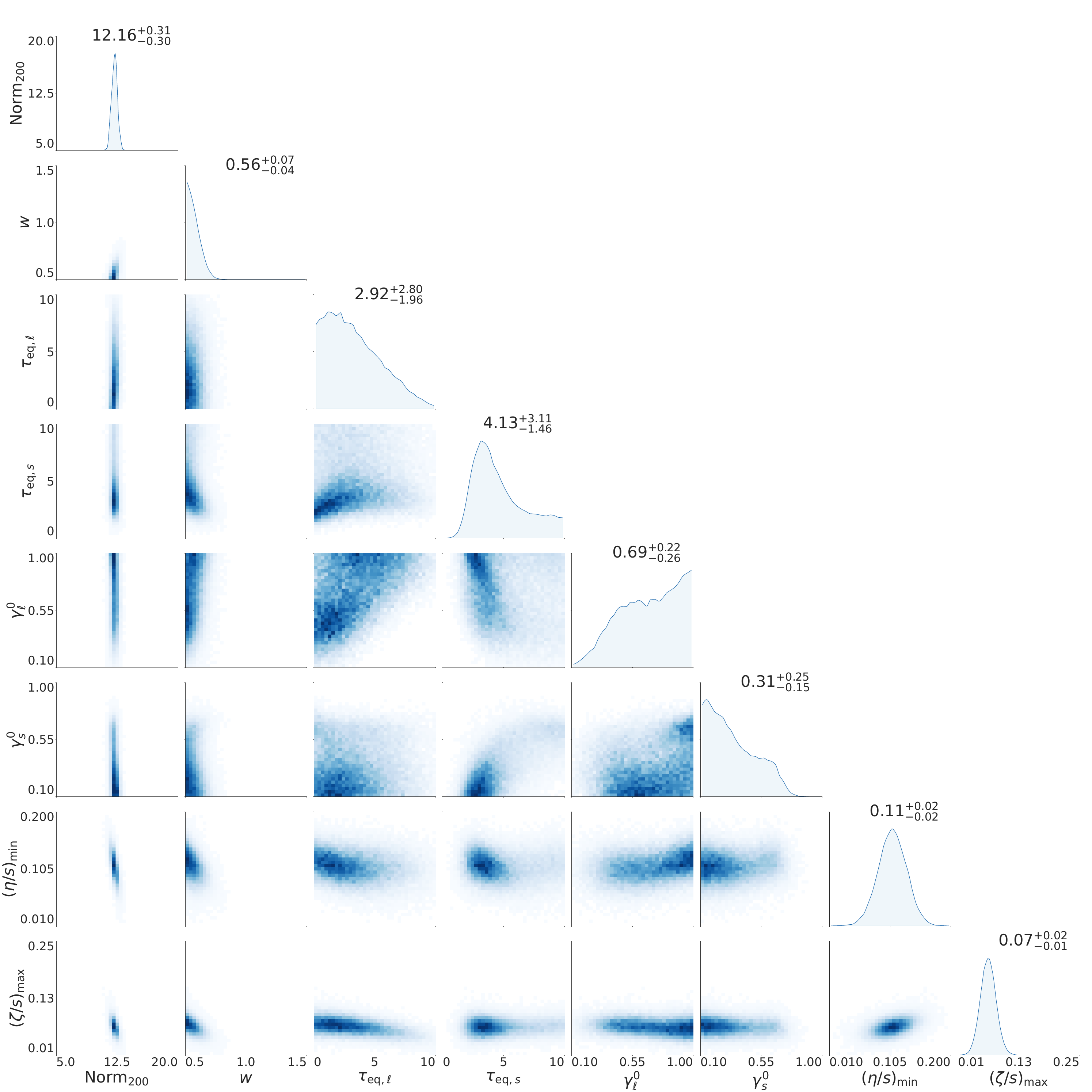}
\caption[Posterior distributions of the calibrated parameters]{Posterior distribution over the eight sampled parameters: the Au+Au normalization $\text{Norm}_{200}$ and nucleon width $w$, the four chemical equilibration parameters $\tau_{\text{eq},l}$, $\tau_{\text{eq},s}$, $\gamma_l^0$, and $\gamma_s^0$ that are the focus of this analysis, and the shear and bulk viscosity magnitudes $(\eta/s)_{\min}$ and $(\zeta/s)_{\max}$. Diagonal panels show the one-dimensional marginal posterior of each parameter, annotated with its median and $68$\% credible interval; off-diagonal panels show the two-dimensional joint posteriors, with darker shading indicating higher posterior density.}
\label{fig:corner_full}
\end{figure}

The calibration yields the joint posterior distribution over the eight sampled parameters. Fig.~\ref{fig:corner_full} shows the marginal and pairwise-joint posteriors for the Au+Au normalization, the nucleon width, the four chemical equilibration parameters that are the focus of this analysis, and the shear and bulk viscosity magnitudes. Quoted uncertainties are the $68$\% credible intervals about the marginal posterior median. A central result of the calibration is the constraint on the two equilibration timescales. The strange quark equilibration time is inferred to be $\tau_{\text{eq},s} = 4.13^{+3.11}_{-1.46}$~fm/$c$ and the light quark time $\tau_{\text{eq},l} = 2.92^{+2.80}_{-1.96}$~fm/$c$. 

Two features bear on the questions posed at the outset. First, instantaneous strange equilibration, $\tau_{\text{eq},s} \to 0$, corresponding to the equilibrium limit, is disfavored by the data, which instead favor a finite time for the strange abundance to approach saturation. Second, the light quark timescale is both smaller in its median and broader, consistent with a comparatively rapid, though less sharply constrained, approach to light quark equilibrium. The ordering $\tau_{\text{eq},s} > \tau_{\text{eq},l}$ favored by the medians is the expected one: the larger strange quark mass and correspondingly smaller production rates make strangeness the slower sector to equilibrate. The marginal intervals overlap substantially, however, so at the present design size and single-system scope this ordering is indicated rather than resolved. It is quantified as a joint posterior probability in section~\ref{subsec:robustness}, and sharpening it is one motivation for the extensions of section~\ref{section55}.

The initial fugacities reinforce this picture. Their credible intervals lie below unity: $\gamma_l^0 = 0.69^{+0.22}_{-0.26}$ and $\gamma_s^0 = 0.31^{+0.25}_{-0.15}$. This is consistent with a plasma that is chemically undersaturated at early times, as expected of an initially gluon-dominated state in which quark abundances have not yet built up. The two are not determined equally well: $\gamma_s^0$ is pinned well below unity by the strangeness data, whereas the light fugacity posterior is broad and only weakly displaced from its prior. With that caveat, the strange sector is the more strongly undersaturated of the two, $\gamma_s^0$ concentrated below $\gamma_s^0$. Together, the initial fugacities and timescales describe an early plasma in which strange quarks are markedly suppressed relative to their equilibrium abundance and equilibrate over a finite time on the order of the fireball lifetime at this energy.

The joint distributions of Fig.~\ref{fig:corner_full} show the correlations among these parameters. Within each flavor sector the initial fugacity and the equilibration time are positively correlated, visible most clearly as the diagonal band in the $\tau_{\text{eq},l}$ -- $\gamma_l^0$ plane. This reflects the expected trade-off at fixed integrated quark production: a lower initial fugacity must be compensated by a more rapid approach to equilibrium, and a higher initial value tolerates a slower one. The strange sector shows the analogous but more pronounced degeneracy between $\tau_{\text{eq},s}$ and $\gamma_s^0$, the curved band along which the measured strange yields are reproduced by trading initial strange content against equilibration time. This degeneracy is the primary residual uncertainty in the strange sector, and because it arises from constraining a single collision system, data from systems of differing size and lifetime would be expected to help break it.

The normalization, nucleon width, and viscosity magnitudes are well constrained by the same observable set. The Au+Au normalization is sharply determined, $\text{Norm}_{200} = 12.16^{+0.31}_{-0.30}$, and the nucleon width is driven toward the lower end of its prior range, $w = 0.56^{+0.07}_{-0.04}$~fm. The shear and bulk viscosity magnitudes are constrained to small positive values, $(\eta/s)_{\min} = 0.11^{+0.02}_{-0.02}$ --- near but above the conjectured lower bound $1/4\pi \approx 0.08$~\cite{Kovtun:2004de} --- and $(\zeta/s)_{\max} = 0.07^{+0.02}_{-0.01}$. Neither viscosity is strongly correlated with the fugacity parameters in the posterior. The degeneracy between bulk viscosity and quark fugacity demonstrated in section~\ref{section32} and anticipated in section~\ref{section52} does not appear to greatly impact the posterior, indicating that the combined yield, spectral, and flow data sufficiently separate the transport and chemical sectors within this analysis.

\subsection{Posterior predictive comparison}
\label{subsec:predictive}

\begin{figure}[!htbp]
\centering
\includegraphics[width=\textwidth]{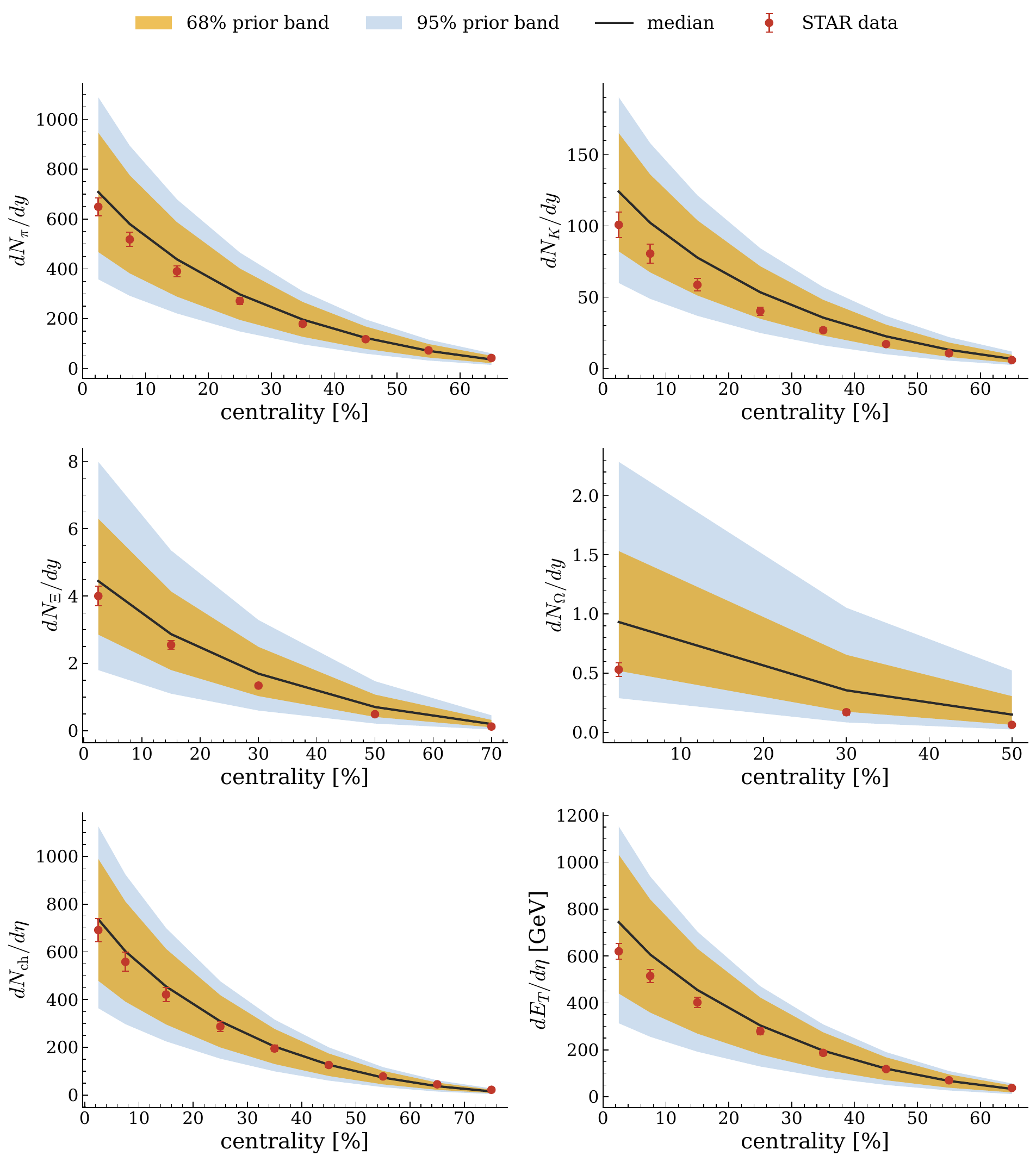}
\caption[Prior predictive yields and transverse energy]{Prior predictive distributions of the integrated yield and transverse energy observables for Au+Au collisions at $\sqrt{s_{NN}} = 200$~GeV: $dN_{\text{ch}}/d\eta$, $dE_T/d\eta$, and the $\pi$, $K$, $\Xi$, and $\Omega$ yields. Each panel shows one observable against collision centrality; the nested shaded bands are the $68$\% and $95$\% credible intervals of the emulated model output sampled over the prior, the solid line is the prior median, and the points are the experimental data. The bands enclose the measured values across the observables shown, confirming that the design spans the region of observable space occupied by the data. The corresponding posterior predictive distributions are shown in Fig.~\ref{fig:posterior_yields}.}
\label{fig:prior_yields}
\end{figure}

\begin{figure}[!htbp]
\centering
\includegraphics[width=\textwidth]{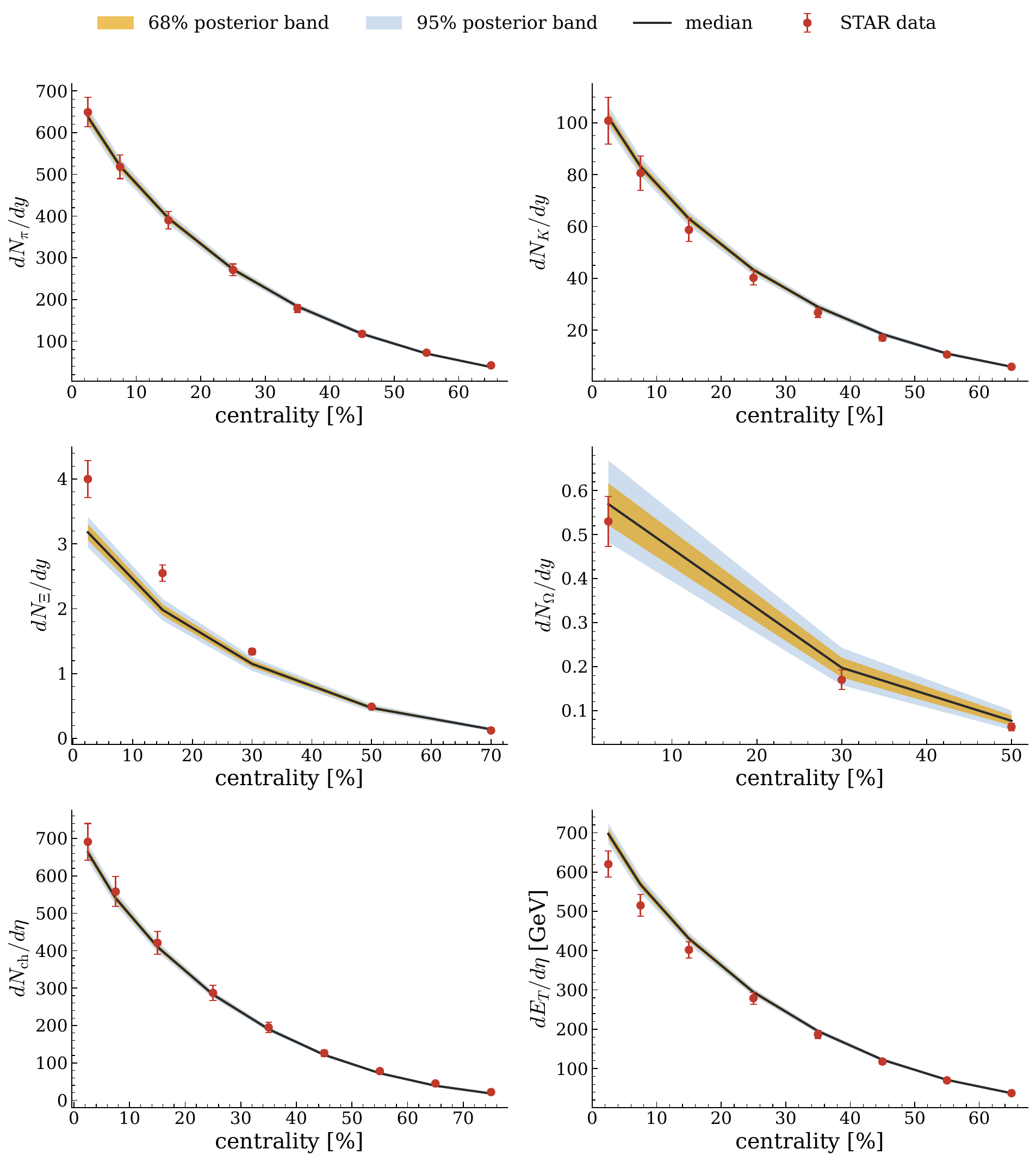}
\caption[Posterior predictive yields and transverse energy]{Posterior predictive distributions of the yield and transverse energy observables, following the panel layout of Fig.~\ref{fig:prior_yields}.}
\label{fig:posterior_yields}
\end{figure}

\begin{figure}[!htbp]
\centering
\includegraphics[width=\textwidth]{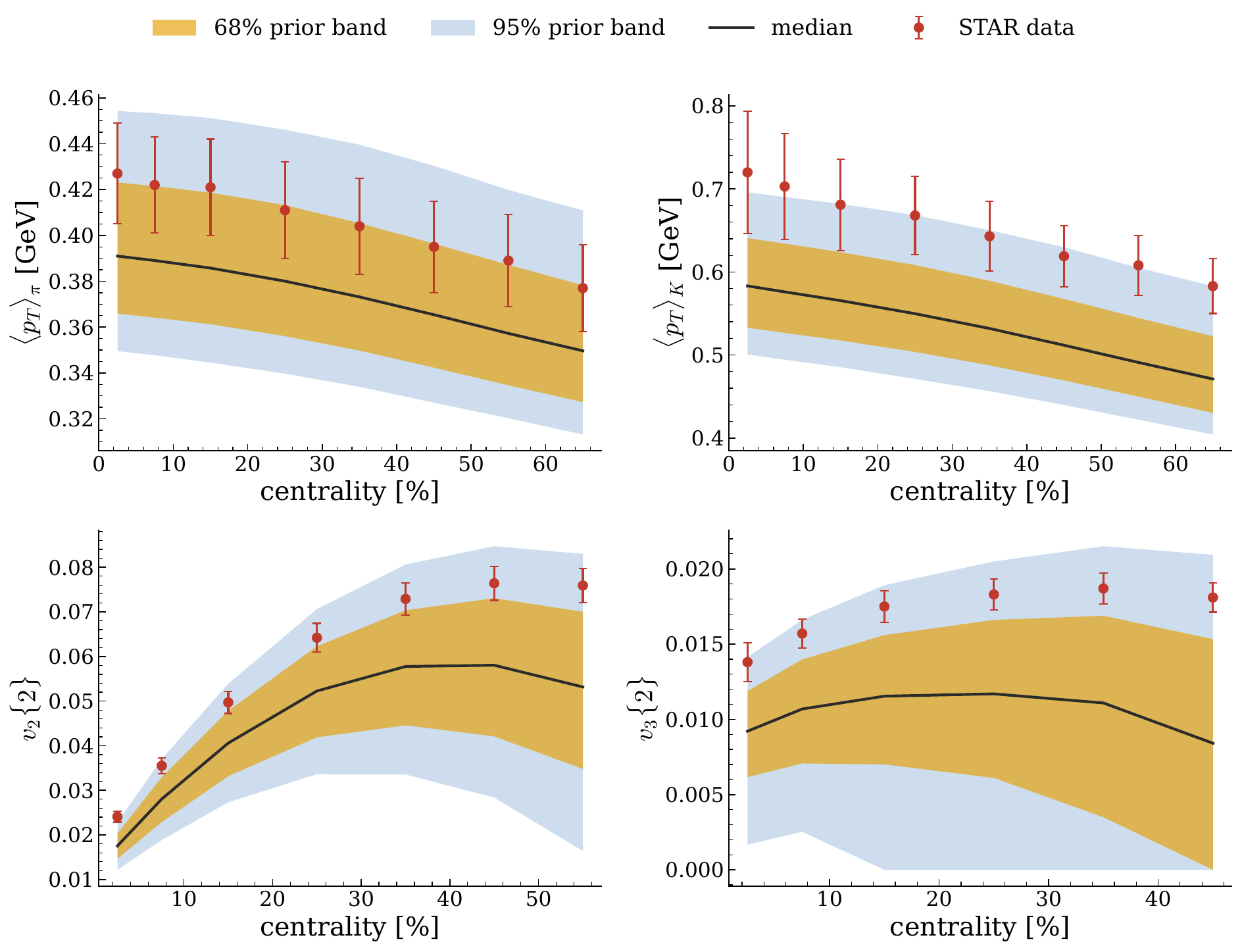}
\caption[Prior predictive mean transverse momenta and flow]{Prior predictive distributions of the mean transverse momenta of pions and kaons and the charged anisotropic flow $v_2\{2\}$ and $v_3\{2\}$, in the same format as Fig.~\ref{fig:prior_yields}. Flow is displayed as $v_n\{2\} = \sqrt{\max(c_n\{2\},\,0)}$, though the calibration is performed in signed-cumulant space (section~\ref{section52}). The corresponding posterior predictive distributions are shown in Fig.~\ref{fig:posterior_ptflow}.}
\label{fig:prior_ptflow}
\end{figure}

\begin{figure}[!htbp]
\centering
\includegraphics[width=\textwidth]{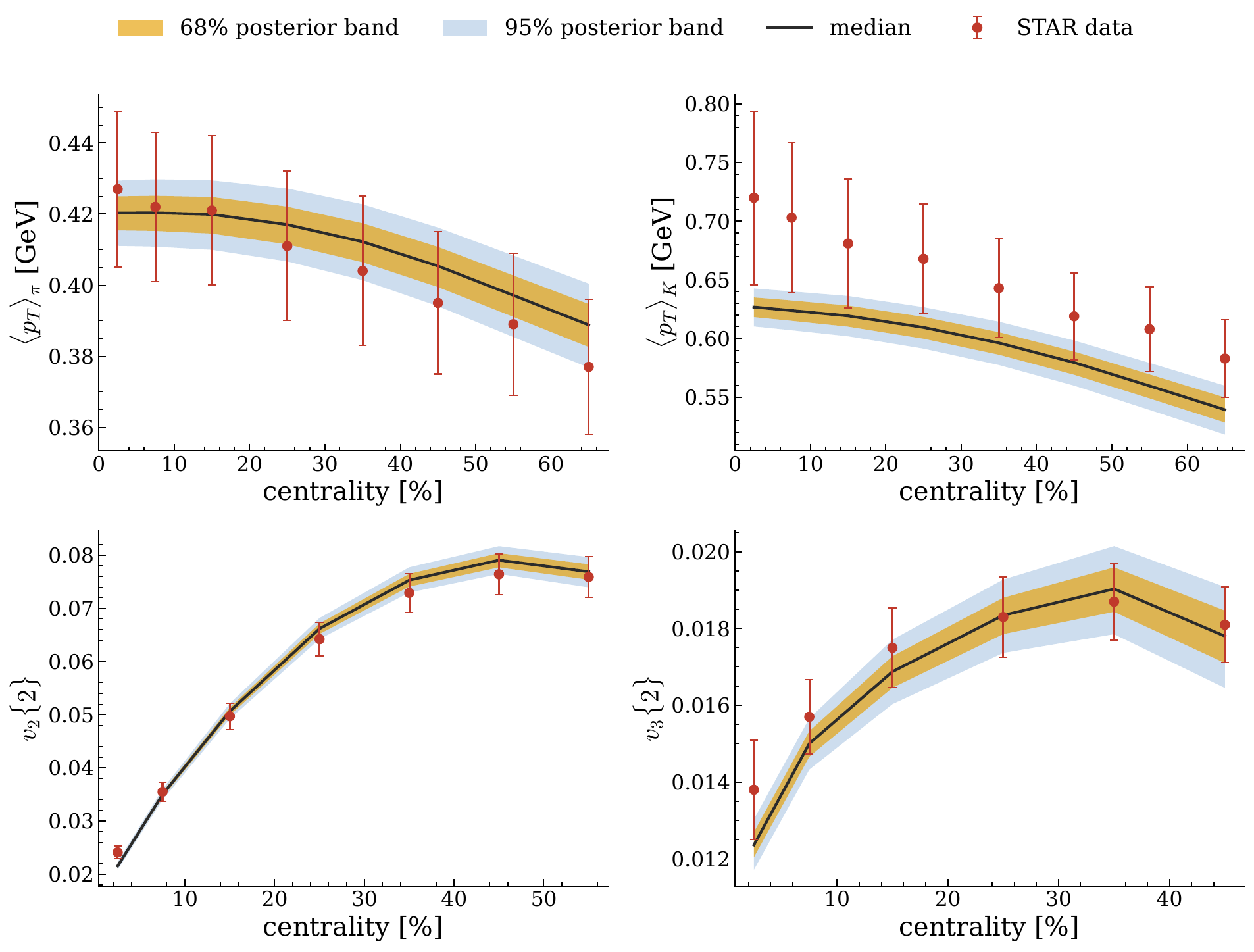}
\caption[Posterior predictive mean transverse momenta and flow]{Posterior predictive distributions of the mean transverse momenta and anisotropic flow, following Fig.~\ref{fig:prior_ptflow}.}
\label{fig:posterior_ptflow}
\end{figure}

The prior and posterior predictive distributions of the fitted observables are shown in Figs.~\ref{fig:prior_yields}--\ref{fig:posterior_ptflow}, split by observable class into the integrated yields and transverse energy, and the mean transverse momenta and anisotropic flow. The prior predictive bands are broad and enclose the measured values for all observables except the kaon mean transverse momentum, whose central and mid-central values sit at or just above the upper edge of the $95$\% prior band. With that exception, the design spans the region of observable space occupied by the data and the calibration is well posed. 

\FloatBarrier

The posterior bands narrow sharply and track the measured centrality dependence of the yields, transverse energy, mean transverse momenta, and flow. The strangeness hierarchy is reproduced across the $\Xi$ and $\Omega$ yields, supporting the strange sector constraints discussed above. The agreement is not perfect: the kaon mean transverse momentum sits about $1\sigma$ below the data in central and mid-central collisions, the $\Xi$ yields are underpredicted for central collisions, and the transverse energy is overpredicted in the most central bins. Apart from these tensions, the calibrated model reproduces most observables entering the fit, and the collapse of the predictive bands from prior to posterior is a direct measure of the constraining power the data exert.

\begin{figure}[!t]
\centering
\includegraphics[width=\textwidth]{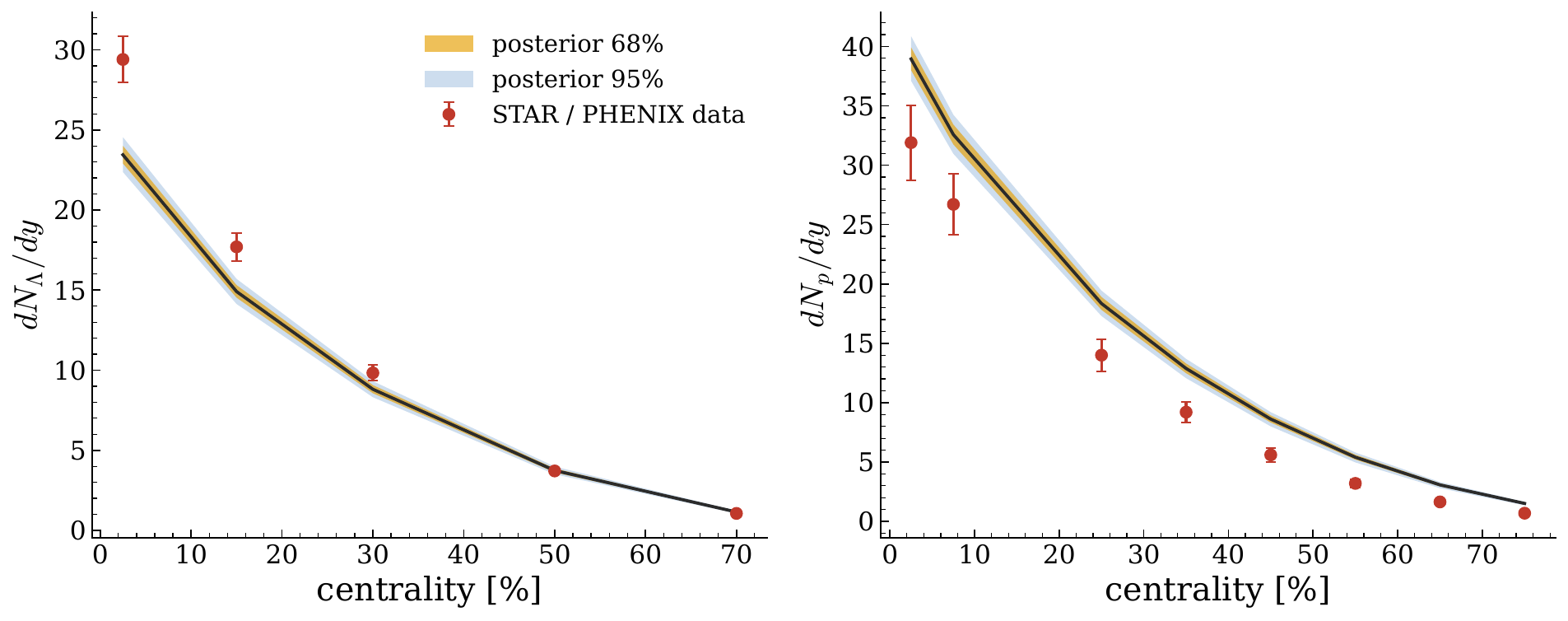}
\caption[Excluded baryons: $\Lambda$ and proton against data]{Posterior predictive comparison for the two excluded baryons, the $\Lambda$, with the $\Sigma^0$ feeddown correction applied, and the proton, compared against the feeddown-corrected PHENIX data. Bands are the posterior credible intervals; neither observable enters the likelihood. The $\Lambda$ is underpredicted in central collisions and the proton overpredicted at all centralities, reflecting the effects discussed in section~\ref{section52}.}
\label{fig:excluded_baryons}
\end{figure}

Two measured baryons excluded from the fit, the $\Lambda$ and the proton, are shown against the posterior in Fig.~\ref{fig:excluded_baryons}, and they depart from the data in opposite directions. The $\Lambda$ falls below the measured yield in central collisions and converges to it toward the periphery, consistent with the density-dependent baryon--antibaryon annihilation discussed in section~\ref{section52}. The proton runs the other way, overpredicted at every centrality, with a posterior median near $39$ against the measured $\approx\!32$ in the most central ($0$--$5$\%) bin. This is the proton surplus discussed in section~\ref{section52}, which annihilation does not explain. That the two miss in opposite directions visually reflects their distinct origins: no single hadronic annihilation rate can bring the proton down to the data without deepening the $\Lambda$ deficit, so the two cannot be reconciled by a single retuning of the hadronic annihilation strength. Their exclusion is intended to protect the primary likelihood from baryon sector effects the fugacity parameters cannot absorb.

\FloatBarrier

\subsection{Robustness of the equilibration constraints}
\label{subsec:robustness}

The magnitudes inferred above are not all constrained to the same degree, and it is worth stating how much each owes to the data rather than the prior. A convenient heuristic is the ratio of the posterior to the prior standard deviation, which approaches unity for a parameter the data do not move and falls toward zero for one they sharply determine. Among the four equilibration parameters, this ratio is $0.75$ for $\gamma_s^0$, $0.78$ for $\tau_{\text{eq},s}$, $0.83$ for $\tau_{\text{eq},l}$, and $0.94$ for $\gamma_l^0$. The initial strange fugacity is thus the chemical parameter most substantially informed by the data; the two timescales are moderately constrained; and the initial light fugacity is very nearly prior-dominated, its posterior scarcely narrower than the prior. This pattern reflects the observable set: the strange-hadron yields tightly constrain the strange fugacity, whereas the bulk yields only weakly constrain the light sector.

\begin{figure}[!htbp]
\centering
\includegraphics[width=\textwidth]{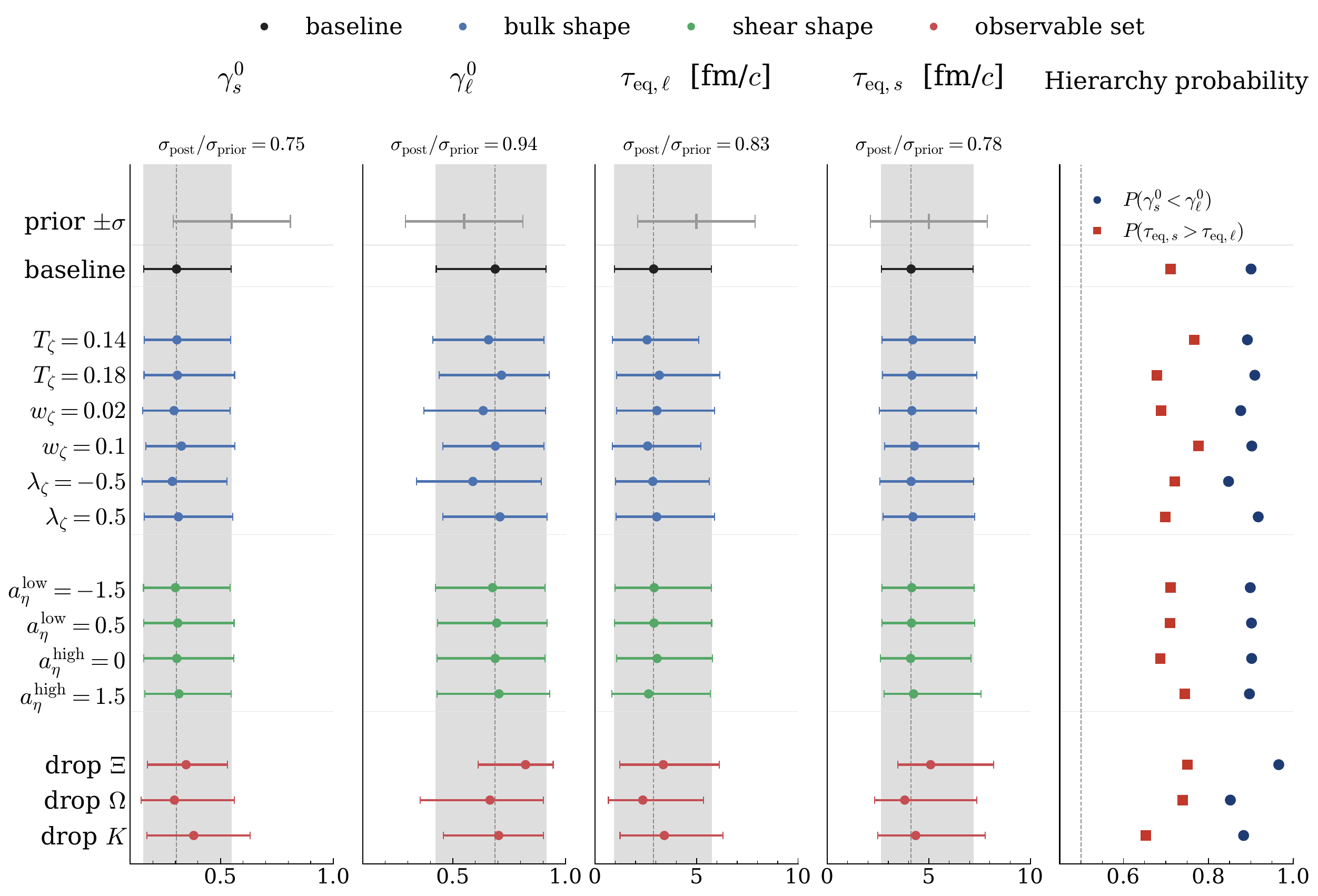}
\caption[Robustness of the equilibration constraints]{Stability of the chemical equilibration constraints across fourteen analysis variants spanning the fixed viscosity-shape parameters and jackknives of the strange observable set. The first four panels show, for $\gamma_s^0$, $\gamma_l^0$, $\tau_{\text{eq},l}$, and $\tau_{\text{eq},s}$, the marginal median and $68$\% credible interval inferred under each variant, each drawn on the full prior range with the baseline result marked for reference. The fifth panel gives the joint posterior probabilities of the two flavor orderings, $P(\gamma_s^0 < \gamma_l^0) \approx 0.90$ and $P(\tau_{\text{eq},s} > \tau_{\text{eq},l}) \approx 0.71$, both favored across all fourteen variants.}
\label{fig:robustness_forest}
\end{figure}

Fig.~\ref{fig:robustness_forest} tests the stability of the equilibration constraints against the analysis choices made in setting up the fit. The calibration is repeated across fourteen configurations --- the baseline and thirteen variants spanning the fixed viscosity shape parameters and jackknives of the strange observable set --- and the resulting marginal constraint on each equilibration parameter is shown as a forest plot. The initial strange fugacity is stable across every variant, its median value ranging over $[0.287, 0.381]$ with a spread of $0.094$, well within its statistical uncertainty of about $0.20$. Varying the fixed viscosity parameters moves $\gamma_s^0$ by only $\sim\!0.04$, confirming directly that fixing them does not dramatically bias the sector of interest. The leading systematic is instead the observable set: the kaon is the primary anchor of $\gamma_s^0$, and removing it shifts the inferred value most (to $0.381$, against $0.348$ and $0.296$ when the $\Xi$ or $\Omega$ is dropped instead). The light fugacity and the two timescales are likewise stable across the variants, though for these the stability is partly of the trivial kind: parameters the data constrain weakly are not strongly pulled by any single analysis choice.

The most robust statements the calibration supports are not the individual magnitudes but the orderings between the two flavor sectors, and these hold across all fourteen variants. The initial strange fugacity lies below the light fugacity with joint posterior probability $P(\gamma_s^0 < \gamma_l^0) \approx 0.90$ (ranging over $0.85$--$0.97$ across the variants): within this model, the plasma begins more strongly undersaturated in strangeness than in light flavor, a stable result across the analysis variants. The timescale ordering is favored in the same sense but far less decisively, $P(\tau_{\text{eq},s} > \tau_{\text{eq},l}) \approx 0.71$ ($0.65$--$0.78$): the data consistently prefer slower strange equilibration, but the preference is weak and should not be read as established. The more defensible conclusion is the hierarchy rather than the numerical values: strangeness starts further from equilibrium and, more tentatively, relaxes toward it more slowly than the light sector.

These constraints are obtained from Au+Au data at a single collision energy. They establish that finite-time flavor equilibration is accessible to this style of analysis and yield a coherent physical picture: an initially undersaturated, gluon-dominated plasma in which strangeness begins markedly more suppressed than the light sector, and relaxes toward equilibrium over a finite time that the data favor to be longer than the light sector time. The principal residual uncertainties are those that a joint calibration across collision systems is positioned to reduce, as discussed in section~\ref{section55}.
\Needspace{4\baselineskip}
\section{Future directions}
\label{section55}

The calibration presented in this chapter should be understood as a first step toward the quantitative characterization of chemical equilibration effects in heavy-ion collisions through systematic model-to-data comparison, rather than as a definitive determination of the equilibration parameters. The framework of a non-equilibrium fugacity sector embedded in a multistage model and constrained by Bayesian inference is, to our knowledge, the first of its kind, and the constraints obtained here establish that such effects are accessible to this style of analysis. At the same time, the specific model calibrated here is deliberately simple, and both the analysis and the underlying physics model allow for substantial extension. This section outlines the most immediate of these, beginning with the limitations most directly bearing on the present results and proceeding to broader extensions of the physics.

The most immediate limitation is the size of the design on which the emulator is trained. The results here rest on the approximately $250$ completed points of the planned $500$-point design, and while the cross-validation of section~\ref{section53} establishes that this sample supports a usable emulator, a denser and more complete design would sharpen the emulator's predictions throughout the parameter space and reduce the contribution of emulator uncertainty to the posterior. This is a purely computational limitation rather than a fundamental one: completing the remaining design points requires only further model runs, and the calibration is expected to tighten as they become available. The parameter constraints reported in section~\ref{section54} should therefore be read with the design size in mind, as constraints that may sharpen or shift as the full design is completed.

The most significant extension of the analysis is the incorporation of additional collision systems. The calibration here is anchored to Au+Au at $\sqrt{s_{NN}} = 200$~GeV, but the forward model is defined for the Pb+Pb and O+O systems as well, and the sensitivity of their observables to the fugacity sector was already established in chapter~\ref{chapter3}. A joint calibration across collision energies and system sizes --- Au+Au and Pb+Pb spanning more than an order of magnitude in $\sqrt{s_{NN}}$, and O+O providing a small collision system at LHC energy --- would test whether a single chemical equilibration picture describes the approach to saturation across widely varying initial temperatures, lifetimes, and system sizes. Such a test is considerably more demanding than a single-system fit and is correspondingly more informative: parameters that are individually constrained by Au+Au data alone, and degeneracies that a single system cannot resolve, may be separated once systems with different spacetime evolution are required to be described simultaneously. The multi-system design underlying this work in Table~\ref{tab:priors} was constructed with this joint calibration in view, and extending the present analysis to it is the natural next step.

Beyond enlarging the design and the set of systems, the physics model itself can be extended in several directions. The present treatment groups the quark flavors into a light and a strange sector; a potential refinement would be to resolve additional flavors. Charm would require a different treatment from the one used here: it is produced primarily in initial hard scatterings and is generally expected to remain far from thermal chemical equilibrium~\cite{Rapp:2009my,Andronic:2015wma}. A conceptually related fugacity or abundance evolution for charm, constrained by open heavy flavor measurements and coupled to heavy quark transport, could therefore provide an independent probe of equilibration dynamics in a regime very different from that of the light and strange sectors.

A second class of refinement concerns the parameterization of the equilibration itself. The fugacities here relax toward equilibrium on a single constant timescale per flavor, taken to be spatially uniform and independent of the local conditions of the medium. In reality the equilibration rate is governed by microscopic processes whose efficiency depends on temperature and density, so that a more realistic model would allow the equilibration timescale to vary with the local temperature, coupling the chemistry to the spacetime evolution of the fireball rather than imposing a single rate throughout. A more flexible parameterization of the timescale, or one tied more directly to an underlying rate calculation, would relax one of the strongest simplifying assumptions of the present model.

Finally, the coupling between the chemical and transport sectors can be made more physical. In the present model the specific shear and bulk viscosities are functions of temperature alone, while the fugacities evolve independently. In reality, the transport properties of the plasma depend on its composition, and a chemically non-equilibrium medium should, in principle, carry viscosities that reflect its instantaneous quark content rather than its temperature alone. Allowing the transport coefficients to depend on the fugacities would couple the two sectors self-consistently and would be a natural way to sharpen the interplay between chemistry and transport that the joint calibration of this chapter is designed to probe. Taken together, these extensions indicate that the non-equilibrium fugacity framework introduced in this work opens a broad program of study, of which the Au+Au calibration presented here is the first step.
\chapter{Conclusions}
\label{chapter6}

Relativistic heavy-ion collisions are the only known means of creating and studying the quark-gluon plasma under controlled conditions. Two decades of measurements at RHIC and the LHC, alongside multistage theoretical modeling, have established that this medium behaves as a strongly coupled fluid of remarkably low specific shear viscosity. This standard modeling picture is usually developed under the additional assumption that the quark and gluon content of the plasma reaches chemical equilibrium by the onset of the hydrodynamic phase. As reviewed in chapter~\ref{chapter1}, that assumption is not obviously justified: gluon-dominated initial states suggest that quark production may continue well into the hydrodynamic stage, and there is no reason to expect light and strange quarks to equilibrate at the same rate. This dissertation set out to relax that assumption, to model the resulting non-equilibrium flavor dynamics, and to confront the model with experimental data.

Chapter~\ref{chapter2} established the multistage framework that underlies modern heavy-ion phenomenology and serves as the baseline for this work: fluctuating initial conditions, viscous relativistic hydrodynamics, particlization, and hadronic transport, together forming a forward model that maps the properties of the medium onto experimental observables. Within this framework, the composition of the plasma enters through the equation of state and the particlization procedure, making these the natural places to implement an incomplete or evolving flavor content.

Chapter~\ref{chapter3} developed the central contribution of this dissertation: a dynamical treatment of quark chemical equilibration. Incomplete equilibration is described through time-dependent quark fugacities that relax toward their equilibrium values over finite timescales, modifying both the equation of state and the particlization of the medium so that its evolving flavor composition shapes both the hydrodynamic evolution and the final-state observables. I first examined the flavor-independent case, in which a single fugacity governs the overall quark content, and then generalized to independent light and strange flavors, allowing the two sectors to equilibrate on separate timescales. Across the two cases, I traced the consequences of delayed equilibration through the hydrodynamic evolution and into hadronic and electromagnetic observables, such as hadron yields, transverse flow, and thermal photon production, identifying those most sensitive to the flavor composition and its rate of equilibration.

Chapter~\ref{chapter4} shifted from the physics model to the statistical framework that connects it to data. Running the model forward, from a choice of parameters to a set of observables, is conceptually straightforward. The more difficult problem is the inverse one: inferring which parameter values produce observables consistent with experiment. Bayesian inference provides a framework for solving it, yielding not a single best-fit point but a posterior distribution over the parameters conditioned on the data. I introduced this framework in general form. Because a model of this complexity is far too expensive to evaluate at the many parameter settings that posterior sampling requires, the approach relies on a fast surrogate: principal component analysis compresses the high-dimensional observable set, and a Gaussian process trained on model evaluations at a designed set of parameter points emulates each retained component across the parameter space. Markov chain Monte Carlo sampling of this emulator then constructs the posterior.

Chapter~\ref{chapter5} brought these elements together in a calibration of the chemical equilibration sector, together with selected initial-state and transport coefficient parameters, against Au+Au collision data at RHIC. Among the chemical equilibration parameters, the calibration places its strongest constraint on the initial strange fugacity, favoring $\gamma_s^0 \approx 0.31$, well below the equilibrium value of unity and robust across the modeling variations tested. The initial light fugacity $\gamma_l^0$ and the two equilibration timescales $\tau_{\mathrm{eq},l}$ and $\tau_{\mathrm{eq},s}$ are individually more weakly determined. Considered jointly, however, the posteriors point to a coherent qualitative picture: the strange sector begins more strongly undersaturated than the light sector, and the data consistently favor its equilibrating more slowly as well, although neither timescale is sharply pinned. Among the transport coefficients, the specific shear viscosity is well constrained by the charged hadron flow harmonics, consistent with the low-viscosity picture established at the outset, while the bulk viscosity peak is constrained to a similarly small value, though for both coefficients the temperature dependence is held fixed rather than inferred. These results represent a first step rather than a final word: the Au-focused calibration presented here is being extended into a larger multi-system analysis, with an expanded set of design points and the simultaneous inclusion of Au+Au, Pb+Pb, and O+O collision data, which should help sharpen these constraints considerably.

Taken together, these chapters demonstrate that chemical equilibration can be incorporated dynamically into a realistic multistage description of heavy-ion collisions and, crucially, that its parameters can be constrained empirically rather than assumed. By extending Bayesian heavy-ion phenomenology beyond the assumption of chemical equilibrium, this work begins to answer, directly from the data, whether the flavor composition of the plasma equilibrates rapidly or continues to evolve throughout the hydrodynamic phase: the present calibration favors a plasma whose strange sector in particular remains undersaturated at early times and approaches equilibrium only gradually.

Several directions follow naturally from this work. The most immediate is the multi-system analysis now under way, which will exploit the differing size and energy of Au+Au, Pb+Pb, and O+O collisions to disentangle the equilibration dynamics from the bulk medium properties with which they are partially degenerate. The calibration can be broadened further, bringing strange hadron flow observables into the likelihood and off-diagonal covariance into the uncertainty model, both deferred in the present analysis. Beyond any single refinement, a quantitative, data-driven account of quark chemical equilibration can tell us how, and how quickly, the quark-gluon plasma reaches equilibrium as it first forms. What emerges is a plasma still approaching chemical equilibrium even as it expands and flows --- not because the model assumes it, but because the data begin to demand it.


\begin{appendices}
  \renewcommand{\chapterlabel}{Appendix}

%

\chapter{Entropy density}
\label{app:entropy}

The entropy density is not required to close the equation of state, but it enters the dissipative sector through the viscosity normalization of section~\ref{subsec:implementation_pce}; we derive it here.

Taking the fundamental thermodynamic relation
\begin{align}
    dU = T dS - P dV + \mu dN
\end{align}
and considering only the total number of (anti)quarks $N_q$, we have
\begin{align}
    \varepsilon = T s - P + \mu_q n_q,
\end{align}
where $\mu_q$ is the (anti)quark chemical potential. Note that in chemical equilibrium, $\mu_q = 0$ and $s = (\varepsilon + P)/T$ follows immediately. Defining the quark fugacity as $\gamma_q = e^{\mu_q/T}$, Maxwell relations give
\begin{align}
    n_q = \left(\frac{\partial N}{\partial V}\right)_{\mu_q, T} = \left(\frac{\partial P}{\partial \mu_q}\right)_{V,T} = \left(\frac{\partial P}{\partial \gamma_q}\right)_{V,T}\left(\frac{\partial \gamma_q}{\partial \mu_q}\right)_{V,T} = \frac{\gamma_q }{T}\left(\frac{\partial P}{\partial \gamma_q}\right)_{V,T},
\end{align}
and thus
\begin{align}
    s = \frac{\varepsilon + P}{T} - \frac{\gamma_q  \ln{\gamma_q}}{T} \left(\frac{\partial P}{\partial \gamma_q}\right)_{V,T}.
    \label{eq:spart}
\end{align}

Using Eq.~\ref{eq:e}:
\begin{align}
\left(\frac{\partial P}{\partial \gamma_q}\right)_{V,T}
= {}&
\left(\frac{T_\text{c}}{T_3}\right)^4
P_3\!\left(T \frac{T_3}{T_\text{c}}\right)
- \left(\frac{T_\text{c}}{T_0}\right)^4
P_0\!\left(T \frac{T_0}{T_\text{c}}\right)
\label{eq:dPdgamma} \\[0.5ex]
&+ \gamma_q
\left(\frac{4 T_\text{c}^3}{T_3^4}\right)
\left(\frac{\partial T_\text{c}}{\partial \gamma_q}\right)
P_3\!\left(T \frac{T_3}{T_\text{c}}\right) \notag \\
&+ (1-\gamma_q)
\left(\frac{4 T_\text{c}^3}{T_0^4}\right)
\left(\frac{\partial T_\text{c}}{\partial \gamma_q}\right)
P_0\!\left(T \frac{T_0}{T_\text{c}}\right) \notag \\
&+ \gamma_q
\left(\frac{T_\text{c}}{T_3}\right)^4
P_3'\!\left(T \frac{T_3}{T_\text{c}}\right)
\left(-\frac{T T_3}{T_\text{c}^2}\right)
\left(\frac{\partial T_\text{c}}{\partial \gamma_q}\right) \notag \\
&+ (1-\gamma_q)
\left(\frac{T_\text{c}}{T_0}\right)^4
P_0'\!\left(T \frac{T_0}{T_\text{c}}\right)
\left(-\frac{T T_0}{T_\text{c}^2}\right)
\left(\frac{\partial T_\text{c}}{\partial \gamma_q}\right).
\notag
\end{align}
Inserting this result into Eq.~\ref{eq:spart} yields
\begin{align}
s(T,\gamma_q) = {}&
\frac{\varepsilon(T,\gamma_q) + P(T,\gamma_q)}{T}
\label{eq:s} \\[0.5ex]
&- \frac{\gamma_q \ln \gamma_q}{T}
\left(
\begin{aligned}
&\left(\frac{T_\text{c}(\gamma_q)}{T_3}\right)^4
P_3\!\left(T \frac{T_3}{T_\text{c}(\gamma_q)}\right) \\
&- \left(\frac{T_\text{c}(\gamma_q)}{T_0}\right)^4
P_0\!\left(T \frac{T_0}{T_\text{c}(\gamma_q)}\right)
\end{aligned}
\right) \notag \\
&- \frac{2 \sqrt{\gamma_q} \ln \gamma_q}{T}
\frac{T_3 - T_0}{T_\text{c}(\gamma_q)}
\left(
\begin{aligned}
&\gamma_q
\left(\frac{T_\text{c}(\gamma_q)}{T_3}\right)^4
P_3\!\left(T \frac{T_3}{T_\text{c}(\gamma_q)}\right) \\
&+ (1-\gamma_q)
\left(\frac{T_\text{c}(\gamma_q)}{T_0}\right)^4
P_0\!\left(T \frac{T_0}{T_\text{c}(\gamma_q)}\right)
\end{aligned}
\right) \notag \\
&+ \frac{\sqrt{\gamma_q} \ln \gamma_q}{2}
\frac{T_3 - T_0}{T_\text{c}(\gamma_q)}
\left(
\begin{aligned}
&\gamma_q
\left(\frac{T_\text{c}(\gamma_q)}{T_3}\right)^3
P_3'\!\left(T \frac{T_3}{T_\text{c}(\gamma_q)}\right) \\
&+ (1-\gamma_q)
\left(\frac{T_\text{c}(\gamma_q)}{T_0}\right)^3
P_0'\!\left(T \frac{T_0}{T_\text{c}(\gamma_q)}\right)
\end{aligned}
\right).
\notag
\end{align}
This result is then computed using the previously defined functions $\varepsilon(T, \gamma_q)$, $P(T, \gamma_q)$, and $T_\text{c}(\gamma_q)$, together with the lattice-derived pressure functions $P_3(T)$ and $P_0(T)$.

In the two-flavor model of section~\ref{section33}, the single fugacity $\gamma_q$ is replaced by independent light and strange fugacities $\gamma_l$ and $\gamma_s$. Repeating the derivation above with two chemical potentials gives
\begin{align}
    s(T,\gamma_l,\gamma_s) = \frac{\varepsilon + P}{T}
    - \frac{1}{T}\sum_{f=l,s} \gamma_f \ln \gamma_f \left(\frac{\partial P}{\partial \gamma_f}\right)_{V,T}.
    \label{eq:s_twoflavor}
\end{align}

Rather than expand the fugacity derivatives in closed form as in Eq.~\ref{eq:s}, we evaluate $(\partial P/\partial \gamma_l)$ and $(\partial P/\partial \gamma_s)$ numerically, by finite differences of the tabulated pressure $P(T,\gamma_l,\gamma_s)$, and obtain the entropy density on the same grid used for the equation of state.
\end{appendices}

\acknowledgements

I thank my advisor, Steffen Bass, for his guidance and support, and the members of my dissertation committee for their time and feedback.

This work was supported by the U.S. Department of Energy under Grant No. DE-FG02-05ER41367. This research used resources of the National Energy Research Scientific Computing Center (NERSC), a U.S. Department of Energy Office of Science User Facility operated under Contract No. DE-AC02-05CH11231 using NERSC award NP-ERCAP0035982. This research was also done using services provided by the OSG Consortium, which is supported by National Science Foundation awards \#2030508 and \#2323298.

\cleardoublepage
\normalbaselines 
\bibliographystyle{apsrev4-1}
\bibliography{./Bibliography/references}






\end{document}